%% file: ms.tex
\RequirePackage{fix-cm}
\documentclass[natbib,smallextended]{svjour3}       
\smartqed  
\usepackage{graphicx}
\usepackage{latexsym}
\usepackage{xspace}
\usepackage{soul}
\usepackage{color}
\usepackage{amsmath}
\usepackage{acronym}
\usepackage{csquotes}
\usepackage{physics}
\usepackage{mathptmx}
\usepackage{amssymb}
\usepackage{hyperref}

\usepackage{enumitem}
\setlist{  
  listparindent=\parindent,
  parsep=0pt,
}

\include{macros}

\begin{document}

\title{Neutron Star Mergers}
\subtitle{and How to Study Them}


\author{Eric Burns}
\institute{E. Burns 
            \at Department of Physics and Astronomy\\
            Louisiana State University\\
            Baton Rouge, LA 70803, USA\\
            \email{ericburns@lsu.edu}
}



\maketitle

\begin{abstract}
Neutron star mergers are the canonical multimessenger events: they have been observed through photons for half a century, gravitational waves since 2017, and are likely to be sources of neutrinos and cosmic rays. Studies of these events enable unique insights into astrophysics, particles in the ultrarelativistic regime, the heavy element enrichment history through cosmic time, cosmology, dense matter, and fundamental physics. Uncovering this science requires vast observational resources, unparalleled coordination, and advancements in theory and simulation, which are constrained by our current understanding of nuclear, atomic, and astroparticle physics. This review begins with a summary of our current knowledge of these events, the expected observational signatures, and estimated detection rates for the next decade. I then present the key observations necessary to advance our understanding of these sources, followed by the broad science this enables. I close with a discussion on the necessary future capabilities to fully utilize these enigmatic sources to understand our universe.

\end{abstract}

\setcounter{tocdepth}{3}
\tableofcontents

\input{acro_list.tex}

\section{Introduction} 
\label{sec:intro} 
\input{s01_intro.tex}


\section{Neutron star mergers}
\label{sec:NSM}

\input{s02_NSM.tex}

\section{Astrophysical inferences}
\label{sec:astrophysics}

\input{s03_astro.tex}

\section{\qJets}
\label{sec:jets}

\input{s04_jets.tex}

\section{\qElem}
\label{sec:elements}
\input{s05_elements.tex}

\section{\qCosmo}
\label{sec:cosmology}

\input{s06_cosmo.tex}

\section{\qNSEOS}
\label{sec:dense}
\input{s07_supranuclear.tex}

\section{\qGRSM}
\label{sec:fundamental_physics}

\input{s08_fundamental.tex}

\section{Recommendations for the future}
\label{sec:future}
\input{s09_recommendations.tex}

\section{Conclusions}
\label{sec:conclusions}

Multimessenger observations of \ac{NS} mergers allow for complementary information on the physics occurring during these events. \acp{GW} and neutrino observations directly probe the central engine. Kilonovae arise from the unbound ejecta released during and after coalescence. The emission of \acp{SGRB} arise from ultralrelativistic jets powered by the accretion torus on the central engine. Together this information could allow \ac{NS} mergers to become the best understood astronomical transient.

In return, we can use them as tools to understand the universe, from cosmology, the origin of heavy elements, extreme particle acceleration, supranuclear matter, and fundamental physics. The science possible with studies of these sources is enormous. We are entering a new era because of the on-set of \ac{GW} astronomy, and are well prepared for the next few years. Beyond that, some necessary capabilities do not exist and are not yet funded. Ensuring these needs are met will maximize what we learn from these unique sources.

\begin{acknowledgements} 
I would like to acknowledge the groups and individuals who I have worked with over many years, which greatly influenced the content of this paper. My close involvement in the \Fermi-\ac{GBM}+\ac{LIGO}/Virgo working group has significantly advanced my understanding of this field. Individuals include, but are not limited to, Nelson Christensen, Regina Caputo, Valerie Connaughton, C.~Michelle Hui, Dan Kocevski, Julie McEnery, Brian Metzger, Judith Racusin, Jay Tasson, Bing Zhang, who I have long worked with and/or provided direct feedback on this manuscript. I led an Astro2020 white paper from the Multimessenger Science Action Group on Multimessenger Science with Neutron Star Mergers. While this manuscript began before that work started, it greatly expanded the science discussed here, and I thank the individuals who contributed to that paper. I would also like to thank all of the scientists who provided feedback on the initial arXiv version. Lastly, I would like to thank both reviewers for providing helpful feedback on such a long manuscript.
\end{acknowledgements}

\addcontentsline{toc}{section}{References}
\bibliographystyle{aasjournal}
\bibliography{references}

\end{document}

%% file: macros.tex
\definecolor{darkpastelgreen}{rgb}{0.01, 0.75, 0.24}

\newcommand{\Fermi}{\emph{Fermi}\xspace}
\newcommand{\Swift}{\emph{Swift}\xspace}

\newcommand{\grbid}{GRB~170817A\xspace}
\newcommand{\gwid}{GW170817\xspace}
\newcommand{\knid}{KN170817\xspace}

\newcommand{\beq}{\begin{equation}}
\newcommand{\eeq}{\end{equation}}
\newcommand{\qGRSM}{Fundamental physics\xspace}
\newcommand{\qCosmo}{Standard sirens and cosmology\xspace}
\newcommand{\qJets}{Short gamma-ray bursts and ultrarelativistic jets\xspace}
\newcommand{\qNSEOS}{Dense matter\xspace}
\newcommand{\qElem}{Kilonovae and the origin of heavy elements\xspace}

\newcommand\aap{Astron. Astrophys.\ }

\newcommand\apjl{Astrophys. J. Lett.\ }
\newcommand\apjs{Astrophys. J. Suppl.\ }

%% file: acro_list.tex
\section*{List of acronyms}

\begin{acronym}[INTEGRAL]
\acro{BAO}[BAO]{Baryon acoustic oscillation}
\acro{BAT}[BAT]{Burst Alert Telescope}
\acro{BATSE}[BATSE]{Burst And Transient Source Experiment}
\acro{BBN}[BBN]{Big Bang Nucleosynthesis}
\acro{BH}[BH]{Black Hole}
\acro{BBH}[BBH]{Binary Black Hole}
\acro{BNS}[BNS]{Binary Neutron Star}
\acro{CBC}[CBC]{Compact Binary Coalescence}
\acro{CMB}[CMB]{Cosmic Microwave Background}
\acro{CCSNe}[CCSNe]{Core-Collapse Supernova explosion}
\acro{CGRO}[CGRO]{Compton Gamma-Ray Observatory}
\acro{CTA}[CTA]{Cherenkov Telescope Array}
\acro{CO}[CO]{Compact Object}
\acro{CR}[CR]{Cosmic Ray}
\acro{EM}[EM]{Electromagnetic}
\acro{FAR}[FAR]{False Alarm Rate}
\acro{FOV}[FoV]{Field of View}
\acro{EOS}[EOS]{Equation of State}
\acro{GBM}[GBM]{Gamma-ray Burst Monitor}
\acro{GCN}[GCN]{Gamma-ray Coordinates Network}
\acro{GR}[GR]{General Relativity}
\acro{GRB}[GRB]{Gamma-ray burst}
\acro{GW}[GW]{Gravitational wave}
\acro{HAWC}[HAWC]{High-Altitude Water Cherenkov}
\acro{HMNS}[HMNS]{HyperMassive Neutron Star}
\acro{IACT}[IACT]{Imaging Atmospheric Cherenkov Telescope}
\acro{INTEGRAL}[INTEGRAL]{INTErnational Gamma-Ray Astrophysics Laboratory}
\acro{IPN}[IPN]{Interplanetary Network}
\acro{IR}[IR]{Infrared}
\acro{ISCO}[ISCO]{Innermost Stable Circular Orbit}
\acro{ISM}[ISM]{Interstellar medium}
\acro{JWST}[JWST]{James Webb Space Telescope}
\acro{KAGRA}[KAGRA]{Kamioka Gravitational Wave Detector}
\acro{KNR}[KNR]{Kilonova remnant}
\acro{LAT}[LAT]{Large Area Telescope}
\acro{LEO}[LEO]{Low Earth Orbit}
\acro{LGRB}[LGRB]{Long Gamma-Ray Burst}
\acro{LIGO}[LIGO]{Laser Interferometer Gravitational-wave Observatory}
\acro{LIV}[LIV]{Lorentz Invariance Violation}
\acro{LSST}[LSST]{Large Synoptic Survey Telescope}
\acro{LVC}[LVC]{The LIGO Scientific Collaboration and Virgo Collaboration}
\acro{LISA}[LISA]{Laser Interferometer Space Antenna}
\acro{NIR}[NIR]{Near infrared}
\acro{MAGIC}[MAGIC]{Major Atmospheric Gamma Imaging Cherenkov Telescopes}
\acro{MCMC}[MCMC]{Markov Chain Monte Carlo}
\acro{MW}[MW]{Milky Way}
\acro{NS}[NS]{Neutron Star}
\acro{NSBH}[NSBH]{Neutron Star--Black Hole}
\acro{PPN}[PPN]{Parametrized Post-Newtonian}
\acro{PTA}[PTA]{Pulsar Timing Arrays}
\acro{QCD}[QCD]{Quantum Chromodynamics}
\acro{QFT}[QFT]{Quantum Field Theory}
\acro{QG}[QG]{Quantum Gravity}
\acro{RAVEN}[RAVEN]{Rapid VOEvent Coincidence Monitor}
\acro{SAA}[SAA]{South Atlantic Anomaly}
\acro{SFR}[SFR]{Star Formation Rate}
\acro{SGRB}[SGRB]{Short Gamma-Ray Burst}
\acro{SMBH}[SMBH]{SuperMassive Black Hole}
\acro{SMNS}[SMNS]{Supramassive Neutron Star}
\acro{SNR}[SNR]{Signal-to-Noise Ratio}
\acro{SME}[SME]{Standard Model Extension}
\acro{SNe}[SNe]{Supernova explosion}
\acro{SNEWS}[SNEWS]{Supernova Early Warning System}
\acro{SPI-ACS}[SPI-ACS]{SPectrometer onboard INTEGRAL - Anti-Coincidence Shield}
\acro{SR}[SR]{Special Relativity}
\acro{SSC}[SSC]{Synchrotron Self Compton}
\acro{TOV}[TOV]{Tolman--Oppenheimer--Volkoff}
\acro{TTE}[TTE]{Time-Tagged Event}
\acro{UHECR}[UHECR]{Ultra-High Energy Cosmic Ray}
\acro{UV}[UV]{Ultraviolet}
\acro{UVOIR}[UVOIR]{Ultraviolet, Optical, and Infrared}
\acro{UVOT}[UVOT]{Ultraviolet/Optical Telescope}
\acro{VHE}[VHE]{Very High Energy}
\acro{WEP}[WEP]{Weak Equivalence Principle}
\acro{XRT}[XRT]{X-ray telescope}
\acro{ZTF}[ZTF]{Zwicky Transient Facility}
\end{acronym}

%% file: s01_intro.tex
Two \acp{NS} from the galaxy NGC 4993 merged, emitting two messengers that traveled together from the age of dinosaurs through the age of civilization. As the messengers neared Sirius the \Fermi Space Telescope was launched; after they passed Alpha Centauri the Advanced \ac{GW} interferometers were turned on for the first time. On August 17th, 2017 the messengers arrived at Earth \citep{GW170817-GRB170817A}: the \acp{GW} observed as \gwid \citep{GW170817-GW} by the Advanced Laser Interferometer Gravitational-Wave Observatory \citep[\acs{LIGO};][]{AdvLIGO} and Advanced Virgo \citep{AdvVirgo} and the gamma rays as \grbid \citep{GBM_only_paper,Savchenko2017} by \Fermi \citep{Meegan2009} and INTEGRAL \citep{SPIACS}. This joint detection resulted in the greatest follow-up observation campaign in the history of transient astrophysics \citep{GW170817-MMAD}, which resulted in six independent detections of AT2017gfo \citep{gw170817_kilonova_first_swope,valenti2017discovery,gw170817_kilonova_tanvir,lipunov2017master,soares2017electromagnetic,arcavi2017optical}, the theoretically predicted radioactively-powered kilonova, whose precise location enabled the identification of \enquote{off-axis} afterglow emission \citep{GW170817_Troja_Xray_discovery,margutti2017electromagnetic,haggard2017deep} that has been detected until more than two years later. \acused{LIGO}

These discoveries culminated in a suite of papers published only two months after the first detection, with contributions from thousands of astronomers and astrophysicists, ushering in the new era of \ac{GW} multimessenger astrophysics. For decades, the scientific promise of these sources has been known, and the first event certainly met expectations with, on average, more than three papers written per day over the first two years.

There have been only three convincing multimessenger detections of individual astrophysical sources: neutrinos and photons from the core-collapse supernova SN 1987A \citep{SN1987A_neutrinos_Hirata_1987}, gravitational waves and photons from a binary neutron star merger \citep[this event;][]{GW170817-MMAD}, and likely neutrinos and photons from a flaring blazar \citep{IceCube170922A_Blazar_2018}. The modern era of time domain, multimessenger astrophysics will hopefully result in multiple detections of multiple source classes with multiple messengers. \ac{BNS} and \ac{NSBH} mergers, collectively referred to here as \ac{NS} mergers, will be important astrophysical multimessenger sources for the foreseeable future.


Several papers and reviews on the astrophysics of \ac{NS} mergers have been written, both before and after \gwid. Several papers have been written on science beyond astrophysics enabled by observations of these events. When available, we reference manuscripts that contain more detailed discussions. This review collates and advances this information into a coherent summary, to ensure the information carried by messengers from \ac{NS} mergers, already long into their journey to Earth, will be captured and utilized to understand our Universe. Our view of these mergers will depend on the ground- and space-based assets available to observe them and our strategies and scientific gains are placed in the context of our current outlook on these future capabilities.


In Sect.~\ref{sec:NSM} we give a broad overview of our current understanding of \ac{NS} mergers and how we observe them. This section contains rough detection rate predictions through the next decade. In Sect.~\ref{sec:astrophysics} we discuss the astrophysical inferences on \ac{NS} mergers that are important for several additional scientific studies and those that are not otherwise discussed. The later science sections are separated into the broad topics: \qJets (Sect.~\ref{sec:jets}), \qElem (Sect.~\ref{sec:elements}), \qCosmo  (Sect.~\ref{sec:cosmology}), \qNSEOS (Sect.~\ref{sec:dense}), and \qGRSM (Sect.~\ref{sec:fundamental_physics}). The individual science sections are, as much as possible, self-contained. Based on the science sections, Sect.~\ref{sec:future} makes recommendations for future capabilities. This discusses both current and funded missions, and identifies where gaps may occur.

Given the broad scope of this paper, particular attention is given to avoid or carefully define field-specific terminology and to use language that should prevent confusion for readers of various backgrounds. We use the astrophysical definition of \enquote{gamma-rays}, referring to all photons with energies $\gtrsim100$\,keV. We will directly state when we are discussing gamma-rays that originate from nuclear processes. We assume, unless otherwise stated, that our general understanding of science is correct, e.g. that \ac{BNS} mergers and (some) \ac{NSBH} mergers are the progenitors of most \acp{SGRB} and all kilonovae, or that the relative propagation of gravity and light is zero. We assume a standard $\Lambda$CDM cosmology, with H$_0$=67.4\,km\,/s\,/Mpc and $\Omega_m$=0.315, from \citet[][]{Planck_2018_cosmo}. Canonical \acp{NS} are those with masses of 1.4M$_{\odot}$; canonical \acp{BH} refer to those with masses of 10M$_{\odot}$. All rates are reported for a calendar year and refer to the prediction of the true rate (i.e. they do not account for Poisson variation). Variables and constants have their usual definition, e.g. $c$ is the speed of light, $G$ the gravitational constant, $M$ represents masses, etc. Subscript $\odot$ denote solar units. When referring to stars in a binary, both massive and compact, the heavier star is always referred to as the primary and is denoted by a subscript 1 and the lighter star is referred to as the secondary with a subscript 2, to match convention. Heavy elements here refers to those beyond-iron.




%% file: s02_NSM.tex
\acp{NS} are the densest matter in the Universe, with \acp{BH} the only known denser object. Binary star systems emit \acp{GW} causing them to slowly inspiral as they lose energy. Tightly bound \ac{BNS} and \ac{NSBH} systems can lose energy fast enough to merge within the age of the Universe. The merging of the two objects can significantly disrupt the \ac{NS}, releasing large amounts of matter and energy that can power the observed \ac{EM} and predicted neutrino signatures.

In Sect.~\ref{sec:overview} we provide a succinct overview of our current understanding of how these systems form, their behavior shortly before, during, and after merger, and potential longer-term signatures. We discuss the intrinsic event rates in Sect.~\ref{sec:intrinsic_rates}, followed by subsections on the canonical signals, their individual detection rates, and what we learn from these observations. Interspersed are subsections on the necessary steps for combining information: Sect.~\ref{sec:association} details the conditions required for robust statistical association, Sect.~\ref{sec:GW-GRB_rates} joint detection rates for independent detections, and Sect.~\ref{sec:followup} methods for follow-up searches. Sect.~\ref{sec:prospects_other_signatures} briefly discusses additional signatures that are expected and prospects of detection. We summarize our predicted future detection rates in Sect.~\ref{sec:detections_summary}.




\subsection{Overview}
\label{sec:overview}
\begin{figure}
    \centering
    \includegraphics[width=1.0\textwidth]{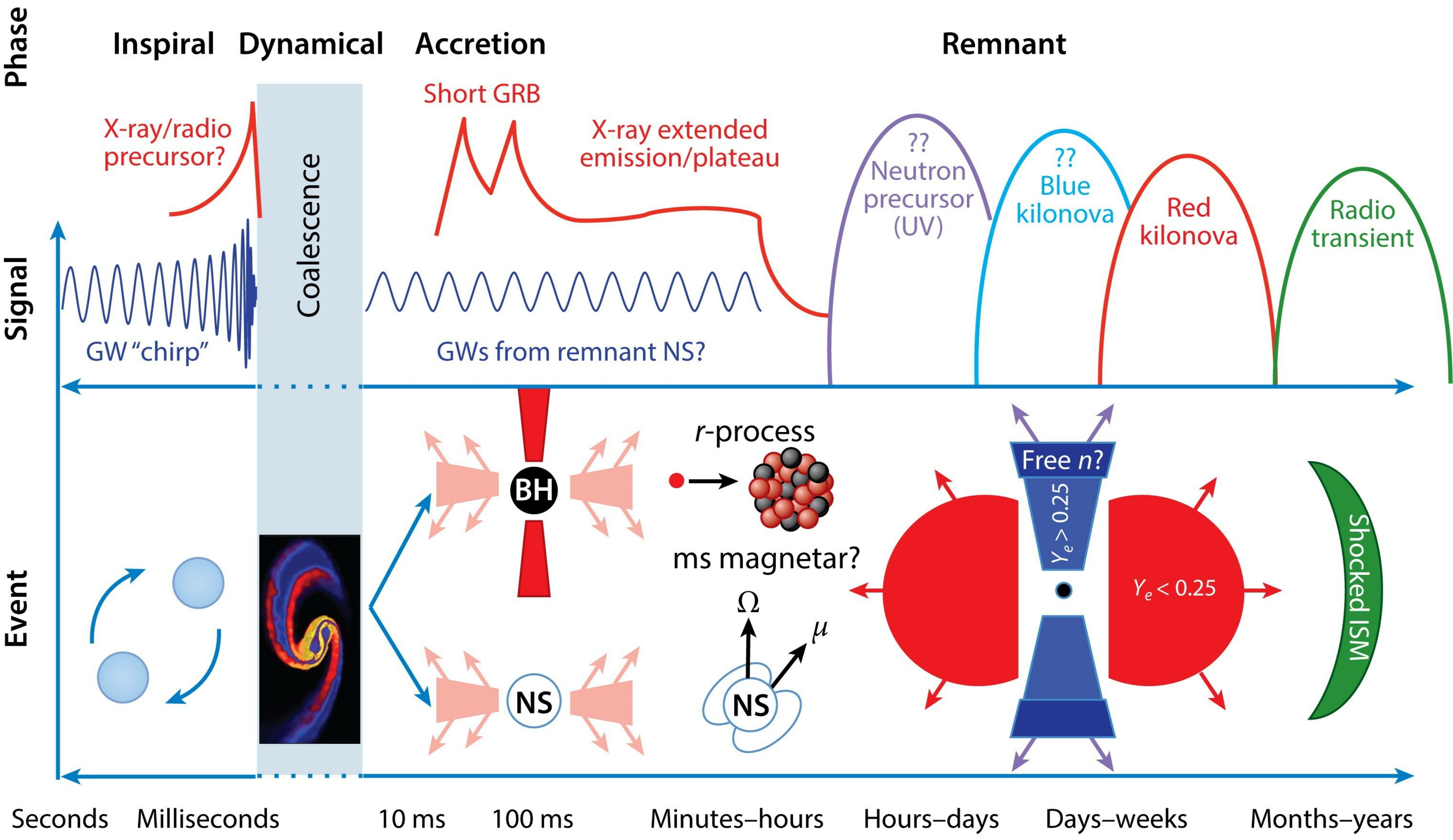}
    \caption{An overview of the expected \ac{GW} and \ac{EM} signatures from minutes before until years after merger, as discussed in Sects.~\ref{sec:overview_inspiral}-\ref{sec:overview_kilonova}.
    The bottom represents what occurs as a function of time with the corresponding observational signature on top. Image reproduced with permission from \citet{overview_fernandez_metzger_2016}}
    \label{fig:signature_overview}
\end{figure}

Information on \ac{NS} mergers can be gleaned from observations of these systems from eons before coalescence to long after merger. This section contains an overview of the lives of these systems; each subsection discusses a stage of their evolution and contains references for further detail. For an in-depth review of the expected \ac{EM} signatures from \ac{NS} mergers see the opening figure of \citet{overview_fernandez_metzger_2016}, which we borrow as Fig.~\ref{fig:signature_overview}. We do not here give an overview on the history of our understanding of these events as we are unlikely to exceed existing literature; for a brief general history we refer the reader to the introduction of \citet{GW170817-MMAD}.

\subsubsection{System formation}
\label{sec:overview_formation}
The formation and evolution of stellar systems is a broad topic in astrophysics. We are focused on the science enabled with \ac{NS} mergers. The events of interest are then \ac{BNS} and \ac{NSBH} systems that will form and merge within the age of the Universe. For relevant reviews see \citet{review_merger_rate_compact_sadowski_2008} and \citet{review_NS_Faber_Rasio}. Before discussing how such systems can form, we show the time until merger as a function of orbital separation radius $R$ for two compact objects inspiraling only through \ac{GW} emission, which is

\begin{equation}
\label{eq:time_to_merger_from_radius}
\begin{split}
t_{\rm merge}(r)
&= \frac{5}{256} \frac{c^5}{G^3} \frac{R^4}{(M_1 M_2)(M_1+M_2)}\\
&\approx 54\,\text{Myr} \Bigg(\frac{1}{q(1+q)}\Bigg) \Bigg(\frac{R}{R_\odot}\Bigg)^4 \Bigg(\frac{1.4\,M_\odot}{M_1}\Bigg)^3
\end{split}
\end{equation}

\noindent individual masses $M_1$ and $M_2$, and mass ratio $q=M_2/M_1$. This equation, and others in this section, assume quasi-circular orbits as compact object systems circularize quickly compared to their total inspiral time \citep{review_NS_Faber_Rasio}. 

A star with mass between $\sim8$--$50\,M_\odot$ will end as a \ac{CCSNe}. Stars on the lower end of this mass range will result in a \ac{NS} and those on the high end will result in a \ac{BH} \citep[see][ and references therein for details]{da2020equation}. Such heavy stars become supergiants near the end of their lives with sizes $R \gtrsim 30\,R_{\odot}$. When two of these stars form already bound together, as a field binary, they can result in compact object binaries once both have undergone supernova. For canonical \ac{BNS} systems with initial separations larger than the size of the progenitor supergiant the \ac{GW}-only inspiral time will be a thousand times the age of the Universe. 

For canonical \ac{BNS} systems to merge within one current age of the Universe, inspiraling only through \ac{GW} radiation, they must have initial separation of $\lesssim5\,R_\odot$. This requires a common envelope stage, where either the two massive stars are not distinct or the primary forms a compact object before being enveloped by the secondary during its supergiant phase. This greatly accelerates the inspiral and results in tighter initial separation of the two compact objects.

If the primary compact object is a \ac{NS} the second is most often also a \ac{NS}. This likely forms a \ac{BNS} system, but could form a \ac{NSBH} system if the primary accretes sufficient mass to collapse into a \ac{BH} during the common envelope phase. If the primary collapses directly to a \ac{BH} the system becomes an \ac{NSBH}\footnote{Some specify \ac{NSBH} or BHNS depending on which object formed first. This is a useful convention for some studies but is not used here as the distinction is beyond the scope of such a general review.} if the secondary is light enough to form a \ac{NS}, otherwise it is a \ac{BBH} system. 

The prior discussion focused on what is thought to be the standard formation channel for \ac{BNS} and \ac{NSBH} systems whose mergers we can observe. It is also believed that a smaller number of systems can be formed dynamically, where two compact objects form separately but become gravitationally bound when they travel close enough to each other. \acp{NS} and \acp{BH} in globular clusters will tend to gravitate towards the center due to dynamical friction, leading to both a higher likelihood of dynamical capture and an accelerated inspiral aided by three-body interactions with other objects. This could contribute $\sim$10\% of merger events \citep[e.g.,][]{belczynski2002comprehensive}. There may be rare head-on collisions that would behave quite differently. These are beyond the scope of this paper, but investigations of their relative importance can be studied from the information relevant for Sect.~\ref{sec:host_redshift_where}.


\subsubsection{Inspiral}
\label{sec:overview_inspiral}
After the \ac{BNS} or \ac{NSBH} system is formed, the two compact objects will lose energy to \acp{GW}, causing the two compact objects to inspiral towards one another. Long before merger this emission is weak and the orbital evolution is slow. Close to merger time the energy released greatly increases and the orbital evolution accelerates. We discuss these two cases and how we can best observe them separately.

Observations of the inspiral long before merger are best performed using \ac{EM} observations of galactic \ac{BNS} systems. An overview of the known galactic \ac{BNS} systems and their observed parameters is available in \citet{Tauris-DNS-Systems}. These \ac{BNS} systems have inspiral times from $\sim$85\,Myr to greater than a Hubble time. There is no known galactic \ac{NSBH} system. 

The discovery of the Hulse-Taylor binary system \citep{Hulse_Taylor_pulsar_discovery} enabled precise measures of the orbital decay of a compact binary system for the first time. Years of careful observation enabled a determination of the properties of the stars and the first proof of \ac{GW} radiation \citep{1982ApJ...253..908T}.

These systems spend only a tiny fraction of their lives in the late inspiral phase, which is roughly hours to minutes before merger. We are unlikely to observe a \ac{NS} system at this phase within the Milky Way, and are thus left to detecting extragalactic events. \ac{BNS} and \ac{NSBH} systems beyond the local group will likely be undetectable in photons during the early inspiral stage. Within the last $\sim$100\,s before merger it is possible that precursor \ac{EM} emission could be detectable for some nearby events. The strongest observational evidence is the claim of precursor activity preceding the main episode of prompt \ac{SGRB} emission \citep{troja2010precursors}; however, this question remains unsettled. There are theoretical models that predict precursor emission in gamma-rays, x-rays, and radio, with typical luminosities $\sim 10^{42}-10^{47}$\,erg/s. These are discussed in Sections \ref{sec:prospects_other_signatures} and \ref{sec:other_high_energy}.

\ac{GW} observations of stellar mass compact object inspirals provide a new method to study these systems at this stage. Because of their extremely dense nature, compact binary inspirals are among the strongest sources of \acp{GW}. As they approach merger time, where the orbital radius is similar to the size of the \acp{NS} themselves, the luminosity of this signal increases and the emitted \ac{GW} frequency enters the band of the ground-based interferometers. Shortly thereafter the objects enter the merger stage.



\subsubsection{Merger}
\label{sec:overview_merger}
The loss of energy to \ac{GW} radiation shrinks the orbital separation, increases the orbital frequency (with $f_{\rm GW}=2f_{\rm orb}$ as the dominant \ac{GW} emission is quadrupolar) and strengthens the \ac{GW} emission. This frequency evolution results in well-known \ac{CBC} chirp signal. The peak \ac{GW} luminosity approaches $10^{56}$\,erg/s around merger time \citep[e.g.,][]{lvc_gwtc1,zappa2018gravitational}. In the surrounding $\sim$seconds the \ac{NS} can be so disrupted that it releases matter which can power ultrarelativistic polar jets (Sect.~\ref{sec:overview_jets}) and mildly relativistic quasi-isotropic outflows (Sect.~\ref{sec:overview_kilonova}) that produce the known \ac{EM} and likely neutrino counterparts.

There are several potential contributions to the matter freed from the \ac{NS}. We follow the discussions from \citet{margalit2019multi,kawaguchi_kilonova_remnants,metzger2020kilonovae}. Dynamical ejecta is released within milliseconds of the merger. The deformation of the \ac{NS} late in the inspiral and efficient angular momentum transport from the remnant can release matter through tidal tails that can become spiral arms, which eject matter predominantly in the equatorial region. Shock-heating occurs at the interface of two \acp{NS}, squeezing out matter through quasi-radial oscillations at the interface region, which can dominate the polar region due to the lower densities in this region and solid angle spin effects.

Additional matter is ejected starting after the dynamical timescale and continuing for up to $\sim$10\,s after merger and is referred to as post-merger or wind ejecta. Disk winds can occur due to several physical processes. Magnetic fields can drive fast outflows with much of the ejection occurring within the first $\sim$1\,s \citep{siegel2017three,fernandez2018long}. Longer term ejection after $\sim$1\,s can occur when viscous heating and nuclear combination dominate over neutrino cooling \citep{metzger2008time,metzger2009neutron}. There can also be significant contributions from a remnant \ac{NS} which can power neutrino winds, magnetically driven outflows, and even strip material from the surface of the remnant itself \citep[e.g.,][]{dessart2008neutrino,overview_fernandez_metzger_2016}. 

The unbound material, or ejecta, is characterized by total mass, average velocity, and electron fraction $Y_e \equiv n_p/(n_n +n_p)$ where $n_n$ and $n_p$ are the number densities for neutrons and protons, respectively. More detailed treatments consider additional behavior, such as the spatial and density distributions. Winds from the central engine can alter these properties, broadening the spatial distributions, accelerating and heating the outflows, providing additional matter, and altering the electron fraction through neutrino irradiation via the charged-current interactions

\begin{equation}
    \label{eq:neutrino_stuff}
    \begin{split}
    p + e^- \leftrightarrow n + \nu_e,\\
    n + e^+ \leftrightarrow p + \bar{\nu_e}.
    \end{split}
\end{equation}

\noindent Given the much larger initial fraction of neutrons to protons, these interactions will drive $Y_e$ to higher values until equilibrium is achieved. The origin of these thermal neutrinos are from the accretion disk or, when one is present, created in pair interactions near the surface of the remnant \ac{NS}

\begin{equation}
    \label{eq:pair_stuff}
    e^+ + e^- \leftrightarrow \nu + \bar{\nu}.
\end{equation}


We expect enormous variation between \ac{NS} mergers. \ac{BNS} and \ac{NSBH} mergers should be quite different. Each of these can be further divided into sub-classes, which are discussed in detail below. Within these sub-classes we expect additional variety depending on the intrinsic parameters of the system.



\ac{NSBH} mergers can be split into two classes. The delineation depends on whether $r_{\rm{tidal}}$, the orbital separation at which the \ac{NS} disrupts, is less than or greater than $r_{\rm{ISCO}}$, the \ac{ISCO} of the \ac{BH} \citep{foucart2012black,foucart2018remnant}.  For a non-spinning \ac{BH} $r_{\rm{ISCO}}=6GM/c^2$. The spin of the \ac{BH} alters this distance, approaching $r_{\rm{ISCO}}=9GM/c^2$ for maximal retrograde spin and approaching the event horizon for maximal prograde spin. The \ac{NS} disruption occurs when tidal acceleration due to the inspiral exceeds the self-gravity of the \ac{NS}, and depends on the properties of the \ac{NS}, including the \ac{NS} \ac{EOS} (Sect.~\ref{sec:NS_EOS}). Disruption is favored for low mass \acp{BH}, for \acp{BH} with high prograde spin, and for large \acp{NS}. When no disruption occurs we refer to these as Heavy \ac{NSBH} mergers; when disruption does occur we refer to them as Light \ac{NSBH} mergers as they have lower mass and should produce bright \ac{EM} radiation.

\begin{itemize}
    \item \textbf{Heavy \ac{NSBH} Mergers}
    
    Heavy \ac{NSBH} mergers swallow the \ac{NS} whole. They will produce significant \ac{GW} emission during inspiral and coalescence, with \ac{BH} ringdown frequencies up to $\sim$1-2\,kHz \citep{pannarale2015gravitational}. Note the frequencies discussed here are the expected maximum values in a given \ac{NS} merger type, not the \ac{ISCO} frequencies. This is likely to be the only observable signal for these events.
    
    
    \item \textbf{Light \ac{NSBH} Mergers}
    
    \ac{NSBH} mergers with tidal disruption can release a sizable fraction of the total \ac{NS} before it enters the \ac{BH}. The \ac{GW} emission from these events is, in general, weaker than the heavy \ac{NSBH} cases due to the lower mass. They will tend to reach higher frequencies, $\sim$3-4\,kHz \citep{pannarale2015gravitational}, owing to the generally smaller \ac{BH} size.
    
    Light \ac{NSBH} mergers are more exciting for traditional (that is, \ac{EM}) and neutrino astronomers. Disruption of the \ac{NS} releases ejecta in the equatorial plane due to tidal effects. This dynamical ejecta moves outward at $\sim0.2-0.3c$, roughly corresponding to the orbital velocity at $r_{\rm{tidal}}$, and is incredibly neutron-rich with $Y_e \lesssim 0.1$ \citep{kiuchi2015high,foucart2014neutron}. The bound material stretches around the \ac{BH} into an accretion disk with a total mass up to $\sim 0.1\,M_\odot$. The disk is initially maintained as neutrino cooling dominates other effects, with peak luminosities approaching $\sim10^{53}$\,erg/s \citep[e.g.,][]{just2016neutron}. The main disk ejection phase can release tens of percent of the total disk mass at $\sim0.1c$; while this material initially also has $Y_e \lesssim 0.1$, neutrino irradiation can significantly raise the electron fraction of polar ejecta due to geometric exposure effects to the disk torus and lower densities in this region \citep[e.g.,][]{fernandez2018long}.
\end{itemize}


The structure of \acp{NS} is determined by the counterbalance of the combination of degeneracy pressure and nuclear forces against gravity. \acp{NS} have a maximum mass, beyond which they will collapse to a \ac{BH}; however, when there are additional mechanisms supporting the star against gravitational collapse this mass threshold can be temporarily altered. The heaviest \acp{NS} that do not immediately collapse to a \ac{BH} are supported against collapse by internal differential rotation, and are referred to as HyperMassive Neutron Stars \citep[\acsp{HMNS};][]{baumgarte1999maximum}. Slightly lighter \acp{NS} can be supported against collapse by uniform rotation, referred to as \acp{SMNS}. \acp{NS} that do not require additional support mechanisms are referred to as Stable \acp{NS}. \acused{HMNS}

\ac{BNS} mergers can be broadly split into four possible outcomes. Cases with the heaviest progenitor \acp{NS} are expected to promptly collapse to a \ac{BH} in $\lesssim10$\,ms. Slightly lighter progenitors should result in a short-lived \ac{HMNS} remnant with typical lifetimes of $\lesssim1$\,s due to efficient energy losses to internal torques \citep{shibata2006merger,GWs_neutrino_BNS_sekiguchi_2011}. At lower masses the remnant object can survive as a \ac{SMNS} with inefficient energy losses through magnetic dipole and quadrupolar \ac{GW} radiation. Shortly after merger the (meta)stable \ac{NS} is expected to have strong magnetic fields, which results in lifetimes as short as hundreds or thousands of seconds \citep{ravi2014birth}. Finally, it may be possible for two low-mass progenitor \acp{NS} to combine into a Stable \ac{NS}. We separate the following paragraphs to discuss our current understanding of these events from the most to least massive cases. Here the Stable \ac{NS} and \ac{SMNS} cases are combined as their lifetimes greatly exceed the merger and ejecta timescales, making these events very similar at this stage.

\begin{itemize}
    \item \textbf{Prompt Collapse}
    
    With sufficiently heavy \acp{NS} the system will collapse to a \ac{BH} within milliseconds. These will be the loudest \ac{BNS} mergers during inspiral due to their higher masses. In this case the \ac{GW} frequencies reach $\sim$6--7\,kHz \citep[e.g.,][]{shibata2006merger,clark2014prospects}, the highest achieved for any \ac{NS} mergers. The inspiral is followed by \ac{BH} ringdown, which has much weaker \ac{GW} emission. 
    
    Near merger, angular momentum transport stretches the \acp{NS}, forming tidal tails in the equatorial plane. Equal-mass binaries have been show to release dynamical ejecta with a low electron fraction $Y_e \lesssim 0.1$ with mass $10^{-4}$--$10^{-3}\,M_\odot$ and outwards velocity $\sim0.3c$ \citep{hotokezaka2013mass,just2015comprehensive}. Asymmetric mass ratios have been shown to achieve $5\times10^-3\,M_\odot$  \citep{kiuchi2019revisiting}. This is far lower total ejecta than the Light \ac{NSBH} merger case as \acp{NS} are larger than similar mass \acp{BH}. The other main dynamical ejecta mechanism in \ac{BNS} mergers is negligible for this case as it is immediately swallowed by \ac{BH} formation.
    
    The tidal tails stretch until they form an accretion disk which can range from $10^{-4}-10^{-2}\,M_\odot$, depending on the \ac{NS} \ac{EOS} \citep[e.g.,][]{shibata2006merger,hotokezaka2013mass,just2015comprehensive,ruiz2017general}. Magnetically-driven outflows and thermally-driven winds can both release up to $20$\% of the disk mass. 
    
    \item \textbf{Hypermassive Neutron Star Remnant}
    
    \ac{BNS} mergers that result in \ac{HMNS} remnants will have similar inspirals as the prompt collapse case, though a bit quieter. During the \ac{HMNS} phase the internal differential rotation releases \acp{GW} about as loud as the peak emission at coalescence, which occurs at $\sim$2-4\,kHz \citep{zhuge1994gravitational,shibata2000simulation,hotokezaka2013mass,maione2017spectral}. When the \ac{HMNS} collapses there is \ac{BH} ringdown emission.
    
    \indent The tidal ejecta for these mergers \citep{hotokezaka2013mass,NS-Bauswein} behave differently than the previously discussed cases. For disks around a \ac{BH} the material accretes in the equatorial region. For a \ac{NS} remnant the presence of a hard surface causes the in-falling matter to envelope the surface, resulting in additional material in the polar regions \citep{metzger2014red}. The unbound tidal ejecta for \ac{BNS} mergers with a \ac{HMNS} remnant will expand outwards at $\sim0.15-0.25c$. These are also the heaviest mergers that will have significant dynamical ejecta from the shock interface between the two \acp{NS}; this ejecta will dominate in the polar regions due to solid angle effects and the lower densities in this region. If the \ac{HMNS} lives for $\gtrsim$50 ms the neutrino luminosity can strip $\sim10^{-3}\,M_\odot$ of material from the surface of the remnant itself \citep{dessart2008neutrino,overview_fernandez_metzger_2016}. 
    
    \indent During these ejection processes the \ac{HMNS} has formed and is of sufficient temperature (few MeV) to produce significant amounts of $e^+e^-$ pairs at its surface. The total MeV neutrino emission can be $10^{53}$\,erg/s with contributions from both the disk and the temporary \ac{NS} \citep[e.g.,][]{GWs_neutrino_BNS_sekiguchi_2011}. The tidal tail ejecta is sufficiently massive, dense, and distant that its electron fraction is largely unchanged ($Y_e \approx 0.1-0.2$). However, the polar material is closer, has lower densities, and a greater geometric exposure to the disk allowing the combined neutrino irradiation to significantly alter the electron fraction of the dynamical material in this region \citep[$Y_e \approx 0.3-0.4$;][]{wanajo2014production}.
    
    \indent Given the larger amount of disruption and the lower overall velocity of the disrupted material, \ac{HMNS} remnants have larger disk masses than the prompt collapse case. The \ac{HMNS} collapses in under a second during the disk wind phase. So long as the \ac{HMNS} lives, the neutrino luminosities will cause an increase in the amount of ejected material and monotonically increase the electron fraction. From \citet{metzger2014red}, the amount of disk wind ejecta can exceed the dynamical ejecta; if the \ac{HMNS} lives for 100 (300) ms the effects of the \ac{HMNS} can eject up to $\sim$10\% ($\sim$30\%) of the total disk mass into the equatorial region and $\sim$5\% ($\sim$10\%) into the polar region. For disk wind ejecta the equatorial material will be distributed between $Y_e \approx0.1-0.5$ and the polar material will be $Y_e \gtrsim 0.3$, and move outwards at up to $\sim 0.1c$. 
    
    \indent The combination of the dynamical and post-merger ejecta and their alteration due to the \ac{HMNS} surface and winds summarizes into a reasonably simple picture. The dynamical ejecta leaves first being lanthanide-rich in the equatorial region and relatively lanthanide-free in the polar region, with a roughly comparable contribution from each component. Behind this is the ejecta from the disk winds which follows a similar spatial distribution of lanthanide-fraction. This combines to the representative Figure\,7 of \citet{metzger2020kilonovae} and our similar representation in Figure\,\ref{fig:remnant_object}.
    
    \item \textbf{Stable and Supramassive Neutron Star Remnants}
    
    \ac{SMNS} remnants survive for \citep[e.g.,][]{ravi2014birth} longer than the ejection phase, meaning they are quite similar to Stable \ac{NS} remnants during merger and ejection. The \ac{GW} emission is similar to the \ac{HMNS} case; the emission is slightly weaker during inspiral, they transition to significant \ac{GW} release to internal differential rotation, but would be followed by secular \ac{GW} radiation \citep[e.g.,][]{foucart2016low} at twice their rotational frequencies for some time. The longevity of this last phase of \ac{GW} emission is not well constrained, but when the \ac{SMNS} collapses there will be weak \ac{BH} ringdown emission. The neutrino flux is similar to the \ac{HMNS} case, but would be significantly greater total irradiation as the cooling time for the full \ac{NS} is longer than the lifetime of \acp{HMNS}.
    
    \indent The initial ejecta is similar to the \ac{HMNS} case, but the longer life of the \ac{NS} provides additional ejecta and wind to the system. This results in greater total ejecta material moving at somewhat larger velocities and the polar dynamical and disk wind ejecta achieving electron fractions approaching the equilibrium value \citep[e.g.,][]{GWs_neutrino_BNS_sekiguchi_2011}. 
    
    \indent The neutrino heating likely causes ejection of the majority of the total disk mass \citep{metzger2014red}. These systems can potentially approach an ejection up to $0.1\,M_\odot$ \citep[e.g.,][]{coughlin2018constraints,margalit2019multi}, with the disk wind ejecta dominating over dynamical ejecta, though large uncertainty remains. Stripping of material from the \ac{NS} surface due to the neutrino-driven wind from the hot \ac{NS} remnant can be more important here than in the \ac{HMNS} case \citep[e.g.,][]{dessart2008neutrino}.
    
    Lastly, the spin-down energy from these remnants should provide massive continued energy injection into the system. This is reviewed in detail in \citet{metzger2020kilonovae}.
    
\end{itemize}

Our understanding of what occurs during \ac{BNS} and \ac{NSBH} mergers comes from detailed simulations accounting for several incredibly complicated, coupled, non-linear effects. Despite the lengthy description in the preceding paragraphs, we have omitted several in-depth investigations into the effects of varying individual parameters, such as eccentricity, mass ratio, total mass, spins, the \ac{NS} \ac{EOS}, etc. The outcome of these variations is not immediately obvious. For a thorough review of these effects we refer to \citet{overview_fernandez_metzger_2016} and \citet{metzger2020kilonovae}. The large uncertainty range in the previously described parameters includes both the intrinsic effects of variation of these parameters and differences in the simulations, which vary their approximations. 

However, some general effects are robust. For \ac{NSBH} mergers there is larger mass ejection for lower mass \acp{BH} with higher values of spin. For \ac{BNS} mergers there is a positive correlation for the total ejecta mass and electron fraction with the lifetime of the \ac{NS}. Combining information from population synthesis models, numerical modeling, and the current constraints on the maximum mass of a \ac{NS} we generally expect to eventually observe all of these cases. The exception might be a \ac{BNS} merger with a Stable \ac{NS}, which may or may not be possible, depending on if the lightest \acp{NS} are less than half the maximum \ac{NS} mass (Sect.~\ref{sec:NS_merger_NS_EOS}).

\subsubsection{Jets}
\label{sec:overview_jets}

The disrupted but still bound material accretes onto the remnant object. In at least some cases, this produces a highly collimated, ultrarelativistic jet that results in a \ac{SGRB}, as confirmed with \gwid and \grbid. As much of this process is still poorly understood we here pull the phenomenological arguments from \citet{overview_fernandez_metzger_2016}. 

These jets have enormous kinetic energies and produce some of the most luminous \ac{EM} events in existence, with each approaching $10^{50}$\,erg \citep{fong2015decade}. These are powered by the accretion disks \citep{oechslin2006torus}, with $10^{-4}$--$0.3\,M_\odot$ available according to simulations (the range includes extreme conditions but neglects heavy \ac{NSBH} mergers with no released matter). The pure conversion of a typical value of $0.1\,M_\odot$ into energy gives $0.1\,M_\odot c^2 \approx 10^{53}$\,erg, which is sufficient to power a \ac{SGRB} with reasonable overall efficiencies.


How this energy reservoir is converted into the jet is somewhat unsettled (Sect.~\ref{sec:jet_formation}). However, it is agreed that an enormous amount of energy, predominantly from the accreting matter, is deposited in the relatively empty polar regions near the surface of the compact object, which launches an ultrarelativistic fireball away from the central engine. This outflow is collimated into a jet by the material encroaching on the polar region, e.g. the thick accretion disk (or torus) and by the magnetic fields emanating from the system. The emission from the collimated ultrarelativistic jet is only detectable for observers within the jet opening angle, $\theta_j$, due to Doppler beaming limiting the visibility region to $1/\Gamma$, where $\Gamma$ is the bulk Lorentz factor with typical value $\sim 100$. The statements here are detailed and referenced in Section\,\ref{sec:jets}.

If there is significant baryonic matter in this region it is expected to sap the available energy and prevent jet launch (Sect.~\ref{sec:central_engines}). If a jet launches and there is ejecta above the launch site in the polar region the jet must propagate through to successfully break-out; otherwise it could, in principle, be choked. The collimation and the jet interaction with polar material imparts structure onto the jet itself (Sect.~\ref{sec:grb_structure}).

For jets that successfully break-out they move outwards at nearly $c$. At $\sim10^{12}$-$10^{13}$\,cm the jet reaches the photospheric radius where light can escape for the first time \citep{beloborodov2017photospheric}. At around the same distance the jet may release the prompt \ac{SGRB} emission due to the occurrence of internal shocks (though there are alternative models with much higher distances, see Sect.~\ref{sec:GRB_production}). The emission is characterized by a total duration of $\sim 0.01-5$\,s predominantly in the $\sim$10\,keV to $\sim$10\,MeV, with peak isotropic luminosities $\sim$10$^{51}$\,erg\,/s \citep[e.g.,][]{von2020fourth,GW170817-GRB170817A}.

After the prompt \ac{SGRB}, the ultrarelativistic jet continues to speed away from the central engine, with a total kinetic energy $\sim 10^{50}$\,erg, and interacts with the surrounding circumburst material with typical densities $\sim 10^{-4}$--$0.1$\,cm$^{-3}$ \citep{fong2015decade}. As the jet interacts its bulk Lorentz factor slows, the observable angle grows, and it emits synchrotron radiation across nearly the entire \ac{EM} spectrum, which has been detected from radio to GeV energies \citep[e.g.,][]{GRB090510_discovery_Fermi,fong2015decade}. This emission is referred to as \ac{GRB} afterglow.

In Sect.~\ref{sec:other_high_energy} we discuss other high energy signatures potentially related to the ultrarelativistic jet. For now it is sufficient to note that observations strongly suggest late-time energy injection into the system from the central engine, which likely has implications for other observable signatures.


\subsubsection{Quasi-isotropic outflows}
\label{sec:overview_kilonova}

The unbound matter from the system evolves far differently than the bound material that powers the ultrarelativistic jet. This ejecta is neutron-rich, contains roughly $\sim10^{-3}$--$10^{-1}\,M_\odot$, and moves outward at a $\sim 0.1-0.3 c$. The rest of this section borrows heavily from \citet{metzger2014red,metzger2014neutron,overview_fernandez_metzger_2016,review_kilonova_tanaka_2016,metzger2020kilonovae}. The merger process significantly raises the temperature of the NS(s). As the ejecta expands and releases energy as thermal neutrinos it rapidly cools, entering relatively slow homologous expansion in only $\sim10$-$100$\,ms. 

At $\lesssim 10^{10}$\,K free nuclei combine into $\alpha$ particles. At $\lesssim 5 \times 10^{9}$\,K the $\alpha$-process forms seed nuclei with $A \sim 90-120$ and $Z \sim 35$ \citep{woosley1992alpha}. The neutron-to-seed ratio results in rapid neutron captures at rates exceeding the $\beta$ decay of the seeds, rapidly synthesizing the heaviest elements. This is the so-called r-process, responsible for half the heavy elements (here meaning beyond iron) in the universe. This continues until the nuclei reach $A \gtrsim 250$ where fission splits the atoms in two, which are subsequently pushed to higher atomic mass in a process referred to as fission recycling. This generically returns peaks near the closed shell numbers $A = 82, 130, 196$, observed in the solar system elemental abundances. A few seconds have passed. 

The heavy nuclei are undergoing heavy radioactive decay, producing copious amounts of neutrinos ($\sim 0.1-10$\,MeV), nuclear gamma-rays (dozens of keV to a few MeV), and elements that approach the line of stability over time \citep[e.g.,][]{kilonova_gamma_rays_hotokezaka_2016}. At early times the overwhelming majority of released energy escapes as neutrinos because the ejecta material is dense and opaque for photons \citep[see Fig.~4, discussion, and references in][]{metzger2020kilonovae}. In base kilonova models, the earliest photons that can escape are the nuclear gamma-rays, beginning on the order of a few hours. Neutrinos escape with $\sim$30--40\% of the energy; gamma-rays carry 20--50\% of the total energy. This significantly lowers the remaining energy in the system before it reaches peak luminosity \citep[e.g.,][]{KN_products_Barnes_2016,kilonova_gamma_rays_hotokezaka_2016}. 

The main frequency range of interested for \ac{EM} observations of kilonova is \ac{UVOIR}. The opacity in this energy range is driven by atomic transitions of bound electrons to another bound energy state. The open $f$ shell for lanthanides ($Z = 58-72$) have angular momentum quantum number of $l=3$, with the number of valence electron states $g=2(2l+1)=14$, where $n$ electrons can be setup in $C = g!/n!(g-n)!$ possible configurations, with bound-bound transitions scaling as $C^2$, resulting in millions of transition lines in the \ac{UVOIR} range. As the ejecta is expanding with a significant velocity gradient \citep[e.g.,][]{NS-Bauswein} all of these lines are Doppler broadened.
This blankets the entire range, preventing this light from escaping at early times. 

As time continues the ejecta loses energy to neutrinos and gamma-rays, cools as it expands, the radioactive heating rate slows, and it transitions to lower densities until eventually the \ac{UVOIR} photons can escape, resulting in a quasi-thermal transient known as a kilonova. The energy deposition rate of most forms of radioactivity of interest here decay as a power law with index $-1.1$ to $-1.4$ \citep[see][and references therein]{metzger2020kilonovae}. In the hours to days post-merger this maintains high temperatures in the ejecta, with values $\sim10^4-10^3$\,K. Ejecta with relatively high initial electron fraction $Y_e \gtrsim 0.3$ will produce mostly lanthanide-free material which will result in a blue kilonova with peak luminosity on the $\sim$1\,day timescale \citep[e.g.,][]{KN_first_named_Metzger_2010}. Ejecta with low electron fraction $Y_e \lesssim 0.3$ will produce lanthanide-rich material (and potentially actinides) that will produce a red kilonova with a peak luminosity timescale of $\sim$1 week \citep[e.g.,][]{Kilonova-Barnes-2013}.

The prior paragraphs in this section discuss the base-kilonova model, but there may be significant additional signals or alteration of these observables from the quasi-isotropic outflows. These include the radioactive decay of neutrons that are not captured into nuclei, the effects of jet interactions on the previously ejected polar material, and late-time energy injection from the central engine. These are summarized in Sect.~\ref{sec:early_uv}, which references detailed works covering each.

\subsubsection{Aftermath}
\label{sec:overview_verylatetime}

After the energy ejection ends and the kilonova cools and fades, the quasi-isotropic ejecta will continue moving outwards. Over the next few months and years the event will transition to the nebular phase. Once it reaches the deceleration radius, where it has swept up a comparable amount of mass from the surrounding environment, the ejecta will transition to a Sedov--Taylor blast wave that releases synchrotron radiation in the radio bands \citep{omni_radio_nakar_2011,omni_radio_KN_Piran_2012,omni_radio_general_Hotokezaka_2015}, analogously described as a kilonova afterglow.

Over decades, centuries, and millennia it forms a \ac{KNR}. These are bound by a shock wave at the interaction of the merger ejecta and surrounding material, providing a transition edge. They may be similar to supernova remnants but have lower total kinetic energies and will tend to occur in regions with lower surrounding material (due to occurring outside of their host galaxies). Even long after merger they will be radioactive, with emission dominated by isotopes with half-lives of similar order to the age of the remnant \citep{wu2019finding,korobkin2019gamma}. Longer still, the kinetic energy will eventually be used up and the shock-front will dissipate. Ejecta that is bound to the host galaxy will eventually return and become part of the diffuse galactic material where long-term mixing distributes the heaviest elements throughout the galaxy \citep{wu2019finding}. Some will eventually join new planets and stars, and a bit may eventually be dug out of the ground by advanced life. Heavy elements unbound from the host galaxy will be lonely for a reasonable part of eternity.

\subsection{Intrinsic event rates}
\label{sec:intrinsic_rates}

The rates of compact object mergers is of interest to several fields. The true value sets how quickly we can achieve specific scientific outcomes, and will determine the necessary devotion of observational resources and prioritization on telescopes with shared time. Estimates have arisen through several means with predicted rates spanning several orders of magnitude. The most direct measurement comes from \ac{GW} observations, calculated from a detection number in a known spacetime volume. These are the basis for our assumed rates, and the large existing uncertainty should rapidly shrink in the next few years. The local volumetric rates assumed in this paper are explained below and summarized in Table \ref{tab:nearby_rates}.

The latest reported local volumetric rate measurements from \ac{LIGO}/Virgo come from the discovery paper in GW190425, the second \ac{GW}-detected \ac{BNS} merger \citep{abbott2020gw190425}. The full 90\% range reported for \ac{BNS} mergers is $250-2810$ Gpc$^{-3}$ yr$^{-1}$. This value is the union of two measurements, one considering a uniform mass prior between $1$--$2\,M_\odot$ for each \ac{NS} in a \ac{BNS} merger and the second adding the sum of the rates of events like GW170817 to those like GW190425. The median value is approximately $1000$\,Gpc$^{-3}$\,yr$^{-1}$. Following the initial release of this paper, which occurred before the publication on GW190425, and to enable for ease of scaling as these reported rates are updated, we chose to use the \ac{BNS} local volumetric rate of $R=1000_{-800}^{+2000}$ (200-3000) Gpc$^{-3}$ yr$^{-1}$.



The rates of \ac{NSBH} mergers are known with less precision. \citet{lvc_gwtc1} bound the local upper limit of \ac{NSBH} mergers as a function of \ac{BH} mass. Since we do not know the distribution of \ac{BH} mass in \ac{NSBH} merger systems we take the least constraining value of $<610$ Gpc$^{-3}$ yr$^{-1}$, which is for $M_{\rm BH}=5\,M_\odot$. The lower and mid-range value come from the merger rates expectations paper prior to the initialization of Advanced \ac{LIGO} \citep{overview_NSMs_GWs_2010}, where the high rate is similar to the constraints reported above. 

The \ac{LVC} has also reported the discovery of a \ac{CBC} with a high mass ratio, GW190814 \citep{abbott2020gw190814}. Owing to the strength of the signal and the large mass asymmetry this allowed for a precise determination of the individual masses, with the secondary being between $2.50$--$2.67\,M_\odot$. This is potentially the first \ac{NSBH} merger identified, but is more likely to be a \ac{BBH} merger. We do not inform our \ac{NSBH} rates with this event. We may expect a directly measured value once a \ac{GW}-detected event is unambiguously classified as an \ac{NSBH} merger.

For comparison, we report the inferred volumetric local \ac{BBH} merger rates with a mass function that is self-consistent with the observed \ac{BBH} mergers from O1 and O2 \citep{LVC_O2_BBH_rates}. This gives a range of 24.4--111.7 Gpc$^{-3}$ yr$^{-1}$ with a central value of 54.4 Gpc$^{-3}$ yr$^{-1}$. This has a factor of four uncertainty. This range is far narrower due to the larger number of detected \ac{BBH} system. As the number of detected \ac{NS} mergers increases the precision of the local rates measure will similarly improve. 

\begin{table}
\caption{The local volumetric merger rates for \ac{BNS}, \ac{NSBH}, and \ac{BBH} mergers. Columns 3--6 contain the nearest event we may expect in a given year, decade, or century. Columns 7 and 8 report the rate per Milky Way-like galaxy per million years and how many millennia we may expect between events.}
\label{tab:nearby_rates}
\centering
\begin{tabular}{| c || c || c | c | c || c | c |}
\hline
 & Local Rates & \multicolumn{3}{c||}{Nearest Event Per (Mpc)} & \multicolumn{2}{c|}{Rate Per MW-Like Galaxy} \\ \cline{2-5} \cline{6-7}
 & (Gpc$^{-3}$ yr$^{-1}$) & year & decade & century & (Myr$^{-1}$) & (Millennia) \\
\hline
BNS &$1000_{-800}^{+2000}$&$60_{-20}^{+40}$&$29_{-9}^{+20}$&$13_{-4}^{+9}$&$100_{-80}^{+200}$&$10_{-7}^{+40}$\\
NSBH &$60_{-59}^{+550}$&$160_{-80}^{+520}$&$70_{-40}^{+250}$&$30_{-20}^{+120}$&$6_{-6}^{+55}$&$170_{-150}^{+16500}$\\
BBH &$53_{-29}^{+58}$&$160_{-30}^{+50}$&$80_{-20}^{+20}$&$35_{-8}^{+10}$&$5_{-3}^{+6}$&$190_{-100}^{+220}$\\
\hline
\end{tabular}
\end{table}

The rates of \ac{NS} mergers vary through cosmic time. Under the standard formation channel, it should track the stellar formation rate modulo their inspiral times. The peak rate of \acp{SGRB} occurred at a redshift of $\sim0.5-0.8$ \citep[e.g.,][]{Berger2013} before declining to the current rate. This is a useful proxy to estimate the largest average inspiral range due to the Malmquist bias in detecting \acp{SGRB}. The furthest known \acp{SGRB} occurred at a redshift of $>2$ and few are expected beyond a redshift of $\sim$5. We do not explicitly account for intrinsic source evolution for our detection rates in this manuscript. The rates of \ac{NS} mergers do not evolve significantly over the distances we can detect these events through \acp{GW}, neutrinos, or as kilonovae for at least a decade. Source evolution does matter for \ac{SGRB} observations, both prompt and afterglow, but our rates for those events are determined from empirical observations and thus source evolution is accounted for intrinsically.

We lastly close with the rates of rare events that may provide unique understanding of these mergers. Particularly nearby events will be able to be characterized to vastly greater detail; as such, we report the nearest event we may expect on fiducial timescales. Assuming the usual number density of \ac{MW}-like galaxies of $\sim$0.01 Mpc$^{-3}$ \citep[e.g.,][]{bns_rates_meta_2018}, we show the rates per Milky Way-like galaxy per million years, and how many millennia we may expect between events in the Milky Way itself. 

From Table \ref{tab:nearby_rates} we can draw a few immediate conclusions. \ac{BNS} mergers are locally more common than \ac{BBH} mergers and likely more common than \ac{NSBH} mergers. We may expect a \ac{BNS} merger to occur within $\sim$30 Mpc about once a decade. Events within $\sim$20 Mpc are rare, occurring about as often as an average human lifetime. We should expect a \ac{BNS} merger in the Milky Way about every 10 millennia.

Strongly lensed events are prize astrophysical occurrences. They provide both complementary and unique tests in cosmology \citep{lens_Refsdal_1964,lens_linder,review_strong_lens} and fundamental physics \citep{strong_lensing_grbs_Biesiada_2009,cgw_strong_lense_Collett_2017,minazzoli2019strong}, and unique studies of transient events \citep[e.g.,][]{lensed_typeIa,lens_anisotropic_PErna_2009}. The detection and successful identification of a strongly lensed \ac{NS} merger would be momentous, which is discussed in more detail in Section\ref{sec:beyond_cosmo} and a few subsections of Section\,\ref{sec:fundamental_physics}. The intrinsic rates of strongly lensed \ac{NS} mergers are likely to be low but likely non-zero \citep[e.g.,][after accounting for new rates estimates]{biesiada2014strong}. These rates could be increased in the future by targeted known strongly lensed systems \citep[see][for these prospects]{surveys_strong_lens_collett_2015}, analogous to the current galaxy targeting approach \ac{EM} follow-up to \ac{GW}-detected \ac{NS} mergers. 


\subsection{Gravitational waves}
\label{sec:GW_observations}

\acp{GW} are detected by measuring their effect on spacetime itself as the strain $h = \Delta L/L$ where $\Delta L$ is the fractional change of length $L$ \citep{abbott2009ligo}. At the reasonably nearby distance of $\sim$100 Mpc (Sect.~\ref{sec:intrinsic_rates}) the strain at Earth for a canonical \ac{BNS} merger is $\sim10^{-21}$. Detection then requires the most sensitive ruler ever built. Weak \acp{GW} can be described by the ordinary plane wave solution. In \ac{GR} \acp{GW} have only two independent polarization modes \citep{Review_Tests_of_GR_overall_Will_2014}. They can be distinguished by a $\pi/4$ rotation in the plane perpendicular to the direction of motion, which, by convention, are referred to as the plus and cross polarization modes. The strain $h$ from these modes are $h_+$ and $h_\times$, respectively. 

Following \citet{GW_patterns_Schutz_2011}, the antenna response function can be written in terms of the two \ac{GR} polarization modes as

\begin{equation}
    \label{eq:strain}
    h(t) = F_+(\theta, \phi, \psi)h_+(t) + F_\times(\theta, \phi, \psi)h_\times(t)
\end{equation}

\noindent where $\theta$ and $\phi$ are spherical coordinates relative to detector normal, and $\psi$ the polarization angle for the merger relative to this same coordinate system. $F_+$ and $F_\times$ are the interferometer response to the two polarization modes

\begin{equation}
    \label{eq:interferometer_response}
    \begin{split}
    F_+ &= \frac{1}{2} (1+\cos^2{\theta}) \cos{2 \phi} \cos{2 \psi} - \cos{\theta} \sin{2 \phi} \sin{2 \psi} \\
    F_\times &= \frac{1}{2} (1+\cos^2{\theta}) \cos{2 \phi} \cos{2 \psi} + \cos{\theta} \sin{2 \phi} \cos{2 \psi}.
    \end{split}
\end{equation}

\noindent The antenna power pattern, which the \ac{SNR} is proportional to, is

\begin{equation}
    \label{eq:antenna_power_pattern}
    \begin{split}
    P(\theta,\phi) &= F_+(\theta,\phi,\psi)^2 + F_\times(\theta,\phi,\psi)^2\\
    &=\frac{1}{4} (1+\cos^2{\theta})^2 \cos^2{2 \phi} + \cos^2{\theta} \sin^2{2\phi}
    \end{split}
\end{equation}

\ac{GW} emission is omnidirectional but not isotropic. For \acp{CBC} we can define the radiated power as a function of inclination angle $\iota$, which goes from 0 to 180 because orientation matters for \ac{GW} observations (as opposed to the 0 to 90 convention used for most \ac{EM} observations). This relation can be represented as as $F_{\rm{rad}}$, referred to as the binary radiation pattern, and is defined as

\begin{equation}
\label{eq:binary_radiation_pattern}
F_{\rm{rad}}(\iota) = \frac{1}{8}\big(1 + 6 \cos^2(\iota) + \cos^4(\iota)\big).
\end{equation}

\noindent It is equivalent to the $\phi$-average of the interferometer antenna pattern. It is strongest along the total angular momentum axis ($\iota$ = 0, 180) and weakest in the orbital plane ($\iota$ = 90). 

The sensitivity of individual ground-based interferometers is usually quoted in terms of detection distances for canonical \ac{BNS} mergers \citep[e.g.,][]{Gen_2_Prospects_2018}. The detection \textit{horizon} is the maximum detection distance, which occurs for face-on events ($\iota\approx$0 or 180, where the rotation axis is oriented towards Earth) that are directly overhead (or under). Converting the total sensitive volume to a spherical equivalent gives a radius referred to as the detection \textit{range}, which is the usual figure of merit for (single) ground-based interferometer sensitivity. The horizon is 2.26 times the range \citep[e.g.,][]{ligo2012sensitivity}.

\ac{NS} mergers are identified in \ac{GW} strain data through \ac{CBC} searches, where \ac{CBC} refers to \ac{BNS}, \ac{NSBH}, and \ac{BBH} mergers for ground-based interferometers, which are found by looking for signals that match waveforms from a template bank of \ac{GW} inspirals \citep[e.g.,][]{usman2016pycbc,messick2017analysis}. Because the signals of interest are so weak and background noise is significant, a \ac{GW} detection generally requires two or more interferometers to jointly trigger on an event. The interferometers are separated by thousands of kilometers, which results in generally uncorrelated background, giving a massive increase in search sensitivity. Signal significance has historically been quantified through the use of a \ac{FAR}, measuring how often an event with a given value of the ranking statistic occurs in background \citep[e.g.,][]{GW150914,GW151226,GW170817-GW}. Recently, the development of $P_{astro}$, the probability that an event is astrophysical in origin, has provided additional information, conveying the chance a given event has an astrophysical origin based on an assumed volumetric event rate against the rate of detector noise in that region of parameter space. This is a more powerful method that should result in increased detection rates, but its effect on detection rates has not been quantified.

Interferometers directly measure amplitude, which falls as $1/d$ \citep[e.g.,][]{AdvLIGO}, rather than the typical $1/d^2$ for most astrophysical instruments. That is, an increase in sensitivity gives a cubic increase in detection rates, rather than the typical $3/2$. For signal-dominated events this corresponds to a cubic increase in detection rates. 

Through kilometer-scale modified Michelson interferometers the direct detection of \acp{GW} has recently been achieved \citep{GW150914}. We first discuss the US-based observatories. The current design sensitivity of the Advanced \ac{LIGO} interferometers is expected to achieve a \ac{BNS} range of 175 Mpc \citep{LIGO_173Mpc} by $\sim$2020\footnote{Note that this is slightly below the historically quoted number, which has been refined due to a greater understanding of the noise from the optical coatings.}. The NSF has funded the Advanced \ac{LIGO}+ upgrade which has a target \ac{BNS} range of 330 Mpc \citep{LIGO_Aplus}. 

Beyond A+, there are proposed concepts. The \ac{LIGO} Voyager upgrade would push the existing interferometers close to their theoretical maximum sensitivity, and we use a representative \ac{BNS} range of 1\,Gpc \citep{LIGO_Voyager}. Lastly, third generation interferometers \citep[e.g.,][]{Cosmic_Explorer,Einstein_Telescope} will detect these events throughout the universe. Converting from values in \citet{reitze2019cosmic}, the early stage Cosmic Explorer ($\sim$2035) would have a \ac{BNS} range of $\sim$12 Gpc and the late-stage version ($\sim$2045) $\sim$60 Gpc. We take $\sim$10 Gpc as a representative value.

\begin{figure}
\centering
\includegraphics[width=1.0\textwidth]{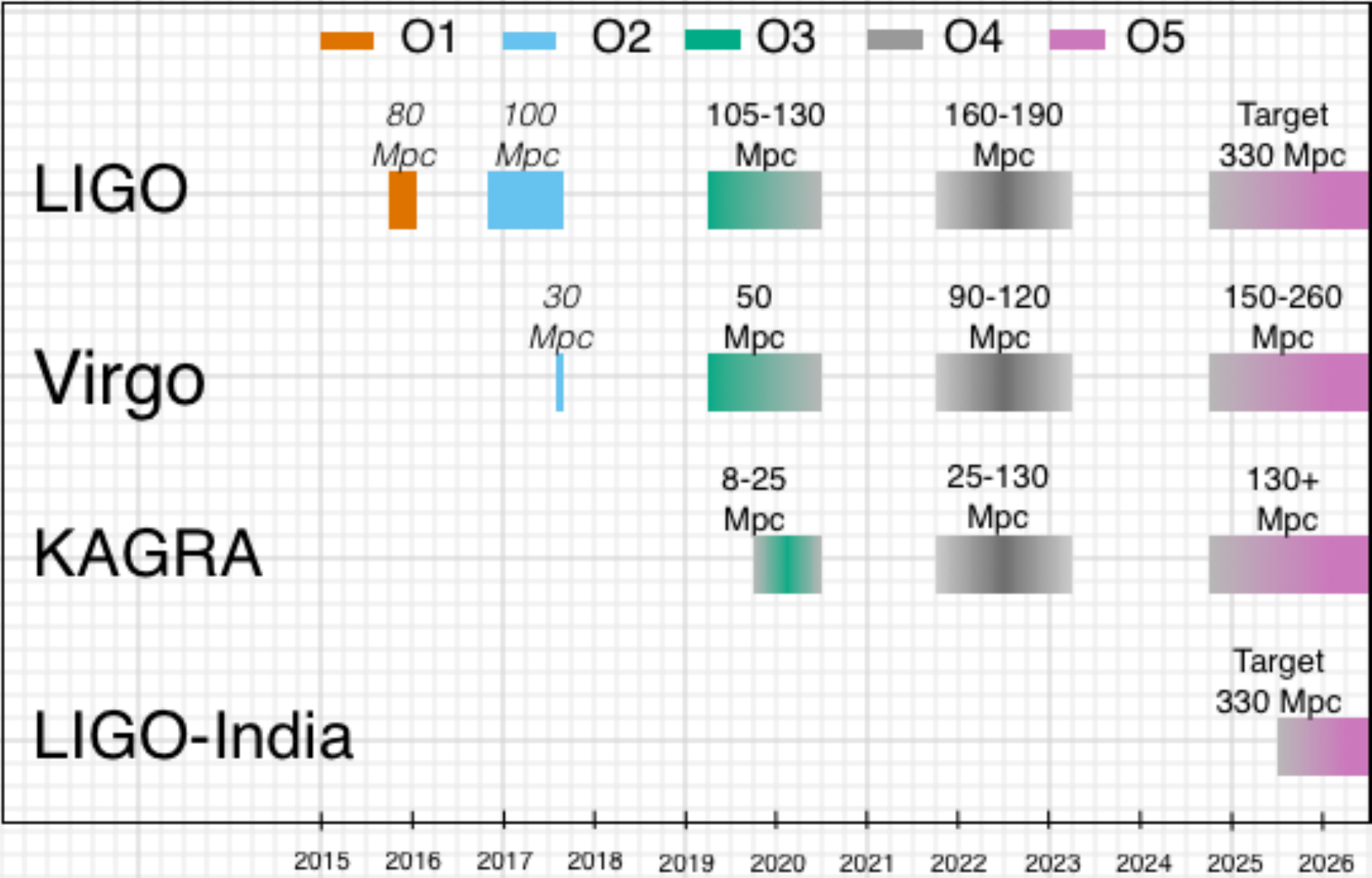}
\caption{The planned ground-based \ac{GW} network observing runs. O1, O2, and about half of O3 have already completed. During O4 the interferometers should approach their Advanced design sensitivity. From 2025+ several interferometers will be upgraded to their advanced configuration.}
\label{fig:GW_obs_plan}
\end{figure}

The \ac{LIGO} interferometers are only part of the ground-based \ac{GW} detection network. The active \ac{GW} detectors are the two Advanced \ac{LIGO} interferometers and the Advanced Virgo \citep{AdvVirgo} interferometer. \ac{LIGO} and Virgo work together as the \ac{LVC}. They are to be joined by the \ac{KAGRA} interferometer \citep{aso2013interferometer} in late 2019 and eventually by LIGO-India which would enter at the A+ version \citep{LIGO_India}. These interferometer sites are generally referred to by letters, H for LIGO-Hanford, L for LIGO-Livingston, V for Virgo, K for KAGRA, and I for LIGO-India. A summary of the  currently expected ground-based \ac{GW} network sensitivity and planned observing runs through $\sim$2026 is shown in Figure \ref{fig:GW_obs_plan}.  The plan updates will be available online\footnote{\url{https://www.ligo.org/scientists/GWEMalerts.php}}. 

In Table \ref{tab:GW_detection_rates} we report reasonable and conservative detection rates for \ac{NS} mergers for the four representative sensitivities. Our base estimate accounts for only two, coaligned interferometers, equivalent to the HL configuration for at least the next decade. This enables easy calculation of a particularly conservative estimate. We also provide a broader network estimate as a function of time based on the network figures of merit in \citet[][which are not directly comparable given the differing interferometer sensitivities]{GW_patterns_Schutz_2011} and simulations in \citet{NS_merger_rates_LVC_2016}. All estimates assume individual interferometer livetime fractions of 70\%, corresponding to 50\% livetime for the HL(-like) configuration(s).

The Advanced and A+ rates are calculated with the intrinsic rates from Table \ref{tab:GW_detection_rates} and their sensitivity volume. Source evolution at these distances are unimportant and neglected. The \ac{NSBH} rates assume they are detected $\sim$2 times further, corresponding to a reasonably light \ac{BH} (giving conservative estimates) which should produce \ac{EM} emission. The Voyager and Gen 3 rates assumes no source evolution, which is a conservative estimate. The Gen 3 rates further only consider events within a redshift of 0.5, providing a very conservative limit. These ranges are 90\% confidence, giving lower limits at 95\% confidence.

\begin{table}
\caption{The expected interferometer sensitivities for the current Advanced interferometers at design sensitivity, the Advanced+ upgrade, the Voyager upgrade, and representative values for third generation interferometers. For each generation we report the \ac{BNS} range and horizon \citep[see][for the interferometer figures of merit that account for cosmological effects]{chen2017distance} in both distance and redshift. For the base detection rate estimates we assume the \ac{GW} network is composed of co-aligned interferometers of identical sensitivity at the Hanford and Livingston sites, giving separate rates for \ac{BNS} and \ac{NSBH} mergers. Based on the scaling and values from \citet{GW_patterns_Schutz_2011} and \citet{review_future_GW_network_2018} we report lower limit rates for a 4 interferometer Advanced network and a 5 interferometer A+ network under by HLV(KI). The frequency range is an approximation and intended only as a rough guideline.
}
\label{tab:GW_detection_rates}
\centering\begin{tabular}{|cc|cccc|}
\hline
 & & \multicolumn{4}{|c|}{GW Interferometer Generation} \\
 & & Advanced & A+ & Voyager & Gen 3 \\
\hline\hline
Frequency Range & Hz & 20-1000 & 10-1000 & 10-3000 & 5-4000 \\
\hline\hline
\multicolumn{2}{|c|}{BNS Merger Detection Distances} & & & & \\ \hline
Range & $D_L$ & 175 Mpc & 330 Mpc & 1 Gpc & $>$10 Gpc \\
 & $z$ & 0.04    & 0.06    & 0.2  & $>$1 \\
Horizon & $D_L$  & 400 Mpc & 750 Mpc & $>$1.5 Gpc & $>$10 Gpc \\
 & $z$  & 0.09     & 0.14     & 0.4 & $>$3 \\
\hline\hline
\multicolumn{2}{|c|}{NSBH Merger Detection Distances} & & & & \\ \hline
NSBH Range   & $D_L$ & 350 Mpc & 660 Mpc & $>$1.5 Gpc & $>$10 Gpc \\
             & $z$   & 0.07    & 0.14    & 0.4   & $>$2 \\
NSBH Horizon & $D_L$ & 800 Mpc & 1.3 Gpc & $>$3 Gpc & $>$10 Gpc \\
             & $z$   & 0.16    & 0.25    & $>$0.5   & $>$2 \\
\hline \hline 
\multicolumn{2}{|c|}{NS Merger Detection Rates} & & & & \\ \hline
BNS - HL & yr$^{-1}$ & 2-32 & 10-200 & $>$Daily & $>$Hourly \\
BNS - HLV(KI) & yr$^{-1}$ & $>$4 & $>$30 & - & - \\
NSBH - HL & yr$^{-1}$ & 0-50 & 0-300 & $>$1 & $>$100 \\
 \hline
\end{tabular}
\end{table}

Beyond just detecting them, characterization of \ac{NS} mergers is an additional priority for design requirements. The high end frequency is set by the wish to directly observe the merger events themselves. From Sect.~\ref{sec:overview_merger} the highest expected maximum frequency is for the \ac{BNS} prompt collapse case reaching $\sim$6-7\,kHz. Sufficiently capturing this range should also enable sensitive searches for \ac{NS} modes above the primary frequency in the \ac{BNS} (meta)stable remnant cases \citep[see][and references therein]{ackley2020neutron}. 

Pushing to lower frequencies has a number of benefits, such as providing vastly improved parameter estimation precision due to a far greater \ac{SNR} for a given event. A canonical \ac{BNS} (NSBH) merger emitting \acp{GW} at 0.1\,Hz will merge in about a decade (a year) \citep[e.g.,][]{graham2017mid}. For \ac{NS} mergers that will merge within an instrument lifetime this provides a reasonable lower frequency goal. This range is also ideal for the best-case \ac{GW} localizations, as we will show. Thus, absent funding or technical considerations, the best range to study these events is $\sim0.1$\,Hz to $\sim$10\,kHz. The rough frequency range for the four ground-based \ac{GW} interferometer sensitivity examples is given in Table \ref{tab:GW_detection_rates}. For the next decade we are largely limited to the $\sim$10-1000\,Hz regime. Achieving higher frequencies may be possible, but pushing lower than 5\,Hz on the ground is nearly impossible. 

Generic \ac{GW} observations of \acp{CBC} measure more than a dozen parameters. The extrinsic system parameters include the location ($\theta$, $\phi$, and the luminosity distance $d_L$), inclination ($\iota$), polarization angle ($\psi$), eccentricity ($e$), coalescence phase ($\phi_0$), and merger time $t_{GW}$. The intrinsic parameters include the mass and spin components of each pre-merger object ($m_1$, $m_2$; $\overrightarrow{S}_1$, $\overrightarrow{S_2}$). Most of these parameters have strong correlations (often referred to as degeneracies). One example is the amplitude dependence on both $\iota$ and $d_L$, contributing to greater uncertainty on both measures \citep{schutz2002lighthouses}. For \ac{NS} mergers matter effects accelerate the late inspiral which can be captured into the tidal deformability parameter ($\Lambda$). 

Eccentricity is generally expected to be zero for these systems, as circularization happens on a shorter interval than the expected inspiral time to merger \citep{peters1963gravitational,review_NS_Faber_Rasio}. The polarization can be constrained for events detected by interferometers that are not coaligned, based on the \acp{SNR} and antenna response as a function of position. These detections will tend to have more precisely measured inclinations, as the parameters are correlated. The merger time and coalescence phase are precisely measured for \ac{NS} mergers given the long inspirals \citep[e.g.,][]{GW170817-GRB170817A}. Tidal deformability is determined by the (non-)detection of accelerated inspirals due to matter effects, and for \ac{NSBH} mergers, by determining the frequency at which tidal disruption occurs, which tends to happen at high frequencies where we currently have insufficient sensitivity.

The remaining \ac{GW}-determined parameters are mass and spin. The masses are determined from the chirp mass

\begin{equation}
    \label{eq:chirp_mass}
    \mathcal{M}_c = \frac{(M_1 M_2)^{3/5}}{(M_1 + M_2)^{1/5}},
\end{equation}

\noindent where $M_1$ and $M_2$ are the masses of the primary and secondary, and the mass ratio $q = M_2/M_1$ which is by definition $q \leq 1$. For \ac{NS} mergers the chirp mass measurement is extremely precise as the \ac{GW} observation covers thousands of cycles, giving a great measure on the frequency evolution of the inspiral. The mass ratio effect on the inspiral is perfectly correlated to first order with one of the spin parameters, requiring high \ac{SNR} near merger to be well constrained. $q$ will be poorly constrained for \ac{BNS} mergers so long as the merger occurs out of band of the \ac{GW} interferometers \citep{lvc_gwtc1}, except for particularly loud events. The spin components are usually written in terms of dimensionless spin $\vec{\chi} \equiv c \vec{S}/(GM^2)$.

A unique aspect of \ac{GW} observations is knowledge of the distance to the source. Both the strain amplitude $h$ and $\dot{f}_{GW}$ depend on the $\mathcal{M}_c$, defined in Equation \ref{eq:chirp_mass}, enabling a determination of the luminosity distance to the source \citep{H0-Schutz-1986,schutz2002lighthouses}. For ground-based interferometers typical distance uncertainty is tens of percent \citep[e.g.,][]{chen2017distance}, with improved uncertainty for higher \ac{SNR} events. Given the distance-inclination correlation, the constraint can be improved when external inclination information is provided \citep[e.g.,][]{guidorzi2017improved}.

The earliest detectable signal for \ac{NS} mergers are \acp{GW}. As such, they play an important role in both the detection and characterization of these events, but also in providing localization information for searches with other instruments. Current ground-based \ac{GW} interferometers can measure \ac{BNS} merger times to sub-ms accuracy. As they are separated by thousands of kilometers and \acp{GW} travel at the speed of light \citep{GW170817-GRB170817A} we can combine pairs of detections into narrow timing annuli on the sky. The narrowness is determined by $\delta t/d_I$ where $d_I$ is the distance between contributing instruments. The precise timing for \ac{BNS} mergers ($\lesssim$ms) enables narrow annuli, despite the (comparatively) short baselines between interferometers.

For two interferometer detections the typical 90\% confidence region is a few hundred square degrees, with large variation in each case \citep[e.g.,][]{singer2014first}. Three interferometer detections decrease to a median of few 10s of square degrees. Additional interferometers improve this accuracy \citep[e.g.,][]{review_future_GW_network_2018}. Table\,\ref{tab:network_livetime} shows the absolute and cumulative livetimes for a number of active interferometers from a network of a given size. Extreme loud single interferometer events can be reported without independent confirmation; in this case the localization will match the antenna pattern of that interferometer, giving a 90\% confidence region of order half the sky. When one interferometer is significantly more sensitive than another the joint detection rate will decrease and two interferometer localizations will be the antenna pattern of the more sensitive instrument, slightly modified by the other, with 90\% confidence region covering several thousand square degrees, as shown by GW190425 \citep{gcn_LVC_S190425z}.

\begin{table}
\caption{The first column varies the number of interferometers contributing to a given observing run. For these rows, the fraction of time a given number of interferometers contribute is given in absolute terms in the central block and cumulative terms in the final block. Each individual interferometer is assumed to have a 70\% livetime, which is a fiducial value based on prior results and future expectations.}
\label{tab:network_livetime}
\centering
\begin{tabular}{|c||ccccc||ccccc|}
\hline
Total Number & \multicolumn{5}{c||}{Active Detectors} & \multicolumn{5}{c|}{Minimum Active Detectors} \\
of Detectors & 1    & 2    & 3    & 4    & 5    & 1     & 2    & 3    & 4    & 5    \\ \hline
1                & 70\% & -    & -    & -    & -    & 70\%  & -    & -    & -    & -    \\
2                & 42\% & 49\% & -    & -    & -    & 91\%  & 49\% & -    & -    & -    \\
3                & 19\% & 44\% & 34\% & -    & -    & 97\%  & 78\% & 34\% & -    & -    \\
4                & 8\%  & 26\% & 41\% & 24\% & -    & 99\%  & 92\% & 65\% & 24\% & -    \\
5                & 3\%  & 13\% & 31\% & 36\% & 17\% & 100\% & 97\% & 84\% & 53\% & 17\% \\ \hline
\end{tabular}
\end{table}


Because inspirals can be detected before merger, \ac{GW} detections can be reported before merger, i.e. act as early warning systems. Knowing the event time in advance can be beneficial for several reasons, such as pointing wide-field telescopes, switching observational modes, increasing temporal resolution, etc, but perhaps the greatest potential outcome would be the pointing of \ac{EM} telescopes to observe the source at merger time, which would uncover vastly greater understanding of these sources. The localizations available before merger using the method discussed above will give typical accuracies about a thousand square degrees a minute before merger \citep[e.g.,][]{cannon2012toward} because the timing uncertainty is not precise until just before merger. Loud events could have improved, but still poor, localizations.

There are additional mechanisms for constraining source position from \ac{GW} observations, relying on the motion of the interferometer. Ground-based interferometers are bound to the surface of Earth and their antenna patterns sweep over the sky as Earth rotates through the day. For signals that are $\sim$hours long this change causes time-dependent exposure that depends primarily on the source position, refining the location. For the recent listed frequency range of Cosmic Explorer, the U.S. third generation proposal, it will achieve 5\,Hz on the low end \citep{reitze2019cosmic}, which would begin to observe \ac{BNS} mergers about an hour before merger. Therefore, even with third generation interferometers we will not be able to rely on additional localization methods and will likely be limited to accuracies of order $\sim$100 square degrees a minute before merger. For comparison, 30 seconds is among the current fastest repoint times (from reception of alert to observation) currently available in time domain astronomy.

Space-based interferometers will localize primarily through measuring Doppler shifts as their orbit moves towards/away the source \citep[e.g.,][]{cutler1998angular}. The longer integration time can give higher \ac{SNR}, providing more precisely determined distances. This is the dominant localization method for the funded satellite constellation mission \ac{LISA}, which would have an Earth-like orbit around the Sun and would cover the $\sim$mHz frequency range. \ac{LISA} may detect \ac{BNS} and \ac{NSBH} systems, but they would be long before merger. 

There are proposed mid-range interferometers, referring to instruments that cover frequencies between \ac{LISA} and the ground-based network, \citep[e.g.,][]{dimopoulos2008atomic,kawamura2011japanese,canuel2018exploring,baker2019space,kuns2019astrophysics}. Such devices would measure \ac{BNS} systems years before merger and are likely the only way to achieve good pre-merger localizations. The details vary, but even conservative instruments/predictions give sub-degree accuracy for at least a few systems per year. These would enable broadband \ac{EM} observations of \ac{NS} mergers during coalescence through the first few hours. There is no funded mission in this range, precluding launch within a decade, but we discuss them as they would enable unique science with \ac{NS} mergers inaccessible through other means. 






\subsection{Prompt gamma-ray bursts}
\label{sec:detect_prompt_grb}
The easiest method to detect \ac{NS} mergers is through their prompt \ac{SGRB} emission. The \ac{GRB} monitors have detected more than a thousand \acp{SGRB}, which is (currently) three orders of magnitude more than \ac{GW} detections of \ac{NS} mergers, two more than claimed kilonovae, and one more than \acp{SGRB} afterglow. These events emit primarily in the $\sim$10\,keV--10\,MeV energy range, which is only observable from space.
There are two classes of \acp{GRB}, short and long, separated in the prompt phase by a duration threshold. These classes have different origins, as proven by follow-up observations. \acp{LGRB} origin from a specific type of core-collapse supernova; \acp{SGRB} originate from \ac{BNS} mergers and likely \ac{NSBH} mergers. Short and long colloquially refer to these separate classes, despite the fact that the duration distributions overlap.

The most prolific active detector of \acp{SGRB} is the \Fermi \ac{GBM} \citep{Meegan09} which identifies more \acp{SGRB} than all other active missions combined. It is this instrument we will use to baseline our rates. \ac{GBM} consists of two types of scintillators to cover $\sim$10\,keV-10\,MeV. The duration threshold where events are equally likely to belong to the short or long distributions for \Fermi \ac{GBM} is 5\,s \citep{GBMBurstCatalog_6Years}. From the combined fit to the short and long log-normal distributions, the weight of each distribution is 20\% and 80\%, respectively. This gives a \Fermi \ac{GBM} \ac{SGRB} detection rate of 48\,\acp{SGRB}/yr. 
The low-energy detectors are oriented to observe different portions of the sky and, to first order, have a cosine response from detector normal. Localization is done by deconvolving the observed counts in each detector with the response of the instrument as a function of energy and constraining the sky region where the event is consistent with a point source origin. The median \ac{GBM} \ac{SGRB} localization, including systematic error, has a 90\% containment region of $\sim$500 deg$^2$. The typical localization accuracy is a few hundred square degrees, comparable to the two-interferometer \ac{GW} localizations, but are quasi-circular blobs rather than narrow arcs.

The \Swift \ac{BAT} consists of an array of gamma-ray scintillators below a partial coding mask, which imparts shadows in a unique pattern \citep{Barthelmy05}. This detector setup trades effective area for localization accuracy, detecting $\sim$8-9\,\acp{SGRB}/yr with localizations to 3' accuracy \citep[e.g.,][]{Lien2016}. \Swift has two narrow-field telescopes, the \ac{XRT} and \ac{UVOT}, which are repointed to the \ac{BAT} localizations for bursts within their field of regard. The \ac{XRT} recovery fraction of \ac{BAT} \acp{SGRB} is 75\%, and is 85\% of those it observes promptly. This enables localization accuracy to a few arcseconds. This is sufficient for follow-up with nearly any telescope, and was the prime mission for \Swift. The \ac{BAT} is sensitive over 15-150\,keV, preventing it from performing broadband spectral studies of \acp{SGRB}.

There are two other instrument types that can promptly detect \acp{SGRB}. The \ac{LAT} is the primary instrument on-board the \Fermi satellite and is a pair-conversion telescope that observe from $\sim$100\,MeV-100\,GeV \citep{atwood2009large}. It detects about $\sim$2\,\acp{SGRB}/yr, though some of these are afterglow-only detections \citep{ajello2019decade}. Compton telescopes are phenomenal \ac{SGRB} detectors that detect photons within the $\sim$100\,keV-10\,MeV energy range, with great sensitivity, wide fields of view, and localization accuracy of order a degree. They can provide a large sample of \acp{SGRB} with localizations sufficient for follow-up with wide-field instruments. 

Beyond autonomous localizations by individual satellites, the \ac{IPN} pioneered using the finite speed of light to constrain events with timing annuli on the sky \citep[see][and references therein]{hurley2010third}. \ac{GRB} temporal evolution is fit by empirical functions and their intrinsic variability is limited to $\gtrsim$50\,ms. That is, to achieve annuli similarly narrow to the \ac{GW} network localizations we require baselines longer than can be achieved in \ac{LEO}. By placing gamma-ray detectors on spacecraft bound for other planets the baseline increases by orders of magnitude, enabling very bright events to be localized to arcminute accuracy. The limitation of the \ac{IPN} is the high data downlink latency, generally too long for the purposes of following \ac{SGRB} afterglow and early kilonova observations. The other issue is the lack of gamma-ray detectors on recent planetary spacecraft, threatening an end to massive baselines for the \ac{IPN}.

The KONUS-Wind instrument has broadband energy coverage comparable to \ac{GBM}, no autonomous localization capability, but sits at the Sun $L_1$ point \citep{aptekar1995konus}. The \ac{INTEGRAL} \ac{SPI-ACS} is an anticoincidence shield sensitive to $\gtrsim 100$\,keV with no energy or spatial information, but has a highly elliptical orbit that brings it up to half a light second from Earth \citep{2003A&A...411L.299V}. With the \ac{LEO} \ac{GRB} monitors they form the backbone of the modern \ac{IPN}, with sufficient distances from Earth and detection rates to regularly constrain the localizations of \acp{GRB} to sub-degree accuracy.

Once a burst is identified it is characterized by its temporal and spectral properties. The \ac{GRB} time is often set to the trigger time, though this definition varies for a given instrument. The on-set time of \ac{GRB} emission can be refined when necessary by fitting a field-specific pulse function and defining the start time as when some amount of the peak height (e.g. 5\% of the maximum) is achieved. The duration of a burst is determined through the $T_{90}$ measure, the time from when 5\% to 95\% of the total fluence is observed, which gives a first assignment as short or long. Out of this analysis comes an estimate of the peak photon and energy flux, and total energy fluence for the event. Spectral analysis of \acp{GRB} is performed with the forward-folding technique, where an empirical functional form is convolved with the detector responses and compared with the data. The usual forms are a basic power law, a smoothly broken power law, or a power law with an exponential cutoff. These functions are not selected with any theoretical motivation. Spectral analysis is often done in a time-integrated manner, which averages out the spectral evolution of the event. Generally a power law fit indicates a burst that is too weak to constrain spectral curvature. When this curvature is constrained it is parameterized as $E_{peak}$, where most of the power is radiated.

When the distance to the source is known (Sect.~\ref{sec:host_redshift_where}) the observed flux and fluence can be converted into the isotropic-equivalent energetics, $L_{iso}$ and $E_{iso}$ for the peak luminosity and total energy released, respectively. These are calculated by assuming the observed brightness is constant over a spherical shell with radius $D_L$ to the source, and are reported in the bolometric range $1$\,keV-$10$\,MeV, after accounting for cosmological redshift through the k-correction factor \citep{bloom2001prompt}. These values can be refined to jet-corrected energetics if the half-jet opening angle is determined through observations of the afterglow \citep{fong2015decade}. 

These are the basic parameters in wide use within the field. There are additional analyses that can be done that are quite useful. Examples include fitting multiple spectral functions simultaneously has provided evidence for additional components \citep[e.g.,][]{guiriec2011detection,tak2019multiple} and a potential spectro-temporal signature indicative of nearby \ac{BNS} mergers \citep{GRB150101B_Burns}. 


\begin{figure}
\centering
\includegraphics[width=0.7\textwidth, hiresbb=true]{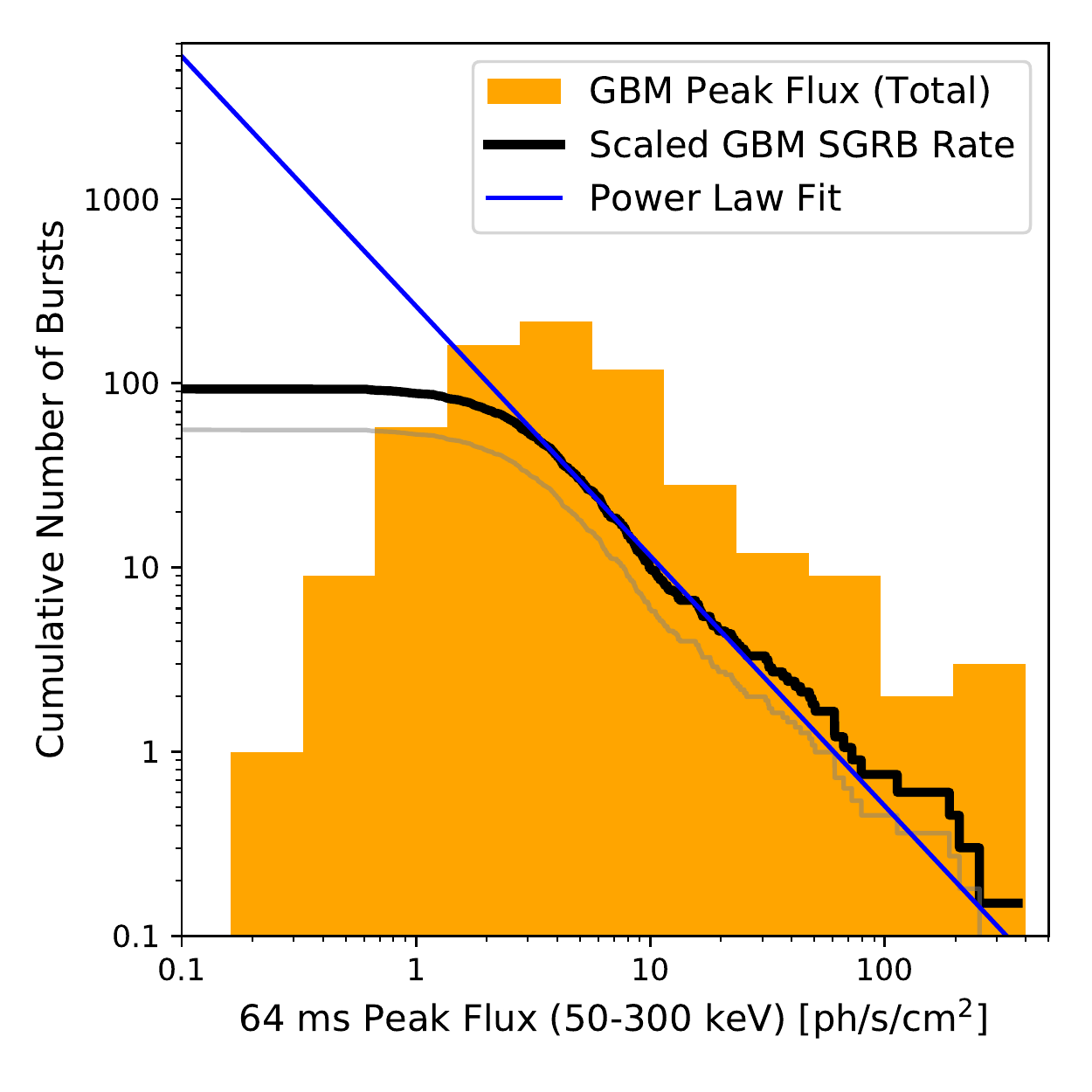}
\caption{The \ac{SGRB} rate as a function of sensitivity. Orange is the histogram of observed 64\,ms peak flux in the 50--300\,keV energy range for \ac{GBM} \acp{SGRB} over an 11-year period. The 64\,ms duration is chosen to encompass most \acp{SGRB} (e.g. the majority of bursts are longer than this timescale) and 50-300\,keV is the dominant triggering range for \ac{GBM}. The grey line is the cumulative logN-logP yearly detection \textit{rate}. GBM has an average exposure of $\sim$60\% (conservatively ignoring sky regions \ac{GBM} observes with poor sensitivity), which is scaled to give the all-sky detection rate of \acp{SGRB} above GBM's on-board trigger sensitivity in black. We fit a power-law to this curve for events above 7\,ph/s/cm$^2$ as this should be a reasonably complete sample. The fit has an index of $-1.3$. 
}
\label{fig:logNlogP}
\end{figure}

Lastly, we discuss how the detection rate of \acp{SGRB} varies with sensitivity, as shown in Figure \ref{fig:logNlogP}. The result is an estimation of the all-sky \ac{SGRB} rate above the on-board trigger threshold for \ac{GBM} of $\sim$80/yr and an extrapolation to higher sensitivity by a logN-logP power-law with an index of -1.3, varying by $\sim$0.1 depending on where the fit threshold is applied. That is, instruments with 2 (10) times \ac{GBM} sensitivity corresponding to a detection rate multiplier of 2.5 (20). Given sensitivity scales as the square root of effective area, to maximize detection rates with a fixed amount of scintillators one should prioritize all-sky coverage over depth in a given direction, though depth is preferred for characterization of individual events.

The \ac{SGRB} detection rates discussed in the previous paragraph were for on-board triggers, which are basic to ensure sample purity, minimize the use of limited bandwidth, and due to the limitations of flight computers. The initial data downlinked after a trigger is limited. Most \ac{GRB} monitors also provide continuous data which is generally binned with somewhat coarse temporal or energy resolution, owing to bandwidth considerations. \Fermi \ac{GBM} is able to downlink continuous \ac{TTE} data, which enables deep searches for additional \acp{SGRB}. There is a blind untargeted search for \ac{SGRB} candidates that reports the results publicly with a few hours delay, limited by the data downlink latency\footnote{\url{https://gcn.gsfc.nasa.gov/gcn/fermi_gbm_subthresh_archive.html}}. The targeted search of \ac{GBM} data \citep{Blackburn2015,Goldstein2016,Targeted_Search_Kocevski_2018} is the most sensitive \ac{SGRB} search ever developed. Based on the maximal detection distance for \grbid with the targeted search against the detection limit of the on-board trigger \citep{Goldstein2017}, the inefficiencies of the on-board trigger due to non-uniform sky coverage, and the logN-logP relation, the \ac{GBM} targeted search should be capable of recovering a few times as many \acp{SGRB} as the on-board trigger, or a few per week.

\subsection{Statistical association and joint searches}
\label{sec:association}
Multimessenger science is incredible. It requires detections in multiple messengers and the robust statistical association of those signals. This is often neglected or totally ignored. As such, we focus on this problem before proceeding to other detections of \ac{NS} mergers. Much work has been done in this endeavor during the past several years, with varied focus and applicability. For example, \citet{ashton2018coincident} developed a general Bayesian framework to associate signals based on commonly measured parameters. For our purposes it is sufficient to use a representative frequentist method using the three dominant parameters that provide association significance: temporal and spatial information, and the rarity of the event itself.

We first discuss time. The rate of \ac{GW}-detected \ac{NS} mergers will remain at less than one per day for the better part of a decade. The rate of \ac{NS} mergers detected as \acp{SGRB} will remain similarly rare. The time offset of these two events is expected to be only seconds long. For example, the chance coincidence of a \ac{GBM} triggered \ac{SGRB} occurring within a few seconds of a \ac{GW} detection of a \ac{NS} merger is $\sim$few$\times 10^{-6}$. Then, with the inclusion of spatial information, even with the independent localizations spanning hundreds of square degrees, the association easily surpasses 5$\sigma$ (see \citealt{GW170817-GRB170817A} and discussions in \citealt{ashton2018coincident}). A pure sample is readily maintained even for large numbers.

Spatial information can be even more powerful. For much of observational astronomy localization alone is sufficient to associate multiwavelength signals because the uncertainty on the localization from radio to X-ray can be a trillionth of the sky, which enables easy association of steady sources. These are so precise that association significance is generally not calculated. We use the nominal \Swift operations as our example here. \Swift has a \ac{GRB} rate (both long and short) of $\sim$100/year which are localized to 3' accuracy with the \ac{BAT}. \Swift autonomously repoints to the majority of these events within about a minute. Fading X-ray signals above the limit of the ROSAT All-Sky Survey \citep{voges2000rosat} within the \ac{BAT} localization are effectively always the \ac{GW} afterglow.

Among the hidden issues exposed by \gwid is the association of kilonovae signals to a \ac{GW} event. For \gwid the last non-detection with sufficient limits was the DLT40 observation 21 days before merger time \citep{GCN_DLT_last_obs}. With our median \ac{BNS} merger rate and the 380 Mpc$^3$ volume from the final \ac{GW} constraint \citep{GW170817-GW}, $P_{\rm chance} \approx (380$\,Mpc$^3) \times (1000$\,Gpc$^{-3}$yr$^{-1}) \times 21 \text{days} \approx 10^{-5}$, which is a reasonably robust association.

To examine a worse-case scenario we can imagine a similar \ac{EM} detection in the follow-up of GW190425 which has a distance estimate of $156 \pm 41$\,Mpc and a 90\% confidence region covering 7461 deg$^2$ \citep{GCN_LVC_190425z}. Then, $P_{\rm chance} \approx 0.5$, a rather questionable association. As the \ac{GW} interferometers improve their reach, events will tend to have similar fractional uncertainty on their distance determination which corresponds to a far larger total localization volumes. Take a middle example with a typical localization region of 500 deg$^2$, distance $200 \pm 50$\,Mpc, and a last (constraining) non-detection a week before, then $P_{\rm chance} \approx 1\%$. So, \textit{even if we know the event is a kilonova, we may not be able to robustly associate it}. This effect is even more important when relatively pure samples are strongly preferred (e.g. standard siren cosmology). This issue can either be solved by increasing the spatial association significance (either through better \ac{GW} or \ac{GRB} localizations) or the temporal association significance. The latter can be accomplished in two ways. More recent non-detections help, but may require sensitivity to $\sim$23--24 Mag \citep{cowperthwaite2015comprehensive}. Alternatively, one can determine the start time to $\sim$1\,day accuracy either by directly constraining the rise or through inferring the age of the kilonova for well-sampled events.

Joint searches for \ac{NS} mergers can be more powerful than individual searches by elevating the significance of a true signal and repressing background. Most work in joint searches for \ac{NS} mergers has focused on \ac{GW}-\ac{GRB} searches. Owing to the rarity of \acp{GRB} and the $\sim$seconds intrinsic time offset, current joint searches can improve the \ac{GW} detection distance by 20--25\% \citep{Joint-Williamson-2014}, which is a corresponding search volume increase of nearly double.\footnote{Note this applies only to the GW-GRB detection volume, and will not significantly affect the GW-only detection rate due to inclination effects. See Sect.~}\ref{sec:grb_structure} Further, for at least the next few years we will have a significant amount of time where only a single \ac{GW} interferometer is active (Table~\ref{tab:network_livetime}, Fig.~\ref{fig:GW_obs_plan}). \acp{SGRB} are so rare that association with a single interferometer trigger could confirm the event. This improves the effective livetime of the \ac{GW} network for \ac{GW}-\ac{GRB} searches. 

In addition to increasing the number of multimessenger detections of \ac{NS} mergers, joint \ac{GW}-\ac{GRB} searches also provide improved localization constraints by combining the two independent, morphologically different localizations. We demonstrate with \gwid. The first localization reported by the \ac{LVC} was the \ac{GBM} localization \citep{gcnLVCGW170817_0}. This was because Virgo data was not immediately available, a massive glitch occurred contemporaneously in \ac{LIGO}-Livingston \citep{gcnLVCGW170817_1}, and the \ac{GBM} localization is more constraining than the single interferometer antenna pattern from \ac{LIGO}-Hanford. The first \ac{GW} network localization (HLV) was reported 5 hours after event time, with a 90\% containment region covering 31 deg$^2$ \citep{gcnLVCGW170817_2}. If we take the HL localization region and combine it with the independent \ac{GBM} localization, the 90\% confidence region covers 60 deg$^2$. These combined localizations also improve the estimate of the distance to the host galaxy. This information was available much earlier than the Virgo information, but was not reported publicly.

Even with the poor localization accuracy of \Fermi \ac{GBM}, the different morphologies of the typical \ac{GBM} and \ac{GW} confidence regions enable greatly improved joint localizations. \ac{GBM} will tend to reduce the 90\% confidence regions for single interferometer events by $\sim$90\%, for double interferometer localizations by $\sim$80\%, but will tend to not improve localizations from three or more interferometers \citep{Burns_Dissertation}. Should a joint \ac{GW}-\ac{GRB} detection occur with \Swift, the \ac{BAT} (or \ac{XRT}) localization would be sufficient for immediate follow-up. \ac{IPN} localizations will be between the two, but with much longer reporting latency (hours-days instead of a minute). 

The other promising joint search is \ac{GW}-neutrino or neutrino-\ac{GRB} searches, for cases where the neutrino emission is nearly immediate \citep[e.g.,][]{van2009joint}, though the prospects for neutrino detections of \ac{NS} mergers are pessimistic or uncertain. Some work has been done on prospects for elevating sub-threshold \ac{GW} detections through association with a kilonova or afterglow. \citet{lynch2018observational} find that to double the number of true \ac{GW} events the \ac{FAR} threshold would increase by five orders of magnitude. They advocate for \ac{LVC} reporting thresholds to be determined by $P_{\rm astro}$, which we support. However, weak events have to overcome the likelihood that the \ac{GW} event is not real for confirmation \citep{ashton2018coincident}. For example, the \ac{LVC}  initial classification for S190718y is 98\% terrestrial (noise) and 2\% \ac{BNS} \citep{GCN_LVC_S190718y}, lowering the claim of a joint detection by more than an order of magnitude. With the prior established difficulty in associating kilonova to \ac{GW} detections, it seems performing follow-up searches of sub-threshold \ac{GW} signals is not a good use of observational resources. Then, for joint searches, the most promising prospect is the identification of a kilonova or afterglow by an optical (or other) survey in its normal operating mode which is then associated to a \ac{GW} or \ac{SGRB} trigger. Such joint searches should be developed and automated.

Because of the importance of this section we summarize the results:
\begin{itemize}
    \item Robust associations are necessary to enable multimessenger astronomy, and are not possible for all events.
    \item Spatial constraints from the discovery instruments are critical for robust statistical association.
    \item Temporal constraints for follow-up instruments are critical for robust statistical association. This can either be through a constraint on rise-time or previous non-detection from wide-field surveys.
    \item Follow-up observations of sub-threshold \ac{GW} signals is ill-advised, but automatically associating signals found in independent surveys should be done.
\end{itemize}

\subsection{Joint GW-GRB detection rates}
\label{sec:GW-GRB_rates}

Prior to \gwid it was considered somewhat unlikely, though possible, for a joint \ac{GW}-\ac{GRB} detection to occur with the Advanced network of interferometers. This belief was continued due to several misconceptions or misunderstandings. We briefly describe these and their resolution:

\begin{figure}
\centering
\includegraphics[width=0.6\textwidth]{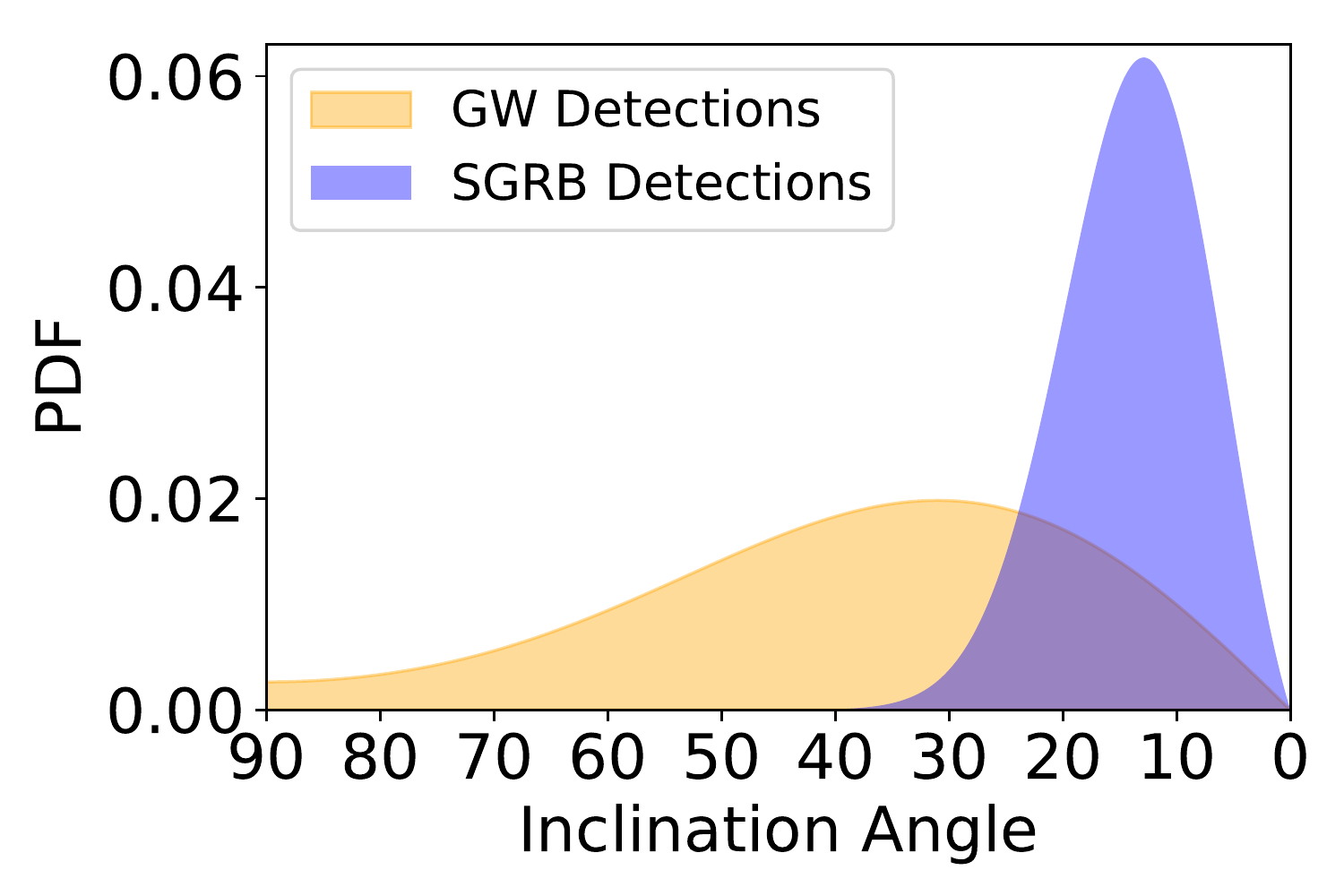}
\caption{The observed inclination angle distributions for \ac{NS} mergers detected through \acp{GW} and prompt \ac{SGRB} observations. The \ac{GW} solution comes from \citet{GW_patterns_Schutz_2011} and the \ac{SGRB} from slight modification (to handle solid angle) from observational results in \citet{fong2015decade}. We use the astrophysical convention of $0 \leq \iota \leq 90$, ignoring handedness relative to Earth. Against the rather naive assumption of a solid-angle distribution, roughly 1 in 8 \ac{GW}-detected \ac{NS} mergers that produce jets will have those jets oriented towards Earth.}
\label{fig:inclination_GW_GRB}
\end{figure}

\begin{figure}
\centering
\includegraphics[width=0.6\textwidth]{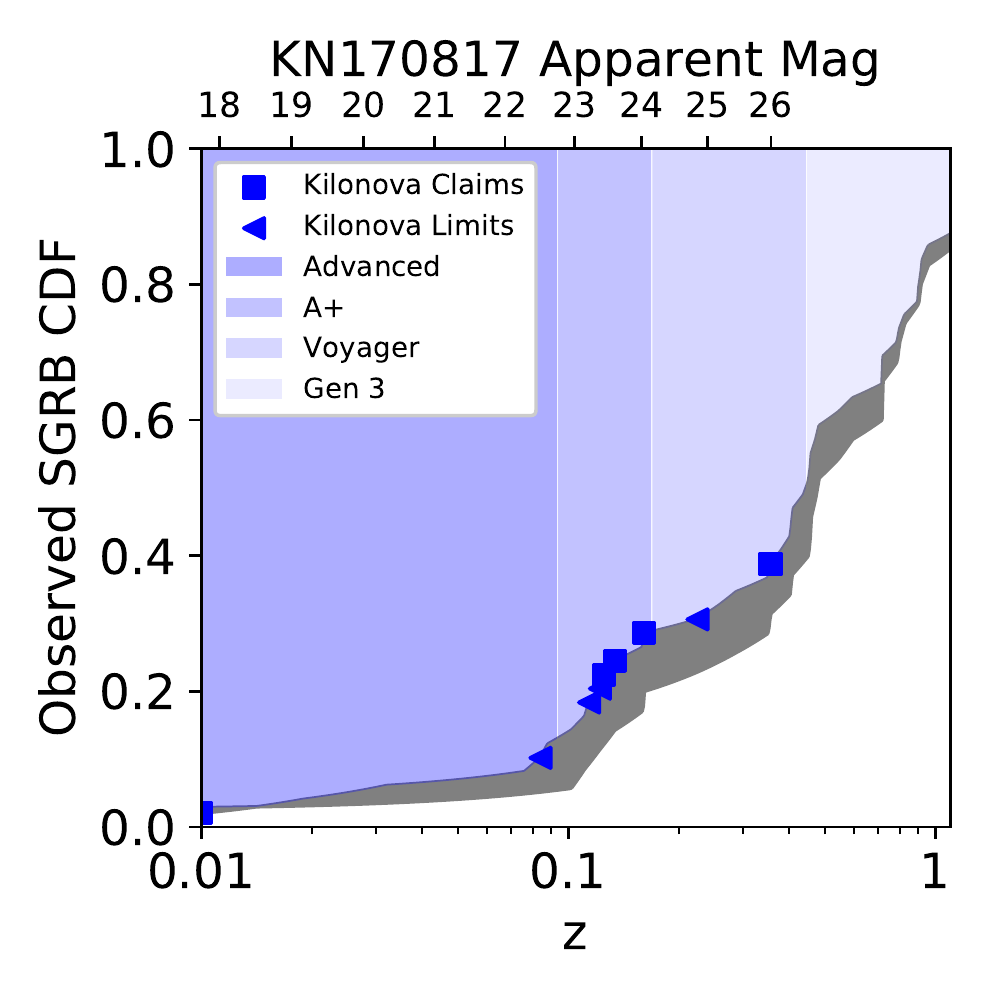}
\caption{Prior observations of \ac{NS} mergers. The grey shaded region is the cumulative redshift distribution observed for \acp{SGRB}, bounded by the pessimistic and optimistic samples from \citet{GW170817-GRB170817A}. Blue squares and triangles are the claimed kilonova and cases with constraining upper limits. The top axis marks the approximate \knid magnitude as a function of distance, based on an assumed 17.5 Mag (within half a Mag of most bands \citealt{villar2017combined}) and neglecting redshift effects. Overlaid are the joint \ac{GW}-\ac{GRB} detection horizons.
}
\label{fig:observed_distances}
\end{figure}

\begin{itemize}
    \item \textbf{Inclination Biases:} \acp{SGRB} have an observed half-jet opening angle distribution of $16^\circ \pm 10^\circ$ \citep{fong2015decade}, which does not include \grbid. Then, from solid angle effects only a few percent of successful \ac{SGRB} jets will be oriented towards Earth. Therefore, the assumption was that only a few percent of \ac{GW}-detected \ac{NS} mergers would have an associated \ac{SGRB} (or less, if not all \ac{NS} mergers produce successful jets).
    
    The emission of \acp{GW} is omnidirectional but not isotropic. It is strongest when the system is face on. Convolving this with solid angle gives an observed inclination angle probability distribution for \ac{GW}-detected \ac{NS} mergers of 

    \begin{equation}
        \label{eq:inclination_gw_obs}
        \rho_{\rm GW-detected}(\iota) = 0.002656\Big(1+6\cos^2(\iota)+\cos^4(\iota)\Big)^{3/2}\sin(\iota)
    \end{equation}
    
    \noindent \citet{GW_patterns_Schutz_2011}. Note that we have altered the distribution to be in terms of degrees (not radians) and removed directionality from $\iota$ (\ac{GW} measures of inclination go from 0 to 180 but \ac{EM} studies of \ac{NS} mergers generally only go to 90).
    
    \indent The effect of this is shown in Fig.~\ref{fig:inclination_GW_GRB}. The \ac{GW} distribution comes from Eq.~(\ref{eq:inclination_gw_obs}). The \acp{SGRB} distribution is a Gaussian convolved with solid angle that roughly recreates the observed distribution compiled in \citet{fong2015decade}, accounting for the intrinsic vs observed differences. The outcome is that roughly 1 in 8 \ac{GW}-detected \ac{NS} mergers that produce \acp{SGRB} will have Earth within the jet angle. 
    
    \item \textbf{The Minimum Luminosity of \acp{SGRB}:} Shifting a typical cosmological \ac{SGRB} with $L_{\rm iso} \approx 10^{52}$\,erg/s within the \ac{GW} detection volume would have an observed flux $\sim$10$^4$ times the typical value. Such a burst has not been observed in half a century of observations.
    
    \indent The implicit assumption is that \acp{SGRB} have a minimum luminosity, which was widely assumed \citep[see, e.g.,][references therein, and references to]{wanderman2015rate}. The was an implicit assumption that \acp{SGRB} arise from top-hat jets, where the jet has uniform properties within its cone, which largely explained observations until \grbid. Structured jets, where there is variation within the jet cone, have now been considered (see Section\,\ref{sec:grb_structure}). For these models the intrinsic luminosity function of \acp{SGRB} refers to the peak luminosity of the jet, generally corresponding to the face-on value. Then, for the same jet, the isotropic-equivalent luminosity as viewed from Earth depends on the inclination angle. Prior to \grbid, there were papers that avoided this implicit assumption, such as \citet{ghirlanda2016short} who predicted joint detect rates without requiring an imposed minimum luminosity.
    
    \indent \citet{evans2015optimization} was the first paper to consider that we may not identify nearby \acp{SGRB} based on flux measurements if they are \enquote{systematically less luminous than those detected to date}. \citet{SGRB_GBM_BAT_Burns_2016} investigated the observed brightness of \acp{SGRB} as a function of redshift and found no relation, empirically showing that we likely had not observed the bottom of the luminosity function, and suggested that subluminous \acp{SGRB} exist. From the knowledge gained from \grbid these subluminous bursts would arise from nearby off-axis events.
    
    \item \textbf{The limited \ac{GW} Detection Distance and the Redshift Distribution of \acp{SGRB}:} There were no known \acp{SGRB} within the Advanced interferometer design \ac{BNS} range of 200\,Mpc. Neglecting the full \ac{GW} network fails to account for the true spacetime volume observed, as shown in Figure \ref{fig:GW_recovery}. Joint \ac{GW}-\ac{GRB} detections will have a restricted inclination angle, giving a sky-averaged \ac{GW}-\ac{GRB} \ac{BNS} range 1.5 times greater than the \ac{GW}-only range. Further, with joint searches we can increase the detection distance by $\sim$25\% (Sect.~\ref{sec:association}). 
    
    \indent The \ac{GW}-\ac{GRB} detection distances and the observed redshift distribution of \acp{SGRB} are shown in Figure \ref{fig:observed_distances} with relevant information in Figure \ref{fig:GW_recovery}. This suggests a few percent of \acp{SGRB} are within the joint detection horizon, corresponding to a few events per year with current sensitivities.
    
\end{itemize}

Combining this information together, \citet{Burns_Dissertation}, published before GW170817, stated that we should expect joint detections with the Advanced network at design sensitivity, and potentially before. With \gwid and \grbid we confirmed that nearby bursts exist, that subluminous \acp{SGRB} exist, and that joint detections should be expected with existing instruments. As a result, in predicting future joint detection rates we use the same underlying principles.

Another issue, that remains unsolved and is not considered in the prior paragraph, is the fraction of observed \acp{SGRB} from \ac{NSBH} mergers. \ac{NSBH} mergers are heavier and can be detected in \acp{GW} roughly an order of magnitude greater volume (for those expected to produce \acp{SGRB}). That is, even a low fraction of detected \acp{SGRB} originating from \ac{NSBH} mergers would result in a sizable fraction of \ac{GW}-\ac{GRB} detections from \ac{NSBH} mergers (as compared to joint detections from \ac{BNS} mergers). The fractional contribution from each progenitor can then significantly alter the expected joint rates.

However, this requires a very important caveat. Since \gwid, several papers have been published that estimate future joint detection rates with the intrinsic \ac{BNS} merger rate, a half-jet opening angle (typically $\sim16^\circ$ from \citealt{fong2015decade}), that all \ac{BNS} mergers produce \acp{SGRB}, and a 100\% recovery efficiency for the \ac{EM} instrument. This last assumption is fundamentally flawed. As a sanity check, applying this calculation to \ac{GBM} vastly overestimates the expected joint detection rate by a factor of several. It is necessary to account for the low recovery fraction of weak \acp{SGRB} due to detection distances like \grbid.

\begin{table}
\caption{The key parameters for joint \ac{GW}-\ac{GRB} detections. We report the \ac{BNS} and \ac{NSBH} ranges and horizons, accounting for the stronger signal for nearly face-on signals (which \ac{SGRB} detections require) and 25\% gain in sensitivity from joint searches \citep[e.g.,][]{Joint-Williamson-2014}. As we are neglecting cosmology in these measures \cite[see][for a full discussion]{chen2017distance} we report lower limits for Voyager and Gen 3 interferometers. For joint detection rates we use \Fermi \ac{GBM} as the baseline, with the joint rates assuming on-board triggers. The last column doubles this value, which is a reasonable estimate for the full set of GRB monitors with the full GW network and the sensitivity gain for joint searches, as discussed in the text.}\label{tab:gw_grb_rates}
\centering
\begin{tabular}{|cc|cccc|}
\hline
 & & \multicolumn{4}{|c|}{GW Interferometer Generation} \\
 & & Advanced & A+ & Voyager & Gen 3 \\
\hline\hline
\multicolumn{2}{|c|}{BNS-GRB Joint Detection Distances} & & & & \\ \hline
Range & $D_L$ & 330 Mpc & 620 Mpc & $>$1.5 Gpc & $>$10 Gpc \\
& $z$ & 0.07    & 0.13    & $>$0.3  & $>$2 \\
Horizon & $D_L$ & 500 Mpc & 900 Mpc & $>$2 Gpc & $>$10 Gpc \\
& $z$ & 0.1     & 0.2     & $>$0.4 & $>$3 \\
\hline \hline
\multicolumn{2}{|c|}{NSBH-GRB Joint Detection Distances} & & & & \\ \hline
Range & $D_L$& 660 Mpc & $>$1 Gpc & $>$3 Gpc & $>$10 Gpc \\
 & $z$ & 0.13 & $>$0.2 & $>$0.5 & $>$3 \\
 Horizon & $D_L$& 1 Gpc & $>$1.5 Gpc & $>$5 Gpc & $>$10 Gpc \\
 & $z$ & 0.2 & $>$0.3 & $>$0.5 & $>$3 \\
 \hline\hline 
\multicolumn{2}{|c|}{GW-GRB Rates} & & & & \\ \hline
GBM+HL & yr$^{-1}$ & $>$0.4--2.2 & $>$1.0--4.8 & $>$4.8--9.6 & $>$Monthly \\
BAT+HL & yr$^{-1}$& $>$0.1--0.4 & $>$0.2--0.8 & $>$0.8--1.7 & $>$Quarterly \\
Realistic & yr$^{-1}$& $>$0.8--4.4 & $>$2.0--9.6 & $>$Monthly & $>$Weekly \\
\hline
\end{tabular}
\end{table}

For the joint rates estimates we use existing literature to determine a reasonable range of the fraction of \acp{SGRB} that will be detected by \ac{NS} mergers, which has the benefit of avoiding the uncertainty on the fraction of \ac{NS} mergers that produce \acp{SGRB}. These rates consider the detections of off-axis events, being built on literature that considers this either explicitly or implicitly. To start, we assume only a two-interferometer network with a 50\% network livetime (70\% each) and that all \acp{SGRB} originate from \ac{BNS} mergers. For the Advanced network at design sensitivity we assume that 0.8--4.5\% of \acp{SGRB} are detected in \acp{GW}. This is consistent with limits on the fraction of nearby \acp{SGRB} from comparing their localizations against galaxy catalogs \citep{mandhai2018rate} and on the inverse fraction of \acp{GW} detections with associated \ac{SGRB} detections \citep{song2019viewing,beniamini2019joint}. These values come from the methods described in \citet{GW170817-GRB170817A} and \citet{ligo2019search}, as well as the simulations from \citet{source_evo_Howell_2018} and \citet{mogushi2019jet}. For the A+ network we take 2--10\%, based on a $\sim$2.5x scaling relative to the Advanced network from \citet{source_evo_Howell_2018}. For Voyager we assume 10--20\% as a representative recovery fraction based on the observed \ac{SGRB} redshift distribution (Figure \ref{fig:observed_distances}). The Gen 3 interferometers have a joint \ac{BNS} range beyond the furthest \ac{SGRB} ever detected; therefore, we assume they recover all events when the network is live.

To calculate an absolute base rate we scale these fractions by the rate of \ac{GBM} on-board triggers.
We note that this is a particularly conservative estimate. It ignores single interferometer \ac{GW} triggers that are confirmed by an associated \ac{SGRB} trigger ($\sim$80\% increase for a two interferometer network), the effects of adding interferometers to the network ($\sim$2--3x for a five interferometer network, with slightly asymmetric sensitives, due to higher network livetime and more uniform coverage), the increase in recovered \acp{SGRB} (a factor of a few, see Sect.~\ref{sec:detect_prompt_grb}), and the contributions from the rest of the active \acp{GRB} monitors ($\sim$30-40\% more than the \ac{GBM} on-board trigger rate). These effects are not fully independent (e.g. a five interferometer network will have negligible single interferometer livetime). As a conservative estimate of the effects of these additional detections we provide the final column in Table~\ref{tab:gw_grb_rates}, which doubles the rate of \ac{GBM}+HL triggers. For Advanced \ac{LIGO} at design sensitivity we should expect a few joint detections per year. With A+ this should happen several times per year.  

We also provide an estimate for \Swift-\ac{BAT}+HL joint detection rates by scaling the \ac{GBM}+HL values. This is reasonable because they have similar detection thresholds. However, this is a lower limit. By reordering the observation list to bias the \ac{BAT} \ac{FOV} to overlap with the \ac{LIGO} sensitivity maximum  the joint detection rates can be increased by several tens of percent. Scaling to instruments with different sensitivities requires accounting for the bias of brighter events being more likely to occur in the nearby universe.


\subsection{Follow-up searches}
\label{sec:followup}
As of the time of this writing, no \ac{NS} merger has ever been discovered without a prompt \ac{SGRB} or \ac{GW} detection. This is not particularly surprising. Using optical as an example, only a few \ac{LGRB} afterglows have been detected without an associated prompt trigger. Detections of \acp{SGRB} are rarer than \acp{LGRB} and have systematically fainter afterglows. Similarly, there are thousands of known supernova identified through optical surveys but they are orders of magnitude brighter and more common than kilonovae.

As such, the dominant mode for finding \ac{SGRB} afterglows, kilonovae, and the other expected \ac{EM} transients from \ac{NS} mergers will be through follow-up observations of prompt \ac{SGRB} and \ac{GW} triggers. This is true at least until the era of \ac{LSST}. These follow-up observations can be performed in a few different ways. The most common method is through follow-up of \Swift-\ac{BAT} \acp{SGRB} with afterglow detections approximately every other month (generally detected by \ac{XRT}).



As previously discussed, \ac{GW} detections of \ac{NS} mergers provide localizations of tens to hundreds, and sometimes thousands, of square degrees. They also provide an estimate of the distance to the event, with typical uncertainty of tens of percent. These 3D localizations are distributed as HEALPix maps \citep{gorski2005healpix} through \ac{GCN}, with the distance reported as a function of position \citep{singer2016rapid}. These localization regions are massive, and difficult to follow-up with the vast majority of telescopes. However, for the initial \ac{GW} era detections will tend to be in the nearby universe ($\lesssim200$\,Mpc), where galaxy catalogs are reasonably complete. That is, narrow-field telescopes can prioritize the position of known galaxies within the \ac{GW}-identified search volume, a technique referred to as galaxy targeting \citep[e.g.,][]{kanner2012seeking,gehrels2016galaxy}.

The other solution to this problem is to build sensitive telescopes with a large \ac{FOV}. When a localization is reported these facilities tile the large error region and rapidly cover the observable containment region to a depth sufficient for a reasonable recovery fraction. This technique can also apply to \ac{GRB} localizations. Such optical facilities identify enormous numbers of transients that have to be down-selected to a small subset of events of interest. A great demonstration of this technique is the \ac{ZTF} follow-up of GW190425, covering $\sim$10\% of the sky on successive nights, in two bands, identifying more than 300,000 candidate transients, and quickly down-selecting to 15 events of interest \citep{coughlin2019growth}.



In estimating follow-up detection rates we should not expect to recover those events that occur near the Sun. The space-based observing constraint is within $\sim$45$^\circ$ of the Sun for many narrow-field space-based telescopes (e.g. \Swift, \textit{Hubble}, \textit{Chandra}). The ground-based limitation is generally a few hours of RA from the Sun, for a comparable exclusion zone size. An exception to this is for events detectable long enough for the Sun to move across the sky, requiring $\sim$months of detectability. We neglect this here, only considering events identified in the first $\sim$week. Either case rules out about 15\% of the sky. We may also not be able to recover \ac{SGRB} afterglow and kilonovae if they occur within about 5$^\circ$ of the galactic plane because of extinction and the insane rate of transients at lower energies. Therefore, follow-up observations could be capable of recovering up to 80\% of \ac{GW} or \ac{GRB} triggers.

We briefly remark on the possibility of separating afterglow and kilonova observations. \ac{SGRB} afterglow can be bright and dominate kilonova emission, or faint and undetectable below a given kilonova. From observations it appears afterglow will dominate in $\sim$25\% of cases \citep{gompertz2018diversity}. When they are of comparable strength, or the observations sufficient, the different spectral signatures and their temporal evolution of these events should enable disentanglement. Further, afterglow will tend to fade away long before the dominant emission of red kilonova.



\subsection{Gamma-ray burst afterglows}
\label{sec:GRB_afterglow}

\Swift identified the first \ac{SGRB} afterglow and has provided a sample\footnote{\url{https://swift.gsfc.nasa.gov/archive/grb_table/}} of about 100. These detections and broadband \ac{EM} observations from radio to GeV have shown afterglow is well described by synchrotron radiation. This radiation spans the \ac{EM} spectrum and is described as power laws with three breaks: the self-absorption break $\nu_a$, the minimum Lorentz factor break $\nu_m$, and the synchrotron cooling break $\nu_c$ \citep{sari1998synchrotron}. 

As summarized in \citet{review_SGRBs_Berger_2014}, broadband observations and closure relations enable determination of these break energies and their temporal evolution allow determination of several parameters. This includes the kinetic energy of the blastwave $E_k$, the half-jet opening angle $\theta_j$ (historically calculated assuming a top-hat jet), the density in the circumburst region $n$ (on $\sim$parsec scales), the power law index of the electron distribution in the jet, and a few microphysical parameters. In response to \grbid excluding the base top-hat jet models, closure relations for structured jet models have been derived \citep{ryan2019gamma}. Afterglow detection also enables arcsecond localizations and thus distance determination (see Sect.~\ref{sec:host_redshift_where}), which allows for the calculation of $E_{iso}$ and $L_{iso}$ of the prompt emission, and the half-jet opening angle allows for the jet-corrected values of these parameters and $E_k$.

\begin{figure}
\centering
\includegraphics[width=1\textwidth]{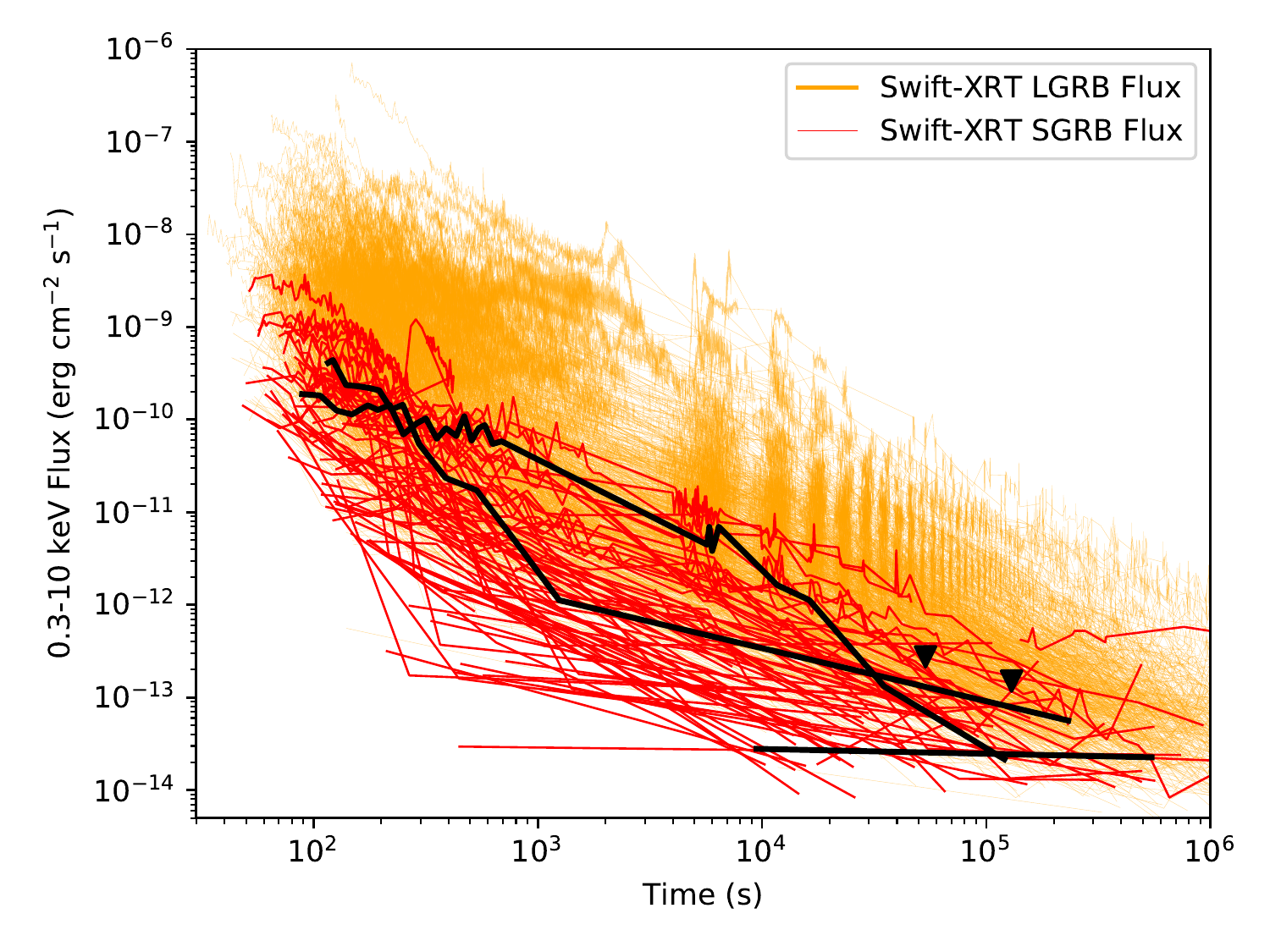}
\caption{The \Swift \ac{XRT} afterglow sample. \acp{LGRB} are orange and \ac{SGRB} in red, showing they are systematically dimmer by $\sim$1-2 orders of magnitudes. \ac{XRT} has an 85\% recovery fraction for \acp{SGRB} it observes in the first 100 seconds. The black markers are the nearest \acp{SGRB} with known redshift. The upper limits (triangles) are for GRBs 170817A and 150101B. The lines are for GRBs 061201, 080905A, and 100628A. Like the prompt \ac{SGRB} emission, they are not brighter at Earth than the full sample.
}
\label{fig:xray_afterglow}
\end{figure}

The rates of \ac{SGRB} afterglow detections is well understood for \Swift bursts. With the rate of \ac{SGRB} detections by \ac{BAT} and the fraction detected in \ac{XRT}, there are $\sim$6-7 X-ray detections of \acp{SGRB}/yr. The \ac{XRT} sample of \ac{GRB} afterglows is shown in Figure \ref{fig:xray_afterglow}. The recovery fraction at other wavelengths is poor. The summary in \citet{fong2015decade} covers observations of 103 \acp{SGRB}; X-rays have a 74\% recovery fraction, optical and \ac{NIR} 34\% and radio 7\%. Note that these pessimistic recovery fractions are for narrow-field telescopes, which are effectively always more sensitive than wide-field telescopes covering the same energy range.


The temporal decay of afterglow is steeper than the sensitivity gain most telescopes get for longer observation times. The faster an observation begins after event time the higher a likelihood of recovery, which was the main technical driver for \Swift. Alternatively, vastly more sensitive telescopes can be pointed at later times and still recover these signals, such as \textit{Chandra} detections days later.



Beyond the typical cosmological \ac{SGRB} afterglows, off-axis afterglows were thought to be promising \ac{EM} counterparts to \ac{GW} detections. From \citet{metzger_berger_2012}, and references therein, when top-hat jets interact with the surrounding material they slow and broaden. Over long enough timescales this emission can become observable to wider angles than the prompt \ac{SGRB} emission, but can still be bright enough to be detected from nearby events. \gwid and \grbid proved that afterglow can be detected significantly off-axis, but it also showed that off-axis afterglows may not be promising \ac{EM} counterparts unless the precise source localization is known through other means (i.e. identification of the kilonova). Fermi-\ac{GBM}, an all-sky monitor that is secondary on its own spacecraft, could detect \grbid nearly as far as the narrow-field X-ray Great Observatory \textit{Chandra}. Indeed without the kilonova determination of the source position the afterglow for \grbid event would not have been identified.



For the previously discussed reasons, searches for blind discovery of \ac{SGRB} afterglow using current wide-field monitors are unlikely to be successful. This is unlikely to change at least until \ac{LSST} operation. The most likely follow-up technique to succeed is then the galaxy targeting technique, as it enables follow-up with more sensitive telescopes; however, this is limited to well-localized and nearby events. The instrument most likely to identify a \ac{SGRB} afterglow following a \ac{GW} detection is the \Swift-\ac{XRT}, as it is the only fast response X-ray instrument. 

Estimating the number of \ac{SGRB} afterglow detections following \ac{NS} mergers is difficult because we do not understand their structure and therefore their brightness distribution. We will lose some events due to Sun constraints, transient contamination Milky Way, or relative sensitivity issues, which we estimate as $\sim$25\% based on the \Swift \ac{XRT} recovery fraction of \ac{BAT} bursts. However, we may also recover some events undetectable by \ac{GBM} due to Earth occultation or livetime considerations. These two effects are likely of similar order. Therefore, we roughly estimate the rates by assuming they have similar recovery fractions as the prompt \ac{GBM} on-board triggers.




\subsection{Kilonovae}
\label{sec:obs_kilonovae}


The first widely discussed claim of a kilonova detection came from follow-up observations of the \Swift \ac{SGRB} 130603B \citep{tanvir2013kilonova}. There are a handful of other claims of kilonova signals in follow-up of \Swift \acp{GRB}, \citep[e.g.,][]{perley2009grb,yang2015possible,jin2016macronova}. Inferred color and luminosity distributions for the claimed events are summarized in \citet{gompertz2018diversity} and \citet{ascenzi2019luminosity}. However, the only well studied kilonova is \knid. This event likely had a \ac{HMNS} remnant (see Sect.~\ref{sec:remnant_determination}), suggesting the brightness was near the middle of the possibilities (with \ac{SMNS} and Stable \ac{NS} being brighter and prompt collapse fainter). However, the early emission was on the bright end of expectations and the exact reason remains a matter of debate (see discussions and references in \citealt{arcavi2018first}, \citealt{metzger2018magnetar}, but see \citealt{kawaguchi_kilonova_remnants}).

If we assume that this unexpected bright behavior is due to our lack of understanding of these sources, rather than being a rare occurrence, we can use it as a representative kilonova, which we do in this paper. \citet{villar2017combined} compiled a large sample of the \ac{UVOIR} observations of \knid. At the distance of $\sim$40\,Mpc the \ac{UV} emission peaked at $\sim$19th Mag (thought it may have peaked before the first observations), blue bands at $\sim$18th Mag, with red and infrared approaching almost $\sim$17th Mag. With a limiting Mag of $\sim$26, within the reach of existing sensitive telescopes, around 30\%-40\% of \Swift \acp{SGRB} occur close enough for a \knid-like event to be detected and studied. The majority of \Swift \acp{SGRB} do not have follow-up at these sensitivities. This is in part because the primary goal of \Swift follow-up was afterglow studies, and \ac{SGRB} afterglow usually fade before the on-set of kilonova emission. With the devotion of sufficient observational resources $\sim$1-2 kilonova per year can be identified by following up \Swift \acp{SGRB}, though we note that many of the nearby bursts have claims of kilonova or interesting upper limits as shown in Figure\,\ref{fig:observed_distances}. 


\knid was independently identified in the follow-up of \gwid through both the wide-field tiling and galaxy targeting techniques \citep[e.g.,][]{gw170817_kilonova_first_swope,soares2017electromagnetic,valenti2017discovery,arcavi2017optical,gw170817_kilonova_tanvir,lipunov2017master}. Both methods will continue to be useful for future events, with the best technique depending on a given event. For events that are nearby (where galaxy catalogs are relatively complete) and reasonably well-localized galaxy targeting will be quite beneficial, with methods that account for galaxy incompleteness being particularly powerful \citep{evans2016swift}. For events that are nearby and poorly localized (e.g. several hundreds of square degrees or more), or events that are further away, the wide-field tiling technique will be dominant, provided the telescopes are sufficiently sensitive. There is no active wide-field \ac{UV} monitor. The band with the current best wide-field telescopes for identifying kilonova are in optical, where instruments like \ac{ZTF} \citep{bellm2014zwicky} can tile a large fraction of the sky to $\sim$21st-22nd Mag in one or two filters in a single night, as demonstrated by the (current) worse-case event \citep{coughlin2019growth}. However, even these depths may be insufficient to recover the majority of kilonova following \ac{GW} detections \citep{carracedo2020detectability}.


Reliably predicting the detection rates of kilonova in follow-up of \ac{GW}-detected \ac{NS} mergers may be a fools errand. The values depend on the volumetric rate of \ac{NS} mergers (each with more than an order of magnitude uncertainty), predictions on the sensitivity of the \ac{GW} network years in advance (that is an attempt to predict how some of the most sensitive machines ever built will change), the color and luminosity distribution of kilonova themselves (and how the intrinsic system parameters affect this, with only a single well-studied event to base our knowledge on), and would have to account for dozens of follow-up instruments scattered over the surface of Earth and teams with different observational strategies. 

Here we bound the rate. To calculate the number of kilonova detected through follow-up of \ac{GW} detected \ac{BNS} mergers we start with the rate of such events within 200\,Mpc. This is estimated using \knid as a baseline, with observations achieving a sensitivity of $\sim$21st Mag, we can recover \knid out to the Advanced design range of \ac{LIGO} and Virgo. At 22nd Mag this reaches to $\sim$250\,Mpc. This is roughly the sensitivity of \ac{ZTF} (depending on the observation time) which has a 47\,deg$^2$ \ac{FOV}, covers the g, r, and i filters (effectively, green, red, and infrared), and observes the northern sky. For our estimate of kilonova detection rates we assume that we can achieve \ac{ZTF}-like depths in the majority of optical filters over the observable night sky, which is a reasonable assumption given active and potential upcoming comparable facilities \citep[e.g.,][]{diehl2012dark,bloemen2015blackgem}. 

\begin{figure}
\centering
\includegraphics[width=1\textwidth]{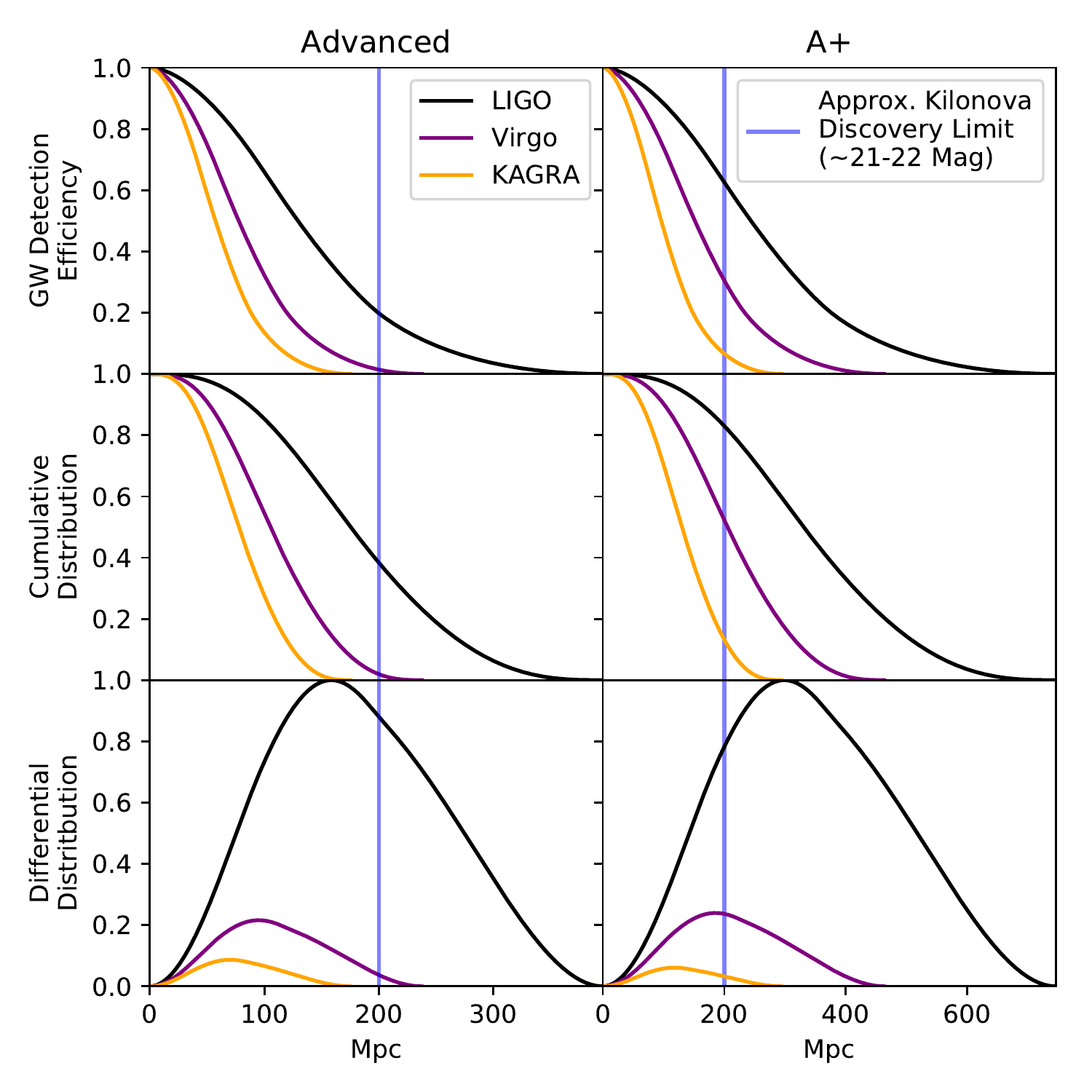}
\caption{The GW detection efficiency and distance distributions for \ac{GW}-detected \ac{NS} mergers by he Advanced and A+ networks. These are constructed with the projection parameter from \citet{finn1993observing}, as used in the literature \citep[e.g.,][]{source_evo_Howell_2018}, and the tables from \citep{dominik2015double}. The left panels are for the Advanced interferometer era and the right for the A+ era. The top panels shows the \ac{GW} detection efficiency for canonical \ac{BNS} mergers as a function of distance. The middle and lower panels scale this by the differential volume to show the cumulative and differential distance distributions for \ac{GW}-detected \ac{NS} mergers. The assumed distances for the different interferometers are the median value for the interferometers (Advanced: LIGO - 175\,Mpc, Virgo - 105\,Mpc, KAGRA - 77.5\,Mpc; A+: LIGO - 330\,Mpc, Virgo - 205\,Mpc, KAGRA - 130\,Mpc) from \citet{Gen_2_Prospects_2018}.}
\label{fig:GW_recovery}
\end{figure}

We do not attempt to estimate the gain from wide-field telescope sensitivity (e.g. \ac{LSST}) as the rate of optical transients becomes too great for this simple method to be accurate. Galaxy-targeting campaigns or smaller field of view telescopes that are more sensitive (e.g. DECam) generally require 3 or more interferometer localizations to succeed. This will not be common for events beyond 200\,Mpc in the Advanced era, but will be in the A+ era where our provided numbers are conservative. This is shown in Figure\,\ref{fig:GW_recovery}.

The first estimation is the \ac{GW}-recovery fraction of these events, which is shown in Figure\,\ref{fig:GW_recovery}. We multiply the GW detection efficiency with differential volume to determine the distance distribution for \ac{GW}-detected \ac{NS} mergers. From this, we can also calculate the recovery fraction of a network for \ac{BNS} mergers within 200\,Mpc, roughly corresponding to the discovery distance for kilonova until \ac{LSST}. This value is calculated by taking the time a network will spend with specific detector combinations and multiplying by the recovery fraction of the second-best live interferometer. We assume 70\% livetime for each individual interferometer and treat Virgo and KAGRA as roughly equivalent (taking the higher recovery fraction). For the Advanced era this suggests the network will recover $\sim$30\% of \ac{BNS} mergers within 200\,Mpc and about 75\% in the A+ HLVKI era. These assumptions neglect the fact that most detectors are not copointed, but this is somewhat counteracted by the additional sensitivity of three and four detector livetimes.

Multiplying this fraction by the 5\% and 95\% bounds on the local volumetric rate of \ac{BNS} mergers gives the expected rates as a function of distance. To estimate the rate of kilonova detections following \ac{GW} detections of \ac{BNS} mergers we account for the 20\% loss of events that occur close to the Sun or in the galactic plane, where follow-up observations are either impossible or likely to be too contaminated to reliably identify as discussed in Section\,\ref{sec:followup}. We calculate reasonable values for pessimistic and optimistic scenarios, as well as a mid-range estimate. For the representative estimate we assume 70\% will be like AT2017gfo or brighter (the remaining 30\% being assumed to be prompt collapse and too faint to detect), and for the high-end estimate we assume 100\% (assuming prompt collapse events are rare). These values come from \citet{margalit2019multi}, which is conservative compared to predicted remnant object fractions from other estimates \citet[e.g.,][]{lu2015millisecond}. 

For the low-end estimate we remove the assumption of kilonova brightness being predominantly determined by the progenitor, e.g. due to properties of the merger or inclination effects, which also removes the assumed mass distributions for \ac{BNS} mergers. \citet{gompertz2018diversity} investigate kilonova brightness based on \acp{SGRB} follow-up. They find that three kilonova candidates would be brighter than \knid and that four events have non-detections with upper limits sufficient to rule out a \knid-like event. We here assume 25\% of kilonova would be as bright as \knid, corresponding to 2 of the 3 candidates being real detections. We caution that this may still prove to be optimistic.

These calculations give a representative estimate of 5.6 \ac{GW}-kilonova detections per year with the Advanced network, with pessimistic and optimistic scenarios estimating between 0.4 and 24 per year. For the Advanced network this is 14/yr in the representative case, and between 1.0 and 60/yr in the other scenarios. For Voyager and Gen 3 we adopt a lower limit of detections of at least once a month, corresponding to the recovery of \ac{BNS} mergers (assuming the 95\% intrinsic lower limit) within 300 Mpc (with the previously mentioned losses). Should wide-field telescopes sufficiently advance in sensitivity, or should \ac{LSST} prioritize the follow-up of \ac{GW} detected \ac{NS} mergers, this rate could greatly increase. The rate of kilonova detected following-up \ac{NSBH} mergers is likely to be low in comparison, due to the generally greater distances and emission peaking in infrared (where wide-field telescopes are much less sensitive), though the intrinsic rates are broadly unknown.

These estimates neglect inclination effects on recovery fraction. As \knid was thought to be oriented for maximal brightness this may suggest the rates are somewhat optimistic. However, this is counteracted by the observed inclination distribution \ac{GW}-detected \ac{NS} mergers. This is discussed in Sections \ref{sec:GW-GRB_rates} and \ref{sec:detections_summary}.

Earlier detections are necessary for characterization of the kilonova and for robust statistical association to the \ac{GW} (or \ac{GRB} signal). The earliest light expected from these events is in \ac{UV}. The only active mission that does \ac{UV} discovery searches is \Swift, which relies on the galaxy targeted technique. Otherwise, observations in b and g filters within about a day (for blue kilonova), and r and i filters on timescales of a week (for red kilonova) are likely the discovery bands \citep{cowperthwaite2015comprehensive}. However, separation of kilonovae from other optical transients must rely on color information, and we likely need detection in multiple bands for discovery. Once the source position is known, either through identification of afterglow or kilonova, broadband study of the kilonova begins. Telescopes covering these wavelengths are abundant, which can make use of both follow-up techniques; however, \ac{NIR} wide-field telescopes are significantly less sensitive than optical ones. 

\ac{UVOIR} observations from the earliest detection until they fade from detectability (in each wavelength) enable us to infer properties of the ejected material. The ejecta mass, velocity, and opacity (or lanthanide fraction, depending on the formulation) can be determined from the broadband evolution of the quasi-thermal signature. This relies on an underlying assumed kilonova model. This is discussed in detail in Sect.~\ref{sec:elements}.

\subsection{Other signatures}
\label{sec:prospects_other_signatures}
\ac{GW} inspirals, prompt \acp{SGRB}, afterglow, and kilonova are the primary signals for detecting and characterizing these events. This section briefly summarizes several other possible signals expected on observational or theoretical grounds. Detecting any of these signatures would provide incredible insight into the physics of \ac{NS} mergers. The discussion here is limited to observational requirements with a base scientific motivation, with more detailed discussion in later sections.

\subsubsection{MeV neutrinos}
As discussed, \ac{BNS} mergers can have neutrino luminosities a few times greater than \ac{CCSNe}. The \ac{SNEWS} was developed to cross-correlate short-duration signal excesses from multiple $\sim$MeV neutrino telescopes to identify and localize nearby \ac{CCSNe} and alert the astronomical community before the first light (from shock break-out) is detectable \citep{antonioli2004snews}. It should also work for \ac{NS} mergers, where the very short intrinsic time offset from a \ac{GW} trigger can enable sensitive joint searches.

To discuss potential detection rates we focus on Hyper-Kamiokande, which is a 0.5 Megaton detector under construction in Japan \citep{mnu_hyperkamiokande_design_report_2018}. It follows the Nobel Prize winning detectors Kamiokande and Super-Kamiokande, will increase our neutrino detection rate of \ac{CCSNe} by an order of magnitude, and provides a potential path forward from the Standard Model. Unfortunately, it will probably not inform our understanding of \ac{NS} mergers as they can only be detected to $\sim$15\,Mpc. The closest \ac{BNS} merger every century should be roughly $13_{-4}^{+9}$\,Mpc, suggesting during a decade run of Hyper-Kamiokande there is a $\lesssim$10\% chance of detecting a \ac{BNS} merger.

\subsubsection{Other observed non-thermal signatures}
Observations of \acp{SGRB} have uncovered several additional non-thermal signatures. These signatures provide unique insight into these events, the possibilities and implications of which are discussed in Sect.~\ref{sec:other_high_energy}. The main peak in prompt emission is sometimes observed with preceding emission referred to as precursor activity and sometimes with extended emission that can last up to $\sim$100\,s. These are reliably identified with the prompt \ac{GRB} monitors. Gamma-ray precursors may require pre-trigger data with high temporal resolution (if the trigger is due to the main emission), and are generally expected to be softer, requiring energy coverage near $\sim$10-100\,keV. There may also be precursor emission at other energies. Clear identification of extended emission requires well-behaved backgrounds after trigger and generally emits at $\lesssim$100\,keV. 

\ac{SGRB} afterglow emission has large variation in addition to the base temporal decay. The \Swift-\ac{XRT} sample of \ac{SGRB} afterglows with X-ray flares and plateau activity in excess of the base temporal decay. These appear to be signatures of late-time energy injection into the jet. They require prompt X-ray observations, generally concluding within 10,000\,s of trigger time.


\subsubsection{High-energy neutrinos}
We may also expect high-energy ($\sim$TeV-EeV) neutrino emission from \ac{NS} mergers. The most sensitive instrument at these energies is the gigaton-class IceCube detector. The prompt and extended emission of \acp{SGRB} and the extra components seen in some \ac{SGRB} afterglow may produce significant amounts of neutrinos \citep[e.g.,][and references therein]{kimura2017high}. These signals are favorable for joint detections given the short time offset and rough localization capability of IceCube. Extended emission appears to be the most favorable signature, but only occurs for a fraction of \ac{NS} mergers. In light of the neutrino search around \gwid \citep{GW170817_neutrinos} approaching interesting limits and the relatively new consideration of the \ac{SGRB} jet interaction with polar kilonova ejecta, new theoretical studies have been performed that suggest we may be able to detect \acp{SGRB} in high energy neutrinos \citep{kimura2018transejecta}. This generally requires a \ac{GW}-\ac{GRB} event within $\sim$50\,Mpc and occurring in the northern hemisphere, where IceCube is far more sensitive. Such an event occurs about once per decade.

\citet{murase2009probing} opened the possibility of observing $\sim$EeV neutrinos over days to weeks after merger from proton acceleration by a new, long-lived \ac{NS} remnant with a high magnetic field, referred to as a magnetar. \citet{fang2017high} applied this to \ac{BNS} mergers and their model was tested in \citet{GW170817_neutrinos}, which suggests we are 2 orders of magnitude away from interesting limits. This high energy neutrino signature is unrelated to the prescence of a jet. The understanding gained through the multimessenger observations of \gwid have led to reevaluation of potential coincident detections \citep[e.g.,][]{kimura2017high} and additional mechanisms for high energy neutrino production, such as choked jet scenarios \citep[e.g.,][]{kimura2018transejecta}. Precise predictions of detection rates are difficult, but are generally expected to be rare.


\subsubsection{Very-high energy electromagnetic detections}
Gamma-rays refers to about half of the electromagnetic spectrum. The primary energy range of \acp{SGRB} ($\sim$keV-MeV energies) are soft gamma-rays. The mid-energy range is covered by the Fermi-\ac{LAT}. In its first decade of observation is has detected 186 \acp{GRB}, 155 of which are with its normal data ($\gtrsim$100\,MeV). The seed information for \ac{LAT} \ac{GRB} searches is usually \ac{GBM} triggers, with about 30\% of \ac{GBM} detections observed within the nominal \ac{LAT} \ac{FOV}, giving a \ac{LAT} recovery efficiency of $\sim$25\%. Of that 25\%, 30\% (2\%) is seen above 5\,GeV (50\,GeV) \citep{ajello2019decade}. Notably, of those with measured redshift 80\% (12\%) have source-frame photons above 5\,GeV (100\,GeV). These detections appear to be a mixture of prompt and afterglow emission, which can occur during the prompt phase even for \acp{SGRB}.

Beyond the reach of \Fermi are \ac{VHE} gamma-rays, roughly defined as $\gtrsim$100\,GeV, that are observed by ground-based facilities utilizing Cherenkov radiation. Detections at these energies are expected observationally from extrapolation of the \ac{LAT} power-law measurements and theoretically, e.g. from synchrotron self-Compton afterglow emission. There are two classes of \ac{VHE} telescopes. Water Cherenkov telescopes like High-Altitude Water Cherenkov Array \citep[\acs{HAWC}][]{HAWC_2} which observe a large fraction of the sky instantaneously (day or night). \acp{IACT} are pointed observations, though by most definitions they are wide-field telescopes ($\sim$few deg$^2$ \ac{FOV}) that are far more sensitive but can only observe at night. \acused{HAWC} 

The first report of a \ac{VHE} detection of a \ac{GRB} occurred earlier this year, with the \ac{MAGIC} detection of LGRB\,190114C \citep{MAGIC_190114C}. The \ac{LAT} observations of this burst are impressive, but within the observed distribution. This suggests that the \ac{MAGIC} observation resulted in detection because it was the first early \ac{VHE} observation of a very bright afterglow. It is sufficiently bright that it could have been detected by \ac{HAWC} in the sensitive region of its \ac{FOV}. There are also two reports from H.E.S.S. of \ac{VHE} detection of afterglow from the LGRBs 180720B and 190829A \citep{HESS_180720B,GCN_HESS_190829A}. This suggests a detection rate of a few \acp{LGRB} per decade with existing telescopes, which is consistent with extrapolation from the \ac{LAT} rates. 

To estimate the detection rate of \acp{SGRB} with \ac{VHE} telescopes we can scale the rate by the fraction of \acp{SGRB} to the total \acp{GRB} rate. The \ac{LGRB}-to-\ac{SGRB} ratio for \ac{GBM} is 4:1. The same ratio for the \ac{LAT} is 10:1. This is not surprising as a large portion of the \ac{LAT} detections are from only afterglow emission (which is fainter for \acp{SGRB}). Then, an optimistic \ac{VHE} detection rate of \ac{NS} mergers with existing instrumentation is $\sim$1/decade. The planned \ac{CTA} is an \ac{IACT} that is roughly an order of magnitude more sensitive than its predecessors. Then, we may expect a \ac{VHE} detection of a \ac{SGRB} every few years. However, we emphasize this is a very rough estimate.


\subsubsection{Neutron precursors to kilonova and additional energy injection}
Among the surprises of \knid that remains unsolved is the origin of the early bright \ac{UV}/blue emission. This topic is discussed in Sect.~\ref{sec:early_uv}. The possibilities range from the decay of free neutrons, shock-heated contributions from jet interactions with polar ejecta, additional heating supplied through a temporary magnetar, etc. In all cases these require observations in \ac{UV} and blue optical wavelengths as early as $\sim$1--2 hours after merger.

\subsubsection{Late-time radio emission}
The quasi-isotropic ejecta will emit late-time radio emission as it interacts with the circumburst material \citep{omni_radio_nakar_2011}. Their estimate of the detectability distances for a representative set of sensitive radio telescope reaches a few hundred Mpc. This signal should therefore be detectable, but we note the assumed densities are higher than most of the observed distribution following \acp{SGRB} \citep{fong2015decade}. We emphasize that this cannot be the only counterpart to a \ac{GW} detection for it to be reliably associated, given the massive delay time preventing robust association.

\subsubsection{Gamma-ray detections of prompt kilonova and kilonova remnants}
Kilonova are nuclear powered transients. Our observational understanding of the properties of the ejecta material comes from indirect, model-dependent inferences. We could directly measure the nuclear yield by detecting the nuclear gamma-rays that emit from $\sim$tens of keV to a few MeV with a flat spectrum across this range due to Doppler broadening of many lines \citep{kilonova_gamma_rays_hotokezaka_2016,korobkin2019gamma}. No existing telescope can detect this emission unless the event occurs within the local group. The current design of the most sensitive proposed instruments \citep[e.g.,][]{mcenery2019all} could detect these signals up to $\sim$15\,Mpc, comparable to the prospects for MeV neutrino detections of these events.


However, another option has recently been identified. Based on fiducial \ac{BNS} merger rates and kilonova ejecta properties both \citet{korobkin2019gamma} and \citet{wu2019finding} discuss the possibility of identifying \acp{KNR} in the Milky Way. They make different assumptions but come to the same conclusion that detecting kilonova remnants in the galaxy may be within reach with next-generation nuclear astrophysics missions. The use of these observations is discussed in Sect.~\ref{sec:heavy_element_milky_way}.

\citet{wu2019finding} also consider potential diffuse emission from ancient \ac{NS} mergers that have fully diffused with the Milky Way. The spatial distribution would likely differ from usual galactic distributions given the natal kicks to these systems, but detection prospects are hopeless for decades.


\subsection{Detections summary}
\label{sec:detections_summary}
\begin{figure}
\centering
\includegraphics[width=1\textwidth]{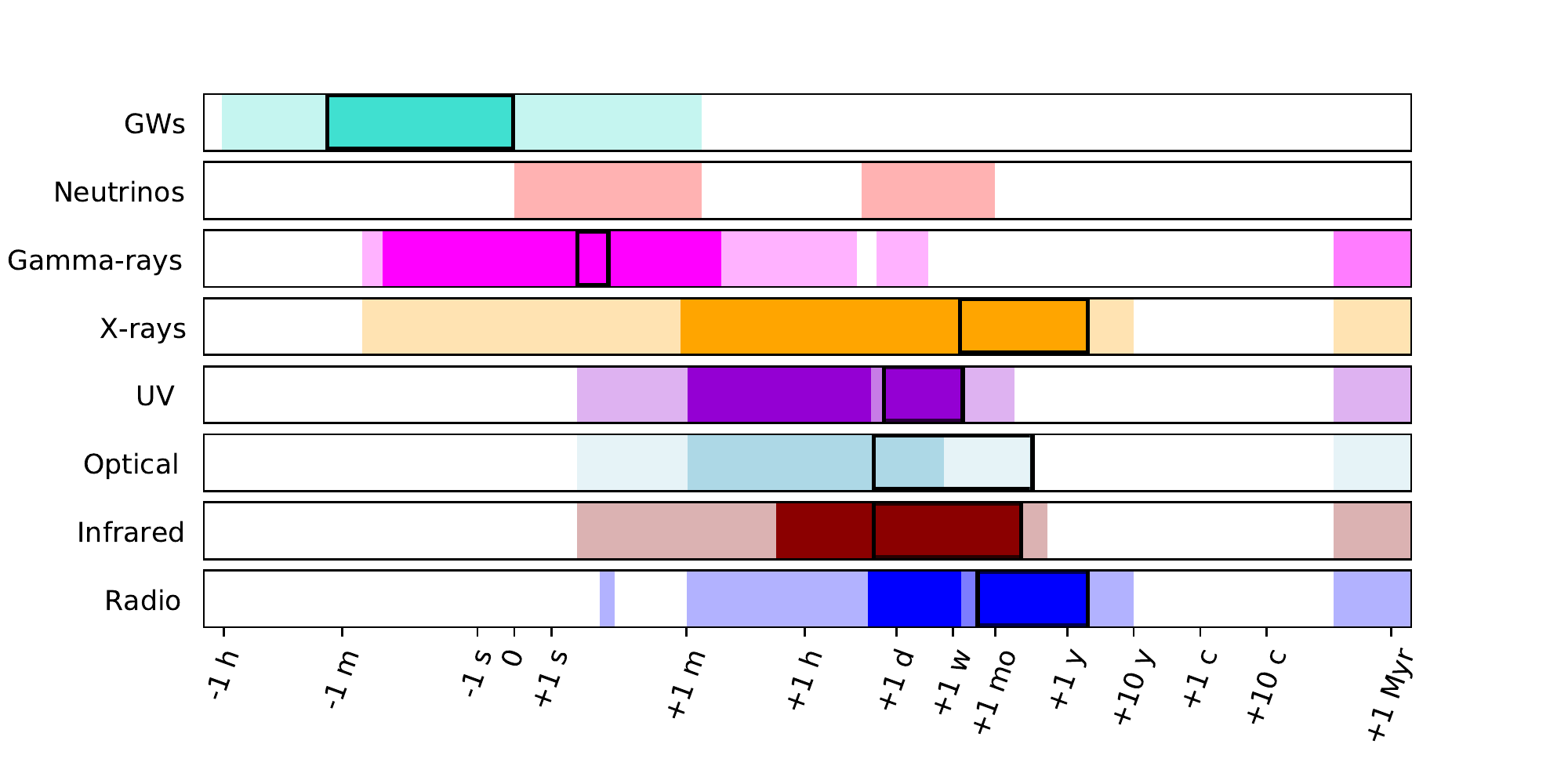}
\caption{The observing timescales when detectable emission is known or expected from \ac{NS} mergers. Because of our greater history (and therefore understanding) of \ac{EM} observations, we divide this messenger into bands. Intervals where signals were detected for \gwid are outlined with black boxes. The full color regions are for times with known observations of other \acp{SGRB}. The shaded regions cover times where we expect to detect signals in the future.}
\label{fig:observing_windows}
\end{figure}

Given the breadth of this total section we provide a short summary tying the observations together. \ac{NS} mergers may produce observable signatures in all astrophysical messengers across wide ranges in energy and time, as shown in Fig.~\ref{fig:observing_windows}. In Tables \ref{tab:total_rates_independent} and \ref{tab:total_rates_dependent} we summarize the rates results of this full section. See the text for a full understanding of the assumptions underlying each number.

We provide a short summary here for convenience. We assume a base intrinsic \ac{BNS} merger rate, neglecting any contribution from \ac{NSBH} mergers. This is used to calculate the \ac{GW} detections where each network assumes only two co-aligned interferometers (corresponding to the two US-based LIGO interferometers for the next several years). Advanced refers to the current design sensitivity, A+ is the funded upgrade, with Voyager and Gen 3 referring to the proposed future interferometers.

The prompt \ac{SGRB} and \ac{SGRB} afterglow rates are based on empirical observations. The joint rates assume a fixed fractional recovery of \acp{SGRB} by \ac{GW} interferometers of a given sensitivity. The \Swift \ac{BAT} joint detection rate comes from scaling the \Fermi-\ac{GBM} values by their relative \acp{SGRB} rates. Note that the GW and GW-GRB rates for \ac{GBM} and \ac{BAT} are for two interferometer GW networks and are therefore a lower bound. See Section\,\ref{sec:GW-GRB_rates} for a broader explanation.

The kilonova rates are very broad bounds, which account for a more complete \ac{GW} network than the \ac{GW} or \ac{GRB} rates shown in this table. The low end is bound by a base recovery fraction of the low end of the \ac{GW} detection rates and on the high end by assuming recovery of the majority of intrinsic event rates within a fixed distance. The detections of kilonova following \acp{SGRB} assume \knid-like events and the fraction of \acp{SGRB} with measured redshift from following within the maximum detection distance for an assumed sensitivity.

\begin{table}
\centering
\begin{tabular}{|cc||cccc|}
\hline
\multicolumn{2}{|c||}{Independent Discovery} & Year & & &\\
\hline
Signal & Instrument & $\sim$2020 & $\sim$2025 & $\sim$2030 & 2035+\\
\hline
GW - BNS	& HL	& 2-32	& 10-200	& $>$Daily	& $>$Hourly\\
GW - NSBH	& HL	& 0-50	& 0-300	& $>$1	& $>$100\\	
SGRB	& GBM &	48	& -	& -	& -\\		
	& BAT	& 8.3	& -	& -	& -	\\	
Afterglow	& Current	& $\sim$0	& -	& -	& -	\\	
	& LSST		& -	& ?	& ? & - \\ 		
Kilonova	& Current	& $\sim$0	& -	& -	& -		\\
	& LSST	& - 	& 5-250	& 5-250	& -	\\
\hline
\end{tabular}
\caption{A summary of the expected individual detection rates of \ac{NS} mergers in their canonical signals. The \ac{GW} rates account for only a two-interferometer network. Several assumptions go into these rates and we strongly caution that these are intended to be representative, not absolute. The details are described throughout this section. Dashes indicate times before/after where relevant instruments are funded. Question marks indicate where estimates are not well characterized.}
\label{tab:total_rates_independent}
\end{table}

\begin{table}
\centering
\begin{tabular}{|cc||cc|cc|}
\hline
\multicolumn{2}{|c||}{Joint} &  &  &	Advanced &	A+ \\	 \hline	
GW-SGRB	& GBM &	& & $>$0.4-2.2 &	$>$1.0-4.8  \\		
 &	BAT & & & 	$>$0.1-0.4 &	$>$0.2-0.8  \\	
 &	Current &	& & $>$0.8-4.4	 &$>$2.0-9.6	  \\
  \hline \hline
\multicolumn{2}{|c||}{Dependent (Follow-up)} & 	BAT & GBM &	Advanced &	A+ \\	 \hline	
Afterglow &	Chandra & $\sim$0 & & 	$>$0.8-4.4 &	$>$2.0-9.6 \\		
	 & XRT & $\sim$0  &	6.2 & 	$>$0.08-0.3 &	$>$0.2-0.6  \\
	 & Radio &$\sim$0 &	$\sim$0.6 &	$>$0.8-4.4 &	$>$2.0-9.6 \\
Kilonova &	~21-22 Mag & 0.2-0.7 &	1-4 &	1.0-21 &	2.7-57 	\\
 &	~24-26 Mag & 1-2	 &	5-10 &	  &	   \\
 \hline
\end{tabular}
    \caption{A summary of the expected joint detection rates of \ac{NS} mergers in their canonical signals. The \ac{GW}-\ac{GRB} and \ac{GW}-kilonova rates roughly account for the full \ac{GW} network. Several assumptions go into these rates and we strongly caution that these are intended to be representative, not absolute. The details are described throughout this section.}
\label{tab:total_rates_dependent}
\end{table}

In broad strokes, all the canonical signals from \ac{NS} mergers are brighter when observed from a polar position than an equatorial one. In Sect.~\ref{sec:GW-GRB_rates} we discuss the effects of inclination bias on joint \ac{GW}-\ac{GRB} detection rates, where \acp{SGRB} are only visible when Earth is within the jet and \acp{GW} are stronger along the total angular momentum axis. Observed kilonova brightness also depends on the inclination angle \citep[e.g.,][]{kasen2017origin}. If polar ejecta is faster moving than the equatorial ejecta then its brightness is fairly constant regardless of the observer angle. If it is slower then its emission is obscured when viewed from an equatorial region \citep[e.g.,][]{kawaguchi_kilonova_remnants}. Equatorial ejecta is brighter when viewed on-axis due to viewing a larger cross section. These conclusions hold for most putative signatures as well (e.g. MeV neutrinos from a thick disk). Overall this may be viewed as a beneficial selection effect for multimessenger astronomy and will result in a larger sample of particularly well characterized events, but will induce biases that must be handled carefully for some science (e.g. standard siren cosmology).

%% file: s03_astro.tex
From the observable parameters for individual events, we may make a number of additional inferences and draw new information from combined information. Sect.~\ref{sec:system_classification} discusses the observations that allow identification of \ac{NS} mergers and classification into \ac{BNS} and \ac{NSBH} mergers; and Sect.~\ref{sec:remnant_determination} discusses how to determine the immediate remnant object formed in \ac{BNS} mergers. The potential contribution to the origin of the observed time delay between the \ac{GW} and \ac{GRB} emission is discussed in Sect.~\ref{sec:GW-GRB_delay}. The origin of the early bright \ac{UV}/blue emission in \knid, and potential contributions to future events, is discussed in Sect.~\ref{sec:early_uv}. Lastly, how to determine where these events occur, both in spatial position and redshift, and the inferences this information allows with respect to formation channels, stellar formation and evolution, and redshift determination for individual events is discussed in Sect.~\ref{sec:host_redshift_where}.


\subsection{Progenitor classification and the existence of neutron star--black hole systems}
\label{sec:system_classification}
There is no known \ac{NSBH} system. These systems are thought to be formed through the same field binary formation channel as \ac{BNS} systems (which we know exist), where instead the primary remnant is either born a \ac{BH} or becomes one through accretion during the common envelope phase. Determining the astrophysical rates and intrinsic properties of these systems has important implications for the science that can be done with \ac{NSBH} mergers. 

As discussed in Sect.~\ref{sec:overview_merger} some \ac{NSBH} mergers are not expected to have \ac{EM} signals. Based on current population synthesis models for intrinsic system parameters, the inferred \ac{BH} spins from LIGO/Virgo observations, and our understanding of which systems will release \ac{NS} material to power the \ac{EM} transients it seems likely that \ac{EM}-dark \ac{NSBH} mergers exist and that \ac{EM}-bright mergers could exist \citep[e.g.][]{foucart2020brief}. Once we have observed them, they provide a separate handle on stellar evolution (Sect.~\ref{sec:host_redshift_where}), may enable a precise determination of \ac{NS} radius in a \ac{NS} merger (Section\,\ref{sec:mr_relation}), and may allow for some more stringent measures of fundamental physics (e.g. speed of gravity) with a given network sensitivity (Sect.~\ref{sec:fundamental_physics}). As they can be detected through \acp{GW} to greater distances and are phenomenologically different, they would require different \ac{EM} capabilities to understand.


Classifying events as \ac{BNS} or \ac{NSBH} mergers is critical to ensure pure samples and understanding how these events differ. \ac{GW} detections of \ac{CBC} provide information on the progenitor masses. Events with the primary constrained to be under the maximum mass of a \ac{NS} can be assumed to be \ac{BNS} systems. Events with the secondary constrained to be over this value can be classified as \ac{BBH} mergers. This value is currently not known (see Sect.~\ref{sec:NS_EOS}) but is almost certainly between 2-3$M_\odot$. Systems with one mass below this value and one above can be classified as \ac{NSBH} mergers. 

These classifications assume that there are no exotic stars in this mass range and that there is a clear separation between \ac{NS} and \ac{BH} masses. For low-mass systems we will tend to precisely measure the chirp mass but poorly measure the mass ratio (unless the event is particularly nearby/loud), so we may expect a significant fraction of events to have inferred individual mass posteriors that cross this boundary. This mass range is particularly difficult to precisely constrain for most events as was shown in \citet{mass_gap_littenberg_2015} who investigate the possibility of probing the existence of the first mass gap of compact objects, i.e. the lack of known \ac{NS} or \ac{BH} between $\sim$2 and $\sim$5 $M_\odot$. Assuming this gap exists would make \ac{GW} classification easier, but this is a strong assumption to make.

Further, the first \ac{GW} detections of \ac{NS} mergers require a higher standard of proof for strong classification claims. \ac{GW} observations can conclusively distinguish between progenitors by finding or ruling out matter effects on the inspiral, characterized by the tidal deformability parameter $\Lambda$. Constraining this value to be non-zero would exclude a \ac{BBH} merger and classify the event as a \ac{NS} merger. Determining between \ac{BNS} and \ac{NSBH} merger would then rely on the mass constraints of the primary.

The difficulty of \ac{GW} measurement of tidal deformability with the current high-frequency sensitivity is demonstrated with \gwid as despite being one of the loudest events detected thus far and utilizing the precise position from the kilonova detection the final \ac{LVC} results cannot rule out a \ac{BBH} merger origin from \ac{GW} observations alone \citep{abbott2020model}. In fact, the \ac{LVC} discovery paper for \gwid comments that the \ac{GW} observations alone do not classify the event as a \ac{BNS} merger, relying on the information provided by the \ac{EM} counterparts and to make the firm claim \citep{GW170817-GW}, which additionally relied upon the assumption that \acp{BH} do not exist in this mass range \citep{hinderer2019distinguishing,coughlin2019can}. For \ac{NSBH} mergers the inspiral can be dominated by the heavier \ac{BH} and appear similar to a \ac{BBH} merger \citep[e.g.][]{foucart2020brief}. \ac{GW}-only classification of these events will not be unambiguous for a large fraction of these events until they achieve sensitivity at higher frequencies.

In the O3 observing run \ac{LIGO} and Virgo reported the \ac{GW} trigger GW190814 \citep{abbott2020gw190814} which demonstrates many of these difficulties. The precisely determined secondary mass requires the object to be either the heaviest known \ac{NS} or the lightest known \ac{BH}, but it cannot be assigned to either class as the boundary between the two is unknown. The precise secondary mass measurement was enabled by a large mass asymmetry and the loud signal, but no evidence for matter effects was observed.



MeV neutrino observations provide another potential direct determinant of the presence of a \ac{NS}, or even determination of a \ac{BNS} progenitor if it observes the (meta-)stable \ac{NS} remnant, but these detections will be very rare for at least a decade (Section\,\ref{sec:prospects_other_signatures}).

Given these difficulties, multimessenger detections provide a solution. If there is an associated \ac{SGRB} we can immediately infer the presence of at least one \ac{NS}. If the inferred \ac{BH} mass is sufficiently heavy then the \ac{GW}-\ac{GRB} observations can classify the event as a \ac{NSBH} merger. Otherwise, they can only conclusively state the system is not a \ac{BBH} merger. This information may be useful in real-time to prioritize follow-up observations once we are in an era where \ac{GW}-detections of \ac{NS} mergers are a regular occurrence. There have been searches for quasi-periodic oscillations in prompt \ac{SGRB} emission \citep{dichiara2013search}, which may occur in \ac{NSBH} mergers if the spin-axis of the \ac{BH} was misaligned with the orbital angular momentum axis \citep{stone2013pulsations}. However, it is unknown if the accretion disk will align with the \ac{BH} equator and precession of the jet may or may not occur \citep{liska2017formation,liska2019bardeen} in \ac{NSBH} mergers. 

Kilonova observations will provide the strongest indirect evidence for system classification. The predictions for the inferred ejecta mass, average velocity, and electron fraction differs for \ac{NSBH} mergers and \ac{BNS} mergers. Delineation between the progenitors and remnants will have to rely on combinations of ejecta mass, velocities, kilonova color, and multimessenger determination of inclination \citep{barbieri2019filling,barbieri2020electromagnetic}. A self-consistent picture with \ac{GW}-determined masses \ac{SGRB} and kilonova observations will strengthen such claims.


\subsection{The immediate remnant object in binary neutron star mergers}
\label{sec:remnant_determination}

\begin{figure}
\centering
\includegraphics[width=1\textwidth]{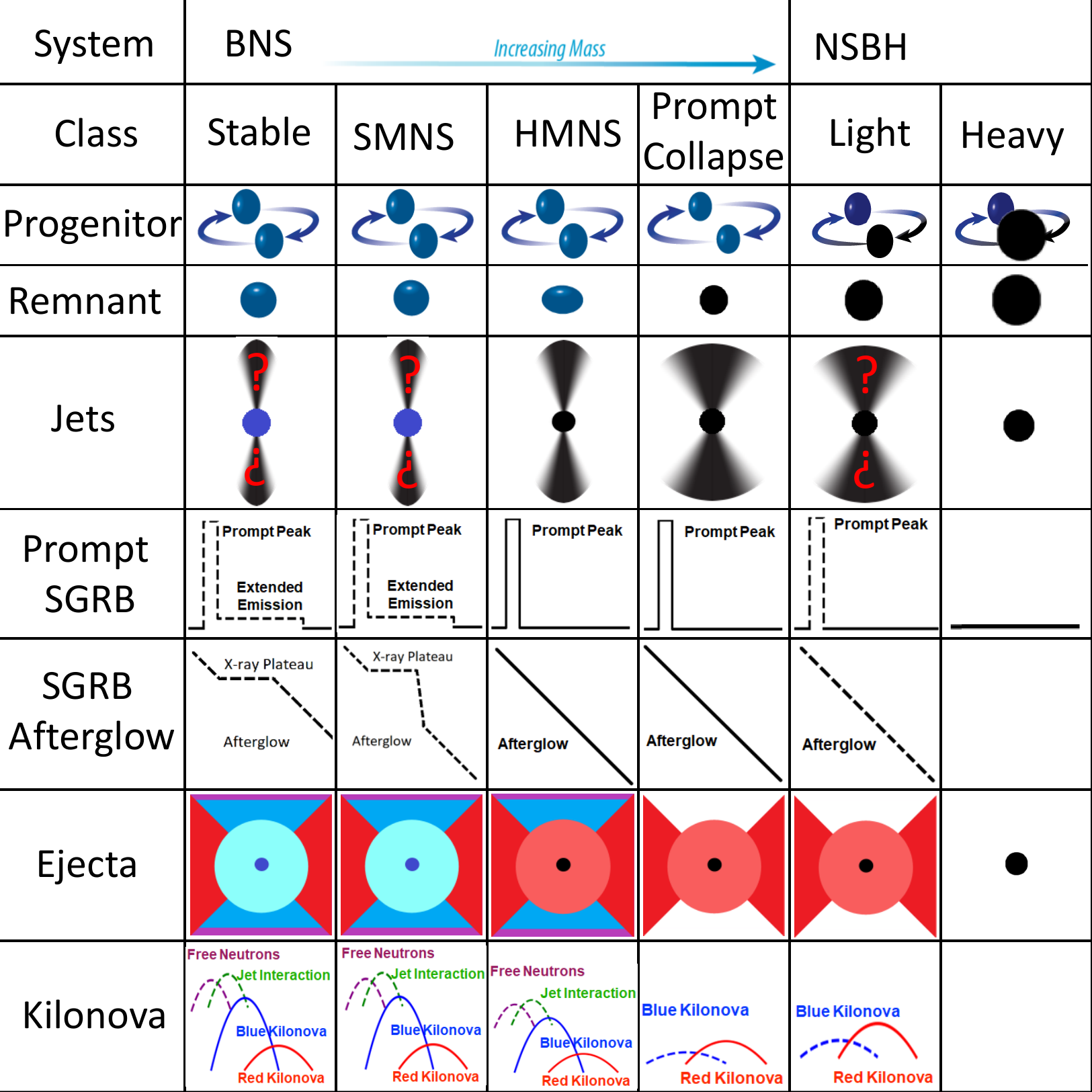}
\caption{The key expected signatures for the different classes of \ac{NS} mergers. Left to right corresponds to increasing mass: \ac{BNS} mergers classed into a Stable \ac{NS}, \ac{SMNS}, \ac{HMNS}, or prompt collapse scenarios, then \ac{EM}-bright \ac{NSBH} mergers and lastly \ac{EM}-dark \ac{NSBH} mergers. The differing prompt \ac{SGRB} and kilonovae signatures are shown for each scenario, providing a potential method to distinguish them. Dashed lines indicate the assignment of this signature to a specific scenario is not yet certain, or that the signature is theoretically expected but not yet confirmed observationally. The geometric representations are approximate and intended only as guidelines.
}
\label{fig:remnant_object}
\end{figure}

In \ac{NSBH} mergers the remnant object will always be a \ac{BH} because one already exists. In \ac{BNS} mergers we have the previously discussed (Sect.~\ref{sec:overview_merger}) four cases: Stable \ac{NS}, \ac{SMNS}, \ac{HMNS}, and prompt collapse to a \ac{BH}. Determining what mergers produce which immediate remnant objects is key to understanding \ac{NS} mergers themselves and informs on the \ac{NS} \ac{EOS} studies, our understanding of the central engines of ultrarelativistic jets, the heavy element yield distribution, and biases in standard siren cosmology. Figure \ref{fig:remnant_object} summarizes the expected differences, collating information from several sections (\ref{sec:overview_inspiral},
\ref{sec:overview_merger},
\ref{sec:overview_jets},
\ref{sec:overview_kilonova}) and is relied upon throughout the paper. While the text and figure represent generally robust expectations and are based on the current understanding of these cases, these will invariably be updated as future multimessenger detections occur and simulations improve. Some current limitations are discussed in Section\,\ref{sec:elements}.

Directly classifying remnants can likely only be done with \ac{GW} or neutrino signals. With neutrino observations we could infer a \ac{NS} remnant because the $\sim$MeV neutrino flux would be in excess of that from the accretion disk. EeV neutrinos should be emitted at late times around long-lived magnetars. Neutrino detections are unlikely to occur with upcoming neutrino telescopes. 

The \ac{GW} merger frequency and strain evolution could reliably differentiate between most of the four cases. For prompt collapse we expect \ac{BH} ringdown at $\sim$6-7\,kHz and an immediate drop in amplitude. For (meta)stable \ac{NS} remnants the merger would occur at $\sim$2-4\,kHz and significant \ac{GW} emission would remain after merger. In the \ac{HMNS} case this emission would cutoff in $\lesssim$1\,s as the object collapsed to a \ac{BH}. For the Stable \ac{NS} and \ac{SMNS} case the amplitude of the \ac{GW} emission would decrease as the remnant transitioned to the isotropic rotation phase where secular \acp{GW} may be released at twice the rotation frequency, which will slowly decrease with time. Distinguishing between Stable \ac{NS} and \ac{SMNS} classes with \ac{GW} observations is unlikely.

With the planned high frequency sensitivity for the Advanced interferometers it may be possible to detect \ac{GW} emission from a \ac{HMNS} remnant at 10-25\,Mpc, where 25\,Mpc is of order a once a decade event \citep{clark2014prospects}. A direct \ac{GW} determination requires improved \ac{GW} sensitivity up to at least $\sim$4\,kHz. The A+ upgrade (and similar upgrades) are currently not aiming to be sensitive beyond $\sim$1\,kHz. Therefore, we are unlikely to have direct determination of the immediate remnant object within a decade. Then indirect determination using \ac{EM} observations is the only viable option. Fortunately, there are expectations for significant \ac{EM} signal variation between remnant classes, guided by theory and simulation. 

Below we summarize how the kilonova, \acp{SGRB}, and other \ac{EM} signatures are expected to vary depending on the immediate merger remnant. Because these rely on model-dependent predictions on the behavior of matter in extreme regimes and the scientific results we wish to claim have incredibly important implications, we require a self-consistent understanding to emerge from these distinct predictions and the \ac{GW} determined masses. 

Kilonovae will be the most common \ac{EM} counterpart, and they should vary significantly between remnant classes. The understanding of the expected differences has come about over the past decade of improvements in simulation and theoretical understanding. The subject was broached with regards to disk winds in \citet{metzger2014red}, refined for general ejecta type in \citet{metzger2020kilonovae}, and well described in \citet{margalit2019multi} and \citet{kawaguchi_kilonova_remnants}. We summarize those arguments here and plot representative early spectra to show how this can be done. We emphasize that these are representative cases and variation on observed emission within a specific remnant class is expected to be significant depending on orientation effects, the mass, mass ratio, spins, etc.

However, the underlying differences are robust. In general, the longer the remnant \ac{NS} lives the more total ejecta will be unbound and it will be systematically bluer. Given enough time, the tidal tails become spiral arms that collide with the dominant \ac{NS} mass and are released. In the \ac{HMNS} case ejection at the shock-interface terminates during collapse, but it can continue in the lower-mass cases. The disk-wind ejecta increases with lifetime as a higher fraction of its total mass is unbound. The massive neutrino luminosities alter the electron fraction for much of this ejecta. These are all generic outcomes.

Second, it also will help resolve the origin of the early \ac{UV} emission. The origin of the early bright \ac{UV}/blue emission in \knid is generally debated. As discussed in detail in Sect.~\ref{sec:early_uv} the resolution to this question should not affect the relative differences between the cases discussed above, as the \ac{UV} brightness expected for different models generally scales with \ac{NS} remnant lifetime. One potential exception is if magnetars cannot power \acp{SGRB}, meaning we would only expect jet interactions in the \ac{HMNS} case.

\begin{figure}
\centering
\includegraphics[width=1\textwidth]{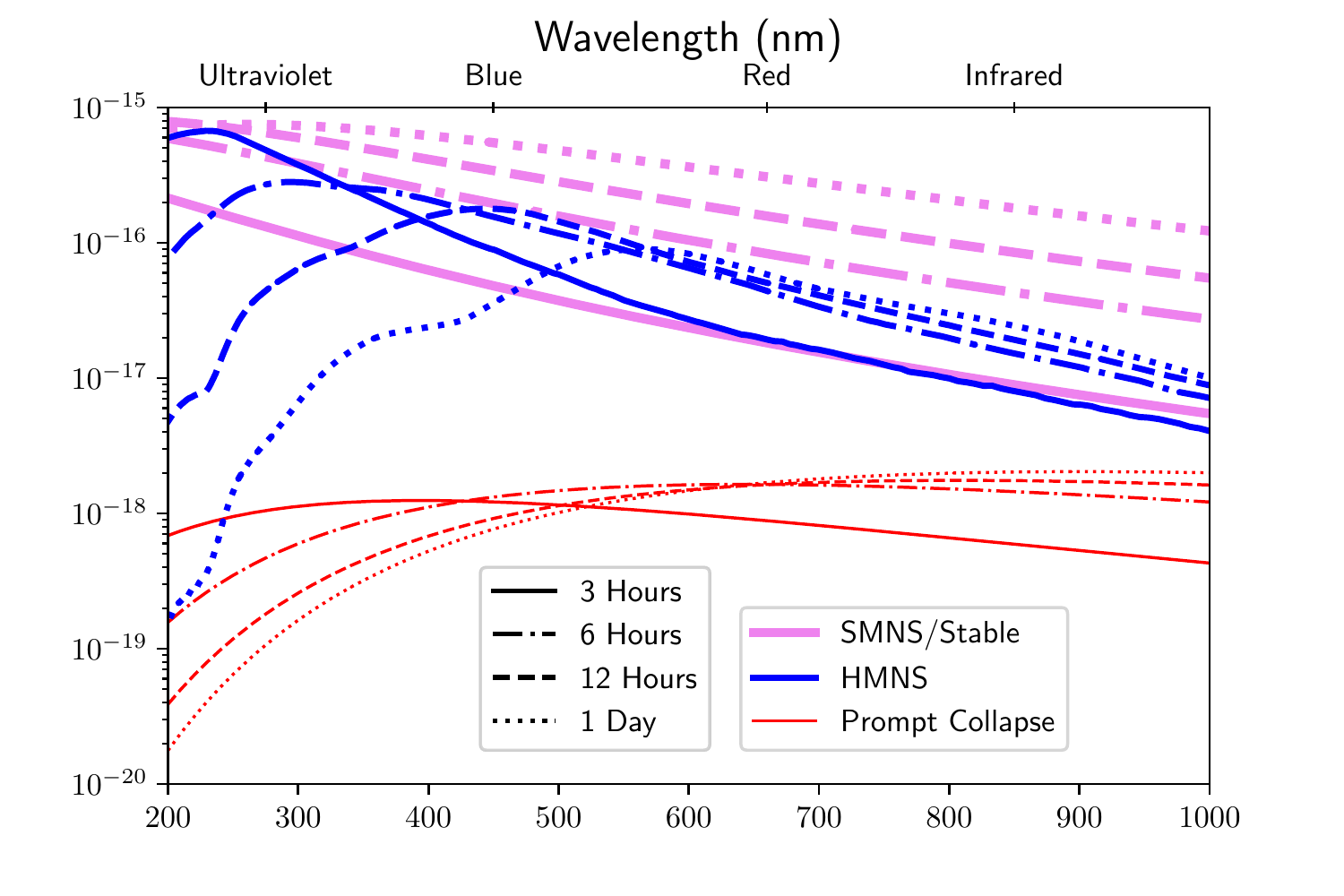}
\caption{Representative early spectra for the Stable \ac{NS} and \ac{SMNS}, \ac{HMNS}, and prompt collapse cases for events at 100\,Mpc. We here assume \knid originated from a \ac{HMNS} remnant and represent this case with the finely tuned model from \citet{kasen2017origin}. The spectra for the prompt collapse and Stable \ac{NS}/\ac{SMNS} cases are generated using the toy kilonova model described in \citet{metzger2020kilonovae}, using the code to generate the lightcurves in \citet{villar2017combined}, and were generated by P.~Cowperthwaite (private communication). The Stable \ac{NS}/\ac{SMNS} case was generated assuming ejecta mass with the properties 
$M_{ej}^{\rm Blue} = 0.1\,M_\odot$, $v_{ej}^{\rm Blue} = 0.3c$, $\kappa^{\rm Blue}=0.1 cm^2/g$ and
$M_{ej}^{\rm Red} = 0.005\,M_\odot$, $v_{ej}^{\rm Red} = 0.25c$, $\kappa^{\rm Red}=10 cm^2/g$. The prompt collapse case has 
$M_{ej}^{\rm Red} = 0.005\,M_\odot$, $v_{ej}^{\rm Red} = 0.25c$, $\kappa^{\rm Red}=10 cm^2/g$, which neglects a potential subdominant blue component.
}
\label{fig:remnant_spectra}
\end{figure}

\citet{kawaguchi_kilonova_remnants} focus on timescales between about a day and a week post-merger and conclude that the peak timescale and luminosity of the infrared emission may enable delineation between the remnant classes. In Figure\,\ref{fig:remnant_spectra} we show early spectra for the different cases using representative parameters from Sect.~\ref{sec:overview_merger}. The early \ac{UV}/blue emission should very easily distinguish prompt collapse from other scenarios for any observation in the first day or so. The fast evolution of the peak in the \ac{HMNS} case can be distinguished from the Stable \ac{NS}/\ac{SMNS} case, as the latter should brighten over time. This method is advantageous as the initial classification can be done relatively soon after merger, allowing for follow-up prioritization and more precise inferences based on the more complete dataset.

\ac{SGRB} observations will provide complementary information on the remnant object, and may provide a key signature to discern between a fully Stable \ac{NS} and a \ac{SMNS} remnant. It is debated if magnetars can power \acp{SGRB} (discussed in Sect.~\ref{sec:central_engines}). 

If magnetars cannot power ultrarelativistic outflows, we would only ever observe \ac{SGRB} emission from mergers that undergo prompt collapse or have a \ac{HMNS}, where in the latter case the jet will not launch until the \ac{NS} has collapsed. Stated another way, there should never be a \ac{SGRB} observed in a \ac{SMNS} or Stable \ac{NS} remnant case. We would also observe the non-thermal plateau emission in the \ac{HMNS} and prompt collapse cases, which would require work to identify its origin.

If magnetars can power ultrarelativistic jets then \ac{SGRB} observations still provide distinguishing characteristics. The non-thermal signatures of extended emission following the prompt peak and X-ray plateaus in the afterglow suggest late-time energy injection into the jet and led to the development of magnetar central engine theory (this is discussed in Sect.~\ref{sec:other_high_energy}). Then, we should expect to detect these signatures only in the Stable \ac{NS} and \ac{SMNS} remnant cases, and will not observe them in the \ac{HMNS} or prompt collapse cases. A sharp drop in X-ray flux at the end of the plateau is thought to occur when the \ac{NS} collapses, providing an observational signature between a Stable \ac{NS} and \ac{SMNS} remnant. The X-ray plateaus have been modeled by late-time fallback accretion, but we should be able to distinguish this from a magnetar central engine (see discussions and references in Sect.~\ref{sec:other_high_energy}).

The time delay from \ac{GW} to \ac{GRB} emission is another key piece of distinguishing information, which may provide another way to disentangle what occurs in these sources. From \citet{zhang2019delay}, the time delay could be up to $\sim$10\,s in cases with a magnetar central engine. This is roughly the timescale for the hot \ac{NS} to cool enough to stop driving baryons from the surface, enabling a clean enough environment for the jet to launch. In other cases the time delay should not exceed a few seconds.

In additional to the potential plateau signature, there are other methods of distinguishing between the Stable \ac{NS} and \ac{SMNS} cases. Long-lived remnants will result in a significantly brightened kilonova signature, which could distinguish the cases as Stable \ac{NS} remnants will be even brighter than \ac{SMNS} remnants \citep[e.g.][]{Bing_5,Bing_6,metzger2018magnetar}. There may be an increase in the radio emission from the quasi-isotropic outflow interactions with the circumburst material $\sim$years after merger \citep{Bing_6,fong2016radio}. There may be differences in the $\sim$EeV neutrino emission weeks after merger \citep[e.g.][]{Bing_7,murase2018double}.

To summarize, with current planned instruments a direct determination of the remnant object for all but the most fortuitous mergers is unlikely for a decade. Until then, we can rely on broadband \ac{EM} observations to characterize these events. Should a self-consistent picture emerge between the observed kilonova, \acp{GRB}, and other signature behavior with the inferred masses from \ac{GW} observations then we can reliably infer the remnant object outcome indirectly. As much of the science from \ac{NS} mergers relies on remnant object classification and it likely has significant effects on the observed signatures, determination of the merger remnant for a sample of events is a key goal of observations of \ac{NS} mergers.

\subsection{The time delay from merger to prompt gamma-ray burst emission}
\label{sec:GW-GRB_delay}

The total observed time offset for two astrophysical messengers is 

\begin{equation}\label{eq:generic_time_offset}
\Delta t_{\rm observed}=\Delta t_{\rm intrinsic}(1+z)+\Delta t_{\rm propagation},
\end{equation}

\noindent with $\Delta t_{\rm intrinsic}$ the intrinsic time delay which is affected by cosmological time dilation $(1+z)$ and $\Delta t_{\rm propagation}$ the induced arrival delays caused during propagation of the messengers from source to observation. Much of the science in this paper relies on $\Delta t_{\rm GRB-GW}$, the observed time offset from the coalescence time as measured by \ac{GW} measurements and the on-set of the prompt gamma-ray emission. Separating the individual contributions to this term could enable us to determine or better constrain the lifetimes of \acp{HMNS}, the speed of gravity, and the emission mechanism of \acp{SGRB}, to name a few.

We will show that possible propagation effects for \ac{GW} to \ac{SGRB} reduce to violations of fundamental physics. So far these all appear to be zero, which simplifies separation of the total individual terms. Should the propagation term be non-zero we can separate them from intrinsic delays as the cosmological redshift effects on the latter should be negligible for the foreseeable future. Alternatively, if they are hard to disentangle we may require future \ac{GW} interferometers to detect \ac{NS} mergers to distances where the redshift will become the dominant term. However, the relevant fundamental physics currently seems rather well supported. We discuss the intrinsic and propagation delay for \ac{GW} to \ac{SGRB} emission separately. We discuss the individual terms to show how we can distinguish the relative contributions of each term, or separately constrain their maximal effects. 

For \ac{BNS} mergers we can write (assuming the standard \ac{GRB} \ac{BH} central engine, relativistic jet, internal shock scenario):

\begin{equation}\label{eq:GW-GRB_time_offset}
\Delta t_{\rm intrinsic} = \Delta t_{\rm collapse} + \Delta t_{\rm formation} + \Delta t_{\rm breakout} + \Delta t_{\Gamma}.
\end{equation}

\noindent $\Delta t_{\rm collapse}$ is the time from coalescence to the formation of the \ac{BH}; $\Delta t_{\rm collapse}\approx0$ if the event undergoes prompt collapse, else $\Delta t_{\rm collapse} \lesssim 1$\,s in the \ac{HMNS} case. $\Delta t_{\rm formation}$ is the time until jet formation once the \ac{BH} has formed, which is expected to be $\lesssim$1\,s (limited by the cooling time in the neutrino powered jet scenario and the accretion timescale in the magnetically powered case; see Sect.~\ref{sec:jet_formation}). If there is previously ejected material in the polar region then the newly formed jet must breakout, where $\Delta t_{\rm breakout} \approx 1$\,s following known closure relations (from \citealt{cocoon_breakout_nakar_2012} as applied to \acp{SGRB} in \citealt{GW170817-GRB170817A}). Lastly, the jet must propagate outwards until the prompt \ac{SGRB} emission. The various \ac{SGRB} emission mechanisms (Sect.~\ref{sec:GRB_production}) usually require at least a few minutes of propagation, but with typical bulk Lorentz factors of $\Gamma \approx 100$ the jet effectively matches the speed of the \acp{GW} and the observed time delay is short (generally of order the duration of the burst).

Equation \ref{eq:GW-GRB_time_offset} can be modified for different \ac{NS} merger cases. Another formulation more useful for \ac{GRB}-specific studies is described in \citet{zhang2019delay}; we do not use this here for clarity with our immediate remnant object discussion. If magnetars can power \acp{SGRB} then $\Delta t_{\rm collapse}$ should be removed from the discussion. In cases with little polar ejecta (prompt collapse \ac{BNS} and \ac{NSBH}) $\Delta t_{\rm breakout}$ is negligible. These differences should enable us to disentangle the relative importance of these individual terms. For example, \ac{NSBH} mergers have $\Delta t_{\rm collapse} = \Delta t_{\rm breakout} = 0$. Should we observe a particularly short \ac{SGRB} then $\Delta t_{\Gamma}$ is small and $\Delta t_{\rm intrinsic} = \Delta t_{\rm formation}$.

One can write a very general equation to capture the total possible propagation effects:

\begin{equation}\label{eq:prop}
\Delta t_{\rm propagation} = \Delta t_{\Delta v} + \Delta t_{\rm LIV} + \Delta t_{\rm WEP} + \Delta t_{\rm massive} + \Delta t_{\rm dispersion} + \Delta t_{\rm deflection} + \Delta t_{\rm other},
\end{equation}

\noindent where each term captures induced relative delay during propagation by different effects: $\Delta t_{\Delta v}$ represents different intrinsic velocities, $\Delta t_{\rm LIV}$ \ac{LIV}, $\Delta t_{\rm WEP}$ relative Shapiro delay, $\Delta t_{\rm massive}$ capturing velocities of massive particles with a given energy according to \ac{SR}, $\Delta t_{\rm dispersion}$ for dispersion, $\Delta t_{\rm deflection}$ the delay induced for magnetic deflection of charged particles, and $\Delta t_{\rm other}$ represents other effects or the unknown. Note that some of these terms are subsets of the other; they are separated in this manner for pedagogical purposes, but see Sect.~\ref{sec:fundamental_physics} for a full explanation.

For \acp{GW} and \acp{SGRB} we can neglect several of these terms. That is, $\Delta t_{\rm deflection}=0$ because (inter)galactic plasma and magnetic fields do not affect $\sim$MeV gamma-rays nor \acp{GW}. The gamma-rays have $\Delta t_{\rm dispersion}=0$, but the \acp{GW} may not. We assume $\Delta t_{\rm other}=0$ for simplicity. This then leaves

\begin{equation}\label{eq:prop_GWGRB}
\Delta t_{\rm propagation} = \Delta t_{\Delta v} + \Delta t_{\rm LIV} + \Delta t_{\rm WEP} + \Delta t_{\rm massive} + \Delta t_{\rm dispersion}.
\end{equation}

\noindent These terms correspond to specific violations of fundamental physics. $\Delta t_{\Delta v}$ is the induced propagation delay for $v_{\rm GW} \neq c$. $\Delta t_{\rm LIV}$ is for different \ac{LIV} by gravity and light; $\Delta t_{\rm WEP}$ is the same except for the \ac{WEP}. $\Delta t_{\rm massive}$ is the delay induced for a graviton with non-zero mass; and $\Delta t_{\rm dispersion}$ capturing other potential forms of \ac{GW} dispersion. Each of these terms and the scientific importance of determining them is discussed in subsections in Sect.~\ref{sec:fundamental_physics}.

By convention, limits on individual fundamental physics terms are set by assuming the other contributions are 0. Should any of these terms be non-zero a sample of events will be required to determine relative contributions, which is possible because the separate possible propagation terms are most strongly dependent on different parameters.

The precision of the tests of fundamental physics (Sect.~\ref{sec:fundamental_physics}) that rely on the \ac{GW}-\ac{GRB} time offset is determined by how accurately we can model the intrinsic time offset for the event of interest, the redshift, and the observed time offset. We can remove the cosmological time dilation of $\Delta t_{\rm intrinsic}$ if we know z, or calculate z from a known distance and an assumed cosmology. Redshift is likely negligible during the Advanced interferometer era. For the A+ era it may begin to be important, and could become the dominant effect for third generation interferometers. 

\begin{figure}
\centering
\includegraphics[width=0.7\textwidth]{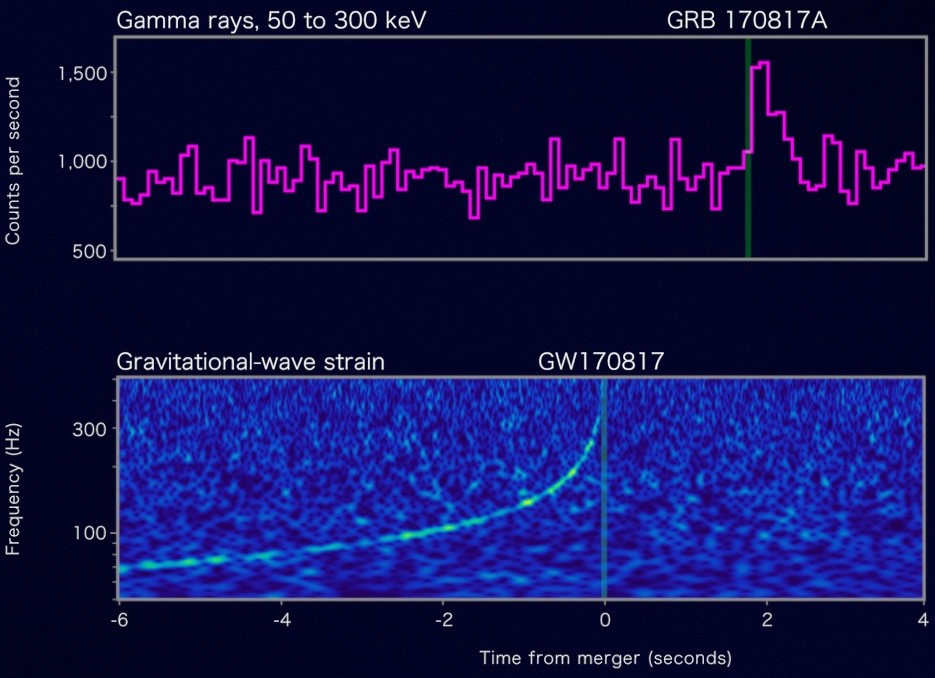}
\caption{The observed time delay from \gwid to \grbid. The top panel is the 50-300\,keV lightcurve from \Fermi \ac{GBM} and the bottom is a time-frequency map from combining the \ac{LIGO} observations. Figure is from NASA Goddard which is modified from Fig.~2 in \citet{GW170817-GRB170817A}.}
\label{fig:time_delay}
\end{figure}

The total intrinsic time delay is expected to be a few seconds \citep[e.g.,][]{zhang2019delay}, and potentially up to 10 seconds in extreme scenarios. Separately constraining the different contributions to the intrinsic time delay unveils great insight into these events \citep[e.g.,][]{li2016grb,GW170817-GRB170817A,zhang2019delay}. The more precisely we can determine the intrinsic time delay the greater our constraints on fundamental physics. Before \gwid the \ac{LVC} prior on this time offset was $[0, +4]$\,s with a 1 second addition on either side for safety (e.g., to account for light travel time from distance spacecraft or differing \ac{GRB} triggering methodologies; see \citealt{GW170817-GRB170817A} and references therein). The time offset from \gwid to \grbid fell right in the middle of this range, as shown in Figure \ref{fig:time_delay}.

When this redshift is accounted for, and allowing for two-sided constraints, we write $\Delta t_{\rm intrinsic,z}^{\pm} = \Delta t_{\rm intrinsic}^{\pm}(1+z)$. Throughout this paper we assume $\delta t_{\rm intrinsic,z} = 2$ s for individual events, giving 1\,s uncertainty for two-sided constraints. This is only twice the precision of the prior set by the \ac{LVC} based only on theory prior to \gwid \citep{lvc_gws_before_grbs}. This assumption also makes the results easily scalable, should this precision be unachievable for some events, given each side of a two-sided constraint are set to 1 s precision. This assumption is used in Sect.~\ref{sec:fundamental_physics}.

\subsection{The origin of early ultraviolet emission}
\label{sec:early_uv}

The observations of \knid were broadly consistent with a two-component kilonova: bright blue emission that peaks on the order of a day which fades to redder emission that peaks on the order of a week before fading out of detectability. It was brighter and bluer than expected, including a somewhat surprising \ac{UV} detection half a day post-merger \citep{gw170817_evans_swift}. The origin of this emission is debated. \citet{metzger2018magnetar} and \citet{arcavi2018first} discuss most of the theoretical explanations that have been invoked, their successes and limitations, and how future early \ac{UV}/blue observations can resolve this question. We summarize the options below but refer to these papers for more detail.

The most basic explanation is a kilonova origin for the emission \citep{villar2017combined}. This is potentially feasible but has some difficulties, which led to the discussion of other models \citep[e.g.][]{metzger2018magnetar,arcavi2018first}. Before proceeding we point out a peculiar outcome of these theoretical models: in all cases we may expect the brightest early \ac{UV}/blue emission to occur for \ac{BNS} mergers with Stable \ac{NS}, \ac{SMNS}, and \ac{HMNS} cases (in expected brightness listed in decreasing order) with little to no early bright \ac{UV}/blue emission from prompt collapse or \ac{NSBH} mergers. This may complicate delineation between these models, but has the benefit that the phenomenological description on indirectly differentiating between \ac{BNS} merger remnants (Sect.~\ref{sec:remnant_determination}) is unaffected by the true origin of the early bright \ac{UV}/blue emission in \knid.

Jet-interaction effects were invoked by a few teams \citep[e.g.][]{gw170817_evans_swift,kasliwal2017illuminating}. In these models the jet is launched after material already exists in the polar region. This can release a portion of very high velocity radioactive ejecta which allows the light to escape much earlier and can provide an additional source of energy through shock-heating. If magnetars can power \acp{SGRB} (Sect.~\ref{sec:central_engines}) then we expect jet interaction effects in the Stable \ac{NS}, \ac{SMNS}, and \ac{HMNS} cases with the amount of polar material related to the lifetime of the \ac{NS}. If magnetars cannot power ultrarelativistic outflows then we only expect jet interactions in the \ac{HMNS} case. The \ac{BNS} prompt collapse and \ac{NSBH} mergers should not have significant material in the polar regions at jet-launch time and should thus have much dimmer blue emission compared to the other cases. However, this model struggles to reproduce the observations of \knid as it would require jet kinetic energies beyond anything previously seen \citep{metzger2018magnetar} but may impart observable signatures in future events.

\citet{metzger2018magnetar} argue the emission can be explained by a neutrino-heated, magnetically accelerated wind from a short-lived magnetar, resulting in mildly-relativistic outflows. Again brightness scales with remnant \ac{NS} lifetime. Observing a more traditional \ac{SGRB} or directly inferring relativistic motion through radio interferometry observations or detection of high energy photons can clearly distinguish between these possibilities. 

Lastly, free neutrons decay as $n^0 \rightarrow p^+ + w^- \rightarrow p^+ + e^- + \bar{\nu}_e$ with a half-life of $\sim$10 minutes. If the fastest moving initial neutrons escape capture they lead the ejecta material, allowing the majority of their decay energy to escape while the photospheric temperatures are still high. This can provide a very blue signature that peaks on the timescale of $\sim$hours, before kilonova emission, with a comparable peak luminosity \citep{metzger2014neutron}. This model is applied to \knid in \citet{metzger2018magnetar}.

Both \citet{arcavi2018first} and \citet{metzger2018magnetar} point out that early \ac{UV} and optical observations should be able to distinguish between these potential contributions as their temporal and spectral evolution differ. \ac{GRB} observations will provide additional information on distinguishing jet interaction effects from the other models. Early \ac{UV} emission may arise from combinations of these potential contributions.

\subsection{Host galaxy, redshift, and where neutron star mergers occur}
\label{sec:host_redshift_where}

Understanding where these events occur determines their source evolution and thus the (volumetric and detection) rates of these events through cosmic time, inform on their formation channels, and provide information on stellar evolution through constraints on rare evolutionary pathways. The best current observational evidence to answer these questions come from observations of \ac{SGRB} afterglows, which provided the strongest evidence tying these events to \ac{NS} mergers prior to \grbid; for an overview see \citet{fong2015decade}. For a study on how \grbid compares to these observed distributions see \citet{fong2017electromagnetic}. For reviews on the formation channels of \ac{BNS} systems see \citet{review_binary_NS_lorimer_2008}, for reviews on compact object binaries see \citet{review_CO_formation_Kalogera_2007,review_CO_evolution_Postnov_2014}. The standard formation channel for \ac{NS} mergers is described in Sect.~\ref{sec:overview_formation}. See \citet{belczynski2002comprehensive} for a discussion on other possible formation channels.

There are two methods to determine the redshift of \acp{GRB}. The first is from direct measurement of redshift from the afterglow itself. This is common for \acp{LGRB} but has only occurred twice for \acp{SGRB}, due to the lower overall brightness. The second method is through statistical association to a host galaxy, and then determining the redshift of that galaxy \citep[e.g.][and references therein]{fong2015decade}.


The natal kicks during supernova explosion send a large fraction of \ac{NS} mergers outside of their host galaxies. More \acp{SGRB} are observed outside of the half-light radius of the inferred host galaxy than within, with typical physical offset $\sim$10\,kpc and the largest inferred of 75\,kpc \citep[see][and references therein]{SGRB_offset_measured_fong_2013}. The assignment of a host galaxy is relatively robust when it is within the half-light radius, and becomes more difficult as the offset increases. The assignment is probabilistic, counting the likelihood of a chance alignment of the source with likely host galaxies. Note that we observe the 2D projection of the 3D offset, e.g. for an event 10\,kpc from the host galaxy we can observe it anywhere from 10\,kpc offset to directly aligned, depending on our viewing geometry. There is no way to directly separate this effect.

\begin{figure}
\centering
\includegraphics[width=1\textwidth]{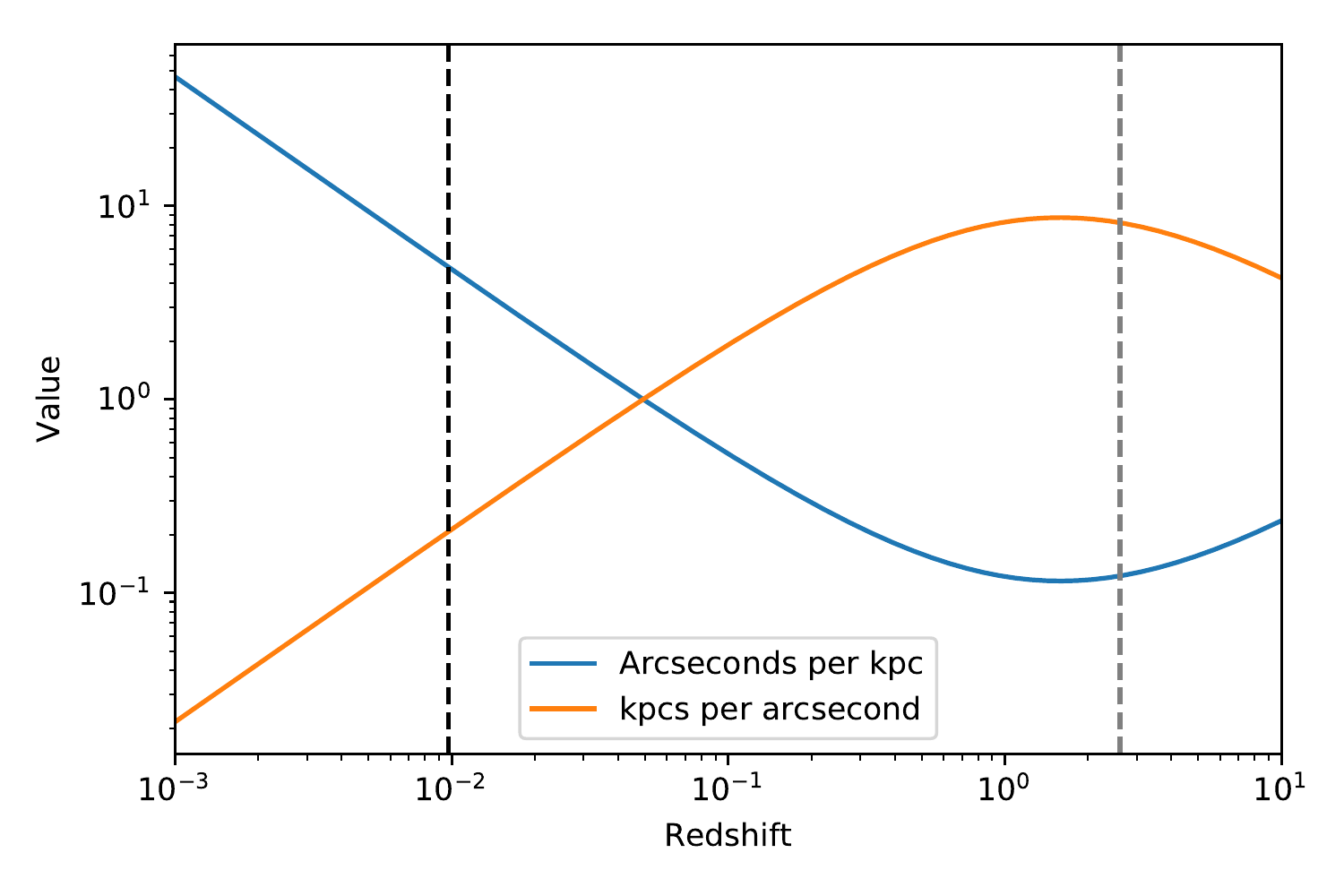}
\caption{Both arcseconds per kpc and kpc per arcsecond as a function of redshift. This figure shows how we require arcsecond precision for distant events to distinguish host from source, and we may fail to associate nearby events as the probability of this depends on the observed 2D offset for a fixed distance. The black dashed line is the distance to NGC 4993, the grey line is the distance to the furthest claimed redshift for a \acp{SGRB}.
}
\label{fig:offset}
\end{figure}

The assignment of a \ac{SGRB} (or kilonova) to a host galaxy requires localizations with $\sim$arcsecond accuracy to be robust. \Swift \ac{XRT} localizations are not sufficient, as the chance alignment of galaxies within typical error regions is non-negligible. There are some \acp{SGRB} where no robust host galaxy assignment can be done as there are no potential hosts very nearby in 2D angular offset ($\lesssim1'$), despite deep observational searches. These hostless \acp{SGRB} have bright galaxies somewhat nearby in 2D offset ($\sim$few arcminutes) in excess of random chance, suggesting at least some belong to these galaxies \citep{SGRB_hostless_offset_Tunnicliffe_2013}. This creates an observational bias against associating some particularly nearby \acp{SGRB} with their true host galaxy, which is shown in Figure \ref{fig:offset}. That is, for a fixed intrinsic offset the maximum observed 2D offset (the vector from host to source being perpendicular to that of host to Earth) can vary by more than an order of magnitude over the observed distance range for \acp{SGRB}. This directly corresponds to the host association probability. There is also the obvious bias of more difficult host galaxy detection for distant events.

The figure also demonstrates that the largest inferred intrinsic offset of 75\,kpc at the distance of \gwid would have a 6' offset from the host galaxy. With \ac{EM}-only observations we could not associate the source to host in this circumstance. Distance determination through \ac{GW} observations will alleviate these issues and will resolve some systematic problems with redshift determination of \acp{SGRB} (and \ac{NS} mergers) These observations require sensitive spectrometers, such as the X-shooter instrument on the VLT.

The cosmic rate evolution of \ac{NS} mergers is not well known. The peak cosmic star formation rate occurred at a redshift of $\sim$1.5-3.0 (e.g. \citealt{cosmic_SFR_Hopkins_2006}, or \citealt{cosmic_SFR_Madau_2014} for a review). We expect the peak rate of \ac{NS} mergers to track the peak star formation rate modulo the average inspiral time (the lifetimes of massive stars that result in \acp{CO} are negligible). \acp{SGRB} will provide the only constraints on the source evolution of \ac{NS} mergers for at least the next decade. The observed median redshift for \acp{SGRB} is inferred to be $<$z$>_{SGRB}\approx$0.5-0.8 \citep{review_SGRBs_Berger_2014}, which corresponds to an average inspiral time of $\lesssim$5 Gyr; this is a lower limit on the average redshift due to the Malmquist bias and the detection threshold of \ac{BAT}.

Population synthesis studies are being provided with much improved data from \ac{EM} observations to test data against \citep[e.g.][]{gaia_overview_2018,bellm2014zwicky,lsst_drivers_Ivezic_2008}. This will be commensurate with \ac{GW} observations from \ac{LIGO}/Virgo and in the future by additional ground-based interferometers and \ac{LISA}. \ac{GW} and \ac{EM} detections of \ac{BNS} and \ac{NSBH} mergers will provide some unique information \citep[e.g.][]{population_synthesis_belczynski_2008} to the overall understanding of how stars form and evolve \citep[e.g.][]{GW170817_LVC_progenitor}. The formation channels can be tested from these studies and by greater understanding of the source evolution of these events. However, results on the inspiral time distribution may require $\sim$100 nearby events \citep{safarzadeh2019measuring}. A detailed description of the input physics and how they are constrained with \ac{GW} observations of \ac{BNS}, \ac{NSBH}, and \ac{NSBH} mergers is given in \citet{kurckow2018}. 

One example is the primary mass gap, i.e. the idea that there is a gap between the heaviest \acp{NS} and the lightest \acp{BH} \citet{belczynski2012missing}. This is borne from the lack of observed compact objects between 2 and 5M$_\odot$. With the \ac{NS} merger determination of the maximum mass of a \ac{NS} (Sect.~\ref{sec:max_NS_mass}) we can set a strict boundary threshold. Then, if the mass gap does not exist we would expect to eventually detect a loud \ac{CBC} merger with a \ac{BH} posterior contained within the putative mass gap. If it does exist then this would never occur. This would have a number of important implications from the \ac{CCSNe} mechanism \citet{belczynski2012missing}, that implication on stellar evolution, and potentially robust \ac{CBC} classification from \ac{GW}-only observations.

This understanding has implications for future detection rates of these events. The \acp{SGRB} detection rate is empirically determined and unaffected, but the \ac{GW} detection rates for Voyager or third generation ground-based interferometers are altered significantly by the evolution of the rates of these events. Further, this has implications for other outstanding questions. Perhaps the best example is the origin of heavy elements, which depends on the source evolution as the modern abundances are determined by the time-integrated history of their creation rate (Sect.~\ref{sec:elements}).

%% file: s04_jets.tex
During the cold war the Vela satellites were launched to monitor Earth for gamma-ray signatures of nuclear detonations in the atmosphere to enforce the Partial Test Ban treaty between the United States and the Soviet Union. The detection of \acp{GRB} in 1967 were initially slightly concerning, before timing annuli placed their origin as outside the solar system, enabling their declassification \citep{GRBs_astrophysical_origin_1973}. This was the beginning of the study of \acp{GRB}, and therefore the beginnings of our observational study of \ac{NS} mergers.

The first three decades of \ac{GRB} study were limited to observations of the prompt phase. The high peak energies, and short duration tended to suggest very energetic phenomena as their source \citep[e.g.,][]{strong1974preliminary,mazets1981catalog}. A cosmological origin for these events would require energetics well beyond anything previously known which strongly suggested a galactic origin. In Euclidean space the observed flux distribution from sources with a homogeneous distribution is $P^{-3/2}$. Deviation from this power law would require a source distribution where space is non-Euclidean, i.e., to cosmological distances \citep{meegan1992spatial,mao1992cosmological}. Sources with a galactic origin have an anisotropic source distribution concentrated in the galactic plane. Data favored an isotropic, inhomogeneous distribution requiring a cosmological origin, with the \ac{BATSE} on-board \ac{CGRO} the first to hit discovery significance \citep{briggs1995batse}. 

Another key result from this era was the discovery of two classes, short and long, separated by their prompt duration and spectral hardness \citep{Dezalay1991,Kouveliotou1993}. This separate has been confirmed by broadband afterglow studies. \acp{LGRB} arise from host galaxies, and regions within those hosts, with high rates of star formation \citep[e.g.][]{fruchter1999hubble,le2003hosts,christensen2004uv,fruchter2006long}. Nearby \acp{LGRB} usually are followed by a \ac{CCSNe} \citep{galama1998unusual,cano2017observer}, giving direct evidence that these events are powered by a subset of \ac{CCSNe} referred to as collapsars \citep{woosley2006supernova}. \acp{SGRB} track older host environments \citep[e.g.,][]{2010ApJ...725.1202L,2013ApJ...769...56F}, occur outside of their host galaxies \citep[e.g.,][]{church2011implications,fong2013locations}, and varied properties of those hosts \citep[e.g.][]{gehrels2005short,fong2013demographics} all matched expectations from \ac{NS} mergers \citep[e.g.,][]{belczynski2006study}. As previously mentioned, the association of \gwid and \grbid directly confirmed that at least some \acp{SGRB} arise from \ac{BNS} mergers. For a review on these properties see \citet{review_SGRBs_Berger_2014}. For a summary of the multiwavelength studies from the first decade of the \Swift mission, which largely confirmed these predictions, see \citep{fong2015decade}.

\begin{figure}
  \centering
  \includegraphics[width=0.5\textwidth]{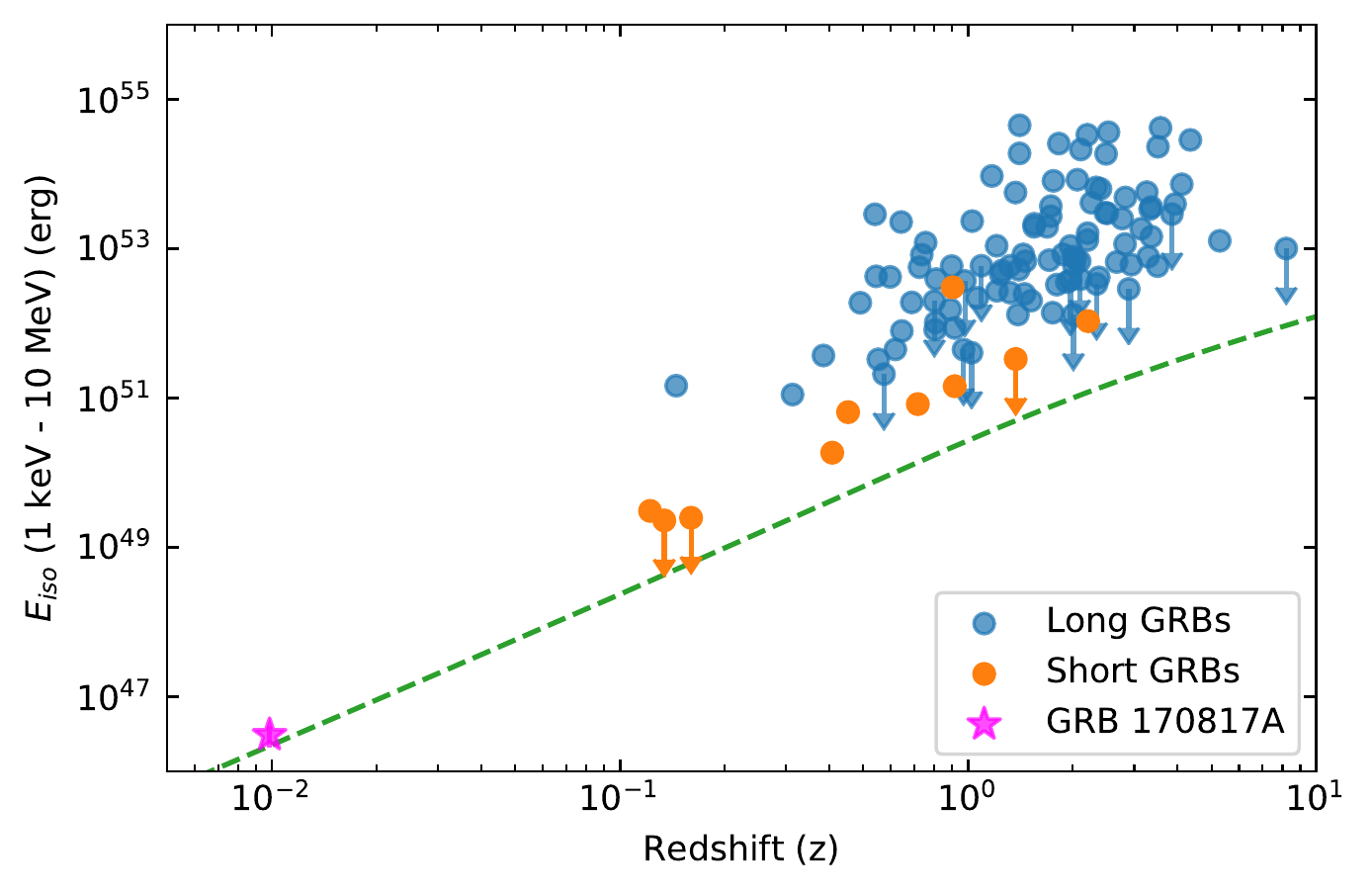}\hfill
  \includegraphics[width=0.5\textwidth]{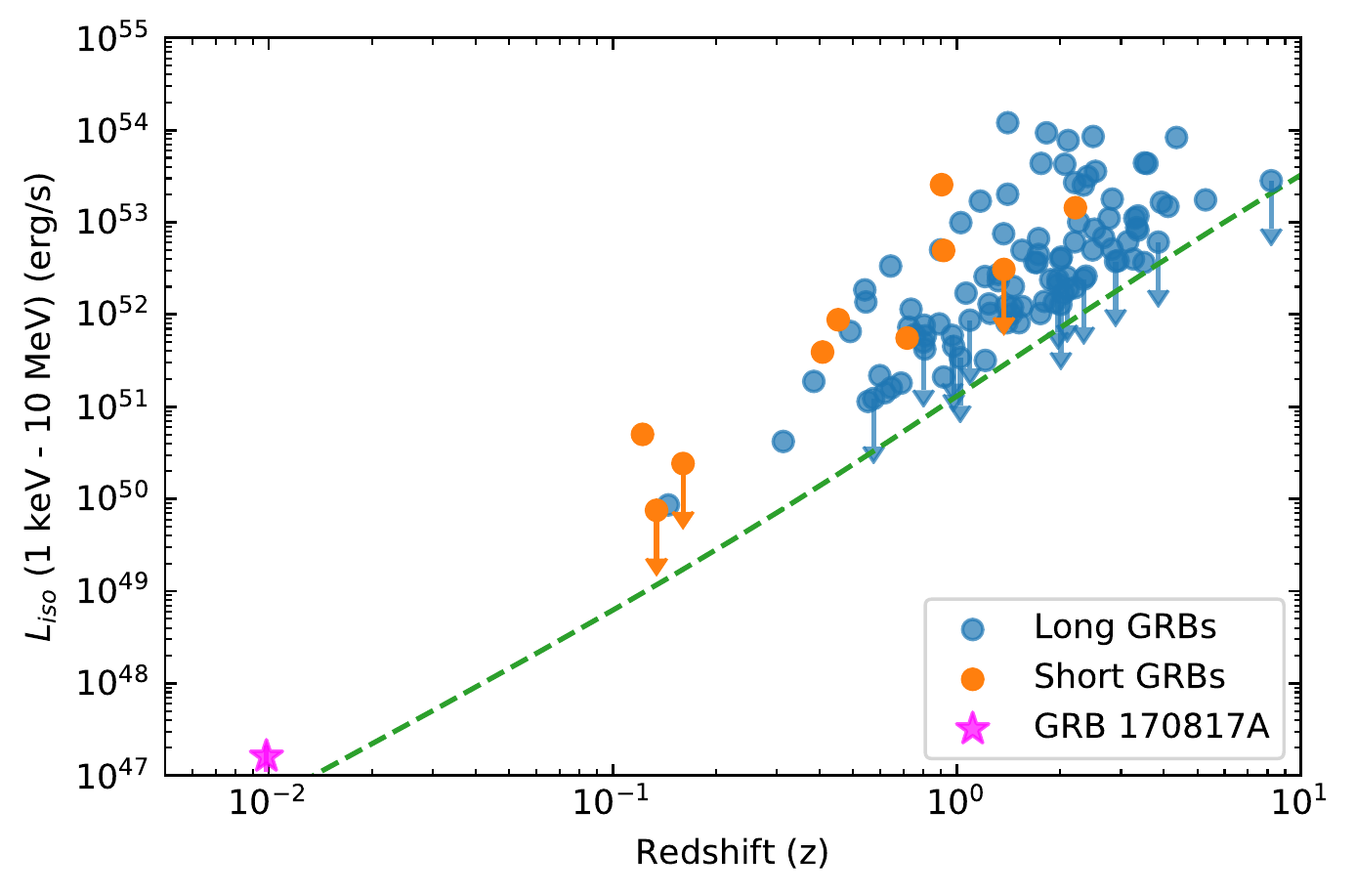}
  \caption{The measured isotropic-equivalent energetics for \ac{GBM} \acp{GRB} with measured redshift. $L_{\rm iso}$ is the peak 64\,ms luminosity; $E_{\rm iso}$ is the total energetics measured over the burst duration. \grbid is both the closest and the faintest by large margins. This figured is modified from Fig.~4 in \citet{GW170817-GRB170817A}.
  }
  \label{fig:GBM_GRB_energetics}
\end{figure}

With base values the observed fluence of \acp{GRB} at Earth from cosmologically distant \acp{GRB} would require intrinsic isotropic-equivalent energetics of $\gtrsim10^{50}$\,erg. The prompt emission of \acp{GRB} have small intrinsic variability timescales, with the most extreme values being sub-millisecond \citep[e.g.][]{bhat1992evidence}. Structure at this timescale constrains the size of the central engine, in the non-relativistic case, to $R \approx c\delta t \lesssim 300$\,km, requiring a compact central engine. This amount of energy being emitted from such a small volume would result in an enormous opacity for $\gtrsim$MeV photons due to pair creation, i.e., $\gamma + \gamma \rightarrow e^+ + e^-$. The resulting spectrum then must be thermal, incompatible with observations. \citet{Paczynski1986,SGRB_progenitor_NS_Goodman_1986} stepped through these issues and determined that \acp{GRB} from cosmological distances require bulk relativistic motion. As calculated through various means, typical values of the bulk Lorentz factor $\Gamma$ are $\sim$100 \citep[e.g.,][]{fenimore1993escape,baring1997escape,lithwick2001lower,hascoet2012fermi}.

It was hypothesized that the fast outflows from \acp{GRB} would interact with the surrounding matter, emitting synchrotron radiation at lower energies \citep{paczynski1993radio}. The first detections of \ac{GRB} afterglow by BeppoSAX confirmed the cosmological origin by localizing events to distant host galaxies \citep[e.g.][]{van1997transient,reichart1998redshift}. Tying specific bursts to specific distances enables a direct determination of the intrinsic isotropic energetics with some approaching a few times $10^{54}$\,ergs (Figure \ref{fig:GBM_GRB_energetics}), which is an energy equivalent to the total mass of the Sun \textit{after} all of the relevant efficiency factors have been accounted for. This strongly suggested that the emission was not isotropic.  

It is now known that the bulk relativistic outflow from \acp{GRB} is not isotropic, but is collimated into jets. The isotropic-equivalent energetics are corrected by the factor $1-\cos(\theta_j)$ where $\theta_j$ is the half-jet opening angle. With a representative values of $\theta_j=1-10$\,deg this reduces the required energetics by $\sim10^2$--$10^4$ \citep[e.g.,][]{sari1999jets,frail2001beaming,panaitescu2001jet,racusin2009jet,cenko2010collimation}. Observational evidence in favor of relativistic jets as the origin of \acp{GRB} includes constraints on the angular size from the detection of radio scintillation \citep{goodman1997radio,frail1997radio} and direct measurement of superluminal motion of compact emitting regions in both long and short \acp{SGRB} \citep{taylor2004angular,mooley2018superluminal}.

We can determine the collimation angle by measuring the jet-break with afterglow studies \citep{rhoads1997tell}. The afterglow undergoes early temporal decay that is somewhat counteracted by the increase in the observable region due to the change in Doppler beaming, $1/\Gamma$, as $\Gamma$ slows due to jet interaction with the circumburst material. Once the beaming angle encompasses the entire jet the temporal decay steepens to the intrinsic value. This signature was first observed in \acp{LGRB} in the late 1990s \citep{kulkarni1999afterglow,fruchter1999hubble,harrison1999optical} while jet-break measurements for \acp{SGRB} required the \Swift era, where a sample now suggests a typical half-jet opening angle for \acp{SGRB} of $\sim16\pm10\,\deg$ \citep[][and references therein]{fong2015decade}. These jet-break measurements relied on a top-hat model, which is consistent with observations for these \acp{GRB}.


The preceding paragraphs discuss the understanding of \acp{GRB} that is generally agreed upon. However, there are many important questions related to \acp{GRB} that remain unresolved. For a thorough and quantitative discussion on \acp{GRB} in general see \citet{kumar2015physics} for a review article and \citet{zhang2018physics} for a book. These publications discuss a number of key outstanding questions relevant to understand \acp{GRB}. Multimessenger studies of \ac{NS} mergers, especially those with detected \acp{SGRB} emission in the prompt or afterglow phase, may provide new pieces of information to answer these questions. Section\,\ref{sec:progenitors_GRBs} focuses how \ac{GW}-\ac{GRB} studies can probe the types and relative contributions of progenitors of \acp{GRB}. Section \ref{sec:central_engines} discusses the possibility of magnetar central engines for ultrarelativistic jets. Determining the possible formation mechanisms for these jets is discussed in Sect.~\ref{sec:jet_formation}. The propagation and structure is explored in Sect.~\ref{sec:grb_structure}. Their role in the production of other high energy particles is discussed in Sect.~\ref{sec:jet_composition}. How \ac{GW} observations may help us understand the prompt emission mechanism is provided in Section \ref{sec:GRB_production}. Lastly, Sect.~\ref{sec:other_high_energy} discusses the other non-thermal signatures seen in \acp{GRB} and how we can uncover their origin. 

\subsection{The progenitors of gamma-gay bursts}
\label{sec:progenitors_GRBs}

As discussed, the circumstantial but convincing evidence enabled by the \Swift mission tied most \acp{SGRB} to a \ac{NS} merger origin. There are two key questions where \ac{GW} observations and multimessenger studies may improve this question. The first is the direct knowledge of the progenitor system for events detected both in in \acp{GW} and as \acp{GRB}, as demonstrated by the association of \gwid to \grbid and the classification as a \ac{BNS} merger \citep{GW170817-GW,GW170817-GRB170817A}. With a larger population of confidently classified events we can constrain the fraction of \acp{SGRB} that arise from \ac{BNS} mergers and those from \ac{NSBH} mergers, providing a method to determine if their \acp{GRB} properties differ.

Second, future \acp{GRB} that are determined via \ac{EM} observations to originate within the \ac{BNS} merger sensitive volume for the \ac{GW} network can be conclusively classified as arising from other sources. There are some events that have some properties of the short class and some of the long class, such as GRB\,060614 \citep[e.g.,][]{zhang2007making}, where a sufficiently sensitive \ac{GW} network will confirm or reject a merger origin, providing some insight into these ambiguous events. 

Further, some fraction of \acp{SGRB} may arise from a different origin. Magnetars are \acp{NS} with magnetic fields of order $\sim$10$^{15}$\,G. Some of them produce soft gamma-ray repeater flares, which are $\sim$10\,ms long and generally softer than \acp{SGRB} \citep[e.g.,][]{lazzati2005soft}. These magnetars sometimes produce a magnetar giant flare, with three having been observed in the Milky Way and its satellite galaxies \citep[e.g.,][]{GMF_790305_LMC,hurley1999giant,GMF_041227_SGR1806_Palmer}. As noted in \citet{hurley2005exceptionally}, observing such flares that originate in other galaxies, out to a few tens of Mpc, would result in temporal and spectral properties largely consistent with cosmological \acp{GRB}. From basic rates estimates it follows that a small fraction of \acp{SGRB}, between $\sim$1-10\%, are from extragalactic giant flares \citep{ofek2007soft,GMF_IPN_Dmitry_search}.

There are two previously published \acp{SGRB} that are strongly suggested to be extragalactic magnetar giant flares \citep[e.g.,][]{GMF_051103_M81,GMF_070201_M31} and an initial report of a third case \citep{gcn_GRB_200415A_IPN}. Non-detections by the \ac{LVC} constrain a \ac{NS} merger origin for the first two to be beyond the likely nearby host galaxy, reducing the options to a giant flare from that galaxy, a \acp{SGRB} from that galaxy with an unknown origin, or chance alignment of a cosmologically distant \acp{SGRB} \citep{abadie2012implications,abbott2008implications}. These measurements were interesting with the previous generation of \ac{LIGO}. As joint observations proceed in the coming years with far more sensitive \ac{GW} interferometers we can more clearly separate \acp{SGRB} arising from different progenitors.

\subsection{The central engines of short gamma-ray bursts}
\label{sec:central_engines}

As summarized in \citet{kumar2015physics}, viable central engines of \acp{GRB} must be able to launch a jet with enormous luminosities, the jet must be relatively clean of baryons to enable ultrarelativistic speeds, and it likely needs to be intermittent to recreate the observed variability timescales and likely able to reactivate to power the later X-ray flares. Based on these criteria, a hyper-accreting stellar-mass \ac{BH} is generally accepted as a viable option \citep[e.g.,][]{woosley1993gamma,popham1999hyperaccreting,lee2000blandford,lei2017hyperaccreting}. Magnetar central engines have also been invoked \citep{uso1992millisecond,thompson1994model,zhang2001gamma} and appear to easily explain observational signatures observed in tens of percent of \acp{SGRB} that require late-time energy injection into the system (Sect.~\ref{sec:other_high_energy}). 

So far, simulations suggest this cannot happen as they may fail to meet the second criterion: if the (meta)stable \ac{NS} remnant lives for $\gtrsim$50 ms the neutrino luminosity strips $\sim10^{-3}\,M_\odot$ of material from the surface of the remnant itself \citep{dessart2008neutrino,overview_fernandez_metzger_2016}. Even with $10^{52}$\,ergs ($\approx 0.1\,M_\odot c^2$, the rough total mass of the accretion disk) to power the jet, this small amount of baryonic material could only be accelerated to $\Gamma \approx 10$, an order of magnitude below the typical values expected for \acp{SGRB} \citep[e.g.,][]{lee2007progenitors,murguia2014necessary}. However, we note the observational signatures requiring ultrarelativistic outflows for \acp{SGRB} is more sparse than in \acp{LGRB}, as demonstrated by \citet{ghirlanda2018bulk} providing a measurement for 1 \acp{SGRB} compared to 67 \acp{LGRB}. While this baryon loading has not been resolved theoretically, there are potential paths forward \citep[see, e.g., discussions in][]{metzger2020kilonovae}.

If magnetars can power ultrarelativistic jets then \acp{SGRB} may be generated in the low-mass \ac{BNS} merger cases so long as the remnant object forms a magnetar \citep{giacomazzo2013formation,giacomazzo2015producing}, could alter the kilonova signatures in the Stable \ac{NS} and \ac{SMNS} cases due to jet interactions with the polar material and enormous energy deposition into the system during spin-down \citep[e.g.,][]{Bing_5,Bing_6,metzger2020kilonovae}. Magnetar central engines have also been studied for a subset of \acp{LGRB} \citep[e.g.][]{bucciantini2008relativistic,ioka2016ultra}; however, the baryon content issues for collapsars is different than for mergers, and it is possible that magnetars may only be viable central engines in the latter case.

If these magnetars cannot power ultrarelativistic jets, then only higher-mass \ac{BNS} mergers produce \acp{SGRB}, we would only expect potential jet interactions in the \ac{HMNS} case and it may suggest magnetars cannot power \acp{LGRB}. Resolving this question is related to confirming the origin of the additional non-thermal emission (Sect.~\ref{sec:other_high_energy}) as originating from magnetar spin-down energy or fall-back accretion, would alter \ac{EM} signatures of the remnant object (Sect.~\ref{sec:remnant_determination}), and has implications on the inferred properties of the ejecta from kilonova observations and therefore their production of the heavy elements (Sect.~\ref{sec:elements}).

Either way we can use this information to classify \ac{BNS} remnant cases, but in different ways (Section \ref{sec:remnant_determination}). Observing a \ac{SGRB}, with the non-thermal plateau emission, from a \ac{BNS} merger confidently classified as a Stable \ac{NS} or \ac{SMNS} merger is suggestive of a magnetar central engine. Given the various possible explanations for the plateau emission we will require several detections with confident classification to prove magnetars can power ultrarelativistic jets (Section \ref{sec:remnant_determination}). 

Otherwise, we will never observe \acp{SGRB} for these events. Approximately 5\% of \ac{GW}-detected \ac{NS} mergers in the Advanced era will have jets oriented towards Earth with \ac{SGRB} emission detectable with current (or funded) missions \citep{song2019viewing}. With the unknown fraction of mergers that result in long-lived magnetars, and the unknown viable viewing angles for the plateau emission, we likely require several tens of \ac{GW} detections of \ac{NS} mergers to confidently rule out this possibility. This will likely be resolved in the A+ era.

\subsection{Ultrarelativistic jet formation}
\label{sec:jet_formation}

A related question to the central engines of \acp{SGRB} is how the jet itself is launched. With the required bulk Lorentz factors and total energetics seen in \acp{GRB}, the jet formation condition requires an enormous energy deposition into environments nearly devoid of baryonic matter, as previously discussed. From Sects.~\ref{sec:overview_merger} and \ref{sec:remnant_determination}, some \ac{NS} mergers have relatively empty polar regions (referenced to the total angular momentum axis) providing a natural jet launching site. The viable jet-launch mechanisms depend on the central engine, intimately tying this question to the previous section. As summarized in \citet{kumar2015physics}, there are three mechanisms thought to be viable for \ac{BH} central engines. We will describe the two most widely discussed options. We refer the reader to that review for more details on all three cases.

One mechanism is through neutrino-antineutrino annihilation \citep{ruffert_janka}, whereby enormous neutrino luminosities interact as $\nu + \bar{\nu} \rightarrow e^+ + e^-$ occurs with moderate efficiency and drives a relativistically expanding fireball away from the central engine \citep[e.g.,][]{katz1996long}. The origin of these neutrinos would generally be the thermal emission from the disk which are geometrically exposed to both polar regions. In cases with, providing 
\begin{equation}
    \dot{E}_{\nu\bar{\nu}} = 1.1 \times 10^{52} \rm{erg/s} \left(\frac{M}{M_\odot}\right)^{-3/2} \left(\frac{\dot{M}}{M\odot/s}\right)^{9/4}
\end{equation}
with $\dot{E}$ the annihilation power, $M$ the \ac{BH} mass, and $\dot{M}$ the accretion rate
\citep{zalamea2011neutrino,lei2013hyperaccreting}. 


The other commonly discussed option is the Blandford-Znajek mechanism which can extract the rotational power of the \ac{BH} from the magnetic field of the disk \citep{blandford1977electromagnetic}. These Poynting flux jets appear capable of recreating \acp{GRB} observations with representative power
\begin{equation}
    \dot{E}_{BZ} = 1.7 \times 10^{50} \rm{erg/s}
    a^2_\star \left(\frac{M}{M_\odot}\right)^2 B^2_{15} F(a_\star)
\end{equation} with $a_\star$ the spin parameter $Jc/GMc^2$ where $J$ is the angular momentum of the \ac{BH} and $F(a_\star)$ is the spin-dependent function that is often approximated \citep{blandford1977electromagnetic,meszaros1997poynting,lee2000blandford,lei2013hyperaccreting}. Note that in this case the neutrino-antineutrino annihilation energy can still be provided to the jet.

\citet{lei2013hyperaccreting} investigate the capability of these two jet-launch mechanisms to reproduce the bulk Lorentz factor and observed intrinsic energetics seen in \acp{GRB} by considering the effects of baryon loading. They find that both mechanisms can produce highly energetic bursts with values spanning order of magnitudes, but that the neutrino-antineutrino case generally results in bulk Lorentz factors lower than has been observed. The high magnetic fields required for the Blandford-Znajek case acts as a barrier preventing protons from entering the jet \citep{2000PhRvD..61h4016L} resulting in a jet with lower numbers of baryons. Given the much larger mass of baryons, compared with electrons, the higher the baryon content the lower the total velocity of the jet (for a given amount of energy).

If magnetars are to be \ac{GRB} central engines then the jet launch mechanism is related to the enormous large-scale magnetic field \citep[e.g.][]{uso1992millisecond,metzger2008short}. From \citet{Metzger2011protomagnetar,kumar2015physics}, the initially hot magnetar drives baryons from the surface, preventing the launch of an ultrarelativistic jet. Once it cools and the baryonic wind stops, the rapid spindown generates magnetic energy via a dynamo mechanism, launching the ultrarelativistic outflow. The total available energy for this case is related to the spin energy of the magnetar, $\sim2\times10^{52}$\,erg, which does not appear violated in the \acp{GRB} with plateau emission suggestive of a magnetar origin \citep{lu2014test}.


Multimessenger observations of \acp{GW} and \acp{SGRB} provide new information to investigate the viable jet launching mechanisms. In the magnetar case you would expect a longer time delay from the \ac{GW}-inferred merger time to the on-set of \acp{GRB} emission \citep{zhang2019delay}. Remnant classification (Section\,\ref{sec:remnant_determination}) may provide conclusive evidence proving the viability of the magnetar mechanism. Delineating between the leading mechanisms to power a \ac{GRB} with a \ac{BH} will benefit from (future) direct \ac{GW} measures on the final \ac{BH} mass and inferred spin, allowing the input of measured instead of assumed values in the above equations. Considering additional information from kilonova and afterglow observations may enable tighter constraints on other parameters, which in the future may favor one method over the other for individual bursts \citep[e.g.,][]{salafia2020accretion}.


\subsection{Propagation and structure}
\label{sec:grb_structure}

Forming a jet also requires some method of collimation. In \acp{SGRB} this can be done by matter surrounding the launch site originating from the expansion of the equatorial ejecta or the dynamical ejecta already in the polar regions \citep[e.g.][]{mochkovitch1993gamma,aloy2005relativistic,nagakura2014jet}.  The observed half-jet opening angle is $\sim 16^\circ \pm 10^\circ$ \citep{fong2015decade}, with a range of observed values from $\sim 3^\circ$ to $\gtrsim 25^\circ$. Given solid angle effects the median observed value is wider than the true value. 

The top-hat jet model refers to a conical emitting regions with uniform parameters as a function of angle \citep{rhoads1997tell}. They have historically been used to model \acp{GRB} because they involve (comparatively) simple math and were capable of reproducing (most) observations. However, structured jets of various forms where the properties vary within the jet opening angle have been considered \citep{meszaros1998viewing,rossi2002afterglow,structure_2,granot2003constraining,kumar2003evolution,perna2003jets,panaitescu2005jets} and have been applied to some particularly well observed bursts \citep[e.g.,][]{berger2003common,starling2005spectroscopy,racusin2008broadband}.

\grbid had a few unusual properties. Its isotropic-equivalent energetics were several orders of magnitude less energetic than the known sample (see Figure\,\ref{fig:GBM_GRB_energetics}), there was no detection of the X-ray afterglow from the earliest observation at $\sim$0.5\,days post-merger \citep{gw170817_evans_swift}, and was first detected in X-rays nine days later by \textit{Chandra} \citep{GW170817_Troja_Xray_discovery}. Some key observations since this time are discussed below. Several models were invoked to explain these characteristics, which can be classified into three options:

\begin{itemize}
    \item \textbf{A Top-hat Jet:} \gwid was so close that we were able to use radio interferometry observations to prove superluminal motion of the main emitting region, confirming compact bulk relativistic motion, proving a successful jet \citep[e.g.][]{mooley2018superluminal,ghirlanda2019compact}. However, in top-hat jets the afterglow fades in time as a power-law, so the lack of detection in X-rays at first observation rules out a top-hat jet origin \citep[e.g.,]{GW170817_Troja_Xray_discovery,margutti2017electromagnetic,2019arXiv190808046F}.
    \item \textbf{A Structured Jets:} A structured jet origin is the leading explanation remaining for \grbid \citep[e.g.,][]{margutti2018binary,alexander2018decline,gcn_170817_latettime,nynka2018fading,lazzati2018late,2019arXiv190808046F,troja2019year}. In this scenario \grbid is usually referred to as off-axis, implying the most luminous section of the \ac{GRB} was oriented away from Earth. The long term monitoring of \grbid has allowed for broadband characterization of the temporal evolution from X-ray to radio over years timescales. This has shown a slow temporal rise to a smooth peak followed by the usually decay rate seen from on-axis jets, as expected once the full jet has slowed enough to be fully visible \citep[see][and references therein]{hajela2019two}. 
    \item \textbf{Cocoon emission} refers to the hot envelope that develops around a jet propagating through dense media \citep{nakar2012relativistic}. Some groups invoked a fully choked jet resulting in cocoon (and shock breakout) emission to simultaneously explain the low luminosity of \grbid, the early afterglow behavior, and the early \ac{UV} emission of the kilonova \citep[e.g.,][]{kasliwal2017illuminating,gottlieb2018cocoon}. Successfully choking a jet generally requires large amounts of material in the path of the jet (i.e., the polar region) in absorb the large kinetic energies and prevent successful propagation of the jet. As such, choked jets in SGRBs was not widely considered before \grbid given the generally low expected densities in that region, as discussed in Section\,\ref{sec:overview_merger}. This has been confirmed by simulations performed after \grbid of jet dynamics in \ac{NS} mergers \citep{duffell2018jet}. While the prompt emission of \grbid is consistent with cocoon closure relations, it would require chance coincidence for this event to occur at the correct distance to produce a burst within all of the normal gamma-ray parameters as measured at Earth \citep{Goldstein2017,GW170817-GRB170817A}. Such emission is expected to produce spectra with peak energies much below those seen in time-resolved analysis of this burst \citep{lazzati2017off,veres2018gamma}. The late-time afterglow emission favors a structured jet origin and disfavor a cocoon origin, as discussed and referenced in the previous bullet. The radio observations of the bulk relativistic motion of the compact emitting region also favors a structured jet \citep{ghirlanda2019compact}. Further, powering the early \ac{UV} emission would require a jet with kinetic energies beyond the previously known sample \citep{metzger2018magnetar}. Together these results strongly suggest that a fully choked jet is incompatible with the broadband observations of \grbid.
\end{itemize}

A major result of \grbid is the exclusion of the top-hat jet model for this burst. While the choked jet cocoon origin for \grbid now appears unlikely, it may be viable for future events \citep{gottlieb2018cocoon} and in cases with prompt detection can be tested by the cocoon closure relations in \citet{nakar2012relativistic} to check for consistency \citep[e.g.,][]{GW170817-GRB170817A,GRB150101B_Burns}. This test can be performed within hours of the merger time, allowing for informed follow-up observations. These studies can confirm the viability of \acp{GRB} originating from a shock breakout origin or exclude this option shortly after event time and may be particularly interesting when tied to investigations of the merger remnant given the different expectations for material in the polar region (Section\ref{sec:remnant_determination}). 



Jets can be collimated by a density gradient in the polar region, preventing particles from expanding too far from the polar region \citep[e.g.,][]{mochkovitch1993gamma,aloy2005relativistic,nagakura2014jet}. Magnetic fields can accelerate charged particles in a preferential direction, where ordered poloidal fields can also contribute to jet collimation \citep[e.g.,][]{rezzolla2011}. These interactions impart structure onto the jet where dependencies on the amount and distribution of polar material and the jet launch time are important \citep[e.g.,][]{xie2018numerical,geng2019propagation,kathirgamaraju2019counterparts,gill2019numerical}. \ac{GW} observations provide new information to investigate this question and additional constraints to be met with future models and simulation. The first is a measure of inclination, $\iota$, where variations in the jet should alter the observed prompt \acp{GRB} properties, such as the observed energetics, and combined study may elucidate their structure or, at least, constrain the properties of assumed functional forms \citep[e.g.,][]{mogushi2019jet,williams2018constraints,song2019viewing,beniamini2019joint,biscoveanu2020constraining}. These studies require detections with non-aligned \ac{GW} interferometers and sensitive sky coverage in gamma-rays, as stringent non-detections are also informative. Combining this information with the jet-opening angle determined from afterglow observations will be particularly powerful. 

The second benefit of \ac{GW} detections for these purposes is immediate identification of particularly nearby \acp{SGRB}; \grbid is so close that it has been observed $\sim$100 times longer than prior \ac{SGRB} afterglows. Among the key parameters to study \ac{GRB} structure is the late-time temporal decay of the afterglow, which can distinguish between jetted and quasi-spherical outflows \citep[e.g.][]{2019arXiv190808046F}. Top-hat jets are predicted to have achromatic jet breaks, but chromatic jet breaks, which are often observed, may allow for inferences on the structure of these outflows \citep[e.g.][]{panaitescu2005models}, though this explanation is not unique \citep[e.g.][]{fox2003early,curran2007grb}.

\ac{GW} detections also constrain the jet launch time from studies of the \ac{GW}-\ac{GRB} time delay \citep[e.g.][]{jet_time_1,gottlieb2018cocoon,jet_time_3}. Jets that are launched earlier will experience less polar material, potentially providing less collimation and would be more likely to breakout. Jets that launch later may be more collimated, given thick disk expansion or additional dynamical polar ejecta. If they launch too late they could potentially be choked and fail. Studies seeking to understand the delay time are strongly tied to understanding the remnant object in the case of a \ac{BNS} merger because of the expected variations in the amount ejecta, the distribution of that ejecta, and the expected time to launch the jet (Sect.~\ref{sec:remnant_determination}).



So far studies of the structure of \grbid and \acp{SGRB} in general have focused on either the prompt or afterglow emission separately. This is for the perfectly understandable reason that it is difficult to address the two together, but a successful, general structured jet model will have to simultaneously explain all observables, including historic constraints. For example, it would need to be capable of recreating the inferred $\Gamma \gtrsim 1000$ observed for GRB 090510 based on \Fermi \ac{LAT} observations of this event \citep{GRB090510_discovery_Fermi}, will also have to reasonably reproduce the observed \ac{SGRB} redshift distribution, reproduce \grbid, and the observed intrinsic energetics distribution \citep{beniamini2018observational}.

\subsection{Gamma-ray burst jet composition and ultra high energy cosmic rays}
\label{sec:jet_composition}
Cosmic rays were first identified more than a century ago, through Victor Hess's high altitude balloon flight \citep{hess1912beobachtungen}. This was before the formulation of \ac{GR} or the postulated existence of the neutrino. These particles carried new information from the Universe to Earth and led to the creation of a new field of study. One of the greatest outstanding questions in astrophysics is the origin of \acp{UHECR}, i.e., cosmic rays with energies in excess of 1\,EeV. For reviews see \citet{nagano2000observations} or \citet{sokolsky2018introduction}

When protons are accelerated to high energies in dense environments they generically undergo photohadronic processes, e.g., $p + \gamma \rightarrow \Delta^+ \rightarrow n + \pi^+$ \citep[e.g.,][]{rachen1998photohadronic}. These can be followed by leptonic decays $\pi^+ \rightarrow \mu^+ + \nu_\mu$ and $\mu^+ \rightarrow e^+ + \nu_e + \bar{\nu}_\mu$, which tie the predicted energies of gamma-rays, neutrinos, and cosmic rays produced in the same interactions; the total observed flux of these messengers are relatively equal, which is suggestive of a common origin \citep[see e.g.][and references therein]{halzen2002high}. Among the problems in determining the origin of \acp{UHECR} is the deflection of charged particles by the (inter)galactic magnetic fields and large gyro radii, causing both a propagation delay and altering the arrival direction. In principle we can reconstruct the source direction for a particle with known properties (e.g. mass, energy), but this relies on our imperfect understanding of the (inter)galactic magnetic fields, obscuring the origin even in the best case. It is for this reason that the quest to detect gamma-rays and neutrinos from a common source, with appropriate relative energies, have been used to search for the origin of \acp{UHECR}.

Given the ultrarelativistic nature of \acp{GRB} and the enormous energetics involved, it is natural to assume they will accelerate some amount of protons to high energies, with simulations showing some level of baryon loading even in the Poynting flux case \citep[e.g.][]{lei2013hyperaccreting}. This led to the suggestion that they may be responsible for \acp{UHECR} \citep{vietri1995acceleration,waxman1995cosmological} and the idea that a large scale neutrino detector could be used to investigate their potential common origin \citep{waxman1997high}. The short intrinsic timescales and external trigger information would make association (after detection) relatively easy.

While IceCube has indeed found an astrophysical flux of high energy neutrinos \citep{aartsen2014observation}, deep searches have never robustly associated these signals with \acp{GRB} \citep{abbasi2011limits,aartsen2015search}. This is somewhat of a puzzling finding, as it suggests a very low baryon loading in \acp{GRB} jets, despite the general expectation that the baryons are present above the jet-launching site and should be accelerated. It could be that these protons are accelerated to high velocities, but the prompt emission radius is significantly larger than the internal shock scenario, where the photohadronic interactions become less likely due to the lower densities and neutrino production is suppressed \citep{ICMART_2}. These non-detections led to suggestions that choked \acp{LGRB}, where the jet fails to breakout through the massive star, may be significant sources of neutrinos \citep[e.g.][]{meszaros2001tev,senno2016choked}.

\acp{LGRB} are generally more favorable for these studies than \acp{SGRB}, as their higher total energetics should produce a higher neutrino flux and their greater total matter above the jet launch site should result in a higher proportion of choked jets. However, the detection of \gwid and \grbid resulted in renewed interest in \acp{SGRB} as neutrino sources. First, among the issues of choked \acp{LGRB} is that they are \ac{EM}-dark (or at least, extremely fainter than successful jets). If there are \ac{NS} mergers with choked jets we can identify nearby events through \ac{GW} detections, which will provide a time and location for joint sub-threshold searches \citep{kimura2018transejecta} as well as inform on the expected \ac{EM} counterparts and their behavior. Second, the inferred structure of \ac{SGRB} jets suggests a higher likelihood of neutrino detection for nearby events identified by \ac{GW} detections \citep[e.g.][]{ahlers2019neutrino}.


It is not known for certain what \acp{GRB} observations are the most likely to produce detectable neutrinos; however, because neutrino telescopes are all-sky monitors we will have observations of nearly all events. Ideally future studies will be able to detect neutrinos from these events and allow us to study baryon presence in the jet. Alternatively more stringent limits may show that the launch of a relativistic outflow in the presence of baryons is not a sufficient condition for the production of \acp{UHECR} or that \ac{UHECR} production may not require significant neutrino production if the emission radius is large enough \citep{ICMART_2}, which may have implications for multimessenger searches for the origin of \acp{UHECR} from other sources. 

\subsection{The prompt emission mechanism(s) of gamma-ray bursts}
\label{sec:GRB_production}

\begin{figure}
\centering
\includegraphics[width=1\textwidth]{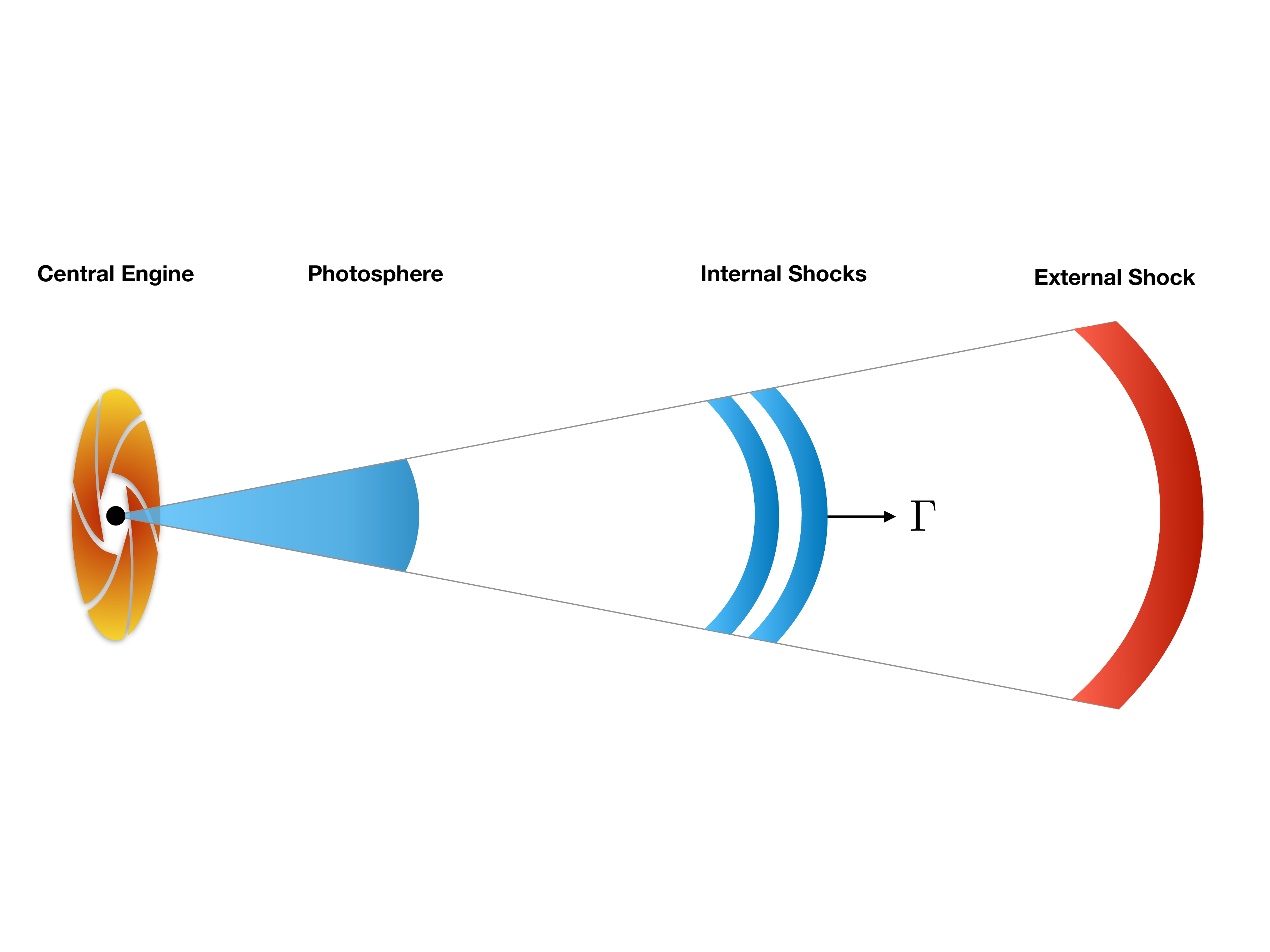}
\caption{A simplified picture of the emission from \acp{GRB}. Thermal emission is possible once the jet has passed the photospheric radius. Internal dissipation of the jet releases the prompt \acp{GRB} signal, shown here with the internal shocks model. Then, the on-set of afterglow emission occurs when the external shock develops as the jet interacts with the surrounding media. This figure is courtesy of Dan Kocevski (private communication).
}
\label{fig:GRB}
\end{figure}

\ac{GRB} jets are often discussed as an ultrarelativistically expanding fireball \citep[e.g.,][]{piran1999gamma,yost2003study,willingale2007testing}. A basic representation of the emission stages is shown in Figure \ref{fig:GRB}. The energy density is truly enormous preventing gamma-rays from escaping until the jet reaches the photospheric radius, where opacity becomes low enough to allow light to escape from within the jet for the first time, at $\sim 10^{11}$--$10^{12}$\,cm \citep{beloborodov2010collisional,kumar2015physics}. Inhomogeneities from the central engine result in shells that propagate outwards with differing bulk Lorentz factors. Fast-moving shells catch slow-moving shells that were emitted at earlier times at $\sim 10^{12}$--$10^{13}$\,cm, releasing the main prompt \ac{GRB} emission through internal shocks \citep{rees1994unsteady}. Lastly, the jet propagates outwards until the interaction with the local environment creates the afterglow emission via synchrotron radiation \citep{kobayashi2007onset}.

Except, maybe not. There are those that argue the dominant emission of \acp{GRB} originates from a photospheric origin \citep[reviewed in][]{beloborodov2017photospheric}. Or that a Poynting flux jet can release the prompt signal once turbulence and magnetic reconnection hit a critical point, at a distance $\sim10^{16}$\,cm from the central engine \citep{zhang2010internal}, which has implications for \acp{GRB} as the origin of \ac{UHECR} (Sect.~\ref{sec:jet_composition}). 

Observations have provided insight, but no full resolution. The broader energy coverage of \Fermi has enabled the study of more complex spectral models. For example, \citet{guiriec2010time} fit the prompt emission of three bright \ac{SGRB} with multiple components, including a thermal component, the main non-thermal component, and an extra power law, which has been seen in additional bursts \citep[e.g.,][]{tak2019multiple}. These components could originate from the three stages (photospheric, internal dissipation, external shock) which can have temporally overlapping signals given the enormous bulk Lorentz factors involved. Alternatively, some explain similar features through synchrotron radiation \citep[e.g.][]{ravasio2018consistency}. The detector response of gamma-ray scintillators is non-linear, requiring a forward-folding spectral analysis method that still (usually) relies on empirical functions rather than theoretically motivated ones, which significantly complicates these studies.

There are two capabilities that are providing new insight into the prompt \ac{GRB} emission mechanism. Polarization probes the existence of large-scale magnetic fields, where significant detection of high polarization implies Poynting flux jets \citep{toma2009statistical}. Population analyses have only recently become available, as these require Compton telescope observations of particularly bright bursts, given the probabilistic scattering angle. Results are not yet conclusive, given the varied results \citep[e.g.,][]{lyutikov2003polarization,yonetoku2012magnetic,chattopadhyay2017prompt,zhang2019detailed,burgess2019time}. Continued advancement in these studies is a promising method to understand the prompt emission mechanism of \acp{GRB}. We note that the lower fluence of \acp{SGRB} implies their polarization will be measured an order of magnitude less often than \acp{LGRB}, but, under the general assumption that \acp{GRB} have the same emission mechanism(s), results from \acp{LGRB} are likely to be informative.

The other new parameter is the time offset from the \ac{GW} to \ac{GRB} emission. These were explored for \gwid and \grbid in \citet{GW170817-GRB170817A} and followed by several wonderful analyses \citep[e.g.,][]{granot2017lessons,GW_GRB_WEP_Shoemaker_2018,zhang2018peculiar}, as well as those that sought to test or distinguish between leading models \citep[e.g.,][]{meng2018origin} or alternative scenarios \citep[e.g.][]{kasliwal2017illuminating}. The separate intrinsic time delay parameters each provide unique information on these events \citep{zhang2019detailed}. With a large enough sample we can independently constrain the separate parameters, providing tighter constraints, e.g., on the jet launch time and the size of the emitting region at emission time. Tying specific bursts to a known central engine type (Section\ref{sec:central_engines}) or potentially constrained to a dominant jet formation mechanism (Section\ref{sec:jet_formation}) will provide additional insights into the viable models.

These studies then require polarization measurements of \acp{GRB}, which will be difficult given there is no active Compton telescope. We also need broadband characterization of the prompt \ac{SGRB} emission in joint \ac{GW}-\ac{GRB} detections. Currently, only KONUS-\textit{Wind} and \Fermi-\ac{GBM} cover the necessary range ($\sim$10\,KeV--10\,MeV). Several proposed SmallSats cover only a restricted energy range ($\sim$50\,keV--2\,MeV), largely due to mass limitations \citep[e.g.][]{racusin2017burstcube,grove2019glowbug}. To constrain the time-resolved $E_{peak}$ in a majority of \acp{SGRB} we require sensitivity to several MeV.

There have been a few detections of the prompt phase of \acp{GRB} by telescopes at lower energies, \citep[e.g.,][]{guiriec2016unified,troja2017significant}. Broadband characterization, beyond the energy range of the \ac{GRB} monitors, of the prompt emission would be phenomenally informative for prompt emission mechanisms \citep[see discussion in][]{kumar2015physics}, so long as their contribution can be separated from a external shock component. This would require either telescopes with massive fields of view, or sufficient early warning from \ac{GW} detectors.

\subsection{The origin of other non-thermal signatures}
\label{sec:other_high_energy}

%

Discussed below are observed or predicted signatures that are likely to be tied to the central engine activity. These includes flares and plateaus in the prompt and early afterglow emission, which are separate from the dominant components. Determining if these events exist and their origin can enable greater understanding of \ac{NS} mergers, as discussed below.


\subsubsection{Short gamma-ray burst precursors}
\label{sec:precursors}
Precursors generally refer to short emission episodes that occur 100\,s or less before the main \ac{GRB} episode. \citet{troja2010precursors} analyzed \Swift data to identify precursor signals, claimed confirmation of these pulses in other instruments, and argue $\sim$10\% of \acp{SGRB} have precursor activity. Other analyses suggest a lower fraction of potential \ac{SGRB} precursors in other instruments \citep[e.g.][]{Zhu_dissertation,Burns_Dissertation,SGRB_precursors_INTEGRAL_2017,li2018insight}.  A similar fraction of \acp{SGRB} have secondary pulses that succeed the main pulse. There is no analysis showing \acp{SGRB} precursors are spectrally distinct from the main emission. As discussed below, the majority of \acp{SGRB} occur at distances beyond where we would theoretically expect to detect precursors. Therefore, it appears feasible that previously observed precursors are just lower-flux \ac{SGRB} pulses. None were observed before \grbid to constraining limits \citep{GW170817-GRB170817A,li2018insight}.

There are theoretical models (mentioned below) that predict precursor emission in gamma-rays, x-rays, and radio, with typical luminosities ($\sim 10^{42}-10^{47}$\,erg s$^{-1}$) and potentially \ac{UHECR} production. Signals at these luminosities would only be detectable by all-sky monitors if the events are particularly nearby, precluding these models as the origin of some claimed precursors (e.g. the precursor for GRB 090510, which occurred at a redshift of 0.9 \citealt{ackermann2010fermi}). Isotropic precursor emission may be expected in these wavelengths from magnetospheric interactions \citep{precursors_hansen_2010,precursors_Metzger_2016,wang2018pre}, disruption of the \ac{NS} crust could produce a short gamma-ray flash \citep{precursors_tsang_resonant}, or emission from the crust can power an \ac{EM} chirp \citep{EM_chirp_schnittman_2018}. These could give unique constraints on the magnetic fields of the progenitors or on the \ac{NS} \ac{EOS} (Section \ref{sec:NS_EOS}). While these signatures would be emitted before merger time, radio precursors may arrive at Earth after merger being delayed by dispersion. 

\ac{GW} observations will enable a resolution to this question. First, they select nearby events where the expected precursor brightness from theory may be detectable by existing or future \ac{GRB} instruments. In some models the precursor emission is more isotropic than the jet, and do not necessarily require an associated prompt \ac{SGRB}. Second, they provide the merger time. This will unambiguously determine if the observed \ac{SGRB} precursors (relative to the main \ac{EM} peak) occur before or after the \ac{GW} merger time, more directly tying precursors to the theoretically-motivated regime or classifying them as prompt \acp{SGRB} pulses.


\subsubsection{Extended emission and X-ray plateaus}
\label{sec:extended_emission}

Extended emission describes an observed behavior of longer, lower flux tails following the main peak of some \acp{SGRB}. While the main peak of \acp{SGRB} is $\lesssim$5\,s, the extended emission can persist for up to $\sim$100 s with the two components having comparable total fluence. This signature was first identified in BATSE data \citep{extended_emission_2001_lazzati,extended_emission_2002_connaughton} and has been found in \ac{BAT} data \citep{extended_emission_2006_norris}. \ac{BAT} allows for the exclusion of extended emission down to stringent flux limits, and suggests it occurs in $\gtrsim$15\% of \acp{SGRB}, but is not ubiquitous \citep{Lien2016}. Extended emission has rapid variability, tying it to late-time energy injection from the central engine. It could be powered by the spin-down energy of a fast-rotating magnetar which can naturally explain the relatively flat emission over the times of interest, corresponding to the Stable \ac{NS} or \ac{SMNS} remnant cases \citep[e.g.][]{dai1998gamma,gao2006short,metzger2008short,bucciantini2011short,fan2013signature,lu2015millisecond}. Matching observations may require significant energy losses to \ac{GW} emission, which would be beneficial for future direct \ac{GW} detections of long-lived remnants.

A somewhat similar plateau signature has been observed on top of the temporal decay of the X-ray afterglow in some \acp{SGRB} \citep[e.g.][]{rowlinson2010unusual} and in \acp{LGRB}. Evidence for which may exist in up to half of \acp{SGRB} afterglows \citep{rowlinson2013signatures}. These signatures can also be reasonably explained by a magnetar central engine \citep[e.g.][]{gompertz2013magnetar}. It may be possible to detect similar signatures from proto-magnetar winds outside of the observable prompt \ac{GRB} line of sight \citep{Bing_2}. There are potentially two such detections already \citep{Bing_3,Bing_4}.

However, there are other models that can result in plateau emission. In the fall-back accretion scenario material is launched with some velocity away from the remnant, but remains gravitationally bound \citep{rosswog2007fallback,2015ApJ...804L..16K}. The variability seen can then arise from interactions of this material during fallback \citep[e.g.][]{coughlin2020variability}. Other models have been considered, such as a two-component jet model \citep[e.g.][]{2011MNRAS.417.2161B,matsumoto2020linking}. Another explanation that arose with the increased consideration of structured jets following \grbid is high latitude emission creating the observed plateaus \citep{oganesyan2020structured,ascenzi2020high}. For each of these models there are additional predictions that will allow for exclusion in some cases, pending sufficient broadband follow-up detections.

Multimessenger observations could provide an unambiguous resolution to the origin of these non-thermal signatures. If magnetars are the origin then we should only expect these signatures following Stable \ac{NS} and \ac{SMNS} cases, corresponding to low-mass \ac{GW} inspirals and bright blue kilonovae (Section \ref{sec:remnant_determination}). If they are observed in other cases, and incompatible with a late-time fall-back origin, then we must search for a different origin. It is also possible that there could be multiple causes for the observed plateau emission, which would require a larger number of multimessenger observations to fully understand.

\subsubsection{X-ray flares in the afterglow}
X-ray flares above the afterglow have also been observed, which differ from plateaus by having a distinct rise and fall \citep{burrows2007x}. Long-lived remnants with high magnetic fields could potentially explain this emission as well \citep[e.g.][]{magnetar_bing_3,gao2006short}; however, these signatures are more often explained via late-time fall-back accretion \citep[e.g.][]{fan2005linearly,rosswog2007fallback,kocevski2007pulse}. Time-resolved multiwavelength observations should be able to distinguish between these models \citep[e.g.][]{lamb2019short}.

There are predicted differences between the progenitor systems, with \ac{NSBH} mergers having up to an order of magnitude more fall-back material than in \ac{BNS} mergers \citep{rosswog2007fallback}. There should also be differences based on the properties of these systems, likely corresponding to the amount of tidal ejecta and being related to the mass ratio of the system. \ac{GW} measurement of these intrinsic parameters and the multimessenger classification of progenitor system and \ac{BNS} remnant type should confirm if observations follow expectations and determine if the X-ray flares are indeed caused by late-time fallback accretion.

\subsubsection{Synchrotron self Compton}
It is generally agreed that the radio to gamma-ray afterglow emission is synchrotron radiation from the external shock \citep{sari1998synchrotron}. From the conditions in \ac{GRB} jets we generically expect \ac{SSC} emission. The first public claim of \ac{VHE} detection of a \ac{GRB} was for GRB\,190114C \citep{MAGIC_190114C}, which has been modeled with a \ac{SSC} origin \citep[e.g.][]{fraija2019analysis,SSC_190114C_2,SSC_190114C_3}. However, no analysis published so far has performed robust multi-instrument spectral analysis showing a statistical preference for a \ac{SSC} origin against a base synchrotron explanation, which may also fit the data. Regardless of this specific burst, the detection of \ac{SSC} emission in \acp{GRB} would give phenomenal constraints on several microphysical parameters which would inform a wide range of \ac{GRB} studies. These would require sensitive \ac{VHE} observations as close to the on-set of prompt emission time as possible. These constraints will likely be most sensitive for \ac{LGRB} observations, but the \ac{GW} identification of nearby \acp{SGRB} and the upcoming \ac{CTA} provide a promising combination to seek \ac{SSC} emission from a \ac{NS} merger. The ideal scenario would be distributed \ac{CTA} coverage of the highest probability region from a \ac{GW} early warning localization.

%% file: s05_elements.tex

The origin of the elements is among the most basic questions in existence. As discussed in Sect.~\ref{sec:cosmology}, Hydrogen, Helium, and Lithium were produced at recombination. Despite 13.8\,Gyr of the production of all other elements, these are still the most common by an overwhelming margin. Some of these atoms coalesced into the first stars. Stellar fusion combine the light elements into heavier elements through well understood nuclear reactions. In massive stars these reactions progress to heavier elements until iron, beyond which fusion becomes endothermic. Eventually the star will explode and release copious amounts of elements from carbon through the $\sim$fifth row of the periodic table. Boron, Beryllium, and nearby elements are created mostly from cosmic ray spallation. 

The heavy elements, those beyond iron, are created by slow and rapid neutron capture processes. The s-process (s for slow) occurs mostly in asymptotic giant branch stars where, over thousands of years, neutrons can be captured into iron seeds from prior supernovae and create heavier elements \citep{johnson2019populating}. Here beta-decay is more rapid than the neutron capture. The reverse is true in the r-process (r for rapid), responsible for the heaviest elements including most of the lanthanides and all of the actinides \citep{burbidge1957synthesis,cameron1957nuclear}, which generally requires material with particularly high neutron density and a low electron fraction. The heaviest (stable) elements must have more neutrons than protons to overcome the massive Coulomb repulsion or else they will radioactively decay to lighter elements. For a recent review on the origin of the heaviest elements see \citet{cowan2019making}.  

In all the universe, the highest neutron density occurs in \acp{NS}. It seems reasonable to investigate the violent births and deaths of \acp{NS} as potential r-process generation sites. For a long time the leading candidate for r-process element production were \ac{CCSNe} \citep[e.g.][]{meyer1992r,woosley1994r}. However, as simulations improved they showed the large neutrino irradiation of the material shifting the electron fraction to higher values, preventing the formation of significant amounts of lanthanides and actinides \citep[e.g.][]{martinez2012charged,roberts2012medium,wanajo2013r}. There are more complicated scenarios that could potentially resolve these issues. For a very nice summary of the current understanding of r-process sites, particularly with respect to common and rare \ac{CCSNe}, we refer to the Supplementary Methods in \citet{siegel2019collapsars}.

The necessary enrichment rate to reproduce the amount of heavy elements ($A > 140$) in the Milky Way, as inferred from the solar system abundances, is $\sim 2 \times 10^{-7}\,M_\odot$/yr \citep{qian2000supernovae}. With a fiducial rate of \ac{CCSNe} per Milky Way-like galaxy of 2.84 per century \citep{li2011nearby}, or 0.0283/yr, if \ac{CCSNe} do produce r-process elements the lanthanide yield of individual events must be low, giving an effective constant enrichment of the heavy elements. There are observational evidence that tend to argue against such a scenario as the dominant r-process site. The first comes from observations of $^{244}$Pu in the ocean floor at two orders of magnitude below the expected value from constant r-process enrichment, favoring a rare process \citep{wallner2015abundance}. Such a measurement relies on using the radionuclide as a natural clock. A second key piece of evidence was the detection of heavy neutron-capture elements in several stars in the dwarf spheroidal galaxy Reticulum II with abundances two orders of magnitude higher than in other such galaxies, again arguing against (relatively) common low-yield events \citep{ji2016r}, with inferred total production capable of reproducing the total r-process production of the Milky Way, suggesting a common origin \citep{beniamini2016r}. The actinide abundances in the early solar system also favor a rare origin \citep{cote2019galactic,bartos2019nearby}.

Ripping apart \acp{NS} promises a neutron-dense, low electron fraction environment. \citet{KN_r_process_Lattimer_1974} were the first to suggest \ac{NSBH} mergers as r-process sites, followed by \citet{symbalisty1982neutron} suggesting \ac{BNS} mergers. \citet{freiburghaus1999r} demonstrated the first simulations showing \ac{NS} mergers could roughly reproduce the observed relative elemental abundances, a result which has been confirmed as simulations have improved. With the apparent r-process production problems in \ac{CCSNe} and observations favoring rare, high-yield sites, \ac{NS} mergers became prime candidates for the dominant r-process sites owing to their much lower rate ($\sim$1 per 10,000 years) in the Milky Way. For a review with a historical discussion on the r-process origin and the role of \ac{NS} mergers see \citet{metzger2020kilonovae}.

The identification of \knid following \gwid with the broadly expected behavior for a kilonova was the first firm detection of r-process nucleosynthesis. With the inferred ejecta mass from \knid and the \ac{GW}-determined local \ac{NS} merger rate it appears that \ac{NS} mergers can be the dominant r-process sites, though large uncertainties remain \citep{cote2018origin}. It appeared then, that we had a reasonably consistent understanding of the origin of heavy elements from theory, simulation, and observation.

Then, \citet{siegel2019collapsars} decided to complicate things, by using knowledge gained from \knid to re-energize an old suggestion \citep[e.g.][]{pruet2004nucleosynthesis}. We briefly summarize their arguments. The observed properties of \knid suggest the dominant ejection method came from accretion disk outflows, from a total disk mass $\sim 0.1\,M_\odot$. \acp{LGRB} originate from collapsars, which are fast-rotating massive stars that undergo core-collapse, and are powered by accretion disks with characteristic mass $\sim 3\,M_\odot$. In short, the thick disk can maintain an electron fraction (in cases with a \ac{BH} central engine) sufficiently low to produce actinides. Despite \ac{CCSNe} being rarer than \ac{NS} mergers (by a factor of a few, based on the inferred \ac{LGRB} and \ac{SGRB} rates) the higher yields (more than an order of magnitude more) of \ac{CCSNe} suggest they have been the dominant r-process production sites over the life-age of the Universe. The viability of this explanation based on current observational evidence is the subject of on-going work \citep[e.g.][]{macias2019constraining,Voort2019}.

As noted in \citet{siegel2019collapsars}, collapsars would be consistent with the observational evidence that support rare, high yield production sites. They predict an infrared signature somewhat similar in evolution to a red kilonova (though from a much larger ejecta mass) that would follow \acp{LGRB} and could be detectable by sufficiently sensitive infrared telescopes such as \ac{JWST}. Should this signature be observationally identified then delineating between the relative importance of collapsars and \ac{NS} mergers will require a more precise yield measurement for each class, their distributions, as well as their relative rates through cosmic time. More generally, there are other suggested rare types of supernova that would produce high yields of the heavy elements, also discussed in \citet{siegel2019collapsars}. We use collapsars as the representative case but note most tests of the two options apply to the larger case of rare supernova.

In Sect.~\ref{sec:nsm_elemental_production} we discuss the nucleosynthetic yield of \ac{NS} mergers, both relative and absolute abundances, and prospects for improving our understanding through simulation and observation of kilonova. Section \ref{sec:heavy_element_milky_way} discusses prospects for determining the current lanthanide and actinide enrichment of our own galaxy. Section \ref{sec:all_r_process} ties these observations to the source evolution of the potential r-process sites and determination of the dominate sites as a function of time through cosmic history.

\subsection{Heavy element production in candidate r-process sites}
\label{sec:nsm_elemental_production}

There are at least three important observational constraints that r-process sites must explain: they must be able to reproduce the relative and absolute heavy element abundances, and they need to be able to explain the varied r-process enrichment in stars. This latter constraint was relied upon to narrow the candidate sites to \ac{NS} mergers and rare types of supernova. The relative values can be inferred from the observed solar system abundances, predicated on the assumption that we do not live in an unusual place. Lower electron fractions may not reproduce the low-mass heavy elements (e.g. iron to lanthanides), and higher electron fractions $Y_e \gtrsim 0.3$ cannot reproduce lanthanides and actinides. \citet{wanajo2014production} first demonstrated with high fidelity simulations that \ac{NS} merger ejecta composed of varying electron fraction successfully reproduce the full range of r-process elements. However, despite the optimal dataset of \knid there were suggestions, but no unambiguous observational proof of production of the heaviest elements (corresponding to the third r-process abundance peak) and an answer may require capabilities that do not yet exist \citep{kasliwal2019spitzer}.

From the arguments suggesting collapsars as potential r-process sites and as checked with initial simulation, collapsars show similar capability to reproduce the observed abundances \citep{siegel2019collapsars}. As stated, most simulations of standard \ac{CCSNe} scenarios appear unable to reproduce the observed relative abundance pattern. It is a reasonable assumption that the dominant r-process production sites should produce these elements with the relative abundances that are observed in the solar system. We assume this is true in ensemble (e.g. the average production from these events, but not necessarily every individual event) when discussing absolute production.

A great deal of simulation work has been performed to tie the observed \ac{UVOIR} behavior to the ejecta properties \citep[e.g.][and references therein]{Kilonova-Barnes-2013,KN_products_Barnes_2016,review_kilonova_tanaka_2016,metzger2020kilonovae}. With prior kilonova candidates \citep[e.g.][]{perley2009grb,tanvir2013kilonova,Berger2013,gompertz2018diversity,ascenzi2019luminosity} the data was insufficient to reliably constrain elemental production of individual \ac{NS} mergers (with some published claimed kilonova signatures relying on a single single data point), especially after accounting for the Malmquist bias towards detecting brighter events (and thus inferring higher average yield per event than the true value).

\begin{figure}
\centering
\includegraphics[width=1\textwidth]{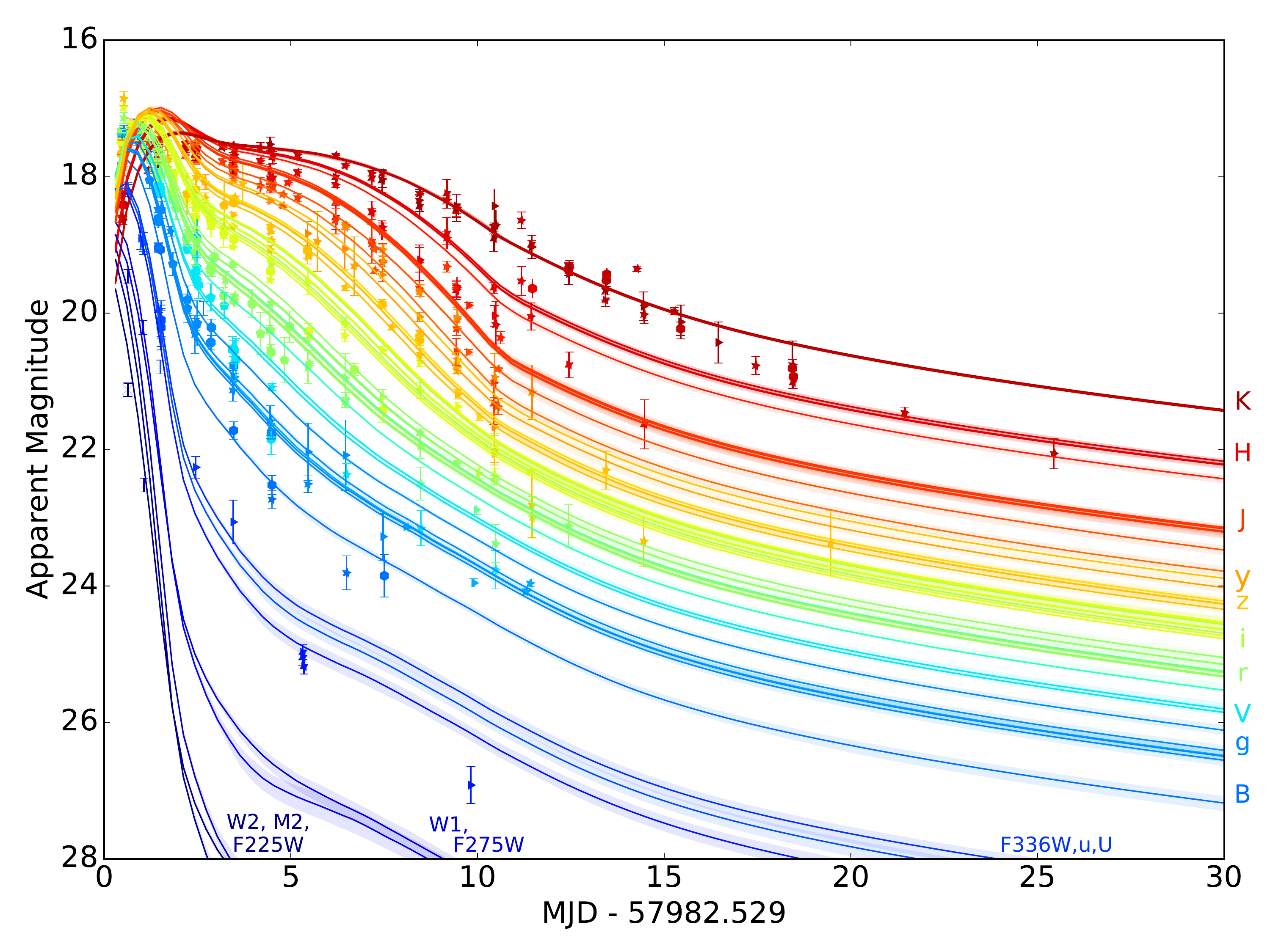}
\caption{The combined \ac{UVOIR} lightcurves for \knid, from \citet{villar2017combined}, with permission. The data comes from several groups. The three component fit using the toy model from \citet{metzger2020kilonovae} is overlaid with solid lines.
}
\label{fig:kn170817_lc}
\end{figure}

In the first detection of a kilonova following a \ac{GW} detection the observers hit the limit of precision of existing models. \citet{villar2017combined} collated the \ac{UVOIR} data reported by various groups for \knid; the results are shown in Figure \ref{fig:kn170817_lc}. Like most authors they identify a red and a blue component, but they favor the addition of a third component with opacity in between the other two\footnote{The color label for this third component has been referred to as \enquote{green} or \enquote{purple}. By physical temperature considerations green is correct. By combination of red and blue, deeply rooted in elementary art classes, purple is the intuitive name. I do not use a name here to prevent confusion, as it does not affect the rest of the paper. Realistically there will be a range from (infra)red to blue, but if three component fits become standard I strongly suggest using a single consistent color name.}. 

We list some of the complications with inferring ejecta properties from the current kilonova models and methods to improve these uncertainties. This is \textit{not} a criticism of these works. They combined several complicated processes into software frameworks that run sets of efficient simulations to predict the signatures of kilonovae before one was ever observed and studied in detail, sometimes hitting the limits of human knowledge itself. The discussion here to show where progress will have to be made in the next few years to determine the true nuclear production in these events. 

In order for kilonova models to be easily utilized to infer ejecta properties from observations they need to provide a range of considered parameter values for comparison. We discuss only two examples out of several options. The kilonova models used in \citet{villar2017combined} are constructed from the toy model presented in \citet{metzger2020kilonovae}, allowing for a broad range of considered ejecta parameters. \citet{kasen2017origin} generated a set of models covering a reasonable parameter space using full radiative transport, reproducing \knid with a specially tailored model, presenting a range of specific models that data can be compared to. They are broadly similar in behavior, but some important differences remain, e.g., the predicted early \ac{UV} flux. \citet{coughlin2018constraints} generated an effective method to interpolate between the available grid models from \citet{kasen2017origin}, providing an important step towards tying observations to simulations.

Laboratory astrophysics is critical. Early work on tying ejecta parameters to lightcurves that predicted a bright blue kilonova with a peak timescale of about a day by assuming iron-like opacities \citep{KN_first_named_Metzger_2010}. Using more realistic opacities for ejecta with lanthanides and actinides results in values orders of magnitude higher, which prevents the quasithermal emission from escaping for longer times, resulting in a redder kilonova with lower peak emission on the timescale of a week \citep{KN_correct_opacities_Kasen_2013,Kilonova-Barnes-2013,KN_correct_opacities_Tanaka_2013}. The dominate contribution to the \ac{UVOIR} opacities are the bound-bound transitions of the lanthanides and actinides. As discussed in Sect.~\ref{sec:overview_kilonova}, the opacities in these papers are calculated from reasonable approximations because we lack the atomic orbital information for these heavy elements which determine the bound-bound transitions. Over the past few years we have improved laboratory and computational determination of these values, work which is critical to improving our estimates of ejecta properties in kilonovae. However, we still do not have key information on individual atoms and much uncertainty remains on how to calculate the ensemble opacities \cite[see discussions in][and references therein]{metzger2020kilonovae}.

Similarly, our current understanding of nuclear physics with regards to the heaviest elements, particularly those far from the region of stability, also limits the accuracy of kilonova lightcurve models. For example, \citet{KN_products_Barnes_2016} investigate a few nuclear mass models to check abundance yields which produce variations in the relative elemental abundances, particularly in the actinides, as well as the fraction of total radioactive energy combined in different decay species as a function of time. The relative $\alpha$, $\beta$, and fission decay differences between nuclear models determines the amount of energy deposited into these products, including neutrinos which can escape very quickly, and gamma-rays which can escape before the peak luminosity time. This alters the thermalization efficiency, i.e. how much energy is converted into heat rather than lost, of the radioactivity as a function of time, which effects the lightcurves and thus our inferences of the total ejecta mass \citep{kilonova_gamma_rays_hotokezaka_2016,KN_products_Barnes_2016}. Fortunately, upcoming atom smashers, particularly the Facility for Rare Isotope Beams \citep{balantekin2014nuclear}, which has astrophysics as a core science goal, will help improve our understanding of the heaviest elements over the next several years.

The simulations themselves make different assumptions and contain different approximations. They vary the assumed velocity gradients of the ejecta, neutron capture fraction, neutrino treatments, radiative transport schemes, nuclear model, opacities, thermalization efficiencies, magnetic fields, entropies, grid formulations and resolution, \ac{NS} \ac{EOS}, etc \citep[e.g.][and references therein]{review_kilonova_tanaka_2016,wollaeger2018impact,kawaguchi_kilonova_remnants,metzger2020kilonovae}. Over the years papers have been published to resolve the importance of these different assumptions which has led to significant improvements in the accuracy of the models and, in general, trends within models based on different input parameters. As examples, that longer-lasting remnants in \ac{BNS} mergers result in more and bluer ejecta, that increasing lanthanide and actinide fraction results in redder kilonova, and that the same kilonova can appear with different color based on the inclination angle. However, uncertainty remains with respect to absolute behavior.

As a particular example we consider the magnetically-driven disk winds. \citet{siegel2017three} investigated these outflows using 3D GRMHD simulations with approximate neutrino transport, running for $\sim$0.4\,s. Extrapolating beyond the end time they conclude that these outflows could ejecta similar amounts of matter as the viscously driven outflows. \citet{fernandez2018long} ran simulations for several seconds, providing direct evidence for those conclusions and additional suggest this ejecta could produce a kilonova precursor signal. \citet{miller2019full} consider full 3D GRMHD simulations with full neutrino transport which significantly altered the electron fraction of the ejected material, suggesting these outflows could power a blue kilonova. 

With each increase in fidelity the conclusions were strengthened or even altered, and this may be expected to continue for some time. As an important example, each still assume idealized initial conditions of the magnetic field. MHD instabilities can significantly amplify magnetic fields \citep[e.g.][]{balbus1991powerful} and their topology is not necessarily simple. \citet{kiuchi2014high} and \citet{kiuchi2015high} study the magnetic fields that develop in \ac{BNS} and \ac{NSBH} mergers respectively, showing strong and complicated fields in both cases, but with different topology. Using the information from careful merger simulations as initial conditions for studies focusing on post-merger effects, as in \citet{nouri2018evolution}, may provide more accurate results. These currently limit our ability to infer ejecta parameters from kilonova lightcurves.

Most kilonova models have assumed spherical symmetry for simplicity, but accounting for more realistic spatial distribution results in inclination effects on the observed lightcurves for the same event \citep[e.g.][]{kasen2017origin,wollaeger2018impact}. In measuring the ejecta properties of the components in \knid nearly every group assumed spherical symmetry for each contributing ejecta region, which is not necessarily a good assumption \cite[e.g.][]{metzger2020kilonovae}. If a red kilonova emitting region is between a blue kilonova emitting region and the observer the blue emission will be blocked by the bound-bound opacity of the intervening material \citep{kasen2017origin}. Even if the view is unobstructed the spatial distribution can alter the inferred ejecta properties. Indeed, accounting for the expected equatorial distribution for the lanthanide-rich material and the polar distribution for the lanthanide-free material for \knid suggests a lower overall yield, removing some of the tension with kilonova simulations \citep{kawaguchi_kilonova_remnants}. This is also considered in \citet{bulla2019possis} with the additional consideration of the polarization which may provide key additional information.

These are further complicated by the intrinsic variations in mergers themselves. The progenitor system and different immediate remnant cases have vastly different ejecta morphology, velocity, opacity, neutrino irradiation, etc. Within each case the mass ratios, spins, and other intrinsic parameters also cause variation in the observational signature. These are further complicated by potential additional sources of energy and heat into the kilonova, like late-time fallback accretion onto the remnant object (see Sect.~\ref{sec:other_high_energy}). Astrophysical observations of a large population of varied \ac{NS} mergers will be particularly helpful in understanding these effects. 

Assuming our general understanding is correct, \ac{NSBH} mergers could release no matter or up to $0.1\,M_\odot$ of lanthanide and actinide-rich ejecta. \ac{BNS} mergers that undergo prompt collapse will produce similar elements, but in lower abundances. Though, these cases could produce the lighter elements if fast magnetically-driven disk outflows occur. \ac{HMNS} could release the full range of beyond-iron elements with higher mass elements from the tidal and disk wind and lower mass elements in the polar ejecta, perhaps up to $\sim0.05\,M_\odot$ based on \knid. Stable \ac{NS} and \ac{SMNS} remnants can release $0.1\,M_\odot$ of the lower mass beyond-iron elements, but only a smaller portion of lanthanides and actinides. To understand the enrichment of heavy elements from \ac{NS} mergers we will likely need to determine the distribution of yield for these different cases, as well as how often these cases occur.

\begin{figure}
\centering
\includegraphics[width=0.5\textwidth]{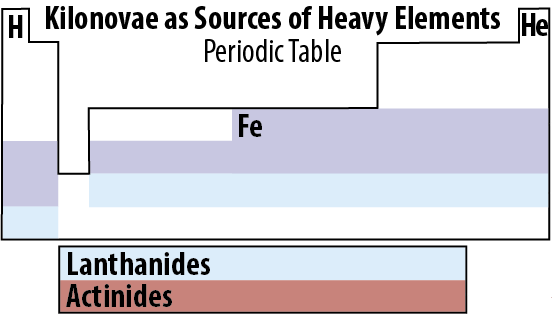}
\caption{The periodic table showing the heavy elements produced in \ac{NS} mergers. The color shading is a simplified representation of the wavelengths that probe production of that element, with violet representing \ac{UV} and near-\ac{UV}, light blue representing optical and some \ac{NIR}, and red showing \ac{NIR} and \ac{IR}. Figure from Judy Racusin (private communication).
}
\label{fig:element_production}
\end{figure}

In order to both precisely test existing models and to accurately infer the ejecta properties for a given event, \ac{UVOIR} observations of \ac{GW}-detected \ac{NS} mergers are absolutely critical. Figure \ref{fig:element_production} shows a basic representation of the elemental yield probed by the different wavelengths. \ac{UV} observations will help understand the unusual excess seen in \knid (Sect.~\ref{sec:early_uv}) which will separate out the contributions of radioactive heating from other potential sources and enable more accurate inferred mass yields. The discovery of the arcsecond position of \ac{EM} counterparts will almost certainly be dominated by optical observations. Infrared uniquely probe the contributions of lanthanide and actinide-rich ejecta, and provide the latest observations of these events. A full understanding of these sources requires the broadband observations from early to late times, noting that limited band observations can be consistent with multiple parameter combinations. \ac{GW} detections provide information on the intrinsic parameters which, with the multimessenger determination of the merger remnant (Sect.~\ref{sec:remnant_determination}), will enable a broad understanding of these sources. The inclination information will be particularly helpful in understanding inclination effects.

Given the complicated nature of these events and our models to understand them, direct determination of nucleosynthetic yield would be helpful. Nuclear gamma-rays can escape beginning a few hours after merger and can carry tens of percent of the total energy of the system \citep[e.g.][]{kilonova_gamma_rays_hotokezaka_2016}. The emission would be concentrated from a few dozen KeV to a few MeV, bright for a few days, and be a relatively flat spectrum due to Doppler broadening. Such a detection would provide another handle on the ejecta properties that is not dependent on a number of assumptions that the \ac{UVOIR} determination is. However, this is beyond the capability of existing instruments, and likely beyond the capability of proposed instruments unless we are lucky \citep{timmes2019catching}.

Alternatively, one could potentially measure yields of individual elements. This can be direct spectroscopic measurements of individual absorption lines with sensitive \ac{IR} telescopes weeks after merger when the ejecta has sufficiently slowed to minimize Doppler broadening. Late-time temporal decay in the infrared may be dominated by the decay of individual (or a few) isotopes. These prospects are reviewed in \citet{metzger2020kilonovae}, who suggest the approaches are promising, though some uncertainty remains.

Observationally measuring or constraining the lanthanide production in collapsars appears phenomenologically similar to that of \ac{NS} mergers. \citet{siegel2019collapsars} argue a late-time infrared signature following \acp{LGRB} detections would arise if they are significant r-process sites. Then, similar modeling to tie the observed light curves to the ejecta properties are required. This may be the only observable signature as the Milky Way is generally too metal-rich for collapsars to occur, preventing study of nearby \ac{LGRB} remnants.


\subsection{On-going heavy element nucleosynthesis in the Milky Way}
\label{sec:heavy_element_milky_way}

Combining yields from individual events with the \ac{GW}-determined volumetric \ac{NS} merger rate measures the local heavy element production from these events. This rate currently has an order of magnitude uncertainty in the 90\% range, which should rapidly shrink over the next few years. With the inferred ejecta for \knid and the merger rates in the Milky Way from Table \ref{tab:nearby_rates}, \ac{BNS} mergers alone can robustly create the r-process elements in the Milky Way at the rate required to be the dominant site of r-process. 

However, as we begin to constrain the yield distribution of \ac{BNS} mergers, better constrain the local rate of \ac{NS} mergers, and determine the relative contribution of \ac{NSBH} mergers, we will have to consider additional effects. From \citet{SGRB_hostless_offset_Tunnicliffe_2013} about 30\% of \acp{SGRB} are hostless, implying no nearby potential galaxy to deep observational limits. From \citet{fong2015decade} some of the \acp{SGRB} with reliable hosts also appear to be significantly outside of the galaxy itself. This implies that a few tens of percent of \ac{NS} mergers are nearly or totally unbound from their host galaxy. Then, the nucleosynthetic yield of these mergers will not contribute to the observed abundances in their galaxies, and we should expect a similar effect for \ac{BNS} and \ac{NSBH} systems born in the Milky Way. This consideration does not apply to either \ac{CCSNe} or collapsars which should track the stellar mass within the galaxy.

The use of radionuclides can uncover recent nucleosynthesis in our own galaxy. That is, explosive nucleosynthesis results in radioactive isotopes. With  nuclear reaction networks we can calculate the expected isotopic ratios of some key elements as a function of time for various initial relative abundances. These natural clocks allow constraints on past explosions in the Milky Way. These studies usually rely on recent \ac{SNe}, supernova remnants, or observations of diffuse radioactive emission. \citet{wu2019finding} consider diffuse emission from \ac{NS} mergers suggesting they are well beyond the capability even of any proposed telescope.

Searches for \ac{KNR} in the galaxy have been proposed \citep{wu2019finding,korobkin2019gamma}, suggesting detections are possible with proposed $\sim$MeV gamma-ray telescopes. The detection of $^{126}$Sn lines would identify a past r-process production site, likely limited to events occurring in the last $\sim$Myr. Detection of additional lines would enable constraints on the age of the remnant and the relative production of actinides. 

Distinguishing between the potential r-process sites could be done through spatial information and yield determination. If the events occur outside of the galactic plane it will favor a \ac{BNS}/\ac{NSBH} merger origin; otherwise a rare \ac{CCSNe} origin. Events with low initial yields will favor basic \ac{CCSNe}. Events with incredible yields ($\sim1\,M_\odot$) would favor a collapsar origin but we do not expect to identify these in the Milky Way. Events with yields $\sim 10^{-2}$--$0.1\,M_\odot$ would favor a \ac{NS} merger origin. Delineating between \ac{BNS} and \ac{NSBH} merger remnants may be difficult unless multiple lines are detected. In general, the inferred actinide fraction will be informative, with \ac{NSBH} mergers generally requiring a high value. Most \ac{BNS} cases do not. The exception is the prompt collapse scenario which may be difficult to distinguish from an \ac{NSBH} merger with low ($\sim0.01\,M_\odot$) initial ejection. Being able to reliably determine what the origin of the r-process site is would require a MeV telescope with line sensitives a factor of a few better than the current advanced proposals.

The other method of direct isotopic determination is through careful cosmic ray studies. \citet{binns2019ultra} argue that uncovering the relative isotopic abundances of the actinides and comparison of their ratios would constrain the rarity of the currently dominant r-process sites, similar to the constraints of observing $^{244}$Pu on the sea floor. They discuss this specifically delineating between base \ac{CCSNe} and \ac{BNS} mergers. 

\subsection{The heavy element enrichment history of the Universe}
\label{sec:all_r_process}

The prior subsection discusses how to resolve the dominant r-process site in the current time. This answer may differ from the site that has produced most of the lanthanides and actinides that now exist. That is, current elemental abundances in the solar system are the cumulative effect of all prior r-process events in the Milky Way. We know that the rates of \ac{BNS} mergers were higher in the past than they are today \citep[e.g.][]{review_SGRBs_Berger_2014}. The peak rates for \ac{CCSNe} occur earlier, and the rates of collapsars earlier still. Then, the relative contributions of each potential source varies through the history of the Universe.

The best understood source evolution of these potential sites is \ac{CCSNe}. Stars that undergo core collapse are massive and have short lifetimes, measured in tens of millions of years, or less than 0.1\% the age of the universe. Their creation should largely track the cosmic star formation history which peaked at roughly $z \approx 1.9$ when the universe was $\sim$3.5\,Gyr old \citep[see, e.g.][]{madau2014cosmic,cosmic_SFR_Hopkins_2006}. The e-folding scale is $\sim$3.9\,Gyr, suggesting half the stellar mass was created before $z \approx 1.3$. These are effectively the source evolution of \ac{CCSNe}, with the normalization determined by the current local rate.

In the early universe the source evolution of collapsars should track that of \ac{CCSNe} (and thus the stellar formation evolution). However, overall, collapsars do not track the environments of \ac{CCSNe} \citep{fruchter2006long}. It is empirical fact that collapsars strongly prefer low metallicity environments. Given the increase in average metallicity as the universe ages due to elemental enrichment from supernovae (and other processes), then the peak collapsar rate should occur earlier than the peak \ac{SFR}. This has been confirmed observationally, suggesting a peak rate before $z \approx2-3$ \citep[e.g.][]{langer2006collapsar,wanderman2010luminosity}.

As previously discussed the formation of \ac{BNS} and \ac{NSBH} mergers likely follows the \ac{SFR} evolution, as they are thought to originate in field binaries of stars that undergo \ac{CCSNe}, but they have long inspirals that delay the merger times. The observed peak rate is around (or greater than) $z \approx 0.5-0.8$ \citep[e.g.][]{review_SGRBs_Berger_2014}, or when the universe was about half its current age.

Then, we can discuss the relative importance of these sites through cosmic time. If collapsars are important r-process producers they are almost certainly the dominant sites for the first several billion years of the Universe. Heavy elements before a redshift of $\sim$3 are likely attributable to these sources. If \ac{CCSNe} are r-process sites then they are likely most important around the times of peak \ac{SFR}, potentially still being sub-dominant during that time. \ac{NS} mergers of either type are likely to be important in the latest half of the universe and currently the dominant sites, with \ac{BNS} and \ac{NSBH} mergers having different yields per event and likely different source evolution.

These studies will have to be done in concert with studies of ancient elemental enrichment \citep[e.g.][]{macias2019constraining,johnson2019origin}, and as we improve our determination of the \ac{SFR}. These are key questions in astrophysics and we can rely on continued investment in these areas. We should seek to determine the \ac{SGRB} source evolution through follow-up observations of prompt signals detected by more sensitive telescopes as a proxy for \ac{NS} merger source evolution. Identical instruments can provide the same for \acp{LGRB} as a proxy for collapsar evolution. We support the use of \ac{JWST} to seek the infrared lanthanide signature in follow-up of \acp{LGRB}.

%% file: s06_cosmo.tex
Cosmology is the study of the Universe on the grandest scales, using observations of the past to understand how it began, how it evolved to its present state, and how it will end. For much of recorded history humanity largely believed in a Geocentric Universe. Copernicus moved us to Heliocentrism through mathematical description. This world view stood until the onset of observational cosmology, little more than a century ago.

\textit{Standard candles} are \ac{EM} sources with known intrinsic luminosities, which enable us to determine their distance from the observed brightness and known $1/d^{2}$ behavior. Cepheid variables were the first known standard candles with luminosities described by the Leavitt Law \citep{cosmo_leavitt_1908, cosmo_leavitt_1912}. Harlow Shapley switched us to Galactocentrism when he used Cepheids to infer the distance to the galactic center \citep{cosmo_shapley_1918}. Soon after, Edwin Hubble used Cepheids to identify other galaxies in the local group as island universes (of Kant's imagination) distinct from the Milky Way \citep{cosmo_hubble_1925, cosmo_hubble_m31_1929}, moving us to Acentrism, and then used them to prove the Universe was expanding in 1929 \citep{cosmo_hubble_1929}. George Lema\^{i}tre found evidence and provided a theoretical explanation for Hubble's results in 1927 \citep{cosmo_lemaitre_1927} and used them to envision the Big Bang \citep{cosmo_lemaitre_1931}. In twenty years the first known standard candle took us from an eternal, static Universe with the solar system at the center to an evolving Universe with a beginning and our galaxy as one of many. As a bit of a cosmic joke, in Hubble's expanding Universe (with a finite propagation speed) we are the center of our \textit{observable} Universe.

Lema\^{i}tre argued that time reversal of an expanding Universe naturally rewound to a single point. Such a Universe would explode outwards, beginning as a super-heated place that cooled as it expanded. Once it was sufficiently cool to allow electrons to bind to nuclei the first atoms were formed, referred to as recombination for historical reasons, which occurred only 380,000 years from the beginning. Careful study of this \ac{BBN} predicts the relative abundances of the light elements \citep[H, He, Li;][]{cosmo_BBN_1948, cosmo_BBN_1999}, which reliably match current measured values \citep[see, e.g.][]{Planck_2018_cosmo}, providing additional support for the theory.

At recombination the Universe became transparent, allowing photons to travel freely for the first time, decoupling radiation and matter. As the Universe expanded these photons were cosmologically redshifted to lower energies until the present time when these photons are microwaves. The prediction of the Cosmic Microwave Background \citep[\acs{CMB};][]{cosmo_alpher_1948} and its accidental discovery \citep{cosmo_CMB_1965} was the third key piece of evidence in favor of the Big Bang. The detection of its blackbody spectrum as the most perfect ever observed added further confirmation \citep{cosmo_mather_1994}. \acused{CMB}

In the 1980s, the idea of inflation (which does not have an agreed upon physical explanation) was developed \citep[e.g.][]{cosmo_inflation_1979, cosmo_inflation_1981, cosmo_inflation_1982}, which resolved a number of outstanding issues such as the lack of magnetic monopoles, the homogeneity and isotropy of the local Universe, and the observed flatness of the Universe. Separately, the inferred baryon matter density in the early Universe is a factor of several below the inferred (total) matter density. This provides strong evidence for the existence of dark matter (\citealt{cosmo_zwicky_1933, cosmo_zwicky_1937}; though there are discussions of the idea back to Lord Kelvin) with similar relative abundances inferred from galaxy rotation curves \citep{cosmo_rubin_1978, cosmo_rubin_1980} and other methods. Altogether, the small anisotropies in the \ac{CMB} \citep{cosmo_smoot_1992} requires inflation (for overall smoothness) and cold dark matter (for some clumping), which made \enquote{Cold Dark Matter} the standard cosmological model at the time. 

The most famous standard candle in astrophysics are type Ia supernovae due to their high intrinsic luminosities and rates enabling distance-redshift studies deep into the Universe (and because explosions are cool). Observations of them gave us the last great surprise in 1998: the expansion rate of the Universe is accelerating \citep{Universe_expanding_Riess_1998, Universe_expanding_Perlmutter_1999}. The currently unknown origin of this acceleration is referred to as dark energy. 

Of the four known fundamental forces, gravity dominates on cosmological scales. Our modern theory of gravity, \ac{GR}, allows for a cosmological constant. A positive cosmological constant (a positive energy density in the vacuum of spacetime) will tend to counteract the pull of gravity. Einstein had originally used it to maintain a static Universe, the need for which was discarded with Hubble's discovery of the expansion of the Universe. Currently, our observations of dark energy are consistent with a positive cosmological constant, $\Lambda$, with sufficient magnitude to accelerate universal expansion. That is, despite the great observational surprise of dark energy, \ac{GR} is still valid on cosmological scales. Adding this to the prior CDM cosmological model gives us $\Lambda$CDM, which has a cosmological constant dark energy, dark matter, and a Big Bang with inflation. For a full understanding of the Standard Model of Cosmology we suggest the reader find a good modern cosmology textbook.

This section is presented with a brief historical overview to make a key point: since the beginning of observational cosmology our understanding of the Universe undergoes a revolution about once a generation as observations achieve the necessary precision to show old models as incomplete. Following this pattern, there are some unsolved issues with $\Lambda$CDM. \ac{QFT} expects a zero-point energy of spacetime with a value one hundred orders of magnitude larger than the observed value; the so-called \enquote{vacuum catastrophe} (also known as the worst prediction in physics) \citep{adler1995vacuum}. The relative amounts of light isotopes in the early Universe formed by \ac{BBN} are in general agreement except for the abundance of Lithium-7 which is significantly rarer than expected \citep[see][for a review]{cosmo_review_BBN_2014}. $\Lambda$CDM has remarkable success at predicting the large scale structure of the Universe, but simulations currently have less success on smaller scales; the so-called \enquote{small scale crisis} \citep[see][for a review]{cosmo_review_smallscalecrisis_2017}. The total amount of baryons in the local Universe is predicted to be much higher from both \ac{BBN} \citep{cosmo_review_BBN_2014} or \ac{CMB} observations \citep{Planck_2018_cosmo} than is actually observed \citep{cosmo_missing_baryon_2012}; the \enquote{Missing Baryon Problem}. The names show that cosmologists have a flare for the dramatic, but the identified problems suggest considering that $\Lambda$CDM is incomplete.

\subsection{The Hubble Constant}
Perhaps the greatest current issue with $\Lambda$CDM is related to the value of the local expansion rate of the Universe, $H_0$, first measured by Hubble generations ago. $H_0$ sets the scale of the Universe, both its age and size, and is one of the fundamental cosmological parameters. In the local Universe one can directly measure $H_0$ through a distance-redshift relation.

In cosmology it is convenient to define the dimensionless scale factor $a$, using the Friedmann equations \citep{1922ZPhy...10..377F}, which grows through time representing the expansion of the Universe with a unity value in the current age. The Hubble Constant is now known to be the local value of the Hubble Parameter which evolves with time and is $H(z) \equiv \dot{a}/a$. Cosmological distance measures are determined by integrating the inverse of this value, e.g. the luminosity distance is
\begin{equation}
    d_L(z) = (1+z) c \int_0^z \frac{dz'}{H(z)},
\end{equation}
where the evolution of $H(z)$ is determined by an assumed cosmological model. With a FLRW metric, $a = 1/(1+z)$. Thus, we can calculate $H_0$ by setting the scale of the Hubble Parameter in the distant Universe and evolved to the present value (the Hubble Constant) by assuming a cosmological model.

Measuring the value from observations of opposite ends of the Universe provides a stringent test of any cosmology. The most precise value of $H_0$ as measured in the nearby (late) Universe is $H_0 = 74.03 \pm 1.42$ \citep{H0-Riess_2019} and the most precise value of $H_0$ from the distance (early) Universe is $H_0 = 67.66 \pm 0.42$ \citep{Planck_2018_cosmo}, which currently disagree at more than 4 $\sigma$ \citep{H0-Riess_2019}. 

In the local Universe ($z \ll 1$), $d=cz/H_0$, where $d$ is the distance (we here neglect peculiar velocities as they are not important to our general conclusions). Therefore, we can use the distance inferred from observations of type Ia supernovae and their associated redshift to measure $H_0$. Redshift is fairly easy to measure (and at cosmological distances the redshift due to the local motion of the galaxies is negligible). 

The cosmological distance ladder is the method used to determine distances to cosmological objects. For an overview of the cosmological ladder and using it to measure $H_0$, see \citet{hubble_constant_Freedman}. Standard candles have known relations, but their zero points (their true intrinsic luminosities) must be properly calibrated for their distance measures to be correct. Galaxies with large numbers of Cepheid variables can have distances determined through the average inferred distance from each variable. The calibration of Cepheid variables can be set through several means, but all require some independent distance measure for the first ladder rung. To move to the distant Universe, a key data set are galaxies where we have observed a type Ia supernova and several Cepheid variables, which enable a calibration of the type Ia distance through comparison. If the zero point is set incorrectly in any of these steps, then a systematic error will be induced in the inferred distance. 

The exact construction of a cosmological ladder can now rely on several different rungs. Regardless of this choice, similar results arise for several other calibration methods. As it is unlikely that all of these calibration methods would be systematically incorrect in the same direction and magnitude, an incorrect distance ladder calibration seems somewhat unlikely to be the origin of the disagreement. For a broad discussion on suggested systematic errors and an investigation into their possible contribution, see \citet{riess2016}. 

The recent value of $H_0$ from the distant Universe come from studies of the CMB with the full data set from the Planck mission \citep{Planck_2018_cosmo} which is connected to the nearby Universe by assuming $\Lambda$CDM. There are several correlations between parameters inferred from CMB data. Density fluctuations at recombination (observed through the \ac{CMB}) result in anisotropies in the large-scale structure of the Universe referred to as \acp{BAO}. \acp{BAO} are \textit{standard rulers}, and combining CMB+BAO observations can break geometric correlations from CMB data alone. The most precise value of $H_0$ from \citet{Planck_2018_cosmo} come from CMB+\ac{BAO} data, with the BAO measures taken from the latest BOSS results \citep{alam2017clustering}. The predecessor to Planck was WMAP, and WMAP+BAO measures give a value of $H_0$ that disagrees with \citet{H0-Riess_2019} at more than 3$\sigma$ significance. \citet{Planck_2018_cosmo} give a thorough discussion of systematics within Planck data, resolve some issues between WMAP and Planck, and conclude that any simple modification to the \ac{CMB}+\ac{BAO} (with $\Lambda$CDM) value of $H_0$ to match the value from \citet{H0-Riess_2018} is disfavored through comparison with other cosmological observations. 

A clever approach to determine the value of $H_0$ sets the calibration of type Ia supernovae from the distant Universe using BAOs, a so-called \enquote{inverted distance ladder} \citep[see, e.g.][]{heavens2014standard}. This is also done assuming $\Lambda$CDM, which results in a value of $H_0$ consistent with the CMB+BAO value \citep{Planck_2018_cosmo}. While the precision of the type Ia and CMB+BAO values of $H_0$ has improved over the past few years, the central value from each method remains largely unchanged. Therefore, it is worth considering that this is evidence against $\Lambda$CDM being the correct cosmological model, which would not be surprising given the historic pace of such advancements.

The most boring outcome is the $H_0$ disagreement is entirely due to statistical chance, but this appears to be unlikely given the disagreement arises through several measures, has persisted for several years, and has become more significant as each measure became more precise. If there is a systematic error in our study of type Ia supernovae or in the calibration of the cosmological ladder, which would not be entirely surprising since we do not understand the explosion mechanism nor the progenitor(s), then it would have implications beyond just the value of $H_0$. Similarly, if there is a systematic error in our study of the CMB then our inferred values for several correlated parameters would be wrong, and the ramifications would be far reaching. If the inferred disagreement is not statistical, and there is no (dominant) systematic error in these studies, then it provides strong evidence that $\Lambda$CDM must be extended.

A quirk of \ac{GR} is that both \ac{GW} amplitude and $\dot{f}_{GW}$ depend on the chirp mass, which enables a determination of the luminosity distance from \ac{GW} observations of chirping binaries \citep{H0-Schutz-1986}, which led to their designation as \textit{standard sirens}. Their importance for cosmology has long been known \citep[see, e.g.][for a review]{schutz2002lighthouses}. With the first \ac{GW} detection of a NS merger as GW170817, the associated redshift enabled the first demonstration of this technique \citep{GW170817-Standard-Siren}. NS mergers will provide a resolution to the disagreement on the value of $H_0$ in the next few to several years (strictly speaking, there is a possibility that standard sirens give a third value inconsistent with the other two). This requires a precision comparable to that of type Ia supernovae, or about 2\%. 

Several people have calculated the number of joint events necessary to hit some level of precision \citep{H0-Dalal-dark-energy-2006, H0-Nissanke-2013, H0_chen_five_years, H0_feeney_2018, H0_170817_VLBI, H0_vitale_NSBH, kyutoku2017gravitational}. There exists a luminosity distance-inclination correlation that limits the precision of the distance estimate. The different results from these papers arise from the different assumptions on how well inclination can be constrained, from the number of active \ac{GW} interferometers (with more giving increased determination of inclination/distance), from the different source classes (with \ac{NSBH} potentially more useful per event than \ac{BNS}), and from using \ac{EM} information to constrain inclination (which is generally done with information from the SGRB jet). 

There are also methods that attempt to infer $H_0$ from \acp{CBC} without associated redshifts \citep[e.g.][]{cosmo_no_EM_messenger_2012,cosmo_no_EM_2012}, but these are inherently model dependent. While they should be useful for constraining parameters in a given cosmology, it would be difficult to falsify a standard cosmology with such a measurement. For comparison, some of the scientists that showed CDM was incomplete \citep{Universe_expanding_Riess_1998} are now using the same sources to suggest $\Lambda$CDM may also be incomplete and have been moving towards discovery significance \citep[e.g.][]{H0-Riess_2019}, but significant works into alternative explanations such as observational biases, source evolution, and underestimates of errors as sources of the disagreement continue \citep[see discussions in][]{riess2016,Planck_2018_cosmo}. We here make generic arguments on prospects for $H_0$ precision with \ac{NS} mergers and, separately, for those with associated \acp{GRB}. We discuss the latter here for organizational purposes, but it is more relevant in the next section.

As the number of \ac{NS} mergers with \ac{GW} measured distances and associated redshift increases, the precision of the $H_0$ measure from standard sirens should scale as $1/\sqrt{N}$. This is only valid outside low-number statistics as the uncertainty on the luminosity distance can vary greatly for individual events (due to \ac{SNR}, orientation, number of contributing interferometers, etc). The scaling from \citet{H0_chen_five_years} for the HLV network is $15/\sqrt{N}$, which we adopt here. To achieve the necessary 2\% precision to prefer either the type Ia or \ac{CMB}+\ac{BAO} $H_0$ value, this would require $\sim$50 \ac{NS} mergers with associated redshift. 

Early discussions of using \ac{NS} mergers as standard sirens focused on \acp{SGRB} as the \ac{EM} counterpart to enable redshift determination \citep[e.g.][]{H0-Dalal-dark-energy-2006, H0-Nissanke-2013}. \ac{GW}-\ac{GRB} detections are advantageous as they can be easily detected deeper into the Universe and the collimation and relativistic beaming of the associated \ac{SGRB} enable inclination constraints that can reduce the distance uncertainty by a factor of a few. If we conservatively say this improvement is a factor of $\sim$2-3 (which is reasonable given uncertainties on the half-jet opening angle distribution and effects of jet structure), then a 2\% determination of $H_0$ could be done with about a dozen events. The \ac{GW}+kilonova approach will likely resolve the $H_0$ tension before the \ac{GW}+\ac{GRB} approach, but the \ac{GW}+\ac{GRB} approach will become more important as the \ac{GW} interferometers search deep into the Universe. In practice, combining the two will provide the best measurement. If our representative estimates are accurate this would be resolved in the A+ era.

Other studies of the distance-redshift relation are all inherently limited by systematic errors, in part because they have achieved sensitivity beyond the statistical uncertainty. These include instrumental calibration uncertainty and zero-point luminosity uncertainties that calibrate the steps of the cosmological ladder. For standard sirens the instrinsic luminosity is determined by well-understood differential geometry, so long as \ac{GR} correctly predicts the inspiral of \acp{CBC}. This leaves the absolute amplitude calibration of the observing intereferometers. The best calibration in a final report of the O2 observing run from Advanced LIGO/Virgo is 2.6\% for LIGO-Hanford \citep{lvc_gwtc1}. Should the calibration uncertainty affect the precision of future standard siren studies there is no obvious technological limitation to achieving sub-percent precision \citep{ALIGO_photon_calibrators,estevez2018first}. Therefore, we do not expect classical systematics to prevent resolution of the $H_0$ controversy with standard sirens. Implicit in this discussion is that standard sirens will also provide a fully independent calibration of the cosmological ladder.
 

However, there is a different potential systematic that has not been discussed in the case of \ac{GW}+kilonova studies. Determination of redshift for \ac{NS} mergers is done by achieving a precise localization with \ac{EM} observations, identifying the host galaxy (section \ref{sec:host_redshift_where}), and measuring the redshift to that host galaxy. The first step requires identification of an \ac{EM} counterpart with follow-up telescopes; if there is an inclination-dependence on the observed \ac{EM} brightness then \ac{GW}-detected mergers with associated redshift will have an inclination bias relative to the total sample of \ac{GW}-detected mergers. This would bias a determination of $H_0$ due to the correlation of distance and inclination on \ac{GW} amplitude. This has been considered in the cases of \ac{GW}-\acp{GRB} because it is an obvious effect. As discussed in Section \ref{sec:detections_summary}, we expect significant inclination effects on \ac{EM} brightness for the majority of counterparts. We support investigations into the level of systematic error this effect can produce.


\subsection{Beyond \texorpdfstring{$\Lambda$}{Lambda}CDM} \label{sec:beyond_cosmo}
Base-$\Lambda$CDM is the standard model of cosmology as it reliably describes our observations of the Universe with only six parameters. As discussed, one of these is the Hubble constant. The expansion rate of the Universe through cosmic time is determined by opposing effects: the pull of gravity opposed by dark energy. These effects are conveniently described through the cosmological density parameters. The density parameter is defined as $\Omega \equiv \rho/\rho_c$ where $\rho_c = 3H^2/8\pi G$ is the critical density necessary for a flat Universe.

The density of matter is $\Omega_M = \Omega_c + \Omega_b + \Omega_\nu$ with the densities of cold dark matter as $\Omega_c$, baryonic matter $\Omega_b$, and neutrinos $\Omega_\nu$. Returning to the use of the dimensionless scale parameter $a$, $\Omega_M$ evolves as $a^{-3}$ as the total amount of matter is largely constant, so the density scales as the inverse volume. The density of radiation, $\Omega_R$, scales as $a^{-4}$ for the same reason with the addition of cosmological redshift (lowering the energy of each photon). The density of dark energy is captured as $\Omega_{DE}$, which is represented as $\Omega_\Lambda$ when it is specifically referring to a cosmological constant. In a flat Universe $\Omega \equiv 1 = \Omega_M+\Omega_{DE}$ (when neglecting terms that are small in the late Universe).

Several extensions to the base $\Lambda$CDM are generally considered. We list a few here and describe the implications of these extensions.
\begin{itemize}
    \item 
\textbf{What is the Shape of the Universe?}\\
Observations of the observable Universe may provide insights into the topology of the Universe as a whole. For ease of use, we can capture curvature as an effective density $\Omega_k = 1-\Omega$ which will scale as $a^{-2}$. If the Universe is flat then $\Omega_k = 0$ and the Universe is infinite (assuming it is simply connected). If the Universe is curved, this is not necessarily true. If $\Omega_k$ is negative then the Universe has positive curvature and is hyperbolic. If $\Omega_k$ is positive then there is negative curvature and we live in a finite, spherical Universe. Allowing this parameter as the only extension to base-$\Lambda$CDM, joint \ac{CMB}+\ac{BAO} observation from Planck and Boss data constrain $\Omega_k = 0.0007 \pm 0.0037$ at 95\% confidence \citep{Planck_2018_cosmo}.
\item \textbf{Is Dark Energy a Cosmological Constant?}\\
We represent dark energy as a cosmological constant because it reliably matches observations, was a simple extension to CDM, and is still consistent with GR. However, it may be that dark energy evolves with time (or, equivalently, $a$). If true, this would prove $\Lambda$CDM incomplete or invalidate GR on cosmological scales, either of which would be a monumental discovery. There are alternative theoretical models to explain dark energy as arising from a dynamic effect, such as Quintessence \citep{quintessence_1, quintessence_2}. 

\indent The \ac{EOS} of dark energy is defined with the dimensionless parameter $w=p/\rho$, the ratio of the pressure to the energy density of a perfect fluid. In $\Lambda$CDM these values are equal and opposite, i.e., $w=-1$. To investigate if dark energy is not a cosmological constant, the dark energy \ac{EOS} is often parameterized as $w = w_0 +w_a (1-a) = w_0 + w_a z/(1+z)$ \citep{chevallier2001accelerating,expansion_linder_2003,Planck_2018_cosmo}. 

\indent For $\Lambda$CDM $w_0=-1$ and $w_a=0$. If $w_0 \neq -1$, the dark energy density is not constant as the Universe expands, and $w_a$ not being equal to 0 would imply a time-varying \ac{EOS}. Fixing $w_a=0$, \citet{Planck_2018_cosmo} combine \ac{CMB} measures with \ac{BAO} measurements from BOSS to limit $w_0 = -1.04 \pm 0.10$ at 95\% confidence. Including \ac{SNe} information allows for constraints on $w_0$ and $w_a$ together, with $w_0 = -0.961 \pm 0.077$ and $w_a = -0.28^{+0.31}_{-0.27}$, both at 68\% confidence, and they are anticorrelated.
\item \textbf{The Mass of Neutrinos}\\
This is discussed in Sect.~\ref{sec:neutrino_mass}.
\end{itemize}

These extensions modify the equations for the cosmological observables in the nearby Universe. Allowing for $\Omega_k$ and $w_0$ to separately deviate from their base values (but not $w_a$ for simplicity), the Hubble Parameter is modified into:
\begin{equation}
    \frac{H^2}{H_0^2} = \Omega_{R}(1+z)^{4} + \Omega_{M}(1+z)^{3} + \Omega_{k}(1+z)^{2} + \Omega_{DE}(1+z)^{3(1+w_0)}
\end{equation}
We here introduce the comoving distance:
\begin{equation}
    d_C(z) = \frac{c}{H_0} \int_0^{z} \frac{dz'}{\sqrt{\Omega_{R}(1+z)^{4} + \Omega_{M}(1+z)^{3} + \Omega_{k}(1+z)^{2} + \Omega_{DE}(1+z)^{3(1+w)}}}.
\end{equation}
Which enables us to write the luminosity distance:
\begin{equation}
    d_L(z) = (1+z)\begin{cases}
               (c/H_0 \sqrt{\Omega_k}) \sinh{(H_0 \sqrt{\Omega_k}d_C(z)/c)}, & \text{for $0 < \Omega_k$} \\
               d_C(z), & \text{for $\Omega_k = 0$} \\
               (c/H_0 \sqrt{| \Omega_k |}) \sin{(H_0 \sqrt{| \Omega_k |} d_C(z) /c)}, & \text{for $\Omega_k < 0$}.
            \end{cases}
\end{equation}
Here the dark energy density is $\Omega_{DE}$ since it may not be a cosmological constant. In the case that $\Omega_k = 0$ and $w_0=-1$ these reduce to the previous equations. 

Answering these questions could change our fundamental understanding of the Universe. The early Universe was radiation dominated. Because of cosmological redshift of photons, it quickly transitioned from radiation dominated to a matter-dominated Universe, referred to as the deceleration era as the pull of gravity slowed the expansion rate of the Universe. Over billions of years $\Omega_M$ was diluted until the effects of dark energy became dominant, bringing us to the dark energy-dominated Universe, also referred to as the acceleration phase. When this transition occurred, and the shape of the transition, is sensitive to the \ac{EOS} of dark energy and the shape of the Universe. 

\begin{table}
\caption{The cosmological parameters ($\Omega_M$, $\Omega_{DE}$, $H_0$) for the variations in $\Omega_k$ and $w_0$, following the procedure from \citep{cosmo_vary_paper,cosmo_gehrels_2015}. The top row are the values from \citet{Planck_2018_cosmo}, whose values we use for the parameters not listed here.}
\label{tab:cosmo_mods}
\centering
\begin{tabular}{ c  c  c  c  c }
\hline 
\hline
$\Omega_k$ & $w_0$ & $\Omega_M$ & $\Omega_{DE}$ & $h$ \\
\hline
0.0     & -1.00  & 0.311  & 0.689  & 67.7 \\
\hline
0.0044  & -1.00  & 0.291  & 0.705  & 70.0 \\
-0.003  & -1.00  & 0.325  & 0.678  & 66.2 \\
0.0     & -0.94  & 0.328  & 0.672  & 65.9 \\
0.0     & -1.14  & 0.275  & 0.725  & 72.0 \\
\hline
\end{tabular}
\end{table}

\begin{figure}
\centering
\includegraphics[width=1\textwidth]{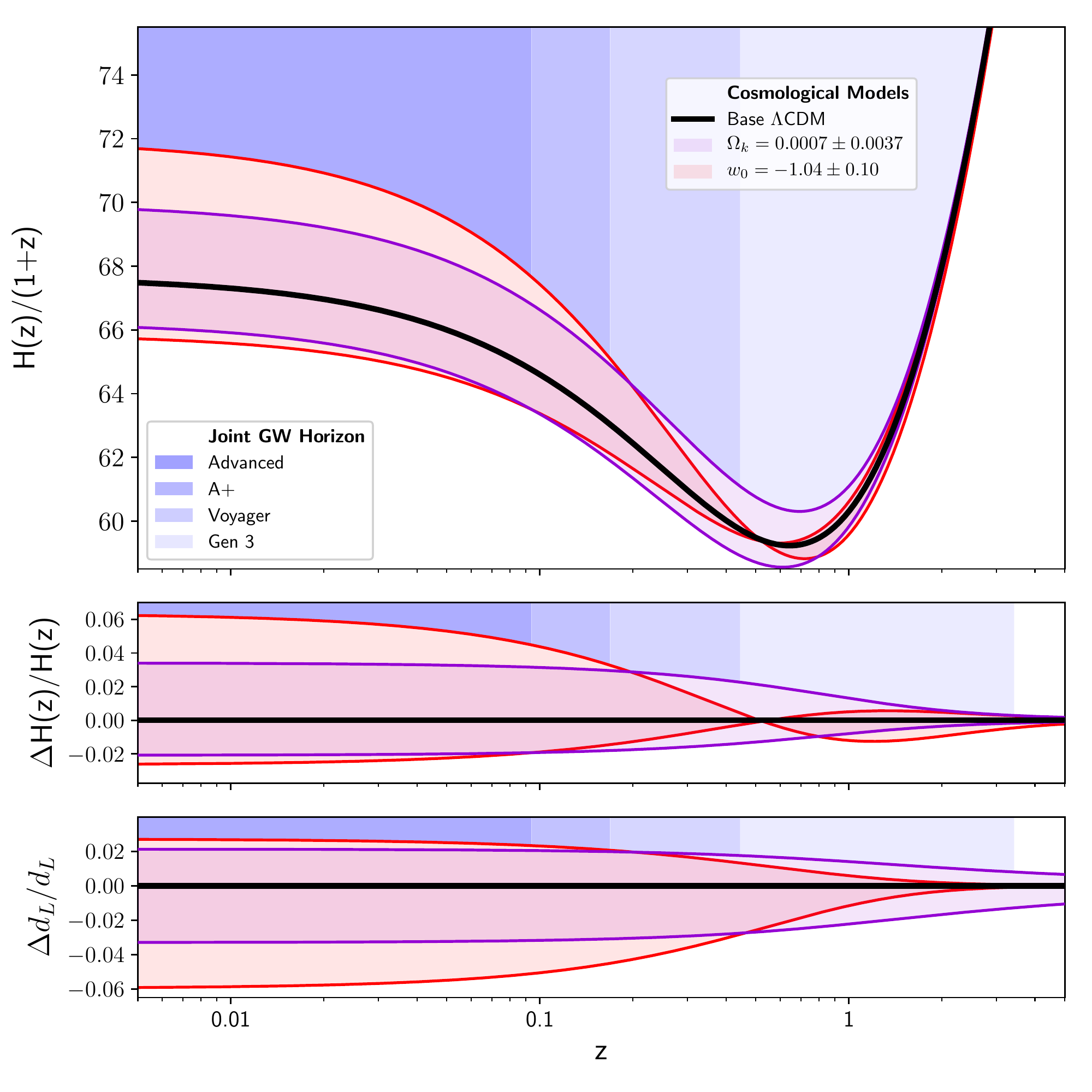}
\caption{The effect on the Hubble Parameter and luminosity distance from allowing $\Omega_k$ and $w_0$ to individually vary, with values taken from Table \ref{tab:cosmo_mods}. We show the base $\Lambda$CDM values from \citet{Planck_2018_cosmo} and the range of the allowed parameter space for a non-flat Universe and non-constant dark energy. The top is for the Hubble parameter, scaled by $(1+z)$, the middle is the fractional deviation of this value, and the bottom the fractional deviation of the luminosity distance, with the latter two corresponding to the necessary measurement precision for informative results on these cases.}
\label{fig:cosmo_mods}
\end{figure}

To demonstrate how changes in these additional parameter models affect observables we match the approach often used for future cosmology experiments \citep[e.g.][]{cosmo_vary_paper,cosmo_gehrels_2015}. We vary the cosmological parameters in a way that the effect on the \ac{CMB} power spectra should be minimized. Specifically, with the convention of $h = H_0/(100 \text{ km s}^{-1} \text{Mpc}^{-1})$, we maintain the distance to the last scattering surface and we fix both $\Omega_M h^2$ and $\Omega_b h^2$. These values for the various cases considered here are given in Table \ref{tab:cosmo_mods}. The effect of these modifications on the observables is given in Figure \ref{fig:cosmo_mods}. We would like to emphasize that what is shown is two separate 1-parameter extensions to the base $\Lambda$CDM. Considering additional options, such as jointly varying $\Omega_k$ and $w_0$ or allowing $w_a$ to vary as well, opens up a vastly larger range of still-acceptable parameter space with correlated variables. Understanding these effects in the middle-age of the Universe is a prime goal of upcoming cosmology experiments like WFIRST, LSST, and EUCLID. 

It is evident from the figure that constraining $H_0$ is a particularly powerful method to constrain beyond-$\Lambda$CDM models. The precision from \citet{H0-Riess_2019} is sufficient to provide useful degeneracy breaking information (e.g., as discussed in Section \ref{sec:neutrino_mass}, its inclusion could resolve the neutrino mass hierarchy, due to their anticorrelation), but this is not done because the value is inconsistent with the \ac{CMB}+\ac{BAO} measures and the reason why is not known. Enter NS mergers. With current GW interferometers we can study the luminosity distance-redshift relation in the nearby Universe. A 1\% measure of $H_0$ in the local Universe corresponds to a $\sim$7\% measure of the \ac{EOS} of dark energy when combined with Planck data, and a $\sim$3\% measure with future \ac{CMB} experiments \citep{riess2016}. If standard siren studies show that other nearby measures of $H_0$ are flawed, then combining \ac{CMB}+\ac{BAO} and standard siren information would be useful for exploring multi-parameter extensions to $\Lambda$CDM\citep[e.g.][]{mnu_cosmo_improvements_2018}. Otherwise, the combined nearby measurement of $H_0$ can be used to study beyond-$\Lambda$CDM models or to further investigate potential issues with the early universe measures. 

Further, this also shows why we require a resolution of the $H_0$ tension using standard sirens even with upcoming cosmological experiments: in order to jointly constrain $\Omega_k$ and both $w_0$ and $w_a$ for the time-varying dark energy \ac{EOS}, as well as other additional parameters, we need precise measures of these observables throughout the Universe \citep{curved_linder,curvature_knox,H0-Dalal-dark-energy-2006,total_bernstein}. Within the decade, the combination of \ac{CMB} observations with information from LSST, EUCLID, and WFIRST as well as standard siren and other local measures of $H_0$ will provide the greatest test of any proposed cosmological model.

As \ac{GW} detectors peer deeper into the Universe, they will enable the most precise Hubble diagram, with a redshift range rivaling or exceeding even type Ia supernovae. These tests are key goals for third generation ground-based interferometers \citep[e.g.][]{et_2010_cosmography,ET_science_objectives_sathyaprakash_2012}, \ac{LISA} \citep[e.g.][]{h0_LISA_2016}, and mid-range space-based interferometers \citep{cutler2009ultrahigh}. \ac{LISA} is expected to detect \acp{CBC} to greater redshifts than the other options, but prospects for the \ac{EM} emission is more uncertain. Third generation ground-based interferometers will tend to have poorer distance uncertainty on an event by event basis, than would a mid-range interferometer. We note the beneficial property for these studies that the peak \ac{SGRB} merger rate (a proxy for \ac{NS} mergers) is around the transition era \citep{review_SGRBs_Berger_2014}.

Traditional cosmology experiments being constructed to answer these questions seek several methods with orthogonal systematics to maximize precision. This is discussed in every document justifying these experiments, for good reasons. We refer to \citet{cosmo_vary_paper} for an in-depth discussion of these methods. They mention the promising prospects for standard sirens but do not consider them in detail because at the time their rates and our capability to detect \ac{EM} counterparts was not known. It is for this reason that, despite these future interferometers coming online in an era where we already expect precision cosmology in the Universe, we still consider this a strong science driver for \ac{NS} mergers. They provide an entirely independent method of distance determination and will become key sources in cosmology.

In the current era we require the capability to detect kilonovae and measure redshift in the local Universe. As we transition to future \ac{GW} interferometers we will require the capability for localizations sufficient for follow-up searches to identify the \ac{GRB} afterglow (as kilonovae will be too faint at these distances). This can be done with a large-scale gamma-ray mission to jointly detect the associated \ac{GRB}, with the added benefit of restricting the inclination angle in the analysis. With mid-range interferometers the localizations should be sufficient in their own right. In all cases, we require the capability to measure the redshift for host galaxies to a reasonable fraction of the \ac{GW} horizon.

Lastly, we briefly comment on the possibility of using lensed \ac{NS} mergers to measure $H_0$. For lensed standard candles we can directly measures $H_0$ \citep{lens_Refsdal_1964,strong_lens_Blandford_1986}, and potentially additional cosmological parameters \citep{lens_linder}. A lensed type Ia has been found \citep{lensed_typeIa}, but cosmological information is still unavailable due to the correlation with the properties of the lensing system. With a \ac{GW}-\ac{GRB} event or mid-range interferometers we could robustly associated multiple detections of the same event (noting that in the \ac{GW}-\ac{GRB} case the jets are sufficiently small that we would not fall out of the jet in most path cases \citealt{lens_anisotropic_PErna_2009}), which would have a precisely ($\sim$ms) measured event time. This would be several orders of magnitude more precise than, e.g. a type Ia supernova. However, this analysis is incredibly difficult and at the time these detections occur they will likely be more important as independent confirmation rather than discovery cases, and we consider them a nice free bonus rather than a driving capability.

%% file: s07_supranuclear.tex

\ac{QCD} is the \ac{QFT} description of the strong force. It describes the interactions between its force boson known as gluons and the elementary particles named quarks, which come in six flavors and three colors (the origin of the chromo- prefix). Gluons bind quarks into hadrons, which are classified as mesons composed of a (valence) quark antiquark pair or baryons composed of three (valence) quarks. Protons and neutrons are baryons that are the composite particles that constitute atomic nuclei, referred to together as nucleons. For a review focused on the nuclear physics description we refer to \citet{baym2018hadrons}. For a review focused on the astrophysical determination of the \ac{NS} \ac{EOS} we refer to \citet{NS-EOS-Ozel}. 

\ac{QCD} is a reliably well tested theory and a foundational aspect of the Standard Model; however, it is incredibly complicated. Constructing large-scale predictions of \ac{QCD} relies on approximate methods. The nuclear saturation density is where baryons begin to overlap, and occurs at $\rho_0 \approx 2.7 \times 10^{14}$\,g\,cm$^{-3}$ \citep[e.g.][]{baym2018hadrons}. Up to about 2$\rho_0$ nucleon interactions dominate with some additional exchanges. In this regime \ac{QCD} lattice methods provide sufficient description to enable tests of \ac{QCD}. At incredibly high densities, $\rho \gtrsim 10-100\rho_0$, the color confinement of quarks to mesons and baryons breaks down. The resulting quark-gluon plasma is well-described by perturbative \ac{QCD}, which have resulted in the most precise tests of \ac{QCD} to date \citep[e.g.][]{altarelli1989experimental,gyulassy2005new}. 

We know less about the behavior of matter in the range $2\rho_0 \lesssim \rho \lesssim 10\rho_0$. It is not known how matter at these densities behaves, i.e. if there is a firm or smooth phase transition between baryon-dominated and quark-dominated interactions. For example, do baryons begin sharing quarks or does color confinement breakdown quickly at some specific density. Constructing predictions at these densities from \ac{QCD} may not be able to directly rely on the previously discussed methods to sufficient accuracy and cannot be built from first principles as this is beyond any existing computational power, and will continue to be for the foreseeable future. Therefore, we rely on varying approximations, often attempting to adapt the approaches viable at either the lower density or higher density end. A description of these methods is beyond the scope of this work and reviewed in \citet{baym2018hadrons}.

These extreme densities are unobtainable in terrestrial laboratories. \acp{NS} are natural experiments. The collapsing core of massive stars converts electrons and protons into the neutrons, resulting in the densest known matter and the only known cold supranuclear matter in the universe, where cold means temperatures $\lesssim$1\,MeV. \acp{NS} can be hot supranuclear matter for comparatively short times when they are born as \ac{CCSNe} or during merger and coalescence in \ac{NS} mergers. In the former case several layers of a large star are between object and observer, a region that is significantly cleaner in the case of \ac{NS} mergers. Their crusts, while incredibly dense by any reasonable measure, have low enough densities that lattice methods may be applicable. Between the crust and the center the densities fall in the 2-10$\rho_0$ range. Therefore, understanding the intrinsic nature of these enigmatic objects allows for unique constraints on the behavior of supranuclear matter, which nicely complement current and upcoming ground-based facilities like the Facility for Rare Isotope Beams \citep{balantekin2014nuclear}.

The key to tying astrophysics to nuclear physics is the \ac{NS} \ac{EOS}, which prescribes the assumed pressure-density relationship. Such a relation can be constructed from the approximate methods described above. From this, one can make testable predictions. Like any star, the structure of \acp{NS} is described from the balance of gravitational forces against internal processes. \citet{NS_first_models_Oppenheimer_Volkoff_1939} and \citet{tolman1939static} derived the equations to calculate \ac{NS} structure and their mass-radius relation from an assumed \ac{EOS}, which are now referred to as the \ac{TOV} equations.

\begin{figure}
\centering
\includegraphics[width=0.7\textwidth]{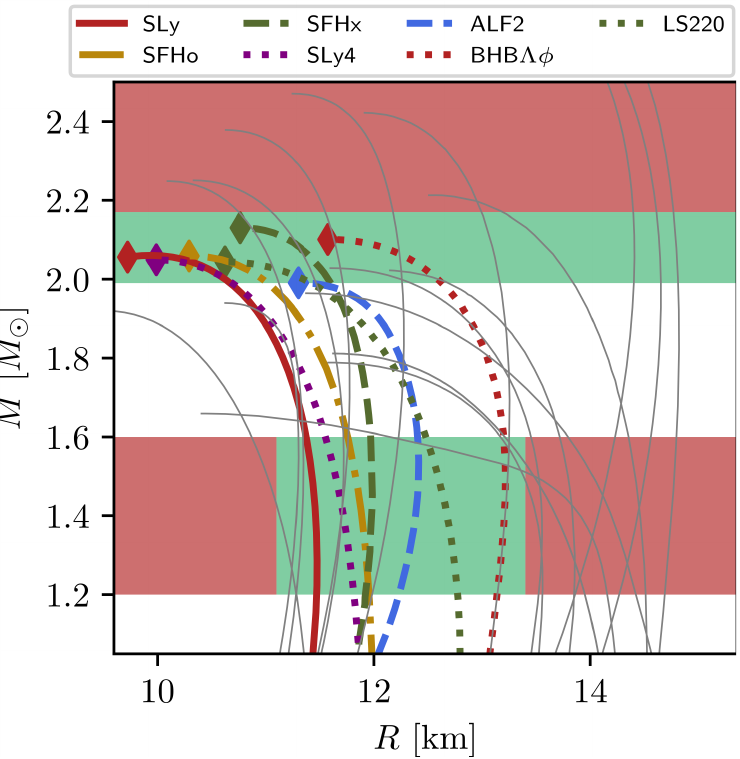}
\caption{Figure from \citet{coughlin2018multi}, reproduced with permission. The mass-radius relations for several representative \acp{EOS} are shown as lines. Constraints are shown in red and green; the viable models must fall into the green regions and must avoid the red regions. The lower band of constraints is set by the mass-radius constraints from multimessenger studies of \gwid. The upper red constraint corresponds to upper limits on $M_{\rm TOV}$ (also referred to in this paper as $M_{\rm Stable}^{\rm Max}$). These values come from the results in \citet{coughlin2018multi}. The lower limit of the upper green constraint is from the lower limit of $M_{\rm TOV}$ \citep{antoniadis2013massive}. Nonviable \ac{NS} \ac{EOS}, according to \citet{coughlin2018multi}, are shown in grey; viable models have thicker, colored lines and the names are labeled at the top. 
}
\label{fig:NS_EOS_mass_radius}
\end{figure}

Soft \ac{EOS} are those with lower pressure for a given mass density, which tend to have lower maximal masses. Stiff \ac{EOS} have higher pressures for a given mass density, and tend to have higher maximal masses. Above $\sim1.5\,M_\odot$ \acp{NS} have a peculiar property: heavier masses correspond to smaller radii, with as asymptotic behavior towards $M_{\rm TOV}$, which is the maximum mass of a stable, non-rotating \ac{NS}. Models exist for more exotic dense stars. We do not discuss these models in detail here, but note that constraints on the parameters relevant for the \ac{NS} \ac{EOS} will constrain viable exotic stars. For example, other stars can have significantly increased mass beyond the range typically considered for \acp{NS} (e.g. the three curves that reach the upper right of Figure \ref{fig:NS_EOS_mass_radius}).

Astrophysical constraints on \ac{NS} \ac{EOS}, prior to \gwid, were set by careful temporal or spectral observations of galactic \acp{NS}, as reviewed in \citet{NS-EOS-Ozel}. A variety of techniques are used, with varying levels of precision. The mass-radius relationship is the most well known prediction that astrophysical observations seek to constrain; examples from a representative set of \ac{NS} \ac{EOS} are shown in Figure \ref{fig:NS_EOS_mass_radius}. Determination of the \ac{NS} \ac{EOS} is important enough to warrant a space-based mission dedicated to this goal. NICER seeks to measure the mass-radius relation of three \acp{NS} through observations of their X-ray emission to 5-10\% precision \citep{gendreau2012neutron}. Current results from NICER begin to approach the 10\% limit in both mass and radius \citep{miller2019psr,riley2019nicer}. 

To be clear, constraining the \ac{NS} \ac{EOS} will \textit{not} provide a new understanding of the fundamentals of \ac{QCD} itself. This is why this section is distinct from Sect.~\ref{sec:fundamental_physics}\footnote{Also because that section is quite long, and this provides a natural split given the different necessary observations.}. Determining the \ac{NS} \ac{EOS} will inform on reliable methods to construct large-scale predictions from \ac{QCD} in an otherwise inaccessible regime, which can be informative for other purposes. 

For example, it is related to the thickness of the neutron skin of heavy nuclei, which are 18 orders of magnitude smaller in size \citep[e.g.][]{horowitz2001neutron}. Neutron skin thickness is probed with terrestrial laboratories \citep[e.g.,][]{horowitz2014electroweak}, providing comparable precision. Constraining these distinct but related properties over such a massive size range gives a particularly stringent test of our understanding of the large-scale behavior of \ac{QCD} in this density regime. The next generation ground-based experiments will provide early results in the mid-2020s.

A full understanding of such densities requires the intersection of nuclear physics and astrophysics. Following discussions in \citet{zhang2018combined}, we can approximate the energy per nucleon $E(\rho, \delta) \approx E_0(\rho) + E_{\rm sym}(\rho)\delta^2$, with the isospin asymmetry $\delta = (\rho_n-\rho_p)/\rho$ and $E_0(\rho)$ the energy in symmetric matter. $E_{\rm sym}$ captures the effects of the neutron-richness of the system. Determining the nuclear symmetry energy density dependence is a key goal for nuclear astrophysics \citep{LRP_US_2015,LRP_EU_2017}. $E_0$ and $E_{\rm sym}$ near the saturation density can be described with characteristic parameters of the \ac{EOS}. Some of these parameters are reasonably well determined from terrestrial experiments and can be used to inform astrophysical inferences, e.g. excluding a Stable \ac{NS} remnant for \gwid \citep{zhang2018combined}. Otherwise, improved measurements on the properties of \acp{NS} and understanding of the \ac{NS} \ac{EOS} can provide new understanding on the nuclear symmetry energy at supranuclear densities \citep{li2019towards}.

\ac{GW} detections of \ac{NS} mergers provide new ways to constrain the \ac{NS} \ac{EOS}. A review of inferences from the observations of \gwid is available in \citet{raithel2019constraints}. We discuss these here, as well as future prospects. The parameters of interest and how observations of \ac{NS} mergers constrain them are discussed in Sect.~\ref{sec:NS_merger_NS_EOS}. Using these measurements together to constrain the \ac{NS} \ac{EOS} is discussed in Sect.~\ref{sec:NS_EOS}. We close with a brief discussion on a potential unique \ac{QCD} measurement from \ac{GW} observations of \ac{BNS} mergers in Section \ref{sec:phase_transition_QCD}.

\subsection{Observables from neutron star mergers}
\label{sec:NS_merger_NS_EOS}
The unique constraints on the \ac{NS} \ac{EOS} utilizing \ac{NS} mergers generally rely on the \ac{GW} measurements. We briefly discuss some the relevant limitations of these measurements here.

Precise knowledge of the \ac{NS} masses is critical. As previously stated, the \ac{GW} measurement of the chirp mass in \ac{NS} mergers is precise, but the measurement of the mass ratio is usually not. From observations of galactic \ac{BNS} systems the individual masses cluster around $1.33 \pm 0.09$ \citep{Tauris-DNS-Systems,lvc_gwtc1}. This also implies that $q \approx 1$ for \ac{BNS} systems, a result generally confirmed by population synthesis models, which would allow for a precise value of the total mass of the system for these events, and much stronger constraints in the masses of the individual progenitors. 

The detection of GW190425 has shown that \ac{GW}-detected merger events do not closely follow the galactic mass distribution, but it is still consistent with $q\approx1$ \citep{abbott2020gw190425}. However, new measurements suggest between 2\% and 30\% of the total population of \ac{BNS} mergers will be asymmetric based on the identification of an asymmetric binary in the Milky Way \citep{ferdman2020asymmetric}. In what follows we generally assume that the majority of \ac{BNS} mergers will be reasonably symmetric, but this absolutely needs to be verified from observations of loud events. Should asymmetric cases be non-negligible all of the following science results will still be possible, but some will require a significantly larger number of events. We note that the low mass of \ac{BNS} systems allows for a reasonable determination of the total mass even if the mass ratio is not precisely known, e.g. for a fixed chirp mass varying q from 0.7 to 1 alters $M_t$ by only 10\%. This may bias measurements if \ac{BNS} mergers do not have $q\approx1$ as expected, but is reasonably accurate for current measurements.

The reason the mass ratio can not be measured for most of these events is its perfect correlation, at leading order, with one of the spin parameters. The highest observed dimensionless spin for \acp{NS} in galactic \ac{BNS} systems are $\vec{\chi} \approx 0.05$. The \ac{GW} parameter estimation can be run assuming this as the maximum allowed spin value for the individual \acp{NS}, referred to as the low-spin prior \citep[e.g.][]{GW170817_GW_updated_parameters}. This assumption allows for tighter constraints on the mass ratio, and therefore tighter constraints on the individual masses and total mass of the system. Again, this assumption is informed from prior \ac{EM} observation, appears to be reasonably valid, but should be tested with particularly loud events. This is demonstrated with GW190814 which has the most precisely measured secondary mass to date as it is both asymmetric and loud \citep{abbott2020gw190814}.

When discussing \ac{BNS} mergers there are two masses of interest: $M_b$ is the baryonic mass and $M_g$ the gravitational mass, which is $M_b$ minus the binding energy. Conservation of mass applies to $M_b$, but $M_g$ is the \ac{GW} observable. To determine the total mass of the remnant object in \ac{BNS} systems one must convert $M_{g,t}$ (with subscript denoting total, as before) into $M_{b,t}$, use mass conservation for $M_{b,\rm remnant} = M_{b,t} - M_{ej}$, then convert back to $M_{g,remnant}$. Below we refer to $M_{g, \rm remnant}$ as $M_{\rm remnant}$. An in-depth discussion of this is presented in \citet{gao2019relation}, who provide \ac{EOS}-insensitive relations to convert between these two.

Quasi-universal relations refer to properties or relations that appear preserved over a wide range of (still viable) \ac{NS} \ac{EOS}. In fact, there are some that exist without an intuitive reason as to why \citep[e.g.][]{yagi2013love}. There are also parametrized forms for \ac{NS} \ac{EOS} \cite[see, e.g.][and references therein]{GW170817_GW_updated_NS_EOS} that are used as phenomenological tools to allow for application of observational constraints to a wide range of \ac{EOS} and are powerful methods to advance our understanding in this complicated, interdisciplinary field of study. However, we note that there are known errors associated with these approaches. Any final assessment on the viability of a \ac{NS} \ac{EOS} may require the direct test of that \ac{EOS} if the measurement is near the level of the quantified errors.

\subsubsection{The maximum mass of neutron stars}\label{sec:max_NS_mass}
For the purposes of this subsection we refer to $M_{\rm TOV}$ as $M_{\rm Stable}^{\rm Max}$ for reasons that will be immediately obvious. We label the maximum mass of a \ac{NS} undergoing only isotropic rotation as $M_{\rm SMNS}^{\rm Max}$ and the maximum mass of a differentially rotating \ac{NS} as $M_{\rm HMNS}^{\rm Max}$. Note that $M_{\rm Stable}^{\rm Max} \leq M_{\rm SMNS}^{\rm Max} \leq M_{\rm HMNS}^{\rm Max}$. \ac{NS} \ac{EOS}-insensitive scalings relate these values \citep[e.g.,][]{cook1994rapidly,gao2019relation,koppel2019general}.

The determination of the immediate remnant object for \ac{BNS} mergers (Sect.~\ref{sec:remnant_determination}) relates to these quantities. If $M_{\rm HMNS}^{\rm Max} < M_{\rm remnant}$ the system will undergo prompt collapse; if $M_{\rm SMNS}^{\rm Max} < M_{\rm remnant} < M_{\rm HMNS}^{\rm Max}$ the remnant will undergo a \ac{HMNS} stage before collapsing to a \ac{BH}; if $M_{\rm Stable}^{\rm Max} < M_{\rm remnant} < M_{\rm SMNS}^{\rm Max}$ the remnant undergoes a phase of internal differential rotation before transitioning to a long-lived \ac{SMNS} stage; otherwise $M_{\rm remnant} < M_{\rm Stable}^{\rm Max}$ and the remnant object is a permanently stable \ac{NS}.

Then, we can use the multimessenger determination of the \ac{BNS} remnant classification to determine $M_{\rm TOV}$. This has been a promising prospect for science with \ac{NS} mergers for some time \citep[e.g.,][]{bauswein2013prompt,rezzolla2018using}. The broad approach discussed here was described in \citet{margalit2019multi}, who argue the detection of $\sim$10\, \ac{BNS} mergers in \acp{GW} with confident remnant classification would constrain $M_{\rm TOV}$ to the level of a few percent, under the assumption that $q\approx1$. This value is largely determined by the pressure at the core of the \ac{NS}.

The current constraints are $1.97\,M_\odot \leq M_{\rm TOV}$ and $M_{\rm TOV} \lesssim 2.3\,M_\odot$. The lower limit comes from observations of a galactic \ac{NS} with mass $2.01 \pm 0.04\,M_\odot$ from \citet{antoniadis2013massive}, which ruled out a significant fraction of soft \ac{EOS} that were otherwise viable \citep[e.g.][]{lattimer2012nuclear}. The maximum value is set by considering several published results \citep[e.g.][]{lu2015millisecond,shibata2019constraint}. We note that more stringent upper limits have been published \citep[e.g.][]{margalit2019multi}, but these results are somewhat in tension. Future observations will resolve this question.

Constraints may improve here with the discovery/characterization of galactic \acp{NS}, especially if they identify \acp{NS} with $M>2.0\,M_\odot$ (with small uncertainties). They will certainly improve with additional \ac{GW} detections of \ac{BNS} mergers (with sufficient \ac{EM} characterization). These multimessenger studies have the advantage that the number of known events will grow rapidly and that we can separately constrain the maximum mass of \ac{SMNS} and \ac{HMNS} with a large enough sample, which can provide additional information to constrain the \ac{EOS}. As shown in Figure \ref{fig:NS_EOS_mass_radius} these limits are powerful constraints on viable \ac{EOS} as they inform on an asymptotic limit (note the maximum $M_{\rm TOV}$ in that figure slightly differs from discussions here). Future constraints at the percent level will provide an incredible constraint on viable \ac{NS} \ac{EOS}.

\subsubsection{The lifetimes of metastable neutron stars}
There is another key parameter that can constrain the \ac{NS} \ac{EOS} that relies on the determination of the merger remnant, that may be unique to multimessenger studies of \ac{BNS} mergers: the lifetimes of \acp{HMNS} or \acp{SMNS}. From simulations, the lifetime of \acp{HMNS} is $\lesssim 1$\,s \citep[e.g.][]{GWs_neutrino_BNS_sekiguchi_2011} and the lifetime of \acp{SMNS} with massive magnetic fields and potential energy losses to \acp{GW} (as may be expected during these mergers) is $\sim10-10^5$\,s \citep[e.g.][]{ravi2014birth}. Though much uncertainty remains \citep[e.g.][]{baiotti2017binary}.

The lifetimes of these metastable \acp{NS} depends on the \ac{NS} \ac{EOS} (though not exclusively, e.g. the mass ratio). There have been studies on the effects of remnant lifetime on observable parameters, especially in the case of neutrino irradiation altering the colors of kilonovae \citep[e.g.][]{metzger2014red,lippuner2017signatures}. The lifetimes may be directly measured by \ac{GW} or neutrino observations, though this will not occur for a decade. We here propose a method to determine the lifetimes of either \acp{HMNS} or \acp{SMNS} (but not both). 

If only \acp{BH} are central engines of \acp{SGRB} (Section \ref{sec:central_engines}) then in the \ac{HMNS} case the jet cannot launch until the collapse of the \ac{NS}. Interesting constraints on this lifetime likely require $\sim1$\,s accuracy. Advancements in understanding the relative contributions of these terms and fortunate events (Section \ref{sec:GW-GRB_delay}) will allow an interesting measure. The most interesting scientific question related to this measure (Section \ref{sec:phase_transition_QCD}) requires precision that is likely impossible without direct measures.

Alternatively, if magnetars can power \acp{SGRB}, then we can directly determine the lifetime of a \ac{SMNS} from the non-thermal observations. This could either be the duration of the extended emission in gamma-rays following the prompt emission or the duration of the X-ray plateaus in the afterglow (Section \ref{sec:other_high_energy}). That is, the sharp drop in flux is expected to correspond to the collapse time of the \ac{SMNS}. Indeed such interpretations have already been applied to a small number of \acp{GRB} \citep[e.g.][]{2001ApJ...552L..35Z,2007ApJ...665..599T}. Multimessenger studies confirming these interpretation will enable they use to study the \ac{NS} \ac{EOS}.

Either option will provide insight on the internal dynamics of the \acp{NS}. The application of these measurements to a broad range of \ac{EOS} is not straightforward. Should an observational measure occur we would expect the necessary simulations to be performed.

\subsubsection{Tidal deformability}
The \ac{GW} determination of the tidal deformability\footnote{This is also referred to as mass quadrupole polarizability or tidal polarizability.} $\Lambda$ of a \ac{NS} is another parameter that can constrain the \ac{EOS} that is uniquely constrained from observations of \ac{NS} mergers, in this case with high-frequency \ac{GW} observations. The inspiral of \ac{NS} mergers are effectively identical to \ac{BBH} mergers with similar intrinsic parameters until very near merger, when the effects of matter and the larger size of \acp{NS} begin to be important. 

\acp{NS} undergo tidal deformation in an \ac{EOS}-dependent manner, which results in an acceleration of the inspiral, generally at $\gtrsim 500$\,Hz. This tidal deformability is often parametrized as

\begin{equation}
    \Lambda \equiv \frac{2}{3}k_2 \Big( \frac{R}{M} \Big)^5
\end{equation}

\noindent where $k_2$ is the quadrupole love number \citep{read2013matter}.  Note that some formulations utilize a parameter that scales as $R^6$ \citep[e.g.][]{de2018tidal}.
Detection of non-zero $\Lambda$ is how \ac{GW} observations alone can infer the presence of a \ac{NS}. Detection of $\Lambda$ is likelier to be measured for \ac{BNS} mergers as these inspirals are slower and the larger mass of the \ac{BH} can dominate the inspiral of \ac{NSBH} mergers, hiding the matter effects signature. In either case, a \ac{NS} with a larger radius will be more strongly deformed.

This parameter was constrained for \gwid. Several analyses provided upper limits on tidal deformability and some also claim lower limits \citep[e.g.][]{GW170817_GW_updated_NS_EOS,most2018new,raithel2018tidal,coughlin2018multi}, providing somewhat varied results. However, it shows the capability of \ac{GW} measurements of this parameter, suggesting the Advanced network can (or already has) constrain this parameter. Providing significantly more accurate constraints likely requires both improved \ac{GW} models that account for matter effects and improved high-frequency response beyond the current funded upgrades.

\subsubsection{The mass-radius relation}\label{sec:mr_relation}
We have already discussed how to infer the masses of the progenitor and remnant object. Then, measurement of the radius, largely determined by the pressure at $\rho \approx 2.5\rho_0$ \citet{lattimer2000nuclear}, enables constraints in the mass-radius plane. There are a few ways to do this with observations of \ac{NS} mergers.

The definition of the tidal deformability being proportional to a high power of the radius relates the parameters, and enables constraints on the \ac{NS} radius from \ac{GW} observations \citep[e.g.][]{GW170817_GW_updated_NS_EOS,omni_radio_optimization_Hotokezaka_2016}. These measurement are of order 10\% accuracy, approaching the precision goal of new \ac{EM} \ac{NS} \ac{EOS} instruments and analyses.

The determination of the merger remnant and the total mass of the remnant enables a constraint on the radius \citep{bauswein2017neutron}. This relies on the \ac{NS}-insensitive relation between $M_{\rm HMNS}^{\rm Max}$ and $M_{\rm Stable}^{\rm Max}$ (labeled differently between these papers), which depends on $R_{1.6}$, the radius of a 1.6M$_\odot$ \ac{NS} for a given \ac{EOS}. Assuming \gwid did not undergo prompt collapse they limit $R_{1.6} > 10.68_{-0.04}^{+0.15}$\,km and demonstrate the identification of a \ac{BNS} merger that undergoes prompt collapse would provide complementary upper bounds. \citet{koppel2019general} expanded this relation to the non-linear regime, giving a tighter constraint.

\ac{NSBH} mergers may also provide a similarly precise determination of the \ac{NS} radius \citep{foucart2012black}. With a reasonable sample of events, or fortunate single events, we can determine the conditions for tidal disruption of the \ac{NS} and the release of the \ac{EM} counterparts. As discussed in Section \ref{sec:overview_merger}, this condition is $r_{\rm{tidal}} > r_{\rm{ISCO}}$. The latter is easily calculated if we can determine $\chi_{\rm eff}$ of the \ac{BH}. $r_{\rm{tidal}}$ depends strongly on the \ac{NS} radius.

Therefore, \ac{BNS} and \ac{NSBH} mergers provide unique methods to constrain the allowable mass-radius relation of \acp{NS}. With \gwid the \ac{GW}-only observations achieved precision of 10--20\% on these parameters \citep{GW170817_GW_updated_NS_EOS}. As the Advanced network approaches design sensitivity this precision will improve.

Beyond even multimessenger studies within astrophysics, interdisciplinary studies in physics can also provide better constraints using astrophysical observations of these events. \citet{capano2019gw170817} combine the astrophysical observations of \gwid with detailed \ac{NS} \ac{EOS} from nuclear theory, which are more advanced than generally considered. They claim a \ac{NS} radius measurement of $11_{-0.6}^{+0.9}$\,km, or $\sim$7\% precision. This suggest future observations of nearby \ac{BNS} mergers will surpass the 5\% precision goal of other methods.

\subsubsection{Ejecta properties}
As described in detail in Section \ref{sec:overview_merger}, the properties of the ejecta that power the kilonova emission are tied to the \ac{NS} \ac{EOS}. This implies that observations of the kilonova ejecta can be tied back to the \ac{NS} \ac{EOS} and provide interesting constraints.

For example, \citet{radice2018gw170817} use the observed properties of \knid to set a lower limit on $\Lambda$, necessary to reproduce the total ejecta mass observed. With the upper limit on $\Lambda$ from the \ac{GW} observations they conclude that we prefer Goldilocks \ac{EOS}, that is, not too soft and not too stiff.

Similar inferences show the promise of multimessenger astronomy, and have been applied by a few groups \citep[e.g.,][]{coughlin2018multi}. We note that these constraints rely on simulations that reliably predicted the broad behavior of \knid before any kilonova had been well-observed and this is evidence supporting their claims. However, we caution against strong inferences based on the current implementations. The existing uncertainty in kilonova modeling (see Section\,\ref{sec:elements}) is not negligible. These methods will prove valuable when modeling has sufficiently advanced (e.g. more realistic initial conditions) and when they have been shown to reliably reproduce a sample of observed events.

\subsubsection{Post-merger gravitational waves}
The \ac{NS} \ac{EOS} should also have an observable imprint on the post-merger \acp{GW}, providing and independent \ac{GW} measure from the pre-merger signal. The most important frequency range is between $\sim$1--4\,kHz where the spectral behavior can be complex. For an overview on our understanding of these signals and prospects for their detection see \citet{clark2016observing}.

The primary frequency depends predominantly on the total mass and the \ac{NS} \ac{EOS} \citep{oechslin2007gravitational}. Detections of post-merger \acp{GW} from \ac{NS} mergers will be limited to particularly nearby events, where we may be able to reliable detector higher order modes in the inspiral and precisely measure the mass ratio. Otherwise we can infer the total mass from the chirp mass, so long as $q\approx1$ is valid. Thus, we can directly tie the primary frequency to tests of the \ac{NS} \ac{EOS}. There are secondary peaks whose prescence or absence will inform on the dynamics of the merger itself \citep{bauswein2015unified}. 

These tests rely on quasi-universal relations that have been confirmed for several \ac{NS} \ac{EOS} over a wide range of parameter space \citep[e.g.][]{2016PhRvD..93l4051R}. These tie the maximum \ac{GW} post-merger frequency to the tidal deformability and thus the radius of the \ac{NS} \citep{2018PhRvL.120c1102B}. These measurements provide additional constraints on parameters also measured through other means, with independent systematics, and in a much higher range of \ac{NS} mass. Therefore, we very strongly encourage an emphasis on high-frequency sensitivity for future \ac{GW} interferometers.


\subsection{Supranuclear matter and the equation of state of neutron stars}
\label{sec:NS_EOS}

Determining the \ac{NS} \ac{EOS} will be greatly aided by the general assumption that all \acp{NS} are governed by the same \ac{EOS}. This requires that any viable \ac{NS} \ac{EOS} has to be consistent with all observed properties of all observed \acp{NS}. It must simultaneously fall into the acceptable range in the mass-radius plane, be capable of producing (stable, non-rotating) \acp{NS} up to $M_{\rm TOV}$, have tidal deformability in the constrained range, live the proper amount of time when metastable, and release approximately the correct amount of ejecta \textit{and it must do so for every observed \ac{NS} merger}. It must also reasonably match the emitted \ac{GW} frequencies during merger for the different \ac{BNS} merger remnant cases, once it is measured. It must do all of this while also satisfying the \ac{NS} \ac{EOS} constraints from other observations, such as the new and upcoming results from NICER \citep{riley2019nicer,miller2019psr}.

We can further apply a few more requirements, as summarized in \citet{GW170817_GW_updated_NS_EOS}. The first is causality, limiting the sound speed at the highest pressure region for a $M_{\rm TOV}$ star to be less than the speed of light. Second, each \ac{EOS} must be self-consistent, e.g., it must be able to create the remnant object with the correct class for the total mass of the system while also be capable of sustaining the inferred properties of the progenitor \acp{NS}. Last, the \ac{NS} must be thermodynamically stable. 

Folding all of these methods into a coherent set of constraints greatly reduces the range of viable \ac{NS} \ac{EOS}. We look forward to the future constraints that the coming years of multimessenger astronomy will provide. As a closing remark on this science, we cite \citet{coughlin2018multi} which creates a multimessenger Bayesian framework to constrain the observed properties of the system based on \ac{GW}, kilonova, and \ac{GRB} constraints. While the results are model-dependent and make a number of assumptions, they provide the first demonstration of a technique that will likely prove invaluable for future studies of \ac{NS} mergers. Additionally, \citet{capano2019gw170817} demonstrate the promise of interdisciplinary work with non-astrophysicists.







\subsection{The phase transitions of quantum chromodynamics}
\label{sec:phase_transition_QCD}
We briefly discuss a prospect that has recently been identified as a potential science goal with \ac{NS} mergers. \ac{QCD} has phase transitions, analogous to the phase transitions of water. That is, regions of parameter space where the behavior of matter changes significantly, such as transition to a quark-gluon plasma at particularly high densities or pressure. 

\ac{GW} observations of \ac{NS} mergers may be able to constrain or measure a phase transition in \ac{QCD} by direct determination of the lifetimes of \acp{HMNS} and the peak \ac{GW} frequencies obtained during merger \citep{most2019signatures,bauswein2019identifying}. Both require \ac{GW} sensitivity at several kHz. The precision required on the merger lifetime likely necessitates direct determination through \ac{GW} (or neutrino) observations, as it is beyond the capability of what we can infer through indirect \ac{EM} methods. As this is potentially the only way to measure \ac{QCD} phase transitions we consider this a critical technical driver for future \ac{GW} interferometer development.































%% file: s08_fundamental.tex
Since 1900, our understanding of the Universe has fundamentally changed. Space and time were thought to be absolute until Einstein showed us space and time were relative and manifestations of spacetime \citep{Einstein_1905_special}. \ac{SR} has now been woven throughout modern physics. Newtonian gravity had stood since the Age of Exploration until it was supplanted by \ac{GR} \citep{einstein_GR_1916}, which has withstood a century of observational and experimental inquiries. We went from no inkling of quantum mechanics to the Standard Model of Particle Physics that brings quantum field theories for three fundamental forces into a single framework. Between them, these theories encompass all known fundamental forces of nature; they have been exquisitely tested, and, so far, observational evidence suggests they are largely correct, though incomplete. As it stands, the two theories are fundamentally incompatible. In a beautiful universe all forces could be described together, which, if possible, is generally thought to require a \ac{QFT} of gravity. However, gravity could be truly distinct, and a not a force in the particle physics definition (i.e., not governed by a gauge boson). For now, true unification eludes us. Observers should strive for ever more stringent tests of these theories until they breakdown and illuminate the path forward.

\ac{NS} mergers have important discovery space, providing insight into the behavior of neutrinos and \acp{GW} according to expectations from the Standard Model and \ac{GR}. The non-zero mass of neutrinos is the clearest evidence for the incompleteness of the Standard Model. The use of \ac{NS} mergers as standard sirens to break geometric correlations from observations of the \ac{CMB} help determine the absolute neutrino mass eigenstates (Sect.~\ref{sec:neutrino_mass}). Small timescales against cosmological baselines enable unmatched timescale ratio tests of fundamental physics. The short intrinsic timescales between messengers for \ac{GW}-\ac{GRB} detections enable for the most precise determination of the speed of gravity relative to the speed of light (Sect.~\ref{sec:speed_of_gravity}). The \ac{GR}-calculated frequency evolution of \acp{GW}, and the time of flight difference for \ac{GW}-\acp{GRB}, allow for precise determinations of dispersion within \acp{GW} (Sect.~\ref{sec:mg}) which also constrains the mass of the graviton (should it exist). A network of several \ac{GW} interferometers allows for searches of beyond-\ac{GR} \ac{GW} polarization modes (Sect.~\ref{sec:GW_polarization}). The comparison of \ac{GW} determined distance and \ac{EM} determined distance allow for searches of large extra dimensions (Sect.~\ref{sec:large_dim}) and, with an associated \ac{GRB}, for searches of gravitational parity violation through enhancement/suppression of the two \ac{GR} polarization modes (Sect.~\ref{sec:GW_parity}). The short intrinsic timescale for prompt \ac{GRB} emission over several decades of energy in \ac{EM} radiation allow for the best constraints on \ac{EM} dispersion (Sect.~\ref{sec:EM_dispersion}). 

Noting several of the other subsections are versions of \ac{LIV}, we discuss general \ac{LIV} in Sect.~\ref{sec:LorentzInvariance}. Lastly, we discuss multimessenger tests of the \ac{WEP} in Sect.~\ref{sec:WeakEquivalencePrinciple}. We also note these last two are two-thirds of the Einstein Equivalence Principle that underlie all metric theories of gravity \citep{Review_Tests_of_GR_overall_Will_2014}. Because more stringent tests will be performed through other means, we do not consider \ac{GW} dispersion as a science driver for \ac{NS} mergers, but discuss it here to show the unique time of flight tests these sources enable will be less constraining than other measures.

For general overviews on the known and expected behavior of particles we refer to the Particle Data Group reviews, \citet{review_particle_physics_Tanabashi_2018}. For a discussion on testing \ac{GR} in general we refer to \citet{Review_Tests_of_GR_overall_Will_2014}. For discussions on the tests of \ac{GR} with ground-based interferometers or \acp{PTA} we refer to \citet{review_GR_tests_with_GWs_yunes_2013} and to \citet{review_GR_tests_with_GWs_LISA_Gair_2013} for \ac{LISA}.

\subsection{The nature of neutrinos}\label{sec:neutrino_mass} 

Neutrinos are bizarre particles. They interact only with the weak nuclear force and gravity, are the lightest (known) particles by a factor of a million, and as a result only rarely interact. Their non-zero mass requires new physics beyond the Standard Model, their role in the primordial plasma have left an imprint in the Universe, and they are important carriers of astrophysical information. Studies of these particles have already resulted in four Nobel Prizes. Below we briefly summarize the evolution of our understanding of these particles and how observations of \ac{NS} mergers complement terrestrial experiments to further elucidate their nature. For a comprehensive overview of our current understanding and expected future results we refer the reader to the Particle Data Group reviews \enquote{Neutrino Masses, Mixing, and Oscillations} and \enquote{Neutrinos in Cosmology} \citep{review_particle_physics_Tanabashi_2018}. See \citet{mnu_general_review_2003} for a review of measuring absolute neutrino masses, \citet{mnu_mass_hierarchy_review_2015} for a review on the neutrino mass hierarchy, and \citet{mnu_right_handed_neutrinos_2013} for a review of right-handed neutrinos and their implications. The Hyper-Kamiokande design report contains a very nice overview for the state of existing, and prospects for upcoming, terrestrial neutrino experiments \citep{mnu_hyperkamiokande_design_report_2018}.

Neutrinos were first proposed by Pauli in 1930 \citep{neutrino_pauli_1930} to maintain energy and momentum conservation during beta decay\footnote{The original letter refers to them as \enquote{neutrons} as the particle now known by that designation was not yet known.}. A few years later Majorana published his relativistic wave equation valid for neutral fermions which are their own antiparticles \citep{neutrino_majorana_1937}. So-called \enquote{Majorana particles} can generate mass in a unique way and the original publication suggested the then-theoretical neutrino as a possible case. Within a few decades the neutrino was observed for the first time \citep{reines_cowan_1953, cowan_reines_1956}.


Particles with left-handed chirality have spin opposite to their direction of motion. For massive particles, helicity (effective chirality) differs for frames of reference moving faster or slower than the particle of interest, as the observed direction of motion inverts while the spin direction does not. Pontecorvo investigated the implications if neutrinos had non-zero masses and left-handed neutrinos oscillate into left-handed antineutrinos (and right-handed antineutrinos into right-handed neutrinos), analogous to neutral kaon oscillations \citep{neutrino_pontecorvo_1957,neutrino_pontecorvo_1958}. These would be two Majorana particles. Only left-handed neutrinos (right-handed antineutrinos) enter the weak interaction Lagrangian, i.e., they only interact with gravity. Therefore, right-handed neutrinos (left-handed antineutrinos) are referred to as \enquote{sterile} neutrinos.

In 1962, two important developments regarding neutrinos occurred. The muon neutrino was directly detected for the first time, confirming a second neutrino type \citep{neutrino_muon_danby_1962}. Neutrino flavors are named to match the charged lepton involved in their usual interactions (e.g., the electron neutrino $\nu_e$ involves interactions with electrons). That same year, Maki, Nakagawa, and Sakata developed the two-neutrino mixing matrix relating neutrino flavor to mass eigenstates \citep{neutrino_maki_1962}. We here describe the modern version of this matrix following \citet{mnu_hyperkamiokande_design_report_2018}. Formally, if we have neutrino flavors $\nu_\alpha$ which are superpositions of the neutrino mass eigenstates $\nu_i$ we have 

\begin{equation}
    \ket{\nu_\alpha} = \sum_i U_{\alpha i}^{\star} \ket{v_i}.
\end{equation}

\noindent $U_{\alpha i}$ is the unitary Pontecorvo--Maki--Nakagawa--Sakata matrix \citep[PMNS matrix; also known as the MNS matrix, lepton mixing matrix, or neutrino mixing matrix;][]{neutrino_maki_1962,mnu_oscillations_original_pontecorvo_1968}. For a modern description of the general matrix see \citet{review_particle_physics_Tanabashi_2018}. Here we discuss the case of three neutrino flavors (electron, muon, tau; $\alpha = e, \mu, \tau$) with three mass eigenstates (i=1,2,3)

\begin{equation}\label{eq:PMNS}
    U_{\alpha i} = 
    \begin{pmatrix}
        1 & 0 & 0 \\
        0 & c_{23} & s_{23} \\ 
        0 & -s_{23} & c_{23} \\
    \end{pmatrix}
    \begin{pmatrix}
        c_{13} & 0 & s_{13} e^{-i \delta_{CP}} \\
        0 & 1 & 0 \\
        -s_{13} e^{-i \delta_{CP}} & 0 & c_{13} \\
    \end{pmatrix}
    \begin{pmatrix}
        c_{12} & s_{12} & 0 \\
        -s_{12} & c_{12} & 0 \\
        0 & 0 & 1 \\
    \end{pmatrix}
    \begin{pmatrix}
        1 & 0 & 0 \\
        0 & e^{i \alpha_{21}/2} & 0 \\
        0 & 0 & e^{i \alpha{31}/2} \\
    \end{pmatrix}
\end{equation}

\noindent with $c_{ij} = \cos(\theta_{ij})$ and $s_{ij} = \sin(\theta_{ij})$. The matrix is characterized by three mixing angles $\theta_{ij}$ and three CP phase terms. The \enquote{Dirac} CP phase term is $\delta$; the Majorana CP phase terms $\alpha_{21}$ and $\alpha_{31}$ only matter if neutrinos are Majorana particles. These phase terms are still unmeasured.

Neutrinos from weak interactions are generated as definite flavors, but, from a basic wave equation, their mass eigenstates propagate at different velocities. As a result, propagating neutrinos are a superposition of flavors and can be detected as a different flavor than their origination. This is referred to as neutrino oscillation and its observation requires non-zero neutrino mass. 

The dominant fusion reactions in the Sun produce vast amounts of electron neutrinos. In 1967 Pontecorvo argued that neutrino oscillations would cause a deficit of solar electron neutrinos at Earth, relative to theoretical flux predictions \citep{neutrino_pontecorvo_1968}. Also in 1967, the Higgs mechanism \citep{Higgs_1,Higgs_2,Higgs_3} was incorporated into the electroweak interaction \citep{SM_1, SM_2}. The Higgs field gives rise to the mass of the Higgs boson itself. The Higgs mechanism explains the origin of the W and Z boson masses. Fermions (leptons and quarks) gain mass through Yukawa-type couplings between left-handed fermions, the Higgs field, and their right-handed counterparts \citep[reviewed in][]{review_particle_physics_Tanabashi_2018}.

The treatment of neutrinos in the Standard Model was constructed in a way that it matched existing observational data. It was thought that individual lepton numbers were conserved within a given flavor (e.g., electrons with electron neutrinos, etc). We have only observed left-handed neutrinos (right-handed antineutrinos). Under the assumption that right-handed neutrinos do not exist, neutrinos cannot gain mass through couplings with the Higgs field, requiring them to be massless in the Standard Model. The suggestion that particles other than gauge bosons could be massless is, in principle, fine, because masses are not predictions of the Standard Model.

So when the first results from the Homestake experiment in the 1970s suggested a deficit of Solar (electron) neutrinos people wanted to believe that either the experiment \citep{mnu_davis_1968} or the theory \citep{mnu_bahcall_1964} was wrong. Three decades later the Sudbury Neutrino Oscillation (SNO) experiment, built to detect all three neutrino flavors, directly confirmed the observed deficit arose from neutrino oscillation \citep{mnu_solar_SNO_2002}. A few years earlier Super-Kamiokande observations of muon neutrinos produced in atmospheric interactions had already proven neutrino oscillations due to matter interaction effects
\citep[][]{mnu_atmospheric_oscillation_SK_1998}. The incompleteness of the Standard Model has now been known for more than two decades. Modern studies of neutrinos observe oscillations from Solar and atmospheric neutrinos, as well as those produced in accelerators or reactors, in order to measure the values of the mixing angles ($\theta_{ij}$) and the squared differences of the neutrino mass eigenstates \citep[$\Delta m_{ij}^2 \equiv m_i^2 - m_j^2$;][]{review_particle_physics_Tanabashi_2018}.

We provide a historical overview as ideas about neutrinos that predate the formulation of the Standard Model may provide a path beyond it. There are several outstanding questions that experimental observations hope to answer \citep[e.g.,][]{review_particle_physics_Tanabashi_2018}:
\begin{itemize}
    \item What is the neutrino mass ordering? 
    \item What gives rise to mass in neutrinos? Are they Dirac or Majorana fermions?
    \item What is the value of the CP phase(s)?
    \item Is there a theory of flavor that encompasses both quarks and leptons?
    \item Why are neutrino masses so much smaller than other particles?
\end{itemize}

Depending on the answers to these questions, studies of neutrinos could uncover solutions to foundational problems. Neutrino parity violation could explain leptogenesis and baryogenesis in the early Universe \citep{neutrino_baryogenesis_fukugita_1986, neutrino_baryogenesis_kuzmin_1985}, explaining why the universe is filled with matter and not antimatter. They may prove Majorana particles exist and require a second method for mass generation. If right-handed \enquote{sterile} neutrinos exist \citep{mnu_right_handed_neutrinos_2013} they should be dark matter constituents. As one example of answering several questions at once, the Type I seesaw mechanism assumes that right-handed neutrinos exist and have Majorana mass terms \citep[see][for a more detailed explanation]{review_particle_physics_Tanabashi_2018}, which could explain the small neutrino mass arising from the difference between the electroweak symmetry breaking scale and the unification scale of the electroweak and strong interactions, baryogenesis, and (at least some) dark matter.

Understanding these particles requires all available information, and answering any one question can resolve or inform future work to answer the others. Ground based experiments have led much of our understanding of these particles and will continue to do so; however, astrophysical observations provide some unique information. The (effective) number of active neutrino species ($N_{\rm eff}$) and the sum of the neutrino mass eigenstates ($m_\nu = \sum_i \nu_i$) can be measured from observations of the \ac{CMB} \citep[e.g.,][]{Planck_2018_cosmo}. The latter is particularly important as the combination of the cosmological measurements of $m_\nu$ and of oscillation experiment measurements of $\Delta m_{ij}^2$ provides a method to determine the absolute values of the neutrino mass eigenstates.

If the values of the neutrino mass eigenstates are larger than the differences between them, then we have the \enquote{quasi-degenerate case} of $m_1 \approx m_2 \approx m_3$. Otherwise the ordering is hierarchical in nature, with either the \enquote{normal hierarchy} ($m_1 < m_2 < m_3$) or the \enquote{inverted hierarchy} ($m_3 < m_1 < m_2$). The resolution to this question will inform on theory \citep[e.g.,][]{mnu_review_2006} and expected results from future experiments (e.g., the search for the 0$\nu \beta \beta$ decay signature of Majorana particles is easier for the inverted case).

\begin{figure}
\centering
\includegraphics[width=1\textwidth]{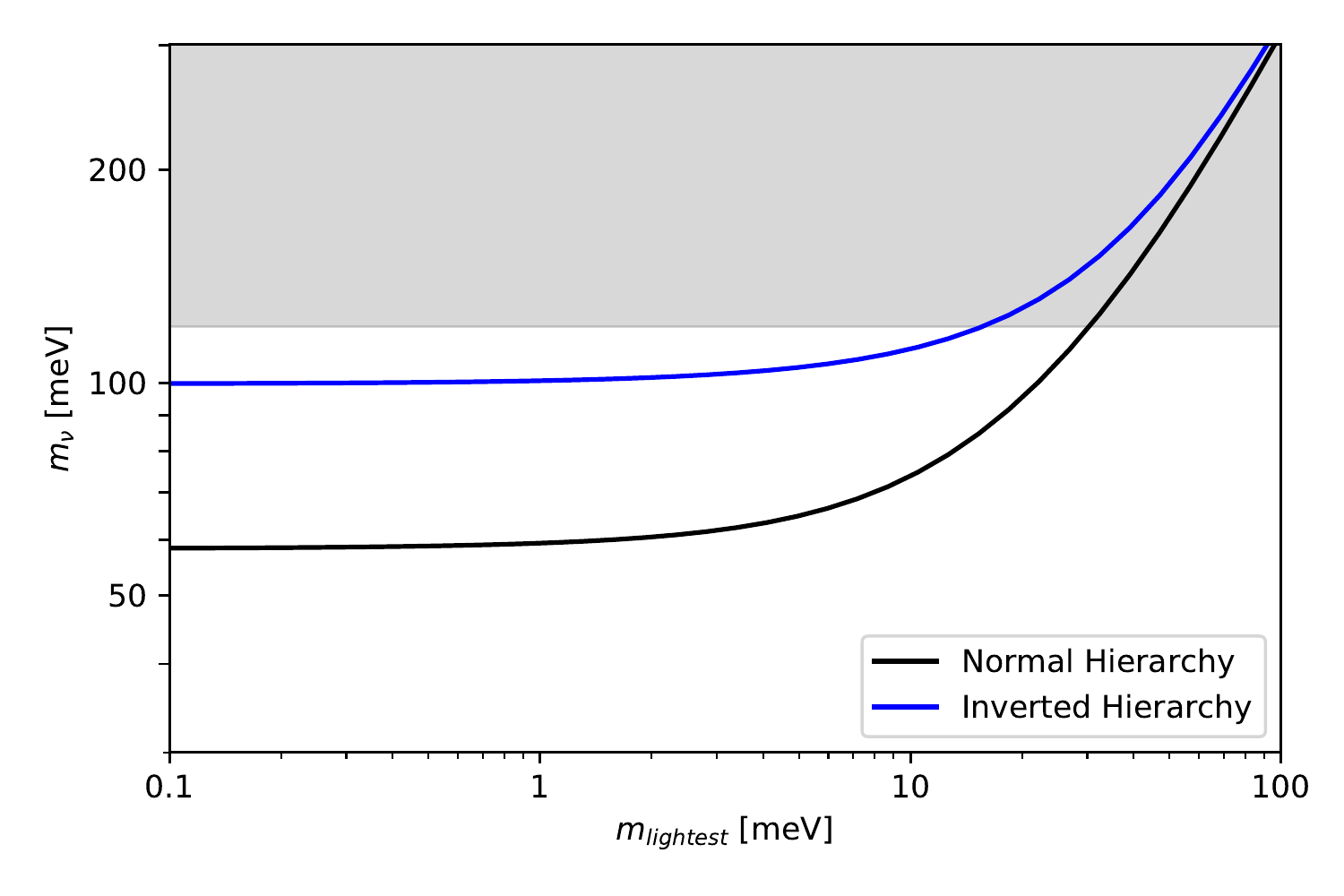}
\caption{The sum of the neutrino mass eigenstates as a function of the lightest eigenstate. The values are shown for both the Normal and Inverted Hierarchy ordering, with the current Planck 2018 results \citep{Planck_2018_cosmo}, which excludes the quasi-degenerate case. Standard siren cosmology will tighten these constraints.}
\label{fig:neutrino_mass_hierarchy}
\end{figure}

From combining Planck measurements, with degeneracy-breaking information and assuming base $\Lambda$CDM, $m_\nu < 0.12$ eV \citep{Planck_2018_cosmo}, which rules out the quasi-degenerate mass ordering. For the squared differences of the mass eigenstates, the current measured values are $\Delta m_{21} \approx 7.5 \times 10^{-5}$\,eV and $\abs{\Delta m_{32}} \approx 2.5 \times 10^{-3}$\,eV \citep{review_particle_physics_Tanabashi_2018}. Therefore, the lowest that $m_\nu$ can be in the inverted hierarchy case is $\sim$0.1 eV and in the normal hierarchy case $\sim$0.06 eV (see Fig.~\ref{fig:neutrino_mass_hierarchy}). 

As the scale of the Hubble parameter and $m_\nu$ and anticorrelated, the inclusion of the \citet{H0-Riess_2019} data may exclude the inverted hierarchy case but this is not done because of the disagreement in the value of $H_0$ \citep{Planck_2018_cosmo}. If the disagreement originates with a systematic error in the measure of $H_0$ from type Ia supernovae then information added by upcoming cosmological experiments will enable $m_\nu$ detection at $>3\sigma$ significance \citep[e.g.,][]{mnu_future_2011,mnu_future_2013,mnu_future_2016}. If the disagreement originates because the base 6-parameter $\Lambda$CDM is incomplete, then standard siren measurements become crucial as they improve $m_\nu$ measurements by $\sim$30\%, even with multi-parameter extensions \citep{mnu_cosmo_improvements_2018}. The timescales for using standard siren cosmology to study neutrino mass are well-suited to complement upcoming direct neutrino mass experiments \citep{mertens2016direct}.

We consider this a science driver for \ac{NS} mergers. As the required observation is the standard siren Hubble diagram, the required observations are the same as discussed in Sect.~\ref{sec:cosmology}. Time of flight tests to directly measure the neutrino mass are briefly discussed in Sect.~\ref{sec:LorentzInvariance}, but the use of \ac{NS} mergers is generally uninteresting as \ac{CCSNe} observed by the same detectors are likely to be more stringent, but still less constraining than the cosmological determination.

\subsection{The speed of gravity}\label{sec:speed_of_gravity}
The speed of gravity is infinite in Newtonian gravity. In the 1800s, attempts were made to include finite propagation speed into theories of gravity. In 1905, \ac{SR} showed that space and time are manifestations of spacetime, and defined {\it c} as the conversion factor between the two, where massless particles do not experience time and travel through space at speed {\it c}. In \ac{GR} \acp{GW} travel at {\it c} but there are some alternative theories of gravity where this may not be true. For example, some explained our evidence for the existence of dark energy as arising from $v_{\rm GW} \neq v_{\rm EM}$, where $v_{\rm EM}$ in a vacuum is $c$ \citep[e.g.,][]{cgw_dark_energy_Felice_2012,cgw_dark_energy_Bellini_2014,cgw_dark_energy_Gleyzes_2015}. Therefore, observational determination of the speed of gravity is critical. In this section we report constraints on $\delta_{\rm GW} \equiv (v_{\rm GW}-v_{\rm EM})/v_{\rm EM}$, the fractional deviation of the speed of gravity to the speed of light. Past, present, and expected future limits on this parameter are given in Table~\ref{tab:c_gw}.

Measuring the speed of gravity has proven difficult. The detection of high energy \acp{CR} at Earth constrains a subluminal graviton, as it would cause energy loss to gravi-Cherenkov radiation \citep{cgw_CR_first_Caves_1980,GR_cgw_CR_limit_Moore_2001}; however, we do not include this in Table~\ref{tab:c_gw} because it presupposes the existence of the graviton and is only valid for very high energy \acp{GW}. Observations of binary pulsars set two-sided limits of $\delta_{\rm GW} \lesssim 10^{-2}$ \citep{speed_of_gravity_pulsars_jimenez_2016}. The first direct measure came from comparing the signal arrival times between the LIGO interferometers from the merging of two black holes \citep{speed_of_gravity_LIGO_GW150914}. These are currently not particularly constraining ($\delta_{\rm GW} \lesssim 0.5$) but could improve to the $\sim$1\% level within a few years with several \ac{BBH} merger detections by a multiple interferometer network \citep{speed_of_gravity_multiple_BBH}. 

It was recognized that the multimessenger detection of a cosmological event in \acp{GW} and \ac{EM} radiation, where the two messengers are emitted close in time, would provide an extremely constraining measure of the speed of gravity relative to the speed of light \citep{GR_mg_Will_1998}. For \ac{NS} mergers we have the coalescence measured by \acp{GW} followed within seconds by a \ac{SGRB} in $\sim$keV-MeV gamma-rays. All other proposed \ac{GW}-\ac{EM} transients either have timescales that are orders of magnitudes larger or are galactic in origin (meaning orders of magnitude smaller baselines); therefore, \ac{GW}-\ac{GRB} observations of \ac{NS} mergers are the ideal multimessenger transient for this test. This method of constraining the speed of gravity was laid out in \citet{GR_mg_Will_1998} for nearby events. It was first (correctly) extended to cosmological distances in \citet{jacob2008lorentz}. \citet{nishizawa_2014_speed_of_gravity} directly applied this equation to \ac{GW}-\ac{GRB} studies. We summarize, modify, and develop arguments from those authors here, with the inclusion of new observational information. 

We are here concerned with the $\Delta t_{v_{\rm GW}}$ term from Eq.~(\ref{eq:prop_GWGRB}), assuming all other propagation terms are 0. We can write $\delta_{\rm GW} \approx v_{\rm EM} \Delta t_{\rm propagation}/d_c$ when $\Delta t_{propagation} \ll d_c/v_{\rm EM}$. The distance can be determined from the \ac{GW} measure of $d_L$ or an \ac{EM} determination of redshift and converted to $d_c$ with an assumed cosmological model. For two-sided constraints on the intrinsic time offset we can set sub- and superluminal constraints as

\begin{equation}
\delta_{\rm GW} < 
9.7 \times 10^{-17} \bigg( \frac{100\text{~Mpc}}{d_c} \bigg) 
\bigg( \frac{\Delta t_{\rm GRB-GW}-\Delta t_{\rm intrinsic,z}^{\pm}}{1 \text{~s}} \bigg). 
\end{equation}
\noindent Where $\Delta t_{\rm intrinsic,z}^{\pm}$ is described in Sect.~\ref{sec:GW-GRB_delay}. We could alternatively replace $1/d_c$ with $(1+z)/d_L$, depending on the specific observables or assumptions used for a given event.

\begin{figure}
\centering
\includegraphics[width=1\textwidth]{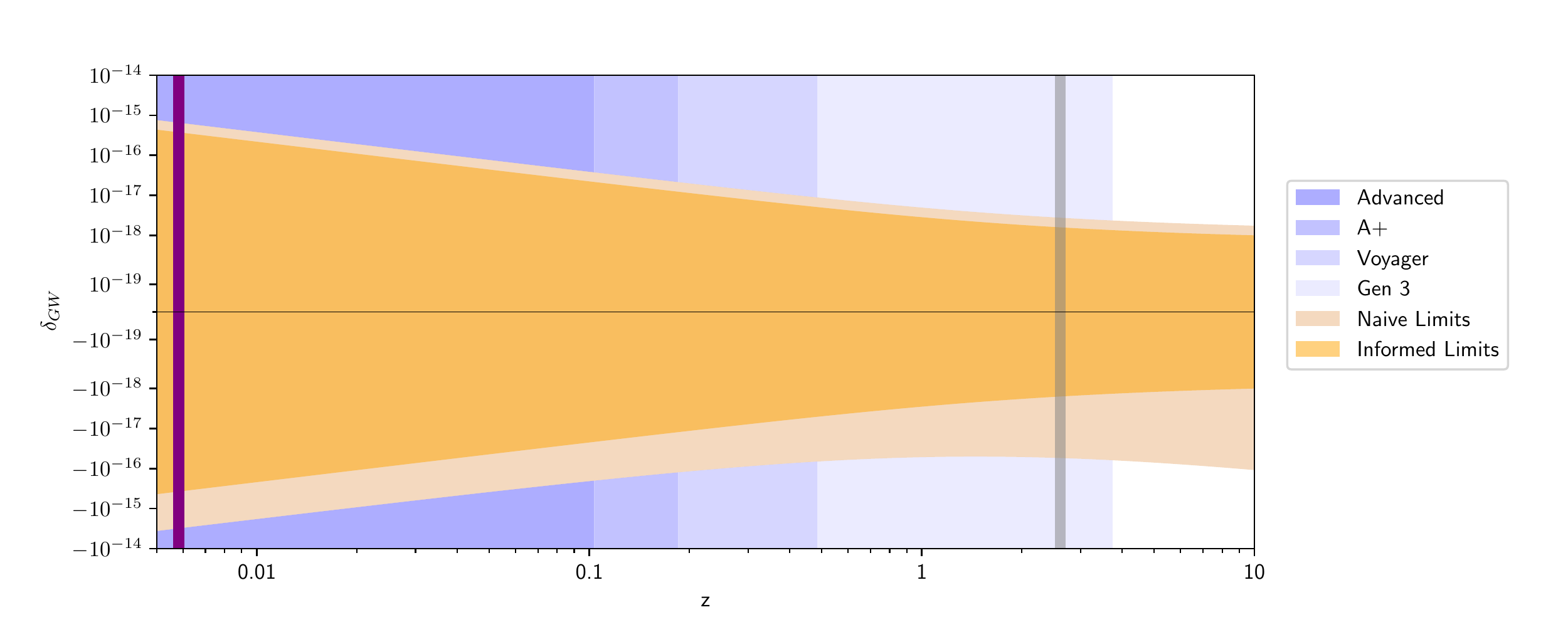}
\caption{Future $\delta_{\rm GW}$ constraints with individual joint \ac{GW}-\ac{GRB} measurements. The sensitivity for a detection at a given distance corresponds to where that redshift interacts with the Naive or Informed limits. The naive (tan) limits are from assuming the conservative intrinsic time offset of $[0, 10]$ s from \citet{GW170817-GRB170817A}. The constraints weaken at high redshift due to cosmological time dilation of the intrinsic time offset. The informed (orange) limits assume 1 s time offset uncertainty on either side and removes the cosmological time dilation of the intrinsic emission offset. The \gwid and \grbid distance is shown with the purple vertical line; the grey line shows the maximum claimed redshift for a \ac{SGRB}. The joint \ac{BNS} detection horizons for the assumed \ac{GW} network are taken from Sect.~\ref{sec:GW_observations}. Note that canonical \ac{NSBH} mergers could be detected to greater distances, corresponding to more stringent limits, within a given network sensitivity.}
\label{fig:speed_of_gravity}
\end{figure}

Indeed, the current best constraints are from the joint detection of \gwid and \grbid \citep{GW170817-GRB170817A}. With a conservative assumed distance, $\Delta t_{\rm intrinsic,z}^{\pm} = [0, 10]$ s, and neglecting redshift effects, we improved the measure by ten orders of magnitude. This single constraint has ruled out large classes of alternative theories of gravity \citep{cgw_beyond_GR_constraints_Creminelli_2017, cgw_beyond_GR_constraints_Baker_2017,cgw_beyond_GR_constraints_Ezquiaga_2017,sakstein2017implications}. There are additional theories that expect deviations at $\delta_{\rm GW} \sim 10^{-40}$, though no yet-imagined test could approach this limit. However, constraining the speed of gravity is intrinsically important as it is a fundamental parameter of the universe. Either we will measure $\delta_{\rm GW} = 0$ ever more precisely and further constrain where deviations from our understanding of fundamental physics occur, or we find one of those deviations. In Fig.~\ref{fig:speed_of_gravity} we show the expected sensitivities of the \ac{GW}-\ac{GRB} time of flight approach as the \ac{GW} detection distance for \ac{BNS} mergers improves, showing both the naive intrinsic time offsets used for GW170817 and GRB 170817A, and the informed $\Delta t_{\rm intrinsic,z}^{\pm}$ method. 

\begin{table}
\centering
\begin{tabular}{| c | c | c |}
\hline
Measurement	& $\delta_{\rm GW}$ Limits (or Sensitivity)	& Citation	\\
\hline
Binary Pulsar Limits & $\delta_{\rm GW} < | 10^{ -2} |$ &  \citet{speed_of_gravity_pulsars_jimenez_2016}  \\
GW150914 IFO Arrival Times & $ \delta_{\rm GW} < 1.7 $ &  \citet{speed_of_gravity_LIGO_GW150914}  \\
3 BBH IFO Arrival Times & $-0.45 < \delta_{\rm GW} < 0.42$ &  \citet{speed_of_gravity_multiple_BBH}  \\
\hline
GW170817-GRB 170817A (Naive) & $ -3 \times 10^{ -15 } < \delta_{\rm GW} <  7 \times 10^{ -16 } $ &  \citet{GW170817-GRB170817A}  \\
\hline
Future GW Network Arrival Times & $\delta_{\rm GW} \lesssim | 10^{-2} |$ &  \citet{speed_of_gravity_multiple_BBH}  \\
Strongly Lensed GW-GRB & $\delta_{\rm GW} \lesssim | 10^{-7} |$ &  \citet{cgw_strong_lense_Collett_2017}  \\
Gen 2 & $\delta_{\rm GW} \lesssim | 2 \times 10^{ -17 } |$ &  This Work  \\
Gen 3 & $\delta_{\rm GW} \lesssim | 2 \times 10^{ -18 } |$ &  This Work  \\
\hline
\end{tabular}
\label{tab:c_gw}
\caption{A summary of previous and potential constraints, or sensitivities, to $\delta_{\rm GW}$. Most of these measures are discussed in the text. The projected time of time values assume 1 s timing offset uncertainty on each side. The expected limits for a given \ac{GW} interferometer sensitivity are calculated for an event at 90\% of the assumed joint detection horizon for canonical \ac{BNS} mergers. As interferometers improve we can expect improvements over current constraints by several orders of magnitude. Not shown is the estimate for this measurement with a mid-range interferometer, which could detect these events to cosmological distances with much more precise determinations of the luminosity distance, perhaps surpassing the constraints of even the Gen 3 ground-based interferometers.}
\end{table}

We consider determination of the speed of gravity to be a science driver for \ac{NS} mergers, as it is fundamental in the universe and best done with these sources. The most important observational capability is increasing the maximum detection distance of ground-based \ac{GW} interferometers, through increasing their low-frequency sensitivity. Beyond the uncertainty on the intrinsic time delay, the dominant source of error is the uncertainty on the luminosity distance, which could be removed with a measured redshift and assumed cosmology, and the on-set of gamma-ray emission which can be tens of ms. The best measurement of the latter will come from prompt \ac{GRB} detectors sensitive to the tens of keV energy range, as these energies tend to precede the harder \ac{EM} emission.

\citet{nishizawa_2014_speed_of_gravity} also raised the prospect of using a population of \ac{GW}-\ac{GRB} detections to jointly determine the intrinsic time offset and further constrain the speed of gravity. In short, the intrinsic time delay will have an additional redshift term, allowing for joint constraints once \ac{GW} interferometers detect events to distances where cosmological redshift is no longer negligible. We add our support to investigations determining how precise these measures can be, but do not perform them here as such work could only improve these limits with the same observational requirements as individual detections. The most precise tests will be enabled through greater understanding of the intrinsic time delay.


\subsection{Gravitational dispersion and the mass of the graviton}\label{sec:mg}
In \ac{GR} \acp{GW} experience no dispersion. But like any aspect of \ac{GR}, there are alternative theories of gravity that expect the opposite. This section is included in the paper because \ac{NS} mergers provide a unique way to search for \ac{GW} dispersion, but as we will show, other investigations into \ac{GW} dispersion will prove superior to what is possible with \ac{NS} mergers. As such, this section contains fewer details. 

We can observationally search for dispersion by modifying the standard energy relation from \ac{SR} to $E^2 = p^2c^2 + A_\alpha p^\alpha c^\alpha$ with $E$, $p$, and $c$, their usual meaning and $A_\alpha$ and $\alpha$ parameters capturing the scale and type of dispersion \citep{mirshekari2012constraining}. This is the phenomenological form that the \ac{LVC} have used in tests of \gwid and the \acp{BBH} from GWTC-1 \citep{LIGO_GR_BBH_GWTC1,LIGO_GR_GW170817}. They consider values of $\alpha$ from 0 to 4 in steps of 0.5, excluding $\alpha=2$ where the effect is an achromatic alteration of the speed of \acp{GW} (see Sect.~\ref{sec:speed_of_gravity}). $\alpha=2.5,3,4$ correspond to specific beyond-\ac{GR} models \citep{mirshekari2012constraining}. For $A_0>0$ this test corresponds to a massive graviton, i.e., $E^2 = p^2c^2+m^2c^4$ from $m_g = A_0^{1/2}/c^2$. This case is useful for pedagogical purposes and projected sensitivities have been reported for several future instruments. We focus on this case, but note our conclusions apply generally (except where stated otherwise). More general formulations of these tests are available in \citet{dispersion_general_2} and \citet{dispersion_general_1} and further discussed in Sect.~\ref{sec:LorentzInvariance}.. 

Because of \ac{GR}, the speed of \acp{GW} being the speed of light, and the effects of gravity being felt on galactic scales it is generally expected that the graviton, if it exists, must be a massless spin-2 gauge boson\footnote{Inversely, it has been shown that if a massless spin-2 gauge boson exists it is a graviton.}. However, there are alternative theories of gravity where the graviton is massive \citep[for a review see][]{review_massive_gravity_Rham_2014}, though some difficulties  have yet to be worked out. Since we are unlikely to be able to directly detect gravitons in the next few years, limits on the mass of the graviton presuppose its existence and come from observations of natural extraterrestrial laboratories.

There are observable effects of a massive graviton. Several of the tests directly limit the Compton wavelength of the graviton, $\lambda_g = h/(m_g c)$, rather than $m_g$ itself. We list prior measurements and predicted future constraints in Table \ref{tab:m_g}. We do not discuss details here (which are available in the relevant citations).

\begin{table}
\caption{Constraints on the mass of the graviton. The top section are constraints from non-\ac{GW} observations, the middle from \ac{GW} observations of \acp{CBC}, the bottom from future expectations. Tests with \ac{NS} mergers are less sensitive than other methods. ToF stands for Time of Flight tests.}
\label{tab:m_g}
\begin{tabular}{| c | c | c | c | c |}
\hline
	&	$m_g$ (${\rm eV}/c^2$)	&	$\lambda_g$ ($km$) & Measurement & Citation	\\
\hline
Non-GW & $< 4.4 \times 10^{ -22 } $ & $> 2.8 \times 10^{ 12 } $ &  Solar System  &  \citet{GR_mg_Will_1998}  \\
 & $< 5.0 \times 10^{ -23 }$ & $> 2.5 \times 10^{ 13 } $ &  SMBH Superradiance  &  \citet{mg_superradiance_Brito_2013}  \\
 & $< 2.0 \times 10^{ -29 }$ & $> 6.2 \times 10^{ 19 } $ &  Clusters  &  \citet{gupta2018limit}  \\
 & $< 1.3 \times 10^{ -30 }$ & $> 9.8 \times 10^{ 20 } $ &  Weak Lensing  &  \citet{mg_weak_lensing_choudhury_2004}  \\
 & $< 7.6 \times 10^{ -20 }$ & $> 1.6 \times 10^{ 10 } $ &  Binary Pulsar  &  \citet{mg_binary_pulsar_Finn_2002}  \\
\hline
\ac{LVC} & $ < 9.5 \times 10^{ -22 }$ & $> 1.6 \times 10^{ 15 } $ &  GW170817  &  \citet{LIGO_GR_GW170817}  \\
 & $< 5.0 \times 10^{ -23 }$ & $> 2.5 \times 10^{ 16 } $ &  GWTC-1 \acp{BBH}  &  \citet{LIGO_GR_BBH_GWTC1}  \\
\hline
Future & $< 3.6 \times 10^{ -21 }$ & $> 3.4 \times 10^{ 11 } $ &  \gwid ToF  &  This Work  \\
 & $< 4.5 \times 10^{ -23 }$ & $> 2.8 \times 10^{ 13 } $ &  \ac{BNS} Merger ToF  &  This Work  \\
 & $< 1.6 \times 10^{ -23 }$ & $> 7.6 \times 10^{ 13 } $ &  ALIGO Future  &  \citet{GR_mg_Keppel_2010}  \\
 & $< 1.7 \times 10^{ -24 }$ & $> 7.1 \times 10^{ 14 } $ &  ET Future  &  \citet{GR_mg_Keppel_2010}  \\
 & $< 2.1 \times 10^{ -27 }$ & $> 5.9 \times 10^{ 17 } $ &  LISA Future &  \citet{GR_mg_Keppel_2010}  \\
\hline
\end{tabular}\end{table}

There are two tests for \ac{GW} dispersion that are of interest for \ac{NS} mergers. Time of flight tests can constrain the mass of a particle through measurement of its arrival time offset from a massless particle (or one with known mass and energy). Taking the energy of a massive particle in the usual form from \ac{SR}, with the definition of group velocity $v \equiv \partial p /\partial E \approx c(1-(mc^2)^2/2E^2)$ for light particles, a massive particle experiences a propagation delay of $\Delta t_{\rm massive} = (d_c/c) (mc^2)^2/2E^2$ compared to a massless particle. This allows for mass constraints of
\begin{equation}\label{eq:mass_delay}
m < E \sqrt{2 \Delta t_{\rm propagation} \frac{c}{d_c}}
\end{equation}
\noindent Using equation \ref{eq:mass_delay} and the time offset and distance values for \gwid and \grbid from \citet{GW170817-GRB170817A} we can constrain the mass of the graviton to $m_g < 3.6 \times 10^{-21}\ {\rm eV}/c^2$. For an exceedingly optimal \ac{GW}-\ac{GRB} conditions ($\sim$1 s known intrinsic time offset uncertainty, a few Gpc source distance, observations starting at 7 Hz) we could achieve $m_g < 5.0 \times 10^{-23}\ {\rm eV}/c^2$.

The second method is through waveform-deviation tests \citep{GR_mg_Will_1998}. This is, in effect, the same test. For a massive graviton, the inspiral of a \ac{CBC} would be altered due to relative propagation delays as a function of energy. That is, waves emitted earlier in the inspiral have a lower energy, and would thus arrive earlier than expected relative to the higher frequency waves emitted later in the inspiral. This method applies to all \ac{CBC}, with the current best limits from \acp{BBH} observed by \ac{LIGO} and Virgo \citep{LIGO_GR_BBH_GWTC1}. 

These limits are already more stringent than the best case option for the multimessenger \ac{GW}-\ac{GRB} test presented above, which is why the method was not performed in \citet{GW170817-GRB170817A}. This is due to the greater timing precision for observations of \ac{GW} inspirals than we can achieve for the \ac{GW}-\ac{GRB} time offsets. For waveform-deviations, the sensitivities depend on the distance to the source, and since \ac{BBH} mergers are more massive than \ac{NS} mergers, they should result in more stringent tests with the same \ac{GW} network, despite being observed over shorter frequencies. \ac{LISA} constraints are also greater than what can be achieved with ground-based interferometers. In the more general \ac{GW} dispersion tests, from \citet{GW_dispersion_future_Samajdar_2017}, ground-based interferometers and space-based interferometers are more sensitive to different $\alpha$ values. In neither case will \ac{NS} mergers be the most sensitive test. We do not consider constraining the mass of the graviton to be a science driver for \ac{NS} mergers. 


\subsection{Gravitational-wave polarization}\label{sec:GW_polarization}
Like \ac{EM} radiation, \acp{GW} are polarized. In \ac{GR} there are \enquote{plus} and \enquote{cross} tensor polarization modes. Such is the faith in \ac{GR} that all waveforms used in \ac{GW} searches and the description of the antenna patterns of \ac{GW} interferometers are constructed from these modes. However, generic metric theories of gravity allow up to six polarization modes: the two tensor modes of \ac{GR}, as well as two vector and two scalar modes \citep{eardley1973gravitational,will2018theory}, with some theories requiring all six \citep[e.g.][]{jacobson2004einstein}. Any detection of non-tensor \ac{GW} polarization would demonstrate a true failure of \ac{GR} while also strictly limiting the allowable beyond-\ac{GR} theories. We here follow the succinct description in \citet{LIGO_GR_GW170817}. For a more thorough summary of \ac{GW} polarization from beyond-\ac{GR} theories we refer to the discussion in \citet{Review_Tests_of_GR_overall_Will_2014} and references therein. 

The tensor modes ($A_+$ and $A_\times$) are transverse to the direction of propagation. The scalar modes are split between a transverse mode referred to as the \enquote{breathing mode} ($A_b$) and a longitudinal mode referred to as the \enquote{longitudinal} mode ($A_l$). Both vector modes ($A_{x}$ and $A_{y}$) are longitudinal. The GW strain measured by a given interferometer can be written as $h(t) = F^A h_A$ with $F^A$ the antenna response to the $h_A$ component of the signal; with all six considered A=($+$, $\times$, x, y, b, l). The response of an individual interferometer to a given polarization is determined by the detector orientation to the source, and we can constrain the contribution of polarization modes by enforcing consistency with observed signals in a network of interferometers \citep{GR_nontensor_polarization_test_PEF_chatziioannou_2012}. $F_+$ and $F_\times$ are the usual response functions; a derivation of the other four is available in \citet{poisson2014gravity}. 

The maximal test of \ac{GW} polarization then has a total of eight unknowns: the six polarization modes and the extrinsic direction to the source. However, the response of quadrupolar antennas to the scalar breathing and longitudinal modes are degenerate and cannot be distinguished, preventing delineation by such \ac{GW} interferometers \citep{Review_Tests_of_GR_overall_Will_2014,LIGO_GR_BBH_GWTC1}. Therefore, the most general possible test of \ac{GW} polarization by the ground-based \ac{GW} network has 7 unknowns. External determination of the source position is particularly powerful for these investigations as it enables precise knowledge of the relative arrival times at the interferometers. \ac{NS} mergers generally provide stronger tests of \ac{GW} polarization because the \ac{EM} counterparts allow for precise, external localizations, resolving two parameters. 


Simulations of studying additional \ac{GW} polarization modes with the ground-based \ac{GW} polarization network confirm these tests are possible \citep{GW_polarization_Takeda_2018}. The authors note a more precise measurement of the chirp mass, dependent on the duration of the signal, enables more powerful \ac{GW} polarization studies, further supporting the use of \ac{NS} mergers over \ac{BBH} mergers. We note that these tests are generally performed using waveforms constructed from \ac{GR} (but see the restricted waveform-deviation tests described in \citealt{arun2012generic}), implicitly assuming additional modes do not alter the behavior of merging compact objects \citep{isi2017probing}. This is a conservative assumption as alteration of the inspiral should produce more obvious deviations from \ac{GR}.

For transient signals, any less than five contributing interferometers results in an underdetermined system for the full test, but interesting tests can be performed with fewer contributing interferometers. The \ac{LVC} have performed basic tests of \ac{GW} polarization modes for mergers detected in LIGO-Hanford, LIGO-Livingston, and Virgo, first performed with the \ac{BBH} detection of GW170814 \citep{GW170814}. These studies compare the agreement of strain data for pure tensor modes against pure vector or pure scalar modes; mixed-mode results have not yet been reported. The current value from \ac{BBH} mergers \citep{LIGO_GR_BBH_GWTC1} is several orders of magnitude less stringent than from GW170817 due to use of the \ac{EM}-determined position \citep{LIGO_GR_GW170817}.

The simulations and current results demonstrate why \ac{NS} mergers with \ac{EM}-determined positions are best suited for tests of beyond-\ac{GR} \ac{GW} polarizations. Table \ref{tab:network_livetime} shows the fraction of time a given network has a number of active interferometers. The maximum distance for a full network detection is determined by the least sensitive interferometer. Additional modes can also be studied with the ground-based \ac{GW} network with continuous waves \citep{isi2017probing} or observations of the stochastic background \citep{callister2017polarization}. They may also be studied with \acp{PTA} \citep{review_GR_tests_with_GWs_yunes_2013}. In these cases the time to first detection is still somewhat uncertain.


As such we consider searches for additional \ac{GW} polarization modes to be science drivers for \ac{NS} mergers. These searches require multiple-interferometer detections, generally resulting in well-constrained \ac{GW}-only localizations within the detection horizon of the least sensitive contributing interferometer. Given this, and the importance of these investigations, we can safely assume an \ac{EM} counterpart will be found for such a detection. The number of (sufficiently sensitive) \ac{GW} interferometers is the only unresolved technical requirement. With current plans we will have LIGO-Hanford, LIGO-Livingston, Virgo, and KAGRA in the next few years, with upgrades and the addition of LIGO-India expected by $\sim$2026. We note LIGO-Hanford and LIGO-Livingston are coaligned (except for the curvature of Earth) to maximize detection prospects, which largely rules out their use as independent interferometers here. Such a network would enable searches for both vector or both scalar modes in addition to the \ac{GR} tensor modes. We support the investment into Virgo, KAGRA, and LIGO-India.

Prospects for a fully-determined test are somewhat pessimistic, unless an additional interferometer is constructed. One option would be to extend the second generation ground-based interferometers into the third generation era, which is an attractive option for several reasons (calibration, maintaining good localization capability for nearby events, etc). Future planned (\ac{LISA}) and proposed (e.g. mid-range interferometers, or the Einstein Telescope) triangular interferometer sets could contribute three independent measures \citep{review_GR_tests_with_GWs_LISA_Gair_2013,ET_science_objectives_sathyaprakash_2012}. These separate instruments could jointly observe sources that emit in the overlap frequencies between them, but such a possibility will not occur for a long time.

\subsection{Extra large dimensions}\label{sec:large_dim}
In \ac{GR} there are four dimensions of spacetime (D=4). Some alternative theories of gravity have a higher number of dimensions. We here discuss tests for additional large dimensions. Observational signatures include effects on the quasinormal modes of \ac{BBH} mergers \citep{large_dim_chakraborty_2018}, additional \ac{GW} polarization modes \citep{large_dim_andriot_2017}, and \enquote{leakage} of \acp{GW} amplitude into the additional dimensions \citep{large_dim_deffayet_2007}. The latter two are well-suited for study with \ac{NS} mergers. Additional polarization modes is discussed in Sect.~\ref{sec:GW_polarization}. Here we focus on \ac{GW} leakage, following \citet{large_dim_deffayet_2007}.

\ac{GW} observations of \acp{CBC} directly measure the luminosity distance to the source, assuming \ac{GR}, where $h \propto d_L^{-1}$. With extra large dimensions conservation of flux dictates that
\begin{equation}
    h \propto \frac{1}{d_L^{-(D-2)/2}}.
\end{equation}
\noindent With studies on \acp{GW} this leakage is generally invoked with a screening mechanism that asymptotes to \ac{GR} in the strong-field regime, maintaining \ac{GR}-predicted waveforms. We adopt the form from \citet{LIGO_GR_GW170817}:
\begin{equation}
    h \propto \frac{1}{d_L^{GW}} = 
    \frac{1}{d_L^{\rm EM} \Big[1+\big(d_L^{\rm EM}/R_c\big)^n \Big]^{(D-4)/2n}},
\end{equation}
\noindent with $R_c$ and $n$ respectively the distance scale and transition steepness of the screening mechanism. When $D > 4$ a given source will appear dimmer than a D=4 equivalent as energy is lost to the higher dimensions, causing the inferred luminosity distance to be greater than the real value. With \ac{GW}-only observations we would systematically measure a higher distance for all sources. With \ac{EM}-determined distances we can compare the two measures. A detectable difference arises only when light and matter propagate in four dimensions of spacetime while gravity may experience more, which is the case in many extra-dimensional theories of gravity \citep{LIGO_GR_GW170817}. 

This test has been performed using $d_L^{GW}$ for GW170817 and $d_L^{\rm EM}$ as from the distance to the host galaxy \citep{large_dim_pardo_2018,LIGO_GR_GW170817}, where the distance between the source and the host galaxy is small compared to the distance to Earth. In both cases they separately constrain parameter space for $D$, $R_c$ or $n$. The results are consistent with the 4 spacetime dimensions of \ac{GR} and constrain the characteristic screening scale as a function of transition steepness, with smooth transitions constraining $R_c$ through Hubble radius scales. With more distant \ac{NS} mergers and a sample of \ac{NS} mergers with both \ac{GW} and \ac{EM} determined distances these constraints will greatly improve. \citet{large_dim_pardo_2018} also limit the graviton lifetime through an amplitude dependent decay-length and test for large dimensions without a screening mechanism.

In \citet{LIGO_GR_GW170817} $d_L^{\rm EM}$ for NGC\,4993 is determined directly through surface brightness fluctuations from \citet{GW170817_distance_cantiello_2018}. This has the advantage of not relying on an assuming $H_0$ or cosmological model, but is limited to mergers in the nearby universe. In \citet{large_dim_pardo_2018} the distance is determined through the redshift measurement and an assumed $H_0$. Given the current disagreement in the value of $H_0$ results assuming each are presented. For future observations it will be necessary to transition to the latter method, which should occur on similar timescales to the standard siren measure of $H_0$ (which assumes $D=4$).

This test is uniquely performed by joint \ac{GW}-\ac{EM} detections. \ac{NS} mergers are the canonical example, and we consider this a science driver. However, we note that \ac{LISA} and partners may perform significantly more stringent measures \citep{large_dim_deffayet_2007}. Possible \ac{LISA} sources can be detected to a redshift of several and the precision of their $d_L^{\rm GW}$ measures \citep{cutler1998angular} is greater than third generation interferometers will achieve for \ac{NS} mergers. However, prospects for \ac{EM} detection, identification, and association of \ac{EM} counterparts to merging supermassive black hole binaries are promising, but speculative. The greatest prospects are for \ac{NS} inspiral observations with mid-range space-based interferometers \citep{cutler2009ultrahigh}.

\subsection{Gravitational parity}\label{sec:GW_parity}
A parity transformation inverts spatial coordinates ($x \rightarrow -x, y \rightarrow -y, z \rightarrow -z)$, creating an effective mirror image. This changes right-handed coordinate systems into left-handed ones (and vice versa). Parity is conserved when a system or process is identical in the original or inverted coordinate system, and violated when not. Such was the belief in the conservation of parity that it was referred to as a law of physics. This tenant remained unchallenged until \citet{lee1956question} suggested on theoretical grounds that weak interactions may not conserve parity, which was very shortly experimentally confirmed \citep{wu1957experimental,garwin1957observations}. In the Standard Model \ac{EM} and strong interactions are parity conserving but the weak interaction has maximal parity violation as its gauge bosons couple only to left-handed particles (right-handed antiparticles). Then, it is worth considering if gravity conserves or violates parity.

In \ac{GR} parity is conserved. Generic gravitational theories that are parity violating and still viable after \gwid and \grbid are now known to reduce to dynamical Chern-Simons gravity \citep{alexander2018gravitational,nishizawa2018parity}, an overview of which is available in \citet{alexander2009chern}. A theoretical motivation for such searches is the requirement of the Chern-Simons parity-violating term in some \acp{QFT} of gravity \citep[e.g.][]{alvarez1984gravitational,ashtekar1989cp,taveras2008barbero,weinberg2008effective} and as a potential explanation for baryogenesis \citep{PhysRevLett.96.081301,alexander2006can}. More fundamentally, we should be certain if parity violation occurs in only one or in two fundamental forces. Testing gravitational parity with direct \ac{GW} detections is reviewed with regards to ground-based interferometers and \acp{PTA} in \citet{review_GR_tests_with_GWs_yunes_2013} and \ac{LISA}-like detectors in \citet{review_GR_tests_with_GWs_LISA_Gair_2013}. We here closely follow the description in \citet{review_GR_tests_with_GWs_yunes_2013}.

Gravitational parity conservation requires left and right-handed circular polarizations to propagate equally. When gravitational parity is violated then this is not true and is referred to as amplitude birefringence\footnote{The term is used in analogy with \ac{EM} birefringence, but \acp{GW} are maintained as a single wave.}. We can write

\begin{equation}
    \begin{pmatrix}
    h_{+,k}(t) \\
    h_{\times,k}(t) \\
    \end{pmatrix}
    =
    e^{-ift/2\pi}
    \begin{pmatrix}
    u & iv \\
    -iv & u \\
    \end{pmatrix}
    \begin{pmatrix}
    h_{+,k}(0) \\
    h_{\times,k}(0) \\
    \end{pmatrix}
\end{equation}
\noindent where $f$ is the \ac{GW} frequency, $t$ is time, \(h_{+,k}\) and \(h_{\times,k}\) are the \ac{GW} Fourier components with wavenumber $k$. \(u\) accounts for curvature effects and is equal to 1 in a flat background; \(v\) captures the degree of birefringence and is equal to 0 in \ac{GR}. With the right and left-circular polarization components \(h_{R,L} = (h_+ \pm h_\times)/\sqrt{2}\), 

\begin{equation}
    \begin{pmatrix}
    h_{R,k}(t) \\
    h_{L,k}(t) \\
    \end{pmatrix}
    =
    e^{-ift/2 \pi}
    \begin{pmatrix}
    u+v & 0 \\
    0 & u-v \\
    \end{pmatrix}
    \begin{pmatrix}
    h_{R,k}(0) \\
    h_{L,k}(0) \\
    \end{pmatrix}.
\end{equation}

\noindent Thus, depending on the sign of \(v\), there is an enhancement of right-handed (left-handed) circularly polarized waves with the suppression of left-handed (right-handed) circularly polarized waves during propagation. The strength of this effect should accumulate based on the number of wavelengths experienced by the \ac{GW} over the full propagation distance, i.e., proportional to \(Df\) with \(D\) the distance to the source and \(f\) the \ac{GW} frequency \citealt{yunes2010testing}.

The importance of \ac{NS} mergers to these tests is evident from the previous paragraph, and was first described in \citet{yunes2010testing} whose conclusions we summarize here. As we are dealing with careful measures of \ac{GW} polarization, a precise determination of the position is extremely beneficial (see Sect.~\ref{sec:GW_polarization}). \ac{NS} mergers occur in the highest \ac{GW} frequencies that we are capable of detecting and are the most distant \ac{EM}-bright sources in that band. Lastly, an ideal observation for testing gravitational parity would be pure left or right-handed circularly polarized waves. Due to collimation and relativistic beaming (and presumed alignment of the jet to the total angular momentum axis) the detection of an associated \ac{SGRB} requires us to be nearly face-on, isolating nearly pure left or right-handed \acp{GW}. Further, such detections have improved constraints on the luminosity distance due to its correlation with inclination.

Gravitational parity violation would manifest as a disagreement in the luminosity distance as measured by \acp{GW}, $d_L^{\rm GW}$ (assuming \ac{GR}), against $d_L^{\rm EM}$ determined through \ac{EM} follow-up (either a direct distance measure of the host galaxy or through a measured redshift and an assumed cosmology). If the waves were the enhanced case then we would measure $d_L^{\rm GW} < d_L^{\rm EM}$  and in the suppressed case $d_L^{\rm GW} > d_L^{\rm EM}$. At least two interferometers, that are not coaligned (i.e. excluding LIGO-Hanford and LIGO-Livingston) are required to determine if the detected \ac{GW} is left or right handed, with additional interferometers providing tighter constraints.

\citet{yunes2010testing} note \ac{NSBH} mergers provide more stringent constraints owing to their greater detection volume\footnote{They consider a 30$M_\odot$ BH which is not expected to result in a \ac{SGRB}, but the general statement is correct.}. They show population-level analysis improves constraints approximately as $1/\sqrt{N}$, where a bimodality in $d_L^{GW}/d_L^{\rm EM}$ would be evident. Lastly, they also consider less than ideal scenarios, such as a wide-angled \ac{SGRB}, and show it only marginally weakens such searches. Full constraints require the separate detection of a left-handed and right-handed events.

To this last discussion we add one further suggestion to the ideal \ac{NS} merger for tests of gravitational parity. From \citet{fong2015decade} the range of half-jet opening angles for \acp{SGRB} goes from a few degrees to $>25^\circ$. These values may not be perfectly valid because they were calculated assuming top-hat jets, but they demonstrate we may have \ac{GW}-\ac{GRB} detections at tens of degrees from the angular momentum axis, further shown with the off-axis detection of \grbid. This event was sub-optimal due to this and the \ac{GW} signal being significant only in coaligned interferometers. Robustly determining the inclination of the system to Earth will fully enable this test which requires detecting a narrowly collimated on-axis burst or, for off-axis bursts, determination of a narrow $\theta_i$. Kilonova observations cannot provide a stringent enough constraint on inclination angle. The best determination will be done with observations of a jet break in the \ac{GRB} afterglow which generally requires initial detections on the order of a few hours and sufficient follow-up to late-times. Therefore, the ideal event is a high \ac{SNR} \ac{GW} detection of a \ac{NS} merger with at least two interferometers, with an associated prompt \ac{GRB}, and confirmation of a small inclination angle. Note the last requirement requires a narrow angle as the measurement of the jet break constrains us only to be within that angle, not our angle within the jet. 

There are other methods to study gravitational parity on cosmological scales. \ac{LISA} will detect massive \ac{BBH} binaries to high redshifts. If gravitational parity is violated then \ac{LISA} will observe a change in the apparent orientation of the system as a function of time \citep{alexander2008gravitational}. Despite the vastly greater detection distances, a face-on \ac{GW}-\ac{GRB} detection would provide more stringent constraints, due to the higher \ac{GW} frequencies involved \citep{yunes2010testing}, and this test will be available a decade sooner. This is true with the current generation of ground-based interferometers and would vastly improve with third generation interferometers. Gravitational parity violation may also cause observable effects in \ac{GW} generation, which could be identified with the observation of a spin-precessing \ac{BBH} or a spinning \ac{NSBH} binary, with a further enhancement for eccentric systems \citep{alexander2018gravitational}. This test would likely be more sensitive, but the detection rates for these systems is unknown. Alternatively tests of gravitational parity can be done with studies of the stochastic background \citep{crowder2013measurement}. We note there are non-\ac{GW} tests for parity violation that could be far more stringent \citep{dyda2012vacuum}. 

However, the technical drivers necessary for this test are already, or soon-to-be, met. This test requires continued improvements to the ground-based \ac{GW} network to increase both distance to which \ac{NS} mergers can be detected in \acp{GW}, an increase in the high-frequency detection range, and an increase in the rate of \ac{GW}-\ac{GRB} detections. It requires all-sky coverage with \ac{GRB} detectors and is greatly aided by the capability to detect \ac{GRB} afterglow emission within about a day in \ac{GW}-\ac{GRB} localization regions. We note that a typical cosmological \ac{GRB} that is face-on and within the detection horizon of the current ground-based interferometers would be sufficiently bright that the prompt \ac{GRB} detectors and follow-up instruments may not need to be particularly sensitive. As we move to more sensitive \ac{GW} interferometers that detect \ac{NS} mergers to distances where even on-axis \acp{GRB} are difficult to detect, the \ac{GW}-\ac{GRB} detection rates will be sufficiently high that we should not require vast improvements to the non-thermal \ac{EM} detection capabilities. However, instruments sensitive to \ac{GRB} emission will improve the population-level constraints.

\subsection{Electromagnetic Dispersion}\label{sec:EM_dispersion}
\ac{QG} may result in observable energy-dependent propagation delay, which is also a signature of \ac{LIV}. \ac{EM} dispersion is then motivated by searches for evidence of \ac{QG}.  The next section contains the scientific importance of these studies in a larger context and is not discussed here. This section is distinct to match the separation of these types of tests in the literature. 

Due to having short intrinsic timescales and detections at cosmological distances, \acp{SGRB} provide stringent tests of \ac{EM} (\textit{in vacuo}) dispersion. For pedagogical purposes we use the effective low-energy field theory formulation from \citet{LIV_090510_Vasileiou_2013}, which assumes $E \ll E_{QG}$ with $E_{QG}$ the scale of \ac{QG} effects. For a massless particle we can perform a series expansion

\begin{equation}\label{eq:LIV_base}
E^2 \approx p^2c^2 \bigg[ 1 - \sum_{n=1}^\infty \xi \Big(\frac{E}{E_{QG}}\Big)^n \bigg]
\end{equation}
\noindent where $E$, $p$, and $c$ have their usual meanings. $\xi$ represents the sign of the violation: for the subluminal case it is equal to +1 (with a negative correlation between photon speed and energy) and -1 for the superluminal case (a positive correlation between photon speed and energy). Such a dispersion would lead to

\begin{equation}\label{eq:LIV_speed}
v_{\rm EM}(E) = \frac{\partial E}{\partial p} \approx c \bigg[1 - \xi \frac{n+1}{2} \Big( \frac{E}{E_{QG}} \Big)^n \bigg].
\end{equation}
\noindent Note that this equation is considering only the dominant term, which is not necessarily for $n=1$ depending on the specific theory considered. Generally, specific terms and signs of violation are considered separately. For two particles of different energies, $E_h > E_l$, from the same source and emitted at the same time, this will induce an arrival delay

\begin{equation}\label{eq:LIV_delay}
\Delta t_{LIV} =  \xi \frac{1+n}{2 H_0} \frac{(E_h^n - E_l^n)}{E_{QG}^n} 
\kappa_n
\end{equation}

\noindent where

\begin{equation}
    \kappa_n \equiv \int_0^z \frac{(1+z')^n}{\sqrt{\Omega_\Lambda + \Omega_M(1+z')^3}} dz'
\end{equation}
\noindent is a comoving distance modified by the order of \ac{LIV} \citep{jacob2008lorentz}. Therefore, the best constraints on $\Delta t_{LIV}$ come events with high energy photons, with low-energy photons as a baseline, that originate at cosmological distances, with small (or well known) intrinsic time offsets. For these tests, often only linear and quadratic \ac{LIV} ($n=1, 2$) are (separately) considered; the small intrinsic time offset is most important for linear tests. 

The use of \acp{GRB} to probe dispersive \ac{LIV} limits was first performed in \citet{GR_quantum_gravity_amelino_1998} using bursts detected by \ac{CGRO}. The detection of high energy photons from cosmological distances and intrinsic impulsive behavior allow for two-sided measures on \ac{LIV}. The primary instrument on the \Fermi Satellite is the \ac{LAT} and the secondary is the \ac{GBM}, which together can detect \acp{GRB} from $\sim$10 keV to $\sim$10s of GeV. A year into its mission, both instruments on-board the \Fermi Satellite detected GRB 090510 \citep{GRB090510_discovery_Fermi}. The burst is a best-case scenario for these tests: it was detected deep into the universe (at a redshift of 0.9), was detected to high energies with a low-energy base, and the impulsive nature constrained the intrinsic emission time offset between low and high energy photons to of order a second. This burst holds the current best limits for linear dispersion, pushing the scale of \ac{QG} beyond the Planck scale, and competitive limits for the quadratic case \citep{LIV_090510_limit_2009,LIV_090510_Vasileiou_2013}. From the duration and hardness of GRB 090510 it is likely a \ac{SGRB}, and likely originates from a \ac{NS} merger.

Significant improvements to these measurements with existing gamma-ray telescopes is unlikely. GRB 090510 is about a once a decade event for \Fermi. \ac{VHE} detections enable improved constraints, but must overcome attenuation for cosmological \ac{VHE} photons. Based on the first detection of a \ac{GRB} at \ac{VHE} with the long GRB 190114C \citep{MAGIC_190114C} and the marginal signal following the short GRB 160821B \citep{MAGIC_160821B} we could expect unambiguous detections of \acp{SGRB} with the upcoming \ac{CTA}. However, these limits would be one-sided as they rely on follow-up detections of non-impulsive \ac{GRB} afterglow emission. Existing wide-field \ac{VHE} instruments have yet to detect any \ac{GRB} (they should be capable, but the necessary events are rare). We likely require either wide-field \ac{VHE} instruments with improved sensitivity or a partnership with \ac{GW} early warning systems and \acp{IACT}.

\subsection{Lorentz Invariance}
\label{sec:LorentzInvariance}

The Standard Model is a quantum description of three of the four (known) fundamental forces. \ac{GR} is a classical description of gravity. If the four forces are to be unified, we almost certainly require a quantum theory of gravity. The scale of interest where \ac{QG} effects may become important ($E_{QG}$) is expected to be of order the Planck Scale, $E_{PL} \equiv \sqrt{(\hslash c^5)/G} \approx 1.2 \times 10^{19}$\,GeV. Lorentz Symmetry is the underlying assumption of Relativity that the laws of physics are the same for all observers with no preferred frame. If there is a fundamental length scale of the Universe, then there is an inertial reference frame where that length is an extrema. Therefore, these two axioms are mutually exclusive and searches for \ac{LIV} are motivated by the quest for \ac{QG}. We note that the unambiguous detection and confirmation of the breaking of Lorentz Symmetry would rewrite textbooks, but setting forever more stringent limits on \ac{LIV} is unlikely to be particularly useful for theoretical development. Therefore, this section is written in terms of the possible sensitivities to \ac{LIV} with a given test, rather than projected future constraints.

For reviews on theory implications of \ac{LIV} we refer the reader to \citet{QG_general_Smolin_2008, QG_LIV_Mattingly_2005, QG_LIV_jacobson_2006}. In brief, \ac{LIV} from \ac{QG} models have been explored for loop quantum gravity \citep[e.g.,][]{QG_LQG_gambini_1999, QG_LQG_rovelli_2008}, string theory \citep[e.g.,][]{QG_string_kostelecky_1989,QG_string_ellis_1999}, and warped brane worlds \citep[e.g.,][]{QG_WBW_Burgess_2002}. \ac{LIV} and Planck-scale effect investigations are also important for non-commutative geometry \citep[e.g.,][]{QG_noncommutative_douglas_2001}, varying speed of light cosmologies \citep[e.g.,][]{LIV_varying_c_moffat_1993, LIV_varying_c_magueijo_2003}, cosmologically varying moduli \citep[e.g.,][]{LIV_varying_moduli_damour_1994}, spacetime-varying couplings \citep[e.g.,][]{LIV_spactime_varying_kostelecky_2003, LIV_spactime_varying_bertolami_2004}, emergent gauge bosons \citep[e.g.,][]{LIV_EGB_kraus_2002}, a consistent theory of \acp{BH} \citep[e.g.,][]{QG_LQG_rovelli_2008}, the prevention of high-energy divergences in \acp{QFT} \citep[e.g.,][]{LIV_BH_solodukhin_2011}, spacetime foam \citep[e.g.,][]{LIV_foam_amelino_1997}, deformed relativity \citep[e.g.,][]{QG_deformed_amelino_2002}, and condensed matter analogues of emergent gravity \citep[e.g.,][]{LIV_superfluid_analogues_volovik_2001}.

There is only one way that Lorentz Invariance can be preserved and numerous methods of violation.  To enable the comparison of a wide range of theories against a wide range of observational tests of \ac{LIV}, the \ac{SME} framework was developed. It is a comprehensive effective field theory description for tests of \ac{LIV}, which includes CPT violation. Its series expansion from \citet{LIV_090510_Vasileiou_2013}, and using $n$ from Sect.~\ref{sec:EM_dispersion} rather than the typical mass dimension ($d=n+4$),

\begin{equation}\label{eq:SME}
\Delta t_{LIV,n} = \frac{1}{H_0} \bigg( \sum_{jm} {}_{0} Y_{jm}(\hat{n}) c_{(I)m}^{(n+4)} \bigg) \kappa_n
\end{equation}

\noindent with $\hat{n}$ the direction of the sources, ${}_{0} Y_{jm}(\hat{n})$ spin-weighted spherical harmonics, and $c_{(I)m}^{(n+4)}$ the framework coefficients representing the \ac{LIV} strength. This expansion encapsulates directional-dependent violation, dispersive and non-dispersive constraints, birefringent and non-birefringent constraints, and it allows for specie-specific tests by separately considering the photon, gravity, neutrino, and matter sectors. \citet{SME_summary} contains a summary of the best constraints of each \ac{SME} parameter, with an annually updated document available on the arXiv\footnote{\url{https://arxiv.org/abs/0801.0287}}, where several of the best existing limits arise from observations of \ac{NS} mergers.

Observationally probing Planck-scale effects is difficult. $E_{PL}$ is several orders of magnitude larger than the highest energy particles ever observed. Astrophysical observations provide some of the best coefficient constraints because very small effects can build up over cosmological baselines into observable effects. Again the short intrinsic time offsets for cosmological detections enable \ac{NS} mergers to provide some of the best discovery space. Second, the multimessenger nature of these sources enable us to use constraints in one sector to probe \ac{LIV} in other sectors.

As discussed in Sect.~\ref{sec:EM_dispersion}, in the \ac{EM} sector \ac{NS} mergers hold the record for linear dispersion and large discovery space for dispersion in general. These limits are non-birefringent, but still theoretically motivated \citep[see discussion in][]{LIV_090510_Vasileiou_2013}. The \ac{EM} birefringent limits are several orders of magnitude more constraining \citep{kostelecky2008astrophysical}. In short, the effect of \ac{EM} birefringence manifests in different propagation speeds for left or right-handed photons, splitting a beam into two components. The detection of linear polarization from a distant source then severely constrains birefringent \ac{LIV}, as the rotation rate over cosmological baselines has to be incredibly tiny. The detection of polarization from \acp{GRB} provide the most stringent \ac{LIV} limits, of which \acp{LGRB} are the better candidate due to their higher fluence.

These constraints allow us to use multimessenger detections to constrain non-birefringent violation in the other sectors by observing the relative arrival times of different messengers. This was done with \gwid and \grbid and improved some non-dispersive constraints in the gravity sector by ten orders of magnitude \citep{GW170817-GRB170817A}, for largely the same reasons as discussed in Sect.~\ref{sec:speed_of_gravity}. \ac{NS} mergers have the greatest discovery space for these kinds of \ac{LIV} tests. As discussed in Sect.~\ref{sec:mg} other observations have larger discovery space for dispersive \ac{LIV} in the gravity sector.

The best, unambiguous, non-dispersive limits on the neutrino sector come from SN 1987A \citep[e.g.][]{neutrinos_speed_1987A_longo_1987,neutrinos_speed_1987A_stodolsky_1988}. Under the assumption that the $\sim$200 TeV neutrino IceCube-170922A association to the Fermi-LAT blazar flare from TXS 0506+056 is true \citep{IceCube170922A_Blazar_2018} and using the gamma-rays as the low-energy baseline, dispersive and non-dispersive (using the gamma-rays as the low energy signal and the high energy neutrino) are improved by orders of magnitude \citep{LIV_neutrino_Ellis} compared to the observations of SN 1987A \citep[e.g.][]{LIV_SN1987a_Ellis}. Should high energy neutrinos be detected from \acp{SGRB} it could shatter these limits given the small timescales and cosmological baselines. The best dispersive limits come from observing the relative arrival times for neutrinos with measured energies in SN 1987A \citep{LIV_SN1987A_Murayama_2001,LIV_neutrinos_Kostelecky_2011}. When MeV neutrino detectors are capable of detecting \ac{NS} mergers they will likely detect at least an order of magnitude more \ac{CCSNe} which can provide a similar test \citep{neutrino_mass_stellar_collapse}. In neither case do we consider \ac{NS} mergers to be critical, given the uncertainties on detection prospects.

Altogether, \ac{NS} mergers have large discovery space for searches of \ac{LIV}. Within the \ac{SME} framework, these sources are critical for dispersive (non-birefringent) measures in the photon sector and non-dispersive measures in the gravity sector. We consider these science drivers for \ac{NS} mergers as detection of \ac{LIV} would usher in a new era of physics.

\subsection{The Weak Equivalence Principle}\label{sec:WeakEquivalencePrinciple}
The \ac{WEP} states that gravitational and inertial masses are identical. It is the outcome of Einstein's famous elevator thought experiment, though similar ideas had been around before the formulation of \ac{GR}. Multimessenger detections provide a unique test of \ac{WEP} by testing if different messengers experience gravity differently. It is a test of the foundation of gravitational theory itself. We first present here the field-standard test method for pedagogical purposes. We close this section with a discussion on the problems with the values from this approach and a new proposed test.

Particles which traverse gravitational potentials undergo a propagation delay due to the warping of spacetime. This was first described by \citet{shapiro_delay_1964} as a fourth test of \ac{GR} by checking for predictions matching the observed time delay due to the gravitational well of the Sun by observing radar pulses reflecting off Venus and Mercury near conjunction. The \ac{PPN} parameter $\gamma$ measures the amount of curvature of space due to unit rest mass \citep{Review_Tests_of_GR_overall_Will_2014}. Its value in \ac{GR} is 1, but this value is not unique to \ac{GR}. Other theories of gravity have $\gamma \neq 1$. Shapiro delay depends on $\gamma$ as:

\begin{equation}\label{eq:shapiro_delay}
\delta t_{s} = - \frac{1 + \gamma}{c^3} 
\int_{r_e}^{r_o} U(r(l))dl
\end{equation}
\noindent where U(r) is the gravitational potential along the path $l$, with integration limits from the distance of emission $r_e$ to observation $r_o$.  

If two particles follow the same path through a gravitational potential but couple to different spacetime metrics (i.e. experience gravity differently) then they would experience different Shapiro de,lays, inducing a relative propagation arrival time. We define the relative Shapiro delay for particles 1 and 2, constraining the term $\Delta t_{\rm WEP}$ from Eq.~(\ref{eq:prop}) which is defined as
\begin{equation}\label{eq:shapiro_delta}
\Delta t_{s1-s2} = 
t_{s1} - t_{s2} 
= \frac{\gamma_2 - \gamma_1}{c^3}
\int_{r_e}^{r_o} U(r(l))dl.
\end{equation}

\noindent For two sided constraints we can rewrite Eq.~(\ref{eq:shapiro_delta})
\begin{equation}\label{eq:shapiro_two_sided}
\gamma_2 - \gamma_1 < 
c^3 \frac{(\Delta t_{\rm GRB-GW}-\Delta t_{\rm intrinsic,z}^{\pm})}{ \int_{r_e}^{r_o} U(r(l))dl }
\end{equation}

Equation~(\ref{eq:shapiro_two_sided}) (or its one-sided version) has been used to constrain deviations between messengers and within messengers. Such observations determine if the trajectory of particles are the same, a test of the \ac{WEP}. The first multimessenger test of \ac{WEP} was between photons and neutrinos with SN 1987A, which showed that neutrinos obey \ac{GR} to the limits of the measurement \citep{WEP_1987A_Krass_1988,WEP_1987A_Longo_1988}. These constraints can be improved using the likely association of IceCube-170922A to the flaring Fermi-LAT blazar \citep{IceCube170922A_Blazar_2018,blazar_WEP_2019a}. Tests have also been performed within photons, \acp{GW}, and neutrinos \citep[e.g.][]{WEP_1987A_Longo_1988,WEP_GRBs_Gao_2015,WEP_FRBs_Wei_2015,WEP_blazars_wei_2016,WEP_GWs_kahya_2016}. Several of these relative constraints exceed the best absolute bounds $\gamma_{\rm EM}-1 = (2.1 \pm 2.3) \times 10^{-5}$ from tracking of the Cassini spacecraft during a close alignment with the Sun \citep{WEP_absolute_Bertotti_2003}.

All of these analyses should be performed: any deviation between or within messengers would have profound implications for the Universe. \enquote{Dark Matter Emulators} were alternative theories of gravity that claimed some of our evidence for dark matter arose from light coupling to a different metric than gravity \citep{EP_dark_matter_eumlators_Desai_2008, EP_dark_matter_emulators_Kahya_2011}. While the evidence for dark matter vastly exceeded what such theories could explain, it would not necessarily be surprising if light and gravity coupled to different spacetime metrics, given the rules that govern force interactions with other particles. \gwid provided the first opportunity to use gravity in a relative test of the \ac{WEP}. \grbid provides the best partner for this test, giving 
\begin{equation}\label{eq:gw_grb_wep_value}
-2.6 \times 10^{-7} 
\leq \gamma_{\rm GW} - \gamma_{\rm EM} \leq
1.2 \times 10^{-6}.
\end{equation}
\noindent \citep{GW170817-GRB170817A}. This measure ruled out most \enquote{Dark Matter Emulators} \citep{cgw_GW170817_dark_matter_Boran_2018}.

\ac{NS} mergers again provide some of the best discovery space. The emission in the first few seconds includes emission over several decades of energy of photons and \ac{GW}, and likely the same for neutrinos. \ac{GW}-\ac{GRB} joint detections are almost certain to set the best relative bounds for \acp{GW} and photons. Should \acp{SGRB} be detected in the prompt phase as neutrinos, then they are also likely to set the best relative constraints for photons and neutrinos. Given the broad energy range within \ac{EM} radiation for prompt \ac{GRB} emission and in \ac{GW} radiation during the strongly chirping inspiral, \ac{NS} mergers will also likely result in the most stringent limits within these messengers. For individual gravitational potentials, these constraints depend weakly on the distance to the events (e.g., improving constraints by a factor of 2 with \gwid and \grbid following the prescription in \citealt{GW170817-GRB170817A} would require a joint detection at 5\,Gpc). Events with smaller intrinsic time offsets, or a greater understanding of that distribution, will provide more stringent tests.

Beyond the observed temporal offset, the dominant parameter for improving these constraints is the total gravitational potential experienced over the paths of the particles of interest. The limits in Equation \ref{eq:gw_grb_wep_value} account only for the Milky Way's gravitational potential, assuming a conservative mass, a Keplerian potential, and integrated from 26 Mpc (the 95\% lower bound of $d_L$ as measured from \gwid) to within 100 kpc. Other papers attempted to account for more of the Milky Way's gravitational potential \citep[e.g.,][]{GW_GRB_WEP_Shoemaker_2018} or contributions from intervening gravitational potentials \citep[e.g. contributions from the Virgo supercluster;][]{GW_GRB_WEP_Wei_2017}. Forecasting future constraints with this method is difficult given the range of possible paths from source to Earth (e.g., contributions from the host galaxy, intervening galaxies or clusters). With our prior note of caution, we do not attempt to provide them here. Improvements to our understanding of the total mass of the Milky Way and the shape of its gravitational potential (e.g., as a result of the \textit{GAIA} or \ac{LISA} missions) will enable more precise statements on relative violations of the \ac{WEP}, and can be applied \textit{ex post facto} to prior joint detections. The most stringent constraints possible would be from a lensed \ac{GW}-\ac{GRB}, where the mass of the lensing system could exceed the Milky Way's by orders of magnitude.

We now discuss the issues with this test, from \citet{minazzoli2019shortcomings}. The formulation of Eq.~(\ref{eq:shapiro_two_sided}) uses an implicit coordinate system that is not gauge-invariant, i.e., depending on the coordinate choice one can obtain positive or negative values \citep{gao2000theorems}. In \citet{minazzoli2019shortcomings} they consider Fermi coordinates associated to an observer, such that the delay is expressed in terms of an observed proper time, which results in a sum of terms with opposing sign. Using reasonable constructions for contributions to the total gravitational well experienced by propagating particles by considering the sum of catalogs they show the induced total (absolute) Shapiro delay is not monotonic, and indeed crosses zero in some cases. Therefore, while we can use the small offsets for, e.g., \ac{GW}-\acp{GRB} to state that we find no evidence for \ac{WEP} violation, we cannot quantify robust limits on relative Shapiro delay. \citet{minazzoli2019shortcomings} suggest this as motivation for the development of tests for specific alternative theories of gravity. We note that if the \ac{WEP} is violated we will find evidence for it, we will just be unable to quantify the degree of violation with this test.

\citet{minazzoli2019strong} propose a new multimessenger test of the \ac{WEP} using strongly lensed events. From the time delay between two images $i$ and $j$ we can apply the usual \ac{PPN} parametrization $\Delta_{ij} \rightarrow (1+\gamma)/2 \Delta_{ij}/2$. By measuring the time delay between these images in two different messengers one can quantify the search for relative \ac{WEP} violation by

\begin{equation}
    \gamma_2-\gamma_1 = 2\frac{\Delta_{ij}^2-\Delta_{ij}^1}{\langle \Delta{ij} \rangle}.
\end{equation}

\noindent The requirement for detecting and identify a strongly lensed \ac{GW}-\ac{GRB} is not an easy one, but the rate should be non-zero with Gen 3 \ac{GW} interferometers and a suitable \ac{GRB} monitor.

%% file: s09_recommendations.tex
The preceding sections have clearly demonstrated astrophysical observations of \ac{NS} mergers enable phenomenal scientific return. Making and reliably interpreting these observations requires input from observers, theorists, and simulation, and advancement in other fields of physics. Below we highlight some recommendations on areas where these needs may not be met, or support existing efforts. In very broad terms these are guided by the following criteria:

\begin{itemize}
    \item A deep understanding of what occurs during \ac{NS} mergers will enable greater scientific return from these sources. Observations of the inspiral, coalescence, and early times from \ac{GW} and neutrino observations nicely complement the \ac{EM} observations of the \ac{SGRB} jet and quasi-isotropic outflows. 
    \item With \gwid and its counterparts the loss of \ac{EM} detection in any energy range would have resulted in significant loss of science. This will also be true for future events. Broadband \ac{EM} coverage is necessary. 
    \item Observations that enable early localizations are crucial to enable sensitive characterization of these events. 
\end{itemize}

We make a number of recommendations on observational resources required for a given messenger (separating the types of photons), the necessary communication improvements to enable time-domain multi-messenger astronomy, and comment on both the necessary work outside of astrophysical observations and the difficulties inherent to interdisciplinary work. 

The previous science and above comments apply generally; however, some of the recommendations below are focused on U.S.\ interests, given the on-going Astro2020 Decadal process (which decides the large mission prioritization of the U.S.) and because I am most familiar with this system. We directly discuss the \ac{GW} interferometers. For \ac{EM} missions we directly discuss some proposed large-scale ($\gtrsim$\$1B) missions proposed to the Astro2020 Decadal, as well as similar scale missions that are in advanced proposal rounds outside of the U.S. For smaller scale missions we make broad recommendations only.

\subsection{Observational resources}
The following sections discuss the existing, planned, proposed, and possible observational capabilities in \acp{GW}, the \ac{EM} spectrum, and neutrinos. To prevent repetition we make the following blanket statements of support:

\begin{itemize}
    \item For missions that have dedicated instrument teams we support sufficient funding to adapt to the new era of \ac{GW} multimessenger astronomy.
    \item For instruments that determine observing time through guest investigator/observer programs we support the allocation of sufficient resources to the observations of \ac{NS} mergers, as well as the necessary prioritization of these observations. When competing proposals for relevant science enter the same round, we suggest the selection of those with community service and prompt open data aspects.
    \item When there are proposed missions that significantly advance the capabilities in a given energy range, we support those missions. We support technological development funding for the cases where sufficient advancements are not yet ready.
\end{itemize}

\subsubsection{Gravitational waves}
\ac{GW} observations are necessary for the majority of the science discussed in this paper, either directly through their own observations or indirectly by identifying \ac{NS} mergers for follow-up. It is widely understood that advancements in \ac{GW} observations are necessary, so we do not summarize why here. The past, present, and funded \ac{GW} network is shown in Fig.~\ref{fig:GW_obs_plan}. Discussions on this and proposed interferometers is in Sect.~\ref{sec:GW_observations}.

We recommend sustained investment into the ground-based gravitational wave interferometers. Improved sensitivity in the $\sim$10--1000\,Hz range will greatly increase the rates of detections of \ac{NS} mergers, enabling population studies. We generally require a network of several interferometers of comparable sensitivity to provide reasonably accurate localizations for most multimessenger studies. This has been directly demonstrated during O3 by the LV detection of GW190425, and a few other events. Under current plans this need will be met in the late Advanced era as Virgo and KAGRA sensitivities become more comparable to Hanford and Livingston. The funded improvements to A+ and similar upgrades for other interferometers are critical. Further, with a sufficiently large number of Advanced/+ interferometers, the downtime can be staggered to allow for continuous \ac{GW} observations. We support consideration of this endeavor to ensure we capture rare, interesting events.

If fewer than three 3rd generation interferometers are funded then the localizations will not be sufficiently constrained for multimessenger studies of these sources. Additionally, the detected events will nearly all be too far to recover kilonova at these distances. One potential solution would be to maintain (sufficiently upgraded) second generation interferometers into the third generation era, allowing for well-constrained localizations for events nearby enough for successful \ac{EM} follow-up. This is also advantageous for other reasons (e.g. early calibration of the new interferometers).

The currently funded upgrades do not significantly broaden the frequency range at which \ac{GW} detections of \ac{NS} mergers can occur. Sensitivity at high frequencies, that is a few to several kHz, is of paramount importance to studies of \ac{NS} mergers. They allow direct observation of merger, and potentially ringdown. This gives some of the greatest tests on the \ac{NS} \ac{EOS}, will allow conclusive classification of more systems from the immediate \ac{GW} detection, the direct determination of the immediate remnant object in \ac{BNS} mergers, and all of the science derived from that knowledge. We support funding to advance the necessary technologies until they are implemented into the existing network. If possible, this upgrade could be included into the A+ network, which would be commensurate with several upcoming facilities (e.g. \ac{EM} upgrades, the nuclear experiment FRIB, and the MeV neutrino experiment Hyper-Kamiokande). Alternatively, building an interferometer focused on this frequency range and utilizing it in partnership with lower frequency interferometers is likely to be a viable solution \citep[e.g.][]{ackley2020neutron}.

For longer-term investment a space-based mid-range interferometer brings unique capabilities. It is the only method to achieve early warning of \ac{NS} mergers with precise localizations far enough in advance to enable broadband \ac{EM} observations of merger time. It is uniquely suited to enable precise standard siren cosmology, broadband studies of the prompt \ac{GRB} emission, population-level studies of the early kilonova emission, and several tests of fundamental physics. These observations would be truly remarkable. We support heavy investment into the technologies necessary for such missions to allow one to be launched as soon as possible.

\subsubsection{Gamma-rays}
We support the extension of the \Fermi mission as \ac{GBM} is the most prolific detector of \acp{SGRB}. We also support the continuation of the \Swift mission as, despite lower joint detection rates, the immediate arcminute localization of a \ac{GW}-detected \ac{NS} merger would allow immediate follow-up across the \ac{EM} spectrum, and an enormously informative dataset. These missions should be extended at least until suitable replacements are launched.

An ideal instrument would provide precise localizations of large numbers of \acp{SGRB}. It is difficult to vastly increase the \ac{SGRB} detection rates with partial coding masks, and the localization accuracy of \ac{GBM}-like instruments is generally limited. Construction of a fourth \ac{IPN} appears unlikely, as it is now difficult to place astrophysics instruments on planetary spacecraft, and the data downlink latency would remain slow. The best option to balance detection rates and localization precision is the construction of a large Compton telescope.

In Astro2020 the only relevant large-scale mission proposal is the Compton+pair conversion telescope AMEGO \citep{mcenery2019all}. AMEGO would have a \ac{SGRB} detection rate roughly an order of magnitude higher than \Swift \ac{BAT}, with $\lesssim$degree accuracy. For events with sufficiently high energy photons, to be detected through pair conversion, these localizations may be smaller. The \ac{LGRB} detection rate is measured in hundreds per year. These would allow greater population studies of these sources and enable study of the prompt, afterglow, and other non-thermal emission in the tens of MeV, where it has not yet been well-studied. If suitable advancements in follow-up instruments can be made to identify \ac{SGRB} afterglow in $\lesssim$deg localizations, which appears reasonable, such an instrument would allow the best determination of the source evolution of \ac{NS} mergers and collapsars, with implications for heavy element enrichment,

With a commensurate ground-based network with A+ sensitivity, AMEGO would be capable of roughly 1 joint \ac{GW}-\ac{GRB} detection per month. The immediate localizations of the multimessenger detections would be of order $\lesssim$degree accuracy for events where we expect afterglow emission, and inform searches for afterglow. This would be roughly an order of magnitude improvement over typical \ac{GW} localizations with several contributing interferometers, and a far greater improvement for most detections at these distances. The joint detections would typically occur within $\sim$500\,Mpc. This guides the necessary capabilities of potential follow-up instruments, e.g. $\sim$23rd Mag for a \knid-like event, which seems possible at that time. Together these would give great constraints on \acp{SGRB} and ultrarelativistic jets, provide a \ac{GW}-\ac{GRB} sample for cosmology, give the most precise measures of the \ac{GW}-\ac{GRB} time delay, be used for tests of fundamental physics, and enable early broadband \ac{EM} observations key for understanding \ac{NS} mergers themselves. This mission would also be beneficial for networks with fewer, but more sensitive \ac{GW} interferometers.

AMEGO may detect early afterglow emission in \acp{SGRB}, though this has not been quantified. The relative balanced priority to the Compton and pair regimes has limited the narrow-line point source sensitivity. With the current design it would require a fortuitous nearby \ac{NS} merger to detect the prompt nuclear gamma-rays from the kilonova. It may be capable of identifying a \ac{KNR} in the Milky Way. We support technology advancement to improve the narrow-line point-source sensitivity. Given the localization method of Compton telescopes, this would have the added benefit of improved prompt \ac{SGRB} localizations.


Should AMEGO be selected in Astro2020 it would not launch for about a decade from now. Until that time, or in the event we get no large-scale gamma-ray mission selected within the decade, we support the launch of sensitive gamma-ray scintillation missions, especially those that exceed the sensitivity of \ac{GBM} from a few keV to several MeV. Even CubeSat missions can provide expanded sky coverage, additional localization constraints, and photon statistics, especially when treated as a network (Section \ref{sec:GRB_together}).

\subsubsection{X-rays}
Given current sensitivities and the relative intrinsic emission, X-rays are the easiest method to detect \ac{SGRB} afterglow emission. They are usually the highest energy detection of this synchrotron emission, enabling inferences like the jet-opening angle, jet structure, and circumburst densities. The temporal evolution in this range also contains several non-thermal signatures, including flares and plateaus, that may have implications for the \ac{NS} \ac{EOS} and jet physics. X-ray observatories can provide the earliest arcsecond localizations, necessary for most \ac{EM} telescopes to observe these events, for robust host association, and some tests of fundamental physics. This is the best wavelength to precisely localize distant events.

\Swift \ac{XRT} utilizes (modified) galaxy targeting for follow-up of \ac{GW}-detected \ac{NS} mergers. It is critical in the current era. \textit{Chandra} provides high spatial resolution in X-rays, enabling host galaxy association for bursts that it detects, and recovery of off-axis afterglows like \grbid. We support the allocation of appropriate \textit{Chandra} time for \ac{NS}-merger follow-up. Future sensitive X-ray observatories with high spatial resolution are helpful for \ac{NS}-merger science. XMM-Newton and ATHENA have spatial resolution that may be problematic to isolate \acp{SGRB} afterglow, as demonstrated with observations of \grbid.

We would ideally launch a time-domain, wide-field X-ray telescope with sufficient sensitivity to recover a reasonable fraction of \ac{SGRB} afterglows. Figure \ref{fig:xray_afterglow} shows the full \Swift \ac{XRT} observations of \ac{SGRB} afterglows, with the nearest known \acp{SGRB} highlighted to demonstrate that they are not brighter than the full sample. Thus, a wide-field X-ray telescope must achieve \ac{XRT}-like sensitivity to recover afterglows at reasonable efficiencies. Recently proposed missions that would utilize lobster eye optics for follow-up observations do not have sufficient sensitivity to recover most off-axis counterparts to most \ac{GW} detections nor on-axis counterparts to \ac{SGRB} detection. With a fiducial lobster sensitivity and an on-target time of 100\,s (which is optimistic given the necessary tiling) only $\sim$15\% of \ac{XRT}-detected afterglows would be recovered. Missions that have lobster eye optics instruments with a \ac{FOV} $\gtrsim$10\% of the sky, like THESEUS or Einstein Probe, may be more promising because they can observe the prompt emission, early afterglow, and potential magnetar plateau. These have somewhat similar detection prospects for low-energy partial coding masks with larger \ac{FOV}, such as STROBE-X. While these latter missions provide useful observations for \ac{NS} mergers, we support relevant technology advancements for significantly more sensitive, wide-field X-ray instruments built for the purposes of follow-up observations of \ac{NS} mergers.


\subsubsection{Ultraviolet}
Bright \ac{UV} emission was among the surprises of \knid. Such observations are key to understanding these sources, by identifying what causes bright \ac{UV} emission, determination of the remnant object, and are the earliest possible light from kilonovae, which allows arcsecond localizations, robust association with \ac{GW} signals, and observations of the rise in optical and \ac{IR} bands.

\Swift \ac{UVOT} is copointed with \ac{XRT}, and follows-up \ac{GW}-detected \ac{NS} mergers with the (modified) galaxy-targeted technique. The only other active \ac{UV} mission is Hubble, which is far more sensitive. We support the allocation of Hubble observing time for \ac{NS} merger studies of \acp{SGRB} afterglow, kilonovae, and to uncover the origin of early \ac{UV} emission. However, the current $\sim$48 hour response time of Hubble is egregiously insufficient and needs to be significantly shortened.

Because \ac{UV} emission is the earliest possible kilonova emission it would be ideal to have a wide-field \ac{UV} telescope to follow-up \ac{GW} detections. A baseline guidance for this mission would achieve $\sim$21st Mag in $\sim$10 minutes, with $\sim$10\,deg$^2$ \ac{FOV}. This should be sufficient to recover \knid-like events to $\sim$150-200\,Mpc with most \ac{GW} localization regions, though greater sensitivity/\ac{FOV} are obviously beneficial. There are several proposed missions that meet these requirements, suggesting the necessary technology already exists. Indeed ULTRASAT exceeds these requirements with a planned launch in 2023.

\subsubsection{Optical}
Optical is likely to be the key discovery wavelength for kilonovae and the most common detections giving arcsecond localizations. They are key to observing the early and middle evolution of kilonovae, and for the spectroscopic determination of redshift.

Of all wavelengths, optical is likely the most prepared for \ac{EM}-counterpart searches of \ac{GW} detections. There are numerous time-domain telescopes that use galaxy-targeting, several wide-field telescopes that can tile the \ac{GW} localizations, and particularly sensitive telescopes to deeply study these events, including those capable of broadband (near-\ac{UV} to \ac{NIR}) spectroscopy.

With the upcoming \ac{LSST} and thirty meter telescopes, these capabilities will continue to improve and meet requirements into the A+ era. These facilities offer unmatched capabilities that are beneficial for \ac{NS}-merger studies. We support the construction of an X-shooter like spectrometer for the thirty meter telescopes.

\subsubsection{Infrared}
\ac{IR} uniquely probes the effects of the lanthanides and actinides. They enable spectroscopic determination of redshift, and are key to doing so for distant events. They may be the discovery wavelengths for particularly (infra)red kilonova, which may be possible for some \ac{NSBH} mergers.

\ac{NIR} can now be reliably observed from the ground, even to deep limits. The existing 10m and upcoming thirty meter telescopes are capable of studying kilonovae weeks after merger. We again support a sensitive spectrometer for these future instruments. Wide-field \ac{NIR} telescopes now exist, but are currently much less sensitive than optical telescopes, which may prevent the detection of some kilonova from \ac{NSBH} mergers. We support technology development to improve the sensitivity of such instruments, with similar technical guidelines to the wide-field \ac{UV} capabilities discussed above.

Hubble provides sensitive \ac{NIR} coverage. Far \ac{IR} can only be observed from space, or near-space. Spitzer will soon end observations. SOFIA observes this energy range. \ac{JWST} will be available in the coming years and its key capabilities enable important study of late-time kilonova emission. It is likely to be joined by WFIRST a few years later which will enable identification of faint (infra)red kilonova due to nearby prompt collapse \ac{BNS} mergers or more distant \ac{NSBH} mergers for events that are well localized but beyond the capabilities of wider-field monitors. Three great observatory-class missions in a row prioritize the \ac{IR} regime, proving reliable coverage for our narrow-field needs. 

\subsubsection{Radio}
Radio observations probe the low end of the synchrotron radiation of \ac{SGRB} afterglow and from the quasi-isotropic outflow interactions long after merger, and are likely to be the latest possible observations of these events. Radio interferometry is capable of directly observing bulk outflow of the jets, and can even do so for events that are not face-on. This is currently limited to particularly nearby events.

Sensitive wide-field radio transient surveys have been developed, and are providing a new aspect to time-domain astronomy. However, another important metric for radio observations to study \ac{NS} mergers is likely narrow-field sensitivity. With improvements in gamma-ray and X-ray sensitivity we would require a commensurate improvement in radio observations to fully study \grbid-like events to greater distances.

\subsubsection{VHE}
\ac{VHE} facilities probe the highest energy emission from these sources which provide stringent tests of \ac{LIV}, and probe the extreme non-thermal emission. Observations of \acp{SGRB} at these energies may be unlikely with current facilities, but possible with the upcoming \ac{CTA} mission. We recommend investigations into staggered observations with the large telescopes to maximize coverage of \ac{GW} early warning localizations before merger, to attempt prompt observations of a \ac{SGRB}.

\subsubsection{Neutrinos}
Neutrino detections of \ac{NS} mergers would be a new messenger from these sources and allow a wealth of new science. MeV detections are likely to be rare, even with the funded Hyper-Kamiokande. High-energy neutrino observations are also likely to be rare with IceCube, but potentially observable. We recommend relevant software and analysis investment from these facilities, and patience until the first neutrino detection of a \ac{NS} merger. 

\subsection{Communication and combined data}
Given the critical information revealed from early observations of \ac{NS} mergers, the necessity of robust associations, the inherent multimessenger and multiwavelength nature, and the need of follow-up observations for most \ac{EM} observations, rapid communication of relevant information is of the utmost importance.


\subsubsection{Combined searches}
\label{sec:GRB_together}
Combining the \ac{GW} interferometers into an effective coherent network for detection has enabled the detection of a far greater number of events, more precise localizations, and the announcement of events quickly after merger. Similar improvements are possible with other types of searches within similar instruments, and between instruments. These are particularly promising outcomes for detections of signals that are prompt (or early). 

The first is the construction of a coherent \ac{GRB} network. These missions are predominantly background-dominated, so joint sub-threshold searches can increase the effective total sensitivity to \acp{SGRB}. Further, the automation of the \ac{IPN} technique would reduce the latency for the annuli to be made available, which may aide in searches for kilonovae. The second are combined \ac{GW}-\ac{GRB} searches, both for independent triggers and sub-threshold searches. The automatic association and combined localizations of \ac{GW}-\ac{GRB} detections can increase the number of \ac{GW} detections and reduce the prompt localization region. The last are combined neutrino-\ac{GW} searches. These joint detections will be rare for the foreseeable future, but when they do occur the science return will be enormous, and the neutrino observations will provide $\sim$deg scale localizations. So despite the likely low chance of success of, investment is warranted.

Pulling out sub-threshold signals requires studies of weaker signals than have been considered before, which is particularly difficult and requires a deep understanding of the instrument, its noise, and its data. These studies likely require heavy investment by the relevant instrument teams. First is the combination of independent detections to aid the follow-up effort, development of shared software, maximizing the likelihood of follow-up success, and rewarding the team investment by credit for discoveries. 



\subsubsection{Reporting systems}
\label{sec:reporting}

Reporting systems are the backbone of multimessenger astronomy. They enable near real-time reporting of transient identification, localization, and initial analysis. This rapid and automated communication is fundamental to the success of the field, which is scientifically obvious. Sociologically, fast reporting allows for the claim of discovery for events of interest while also enabling greater observations and study for key events of interest.

Fully realizing the potential of multimessenger science requires advancements to the existing reporting systems. We comment on some ideal capabilities that these systems should be capable of.
\begin{itemize}
    \item General automated notice types including those for independently discovered transients, potential counterpart and claimed counterpart reporting, classification of those counterparts, planned and actual observations for pointed telescopes, retractions.
    \item It must be capable of distributing alert information for all relevant observatories. For example, radio transients cannot be distributed through some optical alert systems because they cannot provide the observation band and magnitude (which is nonsense for a radio observation). A system that works for all relevant observatories and their relevant information is necessary.
    \item Distribute alerts in real time through various standards to enable ease of access, and maintain an active database that can be polled on demand. With the above capability, this will enable things like requesting all current candidate counterparts to \ac{GW} observations at a given time.
    \item Easy creation of new alert streams, including user-created streams. This would allow for a distributed system for the reporting of value-added information, allowing individuals to contribute their specific expertise. and the minimization of duplicated effort. Examples include combined GRB and joint GW-GRB skymaps, or the convolution of GW skymaps with galaxy catalogs.
    \item It should not attempt to duplicate or supplant the roles of instrument teams. The reporting, individual or combined, should give credit to those who enable the work.
    \item The ranking of candidates to allow for automatic prioritization.
    \item No single point failures, including computers, individuals, and networks. A good option is the use of cloud providers, which provide redundancy and high livetime while also avoiding some potential headaches (e.g. network restrictions due to national security concerns for systems housed within a NASA network).
    \item Be a general multimessenger reporting system, not focused on \ac{NS} mergers.
    \item Induce as little delay as possible to high priority alerts. The most extreme example would be the early warning systems from \ac{GW} observations. Early reports will send alerts on the order of tens of seconds, where transmission delay of the report could prevent successful \ac{EM} observation of the merger time.
    \item Enable private alert streams. This was done by GCN for O1 and O2 alerts from the \ac{LVC} to enable the maturation of \ac{GW} astronomy and could be used for other facilities with private data. It would also allow joint sub-threshold searches to be performed through existing alert streams, without public announcement of all individual sub-threshold searches.
    \item The transient name server is used to provide unique identifiers to astronomical transients, working between the identifiers specific to individual fields or instruments. These two systems must integrate.
    \item Able to access relevant catalogs and use the information accordingly, e.g. report galaxy-weighted localizations.
    \item Be able to promptly determine the recent observations from relevant instruments of a given position to report or help determine the last sensitive non-detection for transients of interest.
    \item The core software has to be written predominantly by professional programmers.
\end{itemize}


There are several proposals for future reporting methods. \ac{LSST} has opted to use Apache Kafka to allow for distributed analysis and reporting through data brokers, which is an architecture that handles several of these wanted capabilities. There is the SCIMMA proposal to the NSF for a multimessenger computing institute. There is the NASA funded GCN upgrade referred to as TACH. AMON is an on-going project that attempts to combine sub-threshold signals to elevate the significance of weak events. Treasure Map provides a method for follow-up teams to report planned observations and coordination. There are several other on-going, proposed, and funded projects with similar ideals.

We must prevent the bifurcation of the time-domain community. As an example, both NSF and NASA have facilities with unique capabilities for multimessenger studies, and each appears open to funding multi-instrument systems as necessary. There should be a single automated organization to distribute alerts of interest for time-domain, multimessenger astronomy. If there is not, then all follow-up groups will have to develop their own software to ingest multiple types of alerts from different systems and combine the information themselves. This is an inefficient allocation of taxpayer money and scientist time. Note that we are not suggesting a single entity be made responsible for multimessenger astronomy. Each sub-group of multimessenger astronomers (e.g. optical surveys, GRB monitors, GW observers) should develop the capabilities necessary to analyze their own data. However, information of interest to the full multimessenger community should be reporting into a unified alert system. This can even be disparate systems that are intentionally built to communicate with each other (which could be, e.g. TACH communicating with the transient name server and Treasure Map allowing for follow-up coordination).


For those who report, automatically or promptly, we list a set of recommendations for best-practices:
\begin{itemize}
    \item A unified skymap format for poor localizations. We suggest the HEALPix format to match \ac{LIGO}/Virgo and \Fermi-GBM. We would support the development of HEALPix maps for things neutrino alerts (both individual MeV, \ac{SNEWS}, and high energy localizations), pair conversion telescopes, Compton telescopes, gamma-ray scintillators, and the \ac{IPN}. The relevant teams should contribute to the shared software, e.g. broaden the use of the multi-resolution maps that will be critical in the next few years.
    \item The development of automated association methods for independent triggers, including (but not limited to) \acp{GW}, prompt \acp{SGRB}, neutrino detections, and optically-identified transients. Further, the automatic combination of localization information.
    \item Full use of the various notice types should be used. For example, the GCN candidate counterpart notice type is substantially underutilized, with the very notable exception of the \Swift-\ac{XRT} team. If this were widely used it would enable fully robotic prioritization.
    \item The assignment of informative names. \knid is a much more useful name for multimessenger studies than AT2017gfo. This is made obvious when you consider having a dozen \ac{GW}-kilonova detections and having to remember which kilonova name belongs to which \ac{GW}. The GWTC-1 catalog \citep{lvc_gwtc1} reports a set of marginal candidates, which are named with the YYMMDD format, with no prefix or suffix marking them as \ac{GW} candidates. This is not helpful for studies that seek to use these signals for future work.
\end{itemize}

\subsubsection{Real-time information}
The \ac{LVC} developed real-time alerts and localizations for the Advanced era. They have heavily improved the information that is distributed in real-time by releasing the initial distance determination (as a function of 2D position), the initial merger classification based on the template mass measurements, and a likelihood of the release of matter from the merger. They have also developed Superevents, to down-select multiple \ac{GW} triggers on the same event due to the multiple search pipelines, which have enabled preliminary notices before any manual selection.

Several astronomers have requested that the \ac{LVC} report additional information in real-time. One usual request is the initial mass measures, which can be determined relatively early during the full parameter estimation. This would enable follow-up observers to make their own inference on the likely system classification, and prioritization to follow-up particularly interesting events when \ac{GW}-detections of \ac{NS} mergers become more routine. Examples may be particularly low or high-mass \ac{BNS} mergers, or \ac{NSBH} mergers near the disruption threshold. In general we support the continued increase in initial reporting information from \ac{GW} detections. 

However, this should be balanced against ensuring fair credit for the \ac{LVC} and its individual members. Two recent results include the heavy (likely) \ac{BNS} merger GW190425 \citep{abbott2020gw190425} and the potential \ac{NSBH} merger where the secondary object falls into the putative primary mass gap range \citep{abbott2020gw190814}. This science is only possible because of years and decades of investment that made \ac{GW} interferometers sensitive enough to bring us into the new era, and much of the science could be inferred from the initial mass measures. We return to this in Sect.~\ref{sec:cultural}.

We support the development of early warning systems for \ac{NS} mergers in the near-future. As discussed, these may not provide particularly well constrained localizations before merger with ground-based interferometers. However, knowing the event time before merger can still be beneficial for several reasons, and there are \ac{EM} facilities that could potentially make use of even rough localizations. These alerts are likely to be complicated, and they must be distributed, received, and reacted to in seconds to be useful. Building this entire system will take heavy investment, and work should begin sooner rather than later.

Improved initial reporting should not be limited to \ac{GW} alerts. For example, the prompt \ac{GRB} monitors should automatically classify events, mark likely \acp{SGRB}, and hopefully combine information in near real-time to support the follow-up effort. This can be broadened to considerations of prompt \ac{GRB} consistency with cocoon origin, which can inform follow-up searches targeting both the quasithermal and non-thermal signals.

The follow-up community is, in general, reporting necessary information in real-time. This includes announcing candidates of interest and their location. This gives the team credit at discovery, while enabling follow-up searches to characterize and classify the transients, and other teams to inform on the last non-detection. This information is also generally reported as soon as possible. Improvements could obviously be made, but the balance of rewarding credit for early reporting of information necessary to maximize science should spread to other aspects of multimessenger astronomy.

\subsubsection{Space-based communication}
Space-based observatories provide key coverage of $\sim$MeV gamma-rays, X-rays, ultraviolet, and infrared wavelengths. They also provide some of the most sensitive and precise observations in optical. There are two communication limitations that matter for existing and proposed missions. 

Data bandwidth is limited given the expensive downlink cost and (in most cases) technical limitations. \Fermi \ac{GBM} can achieve far more sensitive searches because of the downlink of individual time-tagged event data. This was only possible because this continuous data time is small compared to the data requirements of the primary instrument on \Fermi. Enabling missions to downlink more data will allow for increased scientific return through software developments.

Second, prompt communication is key. The prompt downlink of triggers from \ac{GRB} monitors enabled time-domain astronomy and the distribution of \ac{SGRB} localizations within a minute of merger. This capability is not widely accessible, requiring access to the NASA TDRSS satellites or a large network of ground stations as done for INTEGRAL. 

Prompt uplink is currently not possible. One of the main sources of delay to the initial \Swift follow-up of \ac{GW} detections is the time to send the commands up to the spacecraft. We strongly endorse advancements that minimize this requirement for \Swift and other missions, including removal (or minimization) of human-in-the-loop approval. This would also allow for future missions to decouple telescopes to separate spacecraft, maximizing their individual scientific return.

Lastly, the limitation of prompt uplink and in some cases approval of targets of opportunity approval to normal working hours during a weekday is problematic. If a once in a lifetime event goes off at 6pm on a Friday in the US then some of the most sensitive facilities in existence may not even send the repoint command until Monday. This is unacceptable for well-funded missions. In contrast, \Swift is on pace for $\sim$1900 targets of opportunity in 2019 (A. Tohuvavohu, private communication).


\subsection{Theory, simulation, and interdisciplinary studies}
Theory and simulation development enabled the detection and study of \gwid. The advanced numerical relativity waveforms that were constructed to build the \ac{CBC} template banks that enable the real-time searches are a relatively new result. Significant improvements to these templates to fill the existing parameter space and consider additional effects, and improvements to analytic models (calibrated against the numerical waveforms) are warranted.

This is also true of the kilonova simulations that combined several very complicated processes into consistent codes that predicted the broad behavior of \knid. They also created the models that were used to infer the parameters of the ejecta, which so much science relies on. Again improvements are warranted, as discussed in Section \ref{sec:elements}. Similar improvements on the simulations of \ac{SGRB} jets and their interaction with surrounding material are recommended. These recommendations also directly apply for general simulations of \ac{NS} mergers. Lastly, these results rely on knowledge of laboratory astrophysics, particularly atomic and nuclear studies of heavy elements. Also, the inclusion of sophisticated nuclear physics simulations can improve multimessenger results.

We strongly support the necessary funding and allocation of computational and experimental resources to advance theory, improve simulations, and encourage interdisciplinary research. It is critical to nearly all of the potential science with \ac{NS} mergers.





\subsection{Cultural change}
\label{sec:cultural}
We close our recommendations with a somewhat contentious issue. The community did not handle the high pressure situation of \gwid as well as it could have. In the future this will be somewhat alleviated because the open public alerts from the \ac{LVC} and few individual discoveries will be as important. However, we should strive to be better and support individuals and teams that act in good faith.

Interdisciplinary work often does not succeed because it is particularly difficult and the funding mechanisms are often lacking. It appears that the interdisciplinary studies in multimessenger astronomy will succeed because of the great interest from scientists and the funding agencies, and the science that can only be uncovered through such means.  Another potential mismatch is the support for individuals that fall between fields, such as those building the inter-mission software that enables multimessenger astronomy. 

This means that communities that have historically valued different metrics of success must adapt. In very broad strokes, astronomy tends to reward individual or small-group efforts, as evidenced by the metrics for faculty positions or prize fellowships, the awards from the professional societies, and the intense competition in time-domain astronomy. In contrast, nuclear and particle physicists, and related communities like astroparticle (neutrino, cosmic ray, gamma-ray) groups and now the \ac{GW} collaborations, tend to work in very large collaboration out of necessity. For multimessenger science to succeed the judgement on the capability of an individual would ideally consider the metrics of success from their field.



%% file: ms.bbl
\begin{thebibliography}{}
\expandafter\ifx\csname natexlab\endcsname\relax\def\natexlab#1{#1}\fi
\providecommand{\url}[1]{\href{#1}{#1}}

\bibitem[{Aartsen {et~al.}(2018)Aartsen, Ackermann, Adams, Aguilar, Ahlers, \&
  et~al.}]{IceCube170922A_Blazar_2018}
Aartsen, M., Ackermann, M., Adams, J., {et~al.} 2018, Science, 361,
  doi:10.1126/science.aat1378

\bibitem[{Aartsen {et~al.}(2014)Aartsen, Ackermann, Adams, Aguilar, Ahlers,
  Ahrens, Altmann, Anderson, Arguelles, Arlen,
  {et~al.}}]{aartsen2014observation}
Aartsen, M.~G., Ackermann, M., Adams, J., {et~al.} 2014, \prl, 113, 101101

\bibitem[{Aartsen {et~al.}(2015)Aartsen, Ackermann, Adams, Aguilar, Ahlers,
  Ahrens, Altmann, Anderson, Arguelles, Arlen, {et~al.}}]{aartsen2015search}
---. 2015, \apjl, 805, L5

\bibitem[{Aasi {et~al.}(2015)Aasi, Abbott, Abbott, {et~al.}}]{AdvLIGO}
Aasi, J., Abbott, B.~P., Abbott, R., {et~al.} 2015, Class. Quantum Grav., 32,
  074001

\bibitem[{{Abadie} {et~al.}(2010){Abadie}, {Abbott}, {Abbott}, {Abernathy},
  {Accadia}, {Acernese}, {Adams}, {Adhikari}, {Ajith}, {Allen}, \&
  et~al.}]{overview_NSMs_GWs_2010}
{Abadie}, J., {Abbott}, B.~P., {Abbott}, R., {et~al.} 2010, Class. Quantum
  Grav., 27, 173001

\bibitem[{Abadie {et~al.}(2012)Abadie, Abbott, Abbott, Abbott, Abernathy,
  Adams, Adhikari, Affeldt, Ajith, Allen, {et~al.}}]{abadie2012implications}
Abadie, J., Abbott, B., Abbott, T., {et~al.} 2012, \apj, 755, 2

\bibitem[{Abbasi {et~al.}(2011)Abbasi, Abdou, Abu-Zayyad, Adams, Aguilar,
  Ahlers, Andeen, Auffenberg, Bai, Baker, {et~al.}}]{abbasi2011limits}
Abbasi, R., Abdou, Y., Abu-Zayyad, T., {et~al.} 2011, \prl, 106, 141101

\bibitem[{Abbott {et~al.}(2008)Abbott, Abbott, Adhikari, Agresti, Ajith, Allen,
  Amin, Anderson, Anderson, Arain, {et~al.}}]{abbott2008implications}
Abbott, B., Abbott, R., Adhikari, R., {et~al.} 2008, \apj, 681, 1419

\bibitem[{{Abbott} {et~al.}(2016{\natexlab{a}}){Abbott}, others, , {LIGO
  Scientific Collaboration}, \& {Virgo Collaboration}}]{GW150914}
{Abbott}, B.~P., others, , {LIGO Scientific Collaboration}, \& {Virgo
  Collaboration}. 2016{\natexlab{a}}, \prl, 116, 061102

\bibitem[{{Abbott} {et~al.}(2016{\natexlab{b}}){Abbott}, others, , {LIGO
  Scientific Collaboration}, \& {Virgo Collaboration}}]{GW151226}
---. 2016{\natexlab{b}}, \prl, 116, 241103

\bibitem[{{Abbott} {et~al.}(2016{\natexlab{c}}){Abbott}, others, {LIGO
  Scientific Collaboration}, \& {Virgo
  Collaboration}}]{NS_merger_rates_LVC_2016}
{Abbott}, B.~P., others, {LIGO Scientific Collaboration}, \& {Virgo
  Collaboration}. 2016{\natexlab{c}}, \apjl, 832, L21

\bibitem[{{Abbott} {et~al.}(2017{\natexlab{a}}){Abbott}, others, {(LIGO
  Scientific Collaboration}, \& {Virgo
  Collaboration}}]{GW170817_LVC_progenitor}
{Abbott}, B.~P., others, {(LIGO Scientific Collaboration}, \& {Virgo
  Collaboration}. 2017{\natexlab{a}}, \apjl, 850, L40

\bibitem[{{Abbott} {et~al.}(2017{\natexlab{b}}){Abbott}, others, {LIGO
  Scientific Collaboration}, \& {Virgo Collaboration}}]{GW170814}
{Abbott}, B.~P., others, {LIGO Scientific Collaboration}, \& {Virgo
  Collaboration}. 2017{\natexlab{b}}, \prl, 119, 141101

\bibitem[{{Abbott} {et~al.}(2018){Abbott}, others, {LIGO Scientific
  Collaboration}, \& {Virgo Collaboration}}]{GW170817_GW_updated_NS_EOS}
---. 2018, \prl, 121, 161101

\bibitem[{{Abbott} {et~al.}(2019{\natexlab{a}}){Abbott}, others, {LIGO
  Scientific Collaboration}, \& {Virgo Collaboration}}]{lvc_gwtc1}
---. 2019{\natexlab{a}}, Phys. Rev. X, 9, 031040

\bibitem[{{Abbott} {et~al.}(2019{\natexlab{b}}){Abbott}, others, {LIGO
  Scientific Collaboration}, \& {Virgo Collaboration}}]{LVC_O2_BBH_rates}
---. 2019{\natexlab{b}}, \apjl, 882, L24

\bibitem[{{Abbott} {et~al.}(2019{\natexlab{c}}){Abbott}, others, {LIGO
  Scientific Collaboration}, \& {Virgo
  Collaboration}}]{GW170817_GW_updated_parameters}
---. 2019{\natexlab{c}}, Phys. Rev. X, 9, 011001

\bibitem[{{Abbott} {et~al.}(2019{\natexlab{d}}){Abbott}, others, {LIGO
  Scientific Collaboration}, \& {Virgo Collaboration}}]{LIGO_GR_BBH_GWTC1}
---. 2019{\natexlab{d}}, \prd, 100, 104036

\bibitem[{{Abbott} {et~al.}(2019{\natexlab{e}}){Abbott}, others, {LIGO
  Scientific Collaboration}, \& {Virgo Collaboration}}]{LIGO_GR_GW170817}
---. 2019{\natexlab{e}}, \prl, 123, 011102

\bibitem[{{Abbott} {et~al.}(2017{\natexlab{c}}){Abbott}, others, {LIGO
  Scientific Collaboration}, {Virgo Collaboration}, \& {IPN
  Collaboration}}]{lvc_gws_before_grbs}
{Abbott}, B.~P., others, {LIGO Scientific Collaboration}, {Virgo
  Collaboration}, \& {IPN Collaboration}. 2017{\natexlab{c}}, \apj, 841, 89

\bibitem[{Abbott {et~al.}(2009)Abbott, Abbott, Adhikari, Ajith, Allen, Allen,
  Amin, Anderson, Anderson, Arain, {et~al.}}]{abbott2009ligo}
Abbott, B.~P., Abbott, R., Adhikari, R., {et~al.} 2009, Rep. Progr. Phys., 72,
  076901

\bibitem[{Abbott {et~al.}(2017{\natexlab{a}})Abbott, Abbott, Abbott, Acernese,
  Ackley, Adams, Adams, Addesso, Adhikari, Adya,
  {et~al.}}]{GW170817-GRB170817A}
Abbott, B.~P., Abbott, R., Abbott, T.~D., {et~al.} 2017{\natexlab{a}}, \apjl,
  848, L13

\bibitem[{Abbott {et~al.}(2017{\natexlab{b}})Abbott, Abbott, Abbott, Acernese,
  Ackley, Adams, Adams, Addesso, Adhikari, Adya, {et~al.}}]{GW170817-GW}
---. 2017{\natexlab{b}}, \prl, 119, 161101

\bibitem[{Abbott {et~al.}(2017{\natexlab{c}})Abbott, Abbott, Adhikari,
  Ananyeva, Anderson, Appert, Arai, Araya, Barayoga, Barish,
  {et~al.}}]{GW170817-MMAD}
Abbott, B.~P., Abbott, R., Adhikari, R.~X., {et~al.} 2017{\natexlab{c}}, \apjl,
  848, L12

\bibitem[{Abbott {et~al.}(2017{\natexlab{d}})Abbott, Abbott, Abbott, Abernathy,
  Ackley, Adams, Addesso, Adhikari, Adya, Affeldt, {et~al.}}]{Cosmic_Explorer}
Abbott, B.~P., Abbott, R., Abbott, T.~D., {et~al.} 2017{\natexlab{d}}, Class.
  Quantum Grav., 34, 044001

\bibitem[{{Abbott} {et~al.}(2017)}]{GW170817-Standard-Siren}
{Abbott}, B.~P., {et~al.} 2017, \nat, 551, 85

\bibitem[{Abbott {et~al.}(2018{\natexlab{a}})Abbott, Abbott, Abbott, Abernathy,
  Acernese, Ackley, Adams, Adams, Addesso, Adhikari,
  {et~al.}}]{Gen_2_Prospects_2018}
Abbott, B.~P., Abbott, R., Abbott, T.~D., {et~al.} 2018{\natexlab{a}}, Living
  Rev. Relativ., 21, 3

\bibitem[{Abbott {et~al.}(2018{\natexlab{b}})Abbott, Abbott, Abbott, Abernathy,
  Acernese, Ackley, Adams, Adams, Addesso, Adhikari,
  {et~al.}}]{review_future_GW_network_2018}
---. 2018{\natexlab{b}}, Living Rev. Relativ., 21, 3

\bibitem[{{Abbott} {et~al.}(2019{\natexlab{f}})}]{ligo2019search}
{Abbott}, B.~P., {et~al.} 2019{\natexlab{f}}, \apj, 886, 75

\bibitem[{Abbott {et~al.}(2020{\natexlab{a}})Abbott, Abbott, Abbott, Abraham,
  Acernese, Ackley, Adams, Adhikari, Adya, Affeldt,
  {et~al.}}]{abbott2020gw190425}
Abbott, B.~P., Abbott, R., Abbott, T.~D., {et~al.} 2020{\natexlab{a}}, \apjl,
  892, L3

\bibitem[{Abbott {et~al.}(2020{\natexlab{b}})Abbott, Abbott, Abbott, Abraham,
  Acernese, Ackley, Adams, Adya, Affeldt, Agathos, {et~al.}}]{abbott2020model}
---. 2020{\natexlab{b}}, Class. Quantum Grav., 37, 045006

\bibitem[{Abbott {et~al.}(2020{\natexlab{c}})Abbott, Abbott, Abraham, Acernese,
  Ackley, Adams, Adhikari, Adya, Affeldt, Agathos,
  {et~al.}}]{abbott2020gw190814}
Abbott, R., Abbott, T., Abraham, S., {et~al.} 2020{\natexlab{c}}, \apjl, 896,
  L44

\bibitem[{Abdo {et~al.}(2009)Abdo, Ackermann, Ajello, Asano, Atwood, Axelsson,
  Baldini, Ballet, Barbiellini, Baring, {et~al.}}]{LIV_090510_limit_2009}
Abdo, A.~A., Ackermann, M., Ajello, M., {et~al.} 2009, Nature, 462, 331

\bibitem[{Acernese {et~al.}(2015)Acernese, Agathos, Agatsuma,
  {et~al.}}]{AdvVirgo}
Acernese, F., Agathos, M., Agatsuma, K., {et~al.} 2015, Class. Quantum Grav.,
  32, 024001

\bibitem[{Ackermann {et~al.}(2010{\natexlab{a}})Ackermann, Asano, Atwood,
  Axelsson, Baldini, Ballet, Barbiellini, Baring, Bastieri, Bechtol,
  {et~al.}}]{GRB090510_discovery_Fermi}
Ackermann, M., Asano, K., Atwood, W.~B., {et~al.} 2010{\natexlab{a}},
  Astrophys. J., 716, 1178

\bibitem[{Ackermann {et~al.}(2010{\natexlab{b}})Ackermann, Asano, Atwood,
  Axelsson, Baldini, Ballet, Barbiellini, Baring, Bastieri, Bechtol,
  {et~al.}}]{ackermann2010fermi}
---. 2010{\natexlab{b}}, Astrophys. J., 716, 1178

\bibitem[{Ackley {et~al.}(2020)Ackley, Adya, Agrawal, Altin, Ashton, Bailes,
  Baltinas, Barbuio, Beniwal, Blair, {et~al.}}]{ackley2020neutron}
Ackley, K., Adya, V.~B., Agrawal, P., {et~al.} 2020, arXiv e-prints,
  arXiv:2007.03128

\bibitem[{Adler {et~al.}(1995)Adler, Casey, \& Jacob}]{adler1995vacuum}
Adler, R.~J., Casey, B., \& Jacob, O.~C. 1995, American J. Phys., 63, 620

\bibitem[{{Ahlers} \& {Halser}(2019)}]{ahlers2019neutrino}
{Ahlers}, M., \& {Halser}, L. 2019, \mnras, 490, 4935

\bibitem[{Ahmad {et~al.}(2002)Ahmad, Allen, Andersen, Anglin, Barton, Beier,
  Bercovitch, Bigu, Biller, Black, {et~al.}}]{mnu_solar_SNO_2002}
Ahmad, Q.~R., Allen, R.~C., Andersen, T.~C., {et~al.} 2002, \prl, 89, 011301

\bibitem[{Ajello {et~al.}(2019)Ajello, Arimoto, Axelsson, Baldini, Barbiellini,
  Bastieri, Bellazzini, Bhat, Bissaldi, Blandford, {et~al.}}]{ajello2019decade}
Ajello, M., Arimoto, M., Axelsson, M., {et~al.} 2019, Astrophys. J., 878, 52

\bibitem[{Alam {et~al.}(2017)Alam, Ata, Bailey, Beutler, Bizyaev, Blazek,
  Bolton, Brownstein, Burden, Chuang, {et~al.}}]{alam2017clustering}
Alam, S., Ata, M., Bailey, S., {et~al.} 2017, Mon. Not. R. Astron. Soc., 470,
  2617

\bibitem[{Albert {et~al.}(2017)Albert, Andr{\'e}, Anghinolfi, Ardid, Aubert,
  Aublin, Avgitas, Baret, Barrios-Mart{\'\i}, Basa,
  {et~al.}}]{GW170817_neutrinos}
Albert, A., Andr{\'e}, M., Anghinolfi, M., {et~al.} 2017, \apjl, 850, L35

\bibitem[{Alexander {et~al.}(2018)Alexander, Margutti, Blanchard, Fong, Berger,
  Hajela, Eftekhari, Chornock, Cowperthwaite, Giannios,
  {et~al.}}]{alexander2018decline}
Alexander, K.~D., Margutti, R., Blanchard, P.~K., {et~al.} 2018, \apjl, 863,
  L18

\bibitem[{Alexander {et~al.}(2008)Alexander, Finn, \&
  Yunes}]{alexander2008gravitational}
Alexander, S., Finn, L.~S., \& Yunes, N. 2008, \prd, 78, 066005

\bibitem[{Alexander \& Yunes(2009)}]{alexander2009chern}
Alexander, S., \& Yunes, N. 2009, Phys. Rep., 480, 1

\bibitem[{Alexander \& Yunes(2018)}]{alexander2018gravitational}
Alexander, S.~H., \& Yunes, N. 2018, \prd, 97, 064033

\bibitem[{Alexander \& Gates~Jr(2006)}]{alexander2006can}
Alexander, S. H.~S., \& Gates~Jr, S.~J. 2006, JCAP, 2006, 018

\bibitem[{Alexander {et~al.}(2006)Alexander, Peskin, \&
  Sheikh-Jabbari}]{PhysRevLett.96.081301}
Alexander, S. H.~S., Peskin, M.~E., \& Sheikh-Jabbari, M.~M. 2006, \prl, 96,
  081301

\bibitem[{Aloy {et~al.}(2005)Aloy, Janka, \& M{\"u}ller}]{aloy2005relativistic}
Aloy, M.~A., Janka, H.-T., \& M{\"u}ller, E. 2005, Astron. Astrophys., 436, 273

\bibitem[{Alpher {et~al.}(1948)Alpher, Bethe, \& Gamow}]{cosmo_BBN_1948}
Alpher, R., Bethe, H., \& Gamow, G. 1948, Phys. Rev., 73, 803

\bibitem[{Alpher \& Herman(1948)}]{cosmo_alpher_1948}
Alpher, R.~A., \& Herman, R. 1948, Nature, 162, 774

\bibitem[{Altarelli(1989)}]{altarelli1989experimental}
Altarelli, G. 1989, Annu. Rev. Nucl. Part. Sci., 39, 357

\bibitem[{Alvarez-Gaume \& Witten(1984)}]{alvarez1984gravitational}
Alvarez-Gaume, L., \& Witten, E. 1984, Nucl. Phys. B, 234, 269

\bibitem[{Amelino-Camelia(2002)}]{QG_deformed_amelino_2002}
Amelino-Camelia, G. 2002, Int. J. Mod. Phys. D, 11, 35

\bibitem[{Amelino-Camelia {et~al.}(1997)Amelino-Camelia, Ellis, Mavromatos, \&
  Nanopoulos}]{LIV_foam_amelino_1997}
Amelino-Camelia, G., Ellis, J., Mavromatos, N.~E., \& Nanopoulos, D.~V. 1997,
  Int. J. Mod. Phys. A, 12, 607

\bibitem[{Amelino-Camelia {et~al.}(1998)Amelino-Camelia, Ellis, Mavromatos,
  Nanopoulos, \& Sarkar}]{GR_quantum_gravity_amelino_1998}
Amelino-Camelia, G., Ellis, J., Mavromatos, N.~E., Nanopoulos, D.~V., \&
  Sarkar, S. 1998, Nature, 393, 763

\bibitem[{Andriot \& G{\'o}mez(2017)}]{large_dim_andriot_2017}
Andriot, D., \& G{\'o}mez, G.~L. 2017, JCAP, 2017, 048

\bibitem[{Antoniadis {et~al.}(2013)Antoniadis, Freire, Wex, Tauris, Lynch, van
  Kerkwijk, Kramer, Bassa, Dhillon, Driebe, {et~al.}}]{antoniadis2013massive}
Antoniadis, J., Freire, P.~C., Wex, N., {et~al.} 2013, Science, 340, 1233232

\bibitem[{Antonioli {et~al.}(2004)Antonioli, Fienberg, Fleurot, Fukuda,
  Fulgione, Habig, Heise, McDonald, Mills, Namba,
  {et~al.}}]{antonioli2004snews}
Antonioli, P., Fienberg, R.~T., Fleurot, F., {et~al.} 2004, New J. Phys., 6,
  114

\bibitem[{Aprahamian {et~al.}(2015)Aprahamian, Lapi, Mantica, Fatemi,
  Velkovska, Wilkerson, Filippone, Cirigliano, Schukraft, Ji,
  {et~al.}}]{LRP_US_2015}
Aprahamian, A., Lapi, S., Mantica, P., {et~al.} 2015, {Reaching for the
  horizon: The 2015 long range plan for nuclear science}, Tech. rep., U.S.\
  Department of Energy

\bibitem[{Aptekar {et~al.}(1995)Aptekar, Frederiks, Golenetskii, Ilynskii,
  Mazets, Panov, Sokolova, Terekhov, Sheshin, Cline,
  {et~al.}}]{aptekar1995konus}
Aptekar, R.~L., Frederiks, D.~D., Golenetskii, S.~V., {et~al.} 1995, \ssr, 71,
  265

\bibitem[{Arcavi(2018)}]{arcavi2018first}
Arcavi, I. 2018, \apjl, 855, L23

\bibitem[{Arcavi {et~al.}(2017)Arcavi, Hosseinzadeh, Howell, McCully,
  Poznanski, Kasen, Barnes, Zaltzman, Vasylyev, Maoz,
  {et~al.}}]{arcavi2017optical}
Arcavi, I., Hosseinzadeh, G., Howell, D.~A., {et~al.} 2017, Nature, 551, 64

\bibitem[{Arnaud {et~al.}(2002)Arnaud, Barsuglia, Bizouard, Cavalier, Davier,
  Hello, \& Pradier}]{neutrino_mass_stellar_collapse}
Arnaud, N., Barsuglia, M., Bizouard, M.~A., {et~al.} 2002, \prd, 65, 033010

\bibitem[{Arun(2012)}]{arun2012generic}
Arun, K.~G. 2012, Class. Quantum Grav., 29, 075011

\bibitem[{Ascenzi {et~al.}(2020)Ascenzi, Oganesyan, Salafia, Branchesi,
  Ghirlanda, \& Dall'Osso}]{ascenzi2020high}
Ascenzi, S., Oganesyan, G., Salafia, O.~S., {et~al.} 2020, arXiv preprint
  arXiv:2004.12215

\bibitem[{Ascenzi {et~al.}(2019)Ascenzi, Coughlin, Dietrich, Foley,
  Ramirez-Ruiz, Piranomonte, Mockler, Murguia-Berthier, Fryer, Lloyd-Ronning,
  {et~al.}}]{ascenzi2019luminosity}
Ascenzi, S., Coughlin, M.~W., Dietrich, T., {et~al.} 2019, Mon. Not. R. Astron.
  Soc., 486, 672

\bibitem[{Ashtekar {et~al.}(1989)Ashtekar, Balachandran, \&
  Jo}]{ashtekar1989cp}
Ashtekar, A., Balachandran, A.~P., \& Jo, S. 1989, Int. J. Mod. Phys. A, 4,
  1493

\bibitem[{Ashton {et~al.}(2018)Ashton, Burns, Dal~Canton, Dent, Eggenstein,
  Nielsen, Prix, Was, \& Zhu}]{ashton2018coincident}
Ashton, G., Burns, E., Dal~Canton, T., {et~al.} 2018, Astrophys. J., 860, 6

\bibitem[{Aso {et~al.}(2013)Aso, Michimura, Somiya, Ando, Miyakawa, Sekiguchi,
  Tatsumi, Yamamoto, Collaboration, {et~al.}}]{aso2013interferometer}
Aso, Y., Michimura, Y., Somiya, K., {et~al.} 2013, \prd, 88, 043007

\bibitem[{Atwood {et~al.}(2009)Atwood, Abdo, Ackermann, Althouse, Anderson,
  Axelsson, Baldini, Ballet, Band, Barbiellini, {et~al.}}]{atwood2009large}
Atwood, W., Abdo, A.~A., Ackermann, M., {et~al.} 2009, \apj, 697, 1071

\bibitem[{Audren {et~al.}(2013)Audren, Lesgourgues, Bird, Haehnelt, \&
  Viel}]{mnu_future_2013}
Audren, B., Lesgourgues, J., Bird, S., Haehnelt, M.~G., \& Viel, M. 2013, JCAP,
  2013, 026

\bibitem[{Bahcall(1964)}]{mnu_bahcall_1964}
Bahcall, J.~N. 1964, \prl, 12, 300

\bibitem[{Baiotti \& Rezzolla(2017)}]{baiotti2017binary}
Baiotti, L., \& Rezzolla, L. 2017, Rep. Prog. Phys., 80, 096901

\bibitem[{Baker {et~al.}(2017)Baker, Bellini, Ferreira, Lagos, Noller, \&
  Sawicki}]{cgw_beyond_GR_constraints_Baker_2017}
Baker, T., Bellini, E., Ferreira, P.~G., {et~al.} 2017, \prl, 119, 251301

\bibitem[{Balantekin {et~al.}(2014)Balantekin, Carlson, Dean, Fuller,
  Furnstahl, Hjorth-Jensen, Janssens, Li, Nazarewicz, Nunes,
  {et~al.}}]{balantekin2014nuclear}
Balantekin, A.~B., Carlson, J., Dean, D.~J., {et~al.} 2014, Mod. Phys. Lett. A,
  29, 1430010

\bibitem[{Balbus \& Hawley(1991)}]{balbus1991powerful}
Balbus, S.~A., \& Hawley, J.~F. 1991, \apj, 376, 214

\bibitem[{Barbieri {et~al.}(2020)Barbieri, Salafia, Perego, Colpi, \&
  Ghirlanda}]{barbieri2020electromagnetic}
Barbieri, C., Salafia, O., Perego, A., Colpi, M., \& Ghirlanda, G. 2020, The
  European Physical Journal A, 56, 1

\bibitem[{Barbieri {et~al.}(2019)Barbieri, Salafia, Colpi, Ghirlanda, Perego,
  \& Colombo}]{barbieri2019filling}
Barbieri, C., Salafia, O.~S., Colpi, M., {et~al.} 2019, The Astrophysical
  Journal Letters, 887, L35

\bibitem[{Baring \& Harding(1997)}]{baring1997escape}
Baring, M.~G., \& Harding, A.~K. 1997, Astrophys. J., 491, 663

\bibitem[{{Barkov} \& {Pozanenko}(2011)}]{2011MNRAS.417.2161B}
{Barkov}, M.~V., \& {Pozanenko}, A.~S. 2011, \mnras, 417, 2161

\bibitem[{Barnes \& Kasen(2013)}]{Kilonova-Barnes-2013}
Barnes, J., \& Kasen, D. 2013, Astrophys. J., 775, 18

\bibitem[{{Barnes} {et~al.}(2016){Barnes}, {Kasen}, {Wu}, \&
  {Mart{\'{\i}}nez-Pinedo}}]{KN_products_Barnes_2016}
{Barnes}, J., {Kasen}, D., {Wu}, M.-R., \& {Mart{\'{\i}}nez-Pinedo}, G. 2016,
  \apj, 829, 110

\bibitem[{Barsotti {et~al.}(2018)Barsotti, Fritschel, Evans, \&
  Gras}]{LIGO_173Mpc}
Barsotti, L., Fritschel, P., Evans, M., \& Gras, S. 2018, {Updated Advanced
  LIGO sensitivity design curve}, Tech. rep., LIGO.
\newblock \url{https://dcc.ligo.org/LIGO-T1800044/public}

\bibitem[{{Barthelmy} {et~al.}(2005){Barthelmy}, {Barbier}, {Cummings},
  {Fenimore}, {Gehrels}, {Hullinger}, {Krimm}, {Markwardt}, {Palmer},
  {Parsons}, {Sato}, {Suzuki}, {Takahashi}, {Tashiro}, \&
  {Tueller}}]{Barthelmy05}
{Barthelmy}, S.~D., {Barbier}, L.~M., {Cummings}, J.~R., {et~al.} 2005, \ssr,
  120, 143

\bibitem[{Bartos \& Marka(2019)}]{bartos2019nearby}
Bartos, I., \& Marka, S. 2019, Nature, 569, 85

\bibitem[{Baumgarte {et~al.}(1999)Baumgarte, Shapiro, \&
  Shibata}]{baumgarte1999maximum}
Baumgarte, T.~W., Shapiro, S.~L., \& Shibata, M. 1999, \apjl, 528, L29

\bibitem[{Bauswein {et~al.}(2019)Bauswein, Bastian, Blaschke, Chatziioannou,
  Clark, Fischer, \& Oertel}]{bauswein2019identifying}
Bauswein, A., Bastian, N.-U.~F., Blaschke, D.~B., {et~al.} 2019, \prl, 122,
  061102

\bibitem[{Bauswein {et~al.}(2013{\natexlab{a}})Bauswein, Baumgarte, \&
  Janka}]{bauswein2013prompt}
Bauswein, A., Baumgarte, T.~W., \& Janka, H.-T. 2013{\natexlab{a}}, \prl, 111,
  131101

\bibitem[{Bauswein {et~al.}(2013{\natexlab{b}})Bauswein, Goriely, \&
  Janka}]{NS-Bauswein}
Bauswein, A., Goriely, S., \& Janka, H.-T. 2013{\natexlab{b}}, Astrophys. J.,
  773, 78

\bibitem[{{Bauswein} {et~al.}(2017){Bauswein}, {Just}, {Janka}, \&
  {Stergioulas}}]{bauswein2017neutron}
{Bauswein}, A., {Just}, O., {Janka}, H.-T., \& {Stergioulas}, N. 2017, \apjl,
  850, L34

\bibitem[{Bauswein \& Stergioulas(2015)}]{bauswein2015unified}
Bauswein, A., \& Stergioulas, N. 2015, \prd, 91, 124056

\bibitem[{Baym {et~al.}(2018)Baym, Hatsuda, Kojo, Powell, Song, \&
  Takatsuka}]{baym2018hadrons}
Baym, G., Hatsuda, T., Kojo, T., {et~al.} 2018, Rep. Prog. Phys., 81, 056902

\bibitem[{Belczynski {et~al.}(2002)Belczynski, Kalogera, \&
  Bulik}]{belczynski2002comprehensive}
Belczynski, K., Kalogera, V., \& Bulik, T. 2002, Astrophys. J., 572, 407

\bibitem[{Belczynski {et~al.}(2008)Belczynski, Kalogera, Rasio, Taam, Zezas,
  Bulik, Maccarone, \& Ivanova}]{population_synthesis_belczynski_2008}
Belczynski, K., Kalogera, V., Rasio, F.~A., {et~al.} 2008, Astrophys. J. Suppl.
  Ser., 174, 223

\bibitem[{Belczynski {et~al.}(2006)Belczynski, Perna, Bulik, Kalogera, Ivanova,
  \& Lamb}]{belczynski2006study}
Belczynski, K., Perna, R., Bulik, T., {et~al.} 2006, \apj, 648, 1110

\bibitem[{Belczynski {et~al.}(2012)Belczynski, Wiktorowicz, Fryer, Holz, \&
  Kalogera}]{belczynski2012missing}
Belczynski, K., Wiktorowicz, G., Fryer, C.~L., Holz, D.~E., \& Kalogera, V.
  2012, Astrophys. J., 757, 91

\bibitem[{Bellini \& Sawicki(2014)}]{cgw_dark_energy_Bellini_2014}
Bellini, E., \& Sawicki, I. 2014, JCAP, 2014, 050

\bibitem[{Bellm(2014)}]{bellm2014zwicky}
Bellm, E.~C. 2014, in Proceedings of the ThirdHot-Wiring Transient Universe
  Workshop, ed. P.~R. Wo{\'z}niak, M.~J. Graham, A.~A. Mahabal, \& R.~Seaman
  (SLAC)

\bibitem[{Beloborodov(2010)}]{beloborodov2010collisional}
Beloborodov, A.~M. 2010, \mnras, 407, 1033

\bibitem[{Beloborodov \& M{\'e}sz{\'a}ros(2017)}]{beloborodov2017photospheric}
Beloborodov, A.~M., \& M{\'e}sz{\'a}ros, P. 2017, \ssr, 207, 87

\bibitem[{Beniamini {et~al.}(2016)Beniamini, Hotokezaka, \&
  Piran}]{beniamini2016r}
Beniamini, P., Hotokezaka, K., \& Piran, T. 2016, Astrophys. J., 832, 149

\bibitem[{Beniamini \& Nakar(2018)}]{beniamini2018observational}
Beniamini, P., \& Nakar, E. 2018, Mon. Not. R. Astron. Soc., 482, 5430

\bibitem[{{Beniamini} {et~al.}(2019){Beniamini}, {Petropoulou}, {Barniol
  Duran}, \& {Giannios}}]{beniamini2019joint}
{Beniamini}, P., {Petropoulou}, M., {Barniol Duran}, R., \& {Giannios}, D.
  2019, \mnras, 483, 840

\bibitem[{{Berger}(2014)}]{review_SGRBs_Berger_2014}
{Berger}, E. 2014, \araa, 52, 43

\bibitem[{{Berger} {et~al.}(2013){Berger}, {Fong}, \& {Chornock}}]{Berger2013}
{Berger}, E., {Fong}, W., \& {Chornock}, R. 2013, \apjl, 774, L23

\bibitem[{Berger {et~al.}(2003)Berger, Kulkarni, Pooley, Frail, McIntyre, Wark,
  Sari, Soderberg, Fox, Yost, {et~al.}}]{berger2003common}
Berger, E., Kulkarni, S.~R., Pooley, G., {et~al.} 2003, Nature, 426, 154

\bibitem[{Bernstein(2006)}]{total_bernstein}
Bernstein, G. 2006, Astrophys. J., 637, 598

\bibitem[{Bertolami {et~al.}(2004)Bertolami, Lehnert, Potting, \&
  Ribeiro}]{LIV_spactime_varying_bertolami_2004}
Bertolami, O., Lehnert, R., Potting, R., \& Ribeiro, A. 2004, \prd, 69, 083513

\bibitem[{Bertotti {et~al.}(2003)Bertotti, Iess, \&
  Tortora}]{WEP_absolute_Bertotti_2003}
Bertotti, B., Iess, L., \& Tortora, P. 2003, Nature, 425, 374

\bibitem[{Bhat {et~al.}(1992)Bhat, Fishman, Meegan, Wilson, Brock, \&
  Paciesas}]{bhat1992evidence}
Bhat, P., Fishman, G., Meegan, C., {et~al.} 1992, Nature, 359, 217

\bibitem[{Bhat {et~al.}(2016)Bhat, Meegan, von Kienlin, Paciesas, Briggs,
  Burgess, Burns, Chaplin, Cleveland, Collazzi, Connaughton, Diekmann,
  Fitzpatrick, Gibby, Giles, Goldstein, Greiner, Jenke, Kippen, Kouveliotou,
  Mailyan, McBreen, Pelassa, Preece, Roberts, Sparke, Stanbro, Veres,
  Wilson-Hodge, Xiong, Younes, Yu, \& Zhang}]{GBMBurstCatalog_6Years}
Bhat, P.~N., Meegan, C.~A., von Kienlin, A., {et~al.} 2016, \apjs, 223, 28

\bibitem[{Biesiada {et~al.}(2014)Biesiada, Ding, Pi{\'o}rkowska, \&
  Zhu}]{biesiada2014strong}
Biesiada, M., Ding, X., Pi{\'o}rkowska, A., \& Zhu, Z.-H. 2014, JCAP, 2014, 080

\bibitem[{Biesiada \& Pi{\'o}rkowska(2009)}]{strong_lensing_grbs_Biesiada_2009}
Biesiada, M., \& Pi{\'o}rkowska, A. 2009, Mon. Not. R. Astron. Soc., 396, 946

\bibitem[{Bilenky {et~al.}(2003)Bilenky, Giunti, Grifols, \&
  Masso}]{mnu_general_review_2003}
Bilenky, S.~M., Giunti, C., Grifols, J.~A., \& Masso, E. 2003, Phys. Rep., 379,
  69

\bibitem[{{Binns} {et~al.}(2019){Binns}, {Israel}, {Rauch}, {Cummings},
  {Davis}, {Labrador}, {Leske}, {Mewaldt}, {Stone}, {Wiedenbeck}, {Brandt},
  {Christian}, {Link}, {Mitchell}, {de Nolfo}, {von Rosenvinge}, {Sakai},
  {Sasaki}, {Waddington}, {Janka}, {Melott}, {Mason}, {Seo}, {Adams},
  {Thielemann}, {Heger}, {Lugaro}, \& {Westphal}}]{binns2019ultra}
{Binns}, W., {Israel}, M.~H., {Rauch}, B.~F., {et~al.} 2019, \baas, 51, 313

\bibitem[{Biscoveanu {et~al.}(2020)Biscoveanu, Thrane, \&
  Vitale}]{biscoveanu2020constraining}
Biscoveanu, S., Thrane, E., \& Vitale, S. 2020, \apj, 893, 38

\bibitem[{{Blackburn} {et~al.}(2015){Blackburn}, {Briggs}, {Camp},
  {Christensen}, {Connaughton}, {Jenke}, {Remillard}, \&
  {Veitch}}]{Blackburn2015}
{Blackburn}, L., {Briggs}, M.~S., {Camp}, J., {et~al.} 2015, \apjs, 217, 8

\bibitem[{Blandford \& Narayan(1986)}]{strong_lens_Blandford_1986}
Blandford, R., \& Narayan, R. 1986, Astrophys. J., 310, 568

\bibitem[{Blandford \& Narayan(1992)}]{review_strong_lens}
Blandford, R.~D., \& Narayan, R. 1992, Annu. Rev Astron. Astrophys., 30, 311

\bibitem[{Blandford \& Znajek(1977)}]{blandford1977electromagnetic}
Blandford, R.~D., \& Znajek, R.~L. 1977, Mon. Not. R. Astron. Soc., 179, 433

\bibitem[{Blas {et~al.}(2016)Blas, Ivanov, Sawicki, \&
  Sibiryakov}]{speed_of_gravity_LIGO_GW150914}
Blas, D., Ivanov, M.~M., Sawicki, I., \& Sibiryakov, S. 2016, JETP Lett., 103,
  624

\bibitem[{Bloemen {et~al.}(2015)Bloemen, Groot, Nelemans, \&
  Klein-Wolt}]{bloemen2015blackgem}
Bloemen, S., Groot, P., Nelemans, G., \& Klein-Wolt, M. 2015, in Living
  Together: Planets, Host Stars and Binaries, Vol. 496, 254

\bibitem[{Bloom {et~al.}(2001)Bloom, Frail, \& Sari}]{bloom2001prompt}
Bloom, J.~S., Frail, D.~A., \& Sari, R. 2001, Astron. J., 121, 2879

\bibitem[{Boran {et~al.}(2018)Boran, Desai, Kahya, \&
  Woodard}]{cgw_GW170817_dark_matter_Boran_2018}
Boran, S., Desai, S., Kahya, E.~O., \& Woodard, R.~P. 2018, \prd, 97, 041501

\bibitem[{{Bose} {et~al.}(2018){Bose}, {Chakravarti}, {Rezzolla},
  {Sathyaprakash}, \& {Takami}}]{2018PhRvL.120c1102B}
{Bose}, S., {Chakravarti}, K., {Rezzolla}, L., {Sathyaprakash}, B.~S., \&
  {Takami}, K. 2018, \prl, 120, 031102

\bibitem[{Bracco(2017)}]{LRP_EU_2017}
Bracco, A. 2017, Europhysics News, 48, 21

\bibitem[{{Briggs} {et~al.}(1996){Briggs}, {Paciesas}, {Pendleton}, {Meegan},
  {Fishman}, {Horack}, {Brock}, {Kouveliotou}, {Hartmann}, \&
  {Hakkila}}]{briggs1995batse}
{Briggs}, M.~S., {Paciesas}, W.~S., {Pendleton}, G.~N., {et~al.} 1996, \apj,
  459, 40

\bibitem[{Brito {et~al.}(2013)Brito, Cardoso, \&
  Pani}]{mg_superradiance_Brito_2013}
Brito, R., Cardoso, V., \& Pani, P. 2013, \prd, 88, 023514

\bibitem[{Brown {et~al.}(2018)Brown, Vallenari, Prusti, De~Bruijne, Babusiaux,
  Bailer-Jones, Biermann, Evans, Eyer, Jansen, {et~al.}}]{gaia_overview_2018}
Brown, A.~G.~A., Vallenari, A., Prusti, T., {et~al.} 2018, Astron. Astrophys.,
  616, A1

\bibitem[{Bucciantini {et~al.}(2011)Bucciantini, Metzger, Thompson, \&
  Quataert}]{bucciantini2011short}
Bucciantini, N., Metzger, B.~D., Thompson, T.~A., \& Quataert, E. 2011, Mon.
  Not. R. Astron. Soc., 419, 1537

\bibitem[{Bucciantini {et~al.}(2008)Bucciantini, Quataert, Arons, Metzger, \&
  Thompson}]{bucciantini2008relativistic}
Bucciantini, N., Quataert, E., Arons, J., Metzger, B., \& Thompson, T.~A. 2008,
  Mon. Not. R. Astron. Soc. Lett., 383, L25

\bibitem[{{Bulla}(2019)}]{bulla2019possis}
{Bulla}, M. 2019, \mnras, 489, 5037

\bibitem[{Bullock \& Boylan-Kolchin(2017)}]{cosmo_review_smallscalecrisis_2017}
Bullock, J.~S., \& Boylan-Kolchin, M. 2017, Annu. Rev. Astron. Astrophys., 55,
  343

\bibitem[{Burbidge {et~al.}(1957)Burbidge, Burbidge, Fowler, \&
  Hoyle}]{burbidge1957synthesis}
Burbidge, E.~M., Burbidge, G.~R., Fowler, W.~A., \& Hoyle, F. 1957, Rev. Mod.
  Phys., 29, 547

\bibitem[{Burgess {et~al.}(2002)Burgess, Cline, Filotas, Matias, \&
  Moore}]{QG_WBW_Burgess_2002}
Burgess, C.~P., Cline, J.~M., Filotas, E., Matias, J., \& Moore, G.~D. 2002,
  Journal of High Energy Physics, 2002, 043

\bibitem[{Burgess {et~al.}(2019)Burgess, Kole, Berlato, Greiner, Vianello,
  Produit, Li, \& Sun}]{burgess2019time}
Burgess, J.~M., Kole, M., Berlato, F., {et~al.} 2019, Astron. Astrophys., 627,
  A105

\bibitem[{Burles {et~al.}(1999)Burles, Nollett, Truran, \&
  Turner}]{cosmo_BBN_1999}
Burles, S., Nollett, K.~M., Truran, J.~W., \& Turner, M.~S. 1999, \prl, 82,
  4176

\bibitem[{Burns(2017)}]{Burns_Dissertation}
Burns, E. 2017, PhD thesis, The University of Alabama in Huntsville

\bibitem[{{Burns} {et~al.}(2016){Burns}, {Connaughton}, {Zhang}, {Lien},
  {Briggs}, {Goldstein}, {Pelassa}, \& {Troja}}]{SGRB_GBM_BAT_Burns_2016}
{Burns}, E., {Connaughton}, V., {Zhang}, B.-B., {et~al.} 2016, \apj, 818, 110

\bibitem[{Burns {et~al.}(2018)Burns, Veres, Connaughton, Racusin, Briggs,
  Christensen, Goldstein, Hamburg, Kocevski, McEnery,
  {et~al.}}]{GRB150101B_Burns}
Burns, E., Veres, P., Connaughton, V., {et~al.} 2018, \apjl, 863, L34

\bibitem[{Burrows {et~al.}(2007)Burrows, Falcone, Chincarini, Morris, Romano,
  Hill, Godet, Moretti, Krimm, Osborne, {et~al.}}]{burrows2007x}
Burrows, D.~N., Falcone, A., Chincarini, G., {et~al.} 2007, Phil.Trans. R. Soc.
  A: Math. Phys. Eng. Sci., 365, 1213

\bibitem[{Caldwell {et~al.}(1998)Caldwell, Dave, \&
  Steinhardt}]{quintessence_2}
Caldwell, R.~R., Dave, R., \& Steinhardt, P.~J. 1998, \prl, 80, 1582

\bibitem[{Callister {et~al.}(2017)Callister, Biscoveanu, Christensen, Isi,
  Matas, Minazzoli, Regimbau, Sakellariadou, Tasson, \&
  Thrane}]{callister2017polarization}
Callister, T., Biscoveanu, A.~S., Christensen, N., {et~al.} 2017, Phys. Rev. X,
  7, 041058

\bibitem[{Cameron(1957)}]{cameron1957nuclear}
Cameron, A. G.~W. 1957, Publ. Astron. Soc. Pac., 69, 201

\bibitem[{Cannon {et~al.}(2012)Cannon, Cariou, Chapman, Crispin-Ortuzar,
  Fotopoulos, Frei, Hanna, Kara, Keppel, Liao, {et~al.}}]{cannon2012toward}
Cannon, K., Cariou, R., Chapman, A., {et~al.} 2012, Astrophys. J., 748, 136

\bibitem[{Cano {et~al.}(2017)Cano, Wang, Dai, \& Wu}]{cano2017observer}
Cano, Z., Wang, S.-Q., Dai, Z.-G., \& Wu, X.-F. 2017, Adv. Astron., 2017

\bibitem[{Cantiello {et~al.}(2018)Cantiello, Jensen, Blakeslee, Berger, Levan,
  Tanvir, Raimondo, Brocato, Alexander, Blanchard,
  {et~al.}}]{GW170817_distance_cantiello_2018}
Cantiello, M., Jensen, J.~B., Blakeslee, J.~P., {et~al.} 2018, \apjl, 854, L31

\bibitem[{Canuel {et~al.}(2018)Canuel, Bertoldi, Amand, Di~Borgo, Chantrait,
  Danquigny, {\'A}lvarez, Fang, Freise, Geiger, {et~al.}}]{canuel2018exploring}
Canuel, B., Bertoldi, A., Amand, L., {et~al.} 2018, Scientific Reports, 8,
  14064

\bibitem[{{Capano} {et~al.}(2019){Capano}, {Tews}, {Brown}, {Margalit}, {De},
  {Kumar}, {Brown}, {Krishnan}, \& {Reddy}}]{capano2019gw170817}
{Capano}, C.~D., {Tews}, I., {Brown}, S.~M., {et~al.} 2019, arXiv e-prints,
  arXiv:1908.10352

\bibitem[{Carbone {et~al.}(2011)Carbone, Verde, Wang, \&
  Cimatti}]{mnu_future_2011}
Carbone, C., Verde, L., Wang, Y., \& Cimatti, A. 2011, JCAP, 2011, 030

\bibitem[{Carracedo {et~al.}(2020)Carracedo, Bulla, Feindt, \&
  Goobar}]{carracedo2020detectability}
Carracedo, A.~S., Bulla, M., Feindt, U., \& Goobar, A. 2020, arXiv e-prints,
  2004.06137

\bibitem[{Caves(1980)}]{cgw_CR_first_Caves_1980}
Caves, C.~M. 1980, Ann. Phys., 125, 35

\bibitem[{Cenko {et~al.}(2010)Cenko, Frail, Harrison, Kulkarni, Nakar, Chandra,
  Butler, Fox, Gal-Yam, Kasliwal, {et~al.}}]{cenko2010collimation}
Cenko, S., Frail, D., Harrison, F., {et~al.} 2010, \apj, 711, 641

\bibitem[{Chakraborty {et~al.}(2018)Chakraborty, Chakravarti, Bose, \&
  SenGupta}]{large_dim_chakraborty_2018}
Chakraborty, S., Chakravarti, K., Bose, S., \& SenGupta, S. 2018, \prd, 97,
  104053

\bibitem[{Chattopadhyay {et~al.}(2019)Chattopadhyay, Vadawale, Aarthy, Mithun,
  Chand, Basak, Rao, Mate, Sharma, Bhalerao,
  {et~al.}}]{chattopadhyay2017prompt}
Chattopadhyay, T., Vadawale, S.~V., Aarthy, E., {et~al.} 2019, \apj, 884, 123

\bibitem[{Chatziioannou {et~al.}(2012)Chatziioannou, Yunes, \&
  Cornish}]{GR_nontensor_polarization_test_PEF_chatziioannou_2012}
Chatziioannou, K., Yunes, N., \& Cornish, N. 2012, \prd, 86, 022004

\bibitem[{Chen {et~al.}(2018)Chen, Fishbach, \& Holz}]{H0_chen_five_years}
Chen, H.-Y., Fishbach, M., \& Holz, D.~E. 2018, Nature, 1

\bibitem[{{Chen} {et~al.}(2017){Chen}, {Holz}, {Miller}, {Evans}, {Vitale}, \&
  {Creighton}}]{chen2017distance}
{Chen}, H.-Y., {Holz}, D.~E., {Miller}, J., {et~al.} 2017, arXiv e-prints,
  arXiv:1709.08079

\bibitem[{Chevallier \& Polarski(2001)}]{chevallier2001accelerating}
Chevallier, M., \& Polarski, D. 2001, Int. J. Mod. Phys. D, 10, 213

\bibitem[{Choudhury {et~al.}(2004)Choudhury, Joshi, Mahajan, \&
  McKellar}]{mg_weak_lensing_choudhury_2004}
Choudhury, S.~R., Joshi, G.~C., Mahajan, S., \& McKellar, B.~H. 2004,
  Astropart. Phys., 21, 559

\bibitem[{Christensen {et~al.}(2004)Christensen, Hjorth, \&
  Gorosabel}]{christensen2004uv}
Christensen, L., Hjorth, J., \& Gorosabel, J. 2004, Astron. Astrophys., 425,
  913

\bibitem[{Church {et~al.}(2011)Church, Levan, Davies, \&
  Tanvir}]{church2011implications}
Church, R.~P., Levan, A.~J., Davies, M.~B., \& Tanvir, N. 2011, \mnras, 413,
  2004

\bibitem[{Clark {et~al.}(2014)Clark, Bauswein, Cadonati, Janka, Pankow, \&
  Stergioulas}]{clark2014prospects}
Clark, J., Bauswein, A., Cadonati, L., {et~al.} 2014, \prd, 90, 062004

\bibitem[{Clark {et~al.}(2016)Clark, Bauswein, Stergioulas, \&
  Shoemaker}]{clark2016observing}
Clark, J.~A., Bauswein, A., Stergioulas, N., \& Shoemaker, D. 2016, Class.
  Quantum Grav., 33, 085003

\bibitem[{Collaboration {et~al.}(2012)Collaboration, Collaboration,
  {et~al.}}]{ligo2012sensitivity}
Collaboration, L.~S., Collaboration, V., {et~al.} 2012, arXiv e-prints,
  1203.2674

\bibitem[{Collett(2015)}]{surveys_strong_lens_collett_2015}
Collett, T.~E. 2015, Astrophys. J., 811, 20

\bibitem[{Collett \& Bacon(2017)}]{cgw_strong_lense_Collett_2017}
Collett, T.~E., \& Bacon, D. 2017, \prl, 118, 091101

\bibitem[{Connaughton(2002)}]{extended_emission_2002_connaughton}
Connaughton, V. 2002, Astrophys. J., 567, 1028

\bibitem[{Cook {et~al.}(1994)Cook, Shapiro, \& Teukolsky}]{cook1994rapidly}
Cook, G.~B., Shapiro, S.~L., \& Teukolsky, S.~A. 1994, Astrophys. J., 424, 823

\bibitem[{Cornish {et~al.}(2017)Cornish, Blas, \&
  Nardini}]{speed_of_gravity_multiple_BBH}
Cornish, N., Blas, D., \& Nardini, G. 2017, \prl, 119, 161102

\bibitem[{C{\^o}t{\'e} {et~al.}(2019)C{\^o}t{\'e}, Lugaro, Reifarth, Pignatari,
  Vil{\'a}gos, Yag{\"u}e, \& Gibson}]{cote2019galactic}
C{\^o}t{\'e}, B., Lugaro, M., Reifarth, R., {et~al.} 2019, Astrophys. J., 878,
  156

\bibitem[{C{\^o}t{\'e} {et~al.}(2018)C{\^o}t{\'e}, Fryer, Belczynski, Korobkin,
  Chru{\'s}li{\'n}ska, Vassh, Mumpower, Lippuner, Sprouse, Surman,
  {et~al.}}]{cote2018origin}
C{\^o}t{\'e}, B., Fryer, C.~L., Belczynski, K., {et~al.} 2018, Astrophys. J.,
  855, 99

\bibitem[{Coughlin {et~al.}(2020)Coughlin, Nixon, Barnes, Metzger, \&
  Margutti}]{coughlin2020variability}
Coughlin, E.~R., Nixon, C., Barnes, J., Metzger, B.~D., \& Margutti, R. 2020,
  arXiv e-prints, 2006.03174

\bibitem[{Coughlin \& Dietrich(2019)}]{coughlin2019can}
Coughlin, M.~W., \& Dietrich, T. 2019, \prd, 100, 043011

\bibitem[{{Coughlin} {et~al.}(2019){Coughlin}, {Dietrich}, {Margalit}, \&
  {Metzger}}]{coughlin2018multi}
{Coughlin}, M.~W., {Dietrich}, T., {Margalit}, B., \& {Metzger}, B.~D. 2019,
  \mnras, 489, L91

\bibitem[{Coughlin {et~al.}(2018)Coughlin, Dietrich, Doctor, Kasen, Coughlin,
  Jerkstrand, Leloudas, McBrien, Metzger, O'Shaughnessy,
  {et~al.}}]{coughlin2018constraints}
Coughlin, M.~W., Dietrich, T., Doctor, Z., {et~al.} 2018, Mon. Not. R. Astron.
  Soc., 480, 3871

\bibitem[{Coughlin {et~al.}(2019)Coughlin, Ahumada, Anand, De, Hankins,
  Kasliwal, Singer, Bellm, Andreoni, Cenko, {et~al.}}]{coughlin2019growth}
Coughlin, M.~W., Ahumada, T., Anand, S., {et~al.} 2019, arXiv e-prints,
  1907.12645

\bibitem[{Coulter {et~al.}(2017)Coulter, Foley, Kilpatrick, Drout, Piro,
  Shappee, Siebert, Simon, Ulloa, Kasen,
  {et~al.}}]{gw170817_kilonova_first_swope}
Coulter, D.~A., Foley, R.~J., Kilpatrick, C.~D., {et~al.} 2017, Science,
  eaap9811

\bibitem[{{Cowan} {et~al.}(1956){Cowan}, {Reines}, {Harrison}, {Kruse}, \&
  {McGuire}}]{cowan_reines_1956}
{Cowan}, Jr., C.~L., {Reines}, F., {Harrison}, F.~B., {Kruse}, H.~W., \&
  {McGuire}, A.~D. 1956, Science, 124, 103

\bibitem[{Cowan {et~al.}(2019)Cowan, Sneden, Lawler, Aprahamian, Wiescher,
  Langanke, Mart{\'\i}nez-Pinedo, \& Thielemann}]{cowan2019making}
Cowan, J.~J., Sneden, C., Lawler, J.~E., {et~al.} 2019, arXiv e-prints,
  1901.01410

\bibitem[{Cowperthwaite \& Berger(2015)}]{cowperthwaite2015comprehensive}
Cowperthwaite, P.~S., \& Berger, E. 2015, Astrophys. J., 814, 25

\bibitem[{Creminelli \&
  Vernizzi(2017)}]{cgw_beyond_GR_constraints_Creminelli_2017}
Creminelli, P., \& Vernizzi, F. 2017, \prl, 119, 251302

\bibitem[{Crowder {et~al.}(2013)Crowder, Namba, Mandic, Mukohyama, \&
  Peloso}]{crowder2013measurement}
Crowder, S.~G., Namba, R., Mandic, V., Mukohyama, S., \& Peloso, M. 2013, Phys.
  Lett. B, 726, 66

\bibitem[{Cuesta {et~al.}(2016)Cuesta, Niro, \& Verde}]{mnu_future_2016}
Cuesta, A.~J., Niro, V., \& Verde, L. 2016, Phys. Dark Universe, 13, 77

\bibitem[{Curran {et~al.}(2007)Curran, Van~der Horst, Wijers, Starling,
  Castro-Tirado, Fynbo, Gorosabel, J{\"a}rvinen, Malesani, Rol,
  {et~al.}}]{curran2007grb}
Curran, P., Van~der Horst, A., Wijers, R., {et~al.} 2007, Mon. Not. R. Astron.
  Soc. Lett., 381, L65

\bibitem[{Cutler(1998)}]{cutler1998angular}
Cutler, C. 1998, \prd, 57, 7089

\bibitem[{Cutler \& Holz(2009)}]{cutler2009ultrahigh}
Cutler, C., \& Holz, D.~E. 2009, \prd, 80, 104009

\bibitem[{{da Silva Schneider} {et~al.}(2020){da Silva Schneider}, {O'Connor},
  Granqvist, Betranhandy, \& Couch}]{da2020equation}
{da Silva Schneider}, A., {O'Connor}, E., Granqvist, E., Betranhandy, A., \&
  Couch, S.~M. 2020, Astrophys. J., 894, 4

\bibitem[{{Dai} \& {Lu}(1998)}]{dai1998gamma}
{Dai}, Z.~G., \& {Lu}, T. 1998, \aap, 333, L87

\bibitem[{{Dai} {et~al.}(2006){Dai}, {Wang}, {Wu}, \&
  {Zhang}}]{magnetar_bing_3}
{Dai}, Z.~G., {Wang}, X.~Y., {Wu}, X.~F., \& {Zhang}, B. 2006, Science, 311,
  1127

\bibitem[{Dalal {et~al.}(2006)Dalal, Holz, Hughes, \&
  Jain}]{H0-Dalal-dark-energy-2006}
Dalal, N., Holz, D.~E., Hughes, S.~A., \& Jain, B. 2006, \prd, 74, 063006

\bibitem[{Damour \& Polyakov(1994)}]{LIV_varying_moduli_damour_1994}
Damour, T., \& Polyakov, A.~M. 1994, Nucl. Phys. B, 423, 532

\bibitem[{Danby {et~al.}(1962)Danby, Gaillard, Goulianos, Lederman, Mistry,
  Schwartz, \& Steinberger}]{neutrino_muon_danby_1962}
Danby, G., Gaillard, J.~M., Goulianos, K., {et~al.} 1962, \prl, 9, 36

\bibitem[{Davis~Jr {et~al.}(1968)Davis~Jr, Harmer, \& Hoffman}]{mnu_davis_1968}
Davis~Jr, R., Harmer, D.~S., \& Hoffman, K.~C. 1968, \prl, 20, 1205

\bibitem[{De {et~al.}(2018)De, Finstad, Lattimer, Brown, Berger, \&
  Biwer}]{de2018tidal}
De, S., Finstad, D., Lattimer, J.~M., {et~al.} 2018, \prl, 121, 091102

\bibitem[{De~Felice \& Tsujikawa(2012)}]{cgw_dark_energy_Felice_2012}
De~Felice, A., \& Tsujikawa, S. 2012, JCAP, 2012, 007

\bibitem[{de~Rham(2014)}]{review_massive_gravity_Rham_2014}
de~Rham, C. 2014, Living Rev. Relativ., 17, 7

\bibitem[{Deffayet \& Menou(2007)}]{large_dim_deffayet_2007}
Deffayet, C., \& Menou, K. 2007, \apjl, 668, L143

\bibitem[{{Derishev} \& {Piran}(2019)}]{SSC_190114C_2}
{Derishev}, E., \& {Piran}, T. 2019, \apjl, 880, L27

\bibitem[{Desai {et~al.}(2008)Desai, Kahya, \&
  Woodard}]{EP_dark_matter_eumlators_Desai_2008}
Desai, S., Kahya, E.~O., \& Woodard, R.~P. 2008, \prd, 77, 124041

\bibitem[{Dessart {et~al.}(2008)Dessart, Ott, Burrows, Rosswog, \&
  Livne}]{dessart2008neutrino}
Dessart, L., Ott, C.~D., Burrows, A., Rosswog, S., \& Livne, E. 2008,
  Astrophys. J., 690, 1681

\bibitem[{Dezalay {et~al.}(1991)Dezalay, Barat, Talon, Sunyaev, Terekhov, \&
  Kuznetsov}]{Dezalay1991}
Dezalay, J.-P., Barat, C., Talon, R., {et~al.} 1991, in AIP Conference
  Proceedings, Vol. 265, Gamma-ray bursts, ed. W.~Paciesas \& G.~J. Fishman
  (American Institute of Physics), 304--309

\bibitem[{Di~Valentino {et~al.}(2018)Di~Valentino, Holz, Melchiorri, \&
  Renzi}]{mnu_cosmo_improvements_2018}
Di~Valentino, E., Holz, D.~E., Melchiorri, A., \& Renzi, F. 2018, \prd, 98,
  083523

\bibitem[{Dichiara {et~al.}(2013)Dichiara, Guidorzi, Frontera, \&
  Amati}]{dichiara2013search}
Dichiara, S., Guidorzi, C., Frontera, F., \& Amati, L. 2013, Astrophys. J.,
  777, 132

\bibitem[{Diehl {et~al.}(2012)Diehl, Collaboration, {et~al.}}]{diehl2012dark}
Diehl, T., Collaboration, D. E.~S., {et~al.} 2012, Physics Procedia, 37, 1332

\bibitem[{Dimopoulos {et~al.}(2008)Dimopoulos, Graham, Hogan, Kasevich, \&
  Rajendran}]{dimopoulos2008atomic}
Dimopoulos, S., Graham, P.~W., Hogan, J.~M., Kasevich, M.~A., \& Rajendran, S.
  2008, \prd, 78, 122002

\bibitem[{Dominik {et~al.}(2015)Dominik, Berti, O'Shaughnessy, Mandel,
  Belczynski, Fryer, Holz, Bulik, \& Pannarale}]{dominik2015double}
Dominik, M., Berti, E., O'Shaughnessy, R., {et~al.} 2015, Astrophys. J., 806,
  263

\bibitem[{Douglas \& Nekrasov(2001)}]{QG_noncommutative_douglas_2001}
Douglas, M.~R., \& Nekrasov, N.~A. 2001, Rev. Mod. Phys., 73, 977

\bibitem[{Drewes(2013)}]{mnu_right_handed_neutrinos_2013}
Drewes, M. 2013, Int. J. Mod. Phys. E, 22, 1330019

\bibitem[{Duffell {et~al.}(2018)Duffell, Quataert, Kasen, \&
  Klion}]{duffell2018jet}
Duffell, P.~C., Quataert, E., Kasen, D., \& Klion, H. 2018, Astrophy. J., 866,
  3

\bibitem[{Dyda {et~al.}(2012)Dyda, Flanagan, \& Kamionkowski}]{dyda2012vacuum}
Dyda, S., Flanagan, {\'E}.~{\'E}., \& Kamionkowski, M. 2012, \prd, 86, 124031

\bibitem[{Eardley {et~al.}(1973)Eardley, Lee, \&
  Lightman}]{eardley1973gravitational}
Eardley, D.~M., Lee, D.~L., \& Lightman, A.~P. 1973, \prd, 8, 3308

\bibitem[{Einstein(1905)}]{Einstein_1905_special}
Einstein, A. 1905, Ann. Phys., 322, 891

\bibitem[{{Einstein}(1916)}]{einstein_GR_1916}
{Einstein}, A. 1916, Sitzungsber. K{\"o}nigl. Preu{\ss}. Akad. Wiss., 688

\bibitem[{Ellis {et~al.}(2008)Ellis, Harries, Meregaglia, Rubbia, \&
  Sakharov}]{LIV_SN1987a_Ellis}
Ellis, J., Harries, N., Meregaglia, A., Rubbia, A., \& Sakharov, A.~S. 2008,
  \prd, 78, 033013

\bibitem[{{Ellis} {et~al.}(1999){Ellis}, {Mavromatos}, \&
  {Nanopoulos}}]{QG_string_ellis_1999}
{Ellis}, J., {Mavromatos}, N.~E., \& {Nanopoulos}, D.~V. 1999, arXiv e-prints,
  arXiv:gr-qc/9909085

\bibitem[{Ellis {et~al.}(2019)Ellis, Mavromatos, Sakharov, \&
  Sarkisyan-Grinbaum}]{LIV_neutrino_Ellis}
Ellis, J., Mavromatos, N.~E., Sakharov, A.~S., \& Sarkisyan-Grinbaum, E.~K.
  2019, Phys. Lett. B, 789, 352

\bibitem[{Englert \& Brout(1964)}]{Higgs_2}
Englert, F., \& Brout, R. 1964, \prl, 13, 321

\bibitem[{Estevez {et~al.}(2018)Estevez, Lieunard, Marion, Mours, Rolland, \&
  Verkindt}]{estevez2018first}
Estevez, D., Lieunard, B., Marion, F., {et~al.} 2018, Class. Quantum Grav., 35,
  235009

\bibitem[{Evans {et~al.}(2015)Evans, Osborne, Kennea, Campana, O'Brien, Tanvir,
  Racusin, Burrows, Cenko, \& Gehrels}]{evans2015optimization}
Evans, P.~A., Osborne, J.~P., Kennea, J.~A., {et~al.} 2015, Mon. Not. R.
  Astron. Soc., 455, 1522

\bibitem[{Evans {et~al.}(2016)Evans, Kennea, Palmer, Bilicki, Osborne,
  {O'Brien}, Tanvir, Lien, Barthelmy, Burrows, {et~al.}}]{evans2016swift}
Evans, P.~A., Kennea, J.~A., Palmer, D.~M., {et~al.} 2016, Mon. Not. R. Astron.
  Soc., 462, 1591

\bibitem[{Evans {et~al.}(2017)Evans, Cenko, Kennea, Emery, Kuin, Korobkin,
  Wollaeger, Fryer, Madsen, Harrison, {et~al.}}]{gw170817_evans_swift}
Evans, P.~A., Cenko, S.~B., Kennea, J.~A., {et~al.} 2017, Science, 358, 1565

\bibitem[{Ezquiaga \&
  Zumalac{\'a}rregui(2017)}]{cgw_beyond_GR_constraints_Ezquiaga_2017}
Ezquiaga, J.~M., \& Zumalac{\'a}rregui, M. 2017, \prl, 119, 251304

\bibitem[{Faber \& Rasio(2012)}]{review_NS_Faber_Rasio}
Faber, J.~A., \& Rasio, F.~A. 2012, Living Rev. Relativ., 15, 8

\bibitem[{Fan {et~al.}(2013)Fan, Wu, \& Wei}]{fan2013signature}
Fan, Y.-Z., Wu, X.-F., \& Wei, D.-M. 2013, \prd, 88, 067304

\bibitem[{Fan {et~al.}(2005)Fan, Zhang, \& Proga}]{fan2005linearly}
Fan, Y.~Z., Zhang, B., \& Proga, D. 2005, \apjl, 635, L129

\bibitem[{Fang \& Metzger(2017)}]{fang2017high}
Fang, K., \& Metzger, B.~D. 2017, Astrophys. J., 849, 153

\bibitem[{Feeney {et~al.}(2019)Feeney, Peiris, Williamson, Nissanke, Mortlock,
  Alsing, \& Scolnic}]{H0_feeney_2018}
Feeney, S.~M., Peiris, H.~V., Williamson, A.~R., {et~al.} 2019, \prl, 122,
  061105

\bibitem[{Fenimore {et~al.}(1993)Fenimore, Epstein, \& Ho}]{fenimore1993escape}
Fenimore, E.~E., Epstein, R.~I., \& Ho, C. 1993, Astronomy and Astrophysics
  Supplement Series, 97, 59

\bibitem[{Ferdman {et~al.}(2020)Ferdman, Freire, Perera, Pol, Camilo,
  Chatterjee, Cordes, Crawford, Hessels, Kaspi,
  {et~al.}}]{ferdman2020asymmetric}
Ferdman, R.~D., Freire, P.~C.~C., Perera, B.~B.~P., {et~al.} 2020, Nature, 583,
  211

\bibitem[{Fern{\'a}ndez \& Metzger(2016)}]{overview_fernandez_metzger_2016}
Fern{\'a}ndez, R., \& Metzger, B.~D. 2016, Annu. Rev. Nucl. Part. Sci., 66, 23

\bibitem[{Fern{\'a}ndez {et~al.}(2018)Fern{\'a}ndez, Tchekhovskoy, Quataert,
  Foucart, \& Kasen}]{fernandez2018long}
Fern{\'a}ndez, R., Tchekhovskoy, A., Quataert, E., Foucart, F., \& Kasen, D.
  2018, Mon. Not. R. Astron. Soc., 482, 3373

\bibitem[{{Fields} {et~al.}(2014){Fields}, {Molaro}, \&
  {Sarkar}}]{cosmo_review_BBN_2014}
{Fields}, B.~D., {Molaro}, P., \& {Sarkar}, S. 2014, arXiv e-prints,
  arXiv:1412.1408

\bibitem[{Finn \& Chernoff(1993)}]{finn1993observing}
Finn, L.~S., \& Chernoff, D.~F. 1993, \prd, 47, 2198

\bibitem[{Finn \& Sutton(2002)}]{mg_binary_pulsar_Finn_2002}
Finn, L.~S., \& Sutton, P.~J. 2002, \prd, 65, 044022

\bibitem[{{Fong} {et~al.}(2013){Fong}, {Berger}, {Chornock}, {Margutti},
  {Levan}, {Tanvir}, {Tunnicliffe}, {Czekala}, {Fox}, {Perley}, {Cenko},
  {Zauderer}, {Laskar}, {Persson}, {Monson}, {Kelson}, {Birk}, {Murphy},
  {Servillat}, \& {Anglada}}]{2013ApJ...769...56F}
{Fong}, W., {Berger}, E., {Chornock}, R., {et~al.} 2013, \apj, 769, 56

\bibitem[{Fong \& Berger(2013{\natexlab{a}})}]{SGRB_offset_measured_fong_2013}
Fong, W.-f., \& Berger, E. 2013{\natexlab{a}}, Astrophys. J., 776, 18

\bibitem[{Fong \& Berger(2013{\natexlab{b}})}]{fong2013locations}
---. 2013{\natexlab{b}}, \apj, 776, 18

\bibitem[{Fong {et~al.}(2015)Fong, Berger, Margutti, \&
  Zauderer}]{fong2015decade}
Fong, W.-f., Berger, E., Margutti, R., \& Zauderer, B.~A. 2015, Astrophys. J.,
  815, 102

\bibitem[{Fong {et~al.}(2016)Fong, Metzger, Berger, \&
  {\"O}zel}]{fong2016radio}
Fong, W.-f., Metzger, B.~D., Berger, E., \& {\"O}zel, F. 2016, Astrophys. J.,
  831, 141

\bibitem[{Fong {et~al.}(2013)Fong, Berger, Chornock, Margutti, Levan, Tanvir,
  Tunnicliffe, Czekala, Fox, Perley, {et~al.}}]{fong2013demographics}
Fong, W.-f., Berger, E., Chornock, R., {et~al.} 2013, \apj, 769, 56

\bibitem[{Fong {et~al.}(2017)Fong, Berger, Blanchard, Margutti, Cowperthwaite,
  Chornock, Alexander, Metzger, Villar, Nicholl,
  {et~al.}}]{fong2017electromagnetic}
Fong, W.-f., Berger, E., Blanchard, P.~K., {et~al.} 2017, \apjl, 848, L23

\bibitem[{{Fong} {et~al.}(2019){Fong}, {Blanchard}, {Alexander}, {Strader},
  {Margutti}, {Hajela}, {Villar}, {Wu}, {Ye}, {Berger}, {Chornock},
  {Coppejans}, {Cowperthwaite}, {Eftekhari}, {Giannios}, {Guidorzi},
  {Kathirgamaraju}, {Laskar}, {MacFadyen}, {Metzger}, {Nicholl}, {Paterson},
  {Terreran}, {Sand}, {Sironi}, {Williams}, {Xie}, \&
  {Zrake}}]{2019arXiv190808046F}
{Fong}, W.-f., {Blanchard}, P.~K., {Alexander}, K.~D., {et~al.} 2019, arXiv
  e-prints, arXiv:1908.08046

\bibitem[{Foucart(2012)}]{foucart2012black}
Foucart, F. 2012, \prd, 86, 124007

\bibitem[{Foucart(2020)}]{foucart2020brief}
---. 2020, arXiv e-prints, 2006.10570

\bibitem[{{Foucart} {et~al.}(2018){Foucart}, {Hinderer}, \&
  {Nissanke}}]{foucart2018remnant}
{Foucart}, F., {Hinderer}, T., \& {Nissanke}, S. 2018, \prd, 98, 081501

\bibitem[{Foucart {et~al.}(2014)Foucart, Deaton, Duez, O'Connor, Ott, Haas,
  Kidder, Pfeiffer, Scheel, \& Szilagyi}]{foucart2014neutron}
Foucart, F., Deaton, M.~B., Duez, M.~D., {et~al.} 2014, \prd, 90, 024026

\bibitem[{Foucart {et~al.}(2016)Foucart, Haas, Duez, O'Connor, Ott, Roberts,
  Kidder, Lippuner, Pfeiffer, \& Scheel}]{foucart2016low}
Foucart, F., Haas, R., Duez, M.~D., {et~al.} 2016, \prd, 93, 044019

\bibitem[{Fox {et~al.}(2003)Fox, Yost, Kulkarni, Torii, Kato, Yamaoka, Sako,
  Harrison, Sari, Price, {et~al.}}]{fox2003early}
Fox, D., Yost, S., Kulkarni, S.~R., {et~al.} 2003, Nature, 422, 284

\bibitem[{Fraija {et~al.}(2019)Fraija, Dichiara, do~ES~Pedreira, Galvan-Gamez,
  Becerra, Duran, \& Zhang}]{fraija2019analysis}
Fraija, N., Dichiara, S., do~ES~Pedreira, A.~C.~C., {et~al.} 2019, \apjl, 879,
  L26

\bibitem[{Frail {et~al.}(1997)Frail, Kulkarni, Nicastro, Feroci, \&
  Taylor}]{frail1997radio}
Frail, D.~A., Kulkarni, S.~R., Nicastro, L., Feroci, M., \& Taylor, G.~B. 1997,
  Nature, 389, 261

\bibitem[{Frail {et~al.}(2001)Frail, Kulkarni, Sari, Djorgovski, Bloom, Galama,
  Reichart, Berger, Harrison, Price, {et~al.}}]{frail2001beaming}
Frail, D.~A., Kulkarni, S., Sari, R., {et~al.} 2001, \apjl, 562, L55

\bibitem[{{Frederiks} {et~al.}(2007){Frederiks}, {Palshin}, {Aptekar},
  {Golenetskii}, {Cline}, \& {Mazets}}]{GMF_051103_M81}
{Frederiks}, D.~D., {Palshin}, V.~D., {Aptekar}, R.~L., {et~al.} 2007, Astron.
  Lett., 33, 19

\bibitem[{Freedman \& Madore(2010)}]{hubble_constant_Freedman}
Freedman, W.~L., \& Madore, B.~F. 2010, Annu. Rev. Astron. Astrophys., 48, 673

\bibitem[{Freiburghaus {et~al.}(1999)Freiburghaus, Rosswog, \&
  Thielemann}]{freiburghaus1999r}
Freiburghaus, C., Rosswog, S., \& Thielemann, F.-K. 1999, \apjl, 525, L121

\bibitem[{{Friedmann}(1922)}]{1922ZPhy...10..377F}
{Friedmann}, A. 1922, Z. Physik, 10, 377

\bibitem[{Fruchter {et~al.}(1999)Fruchter, Thorsett, Metzger, Sahu, Petro,
  Livio, Ferguson, Pian, Hogg, Galama, {et~al.}}]{fruchter1999hubble}
Fruchter, A.~S., Thorsett, S., Metzger, M.~R., {et~al.} 1999, \apjl, 519, L13

\bibitem[{Fruchter {et~al.}(2006)Fruchter, Levan, Strolger, Vreeswijk,
  Thorsett, Bersier, Burud, Cer{\'o}n, Castro-Tirado, Conselice,
  {et~al.}}]{fruchter2006long}
Fruchter, A.~S., Levan, A.~J., Strolger, L., {et~al.} 2006, Nature, 441, 463

\bibitem[{Fukuda {et~al.}(1998)Fukuda, Hayakawa, Ichihara, Inoue, Ishihara,
  Ishino, Itow, Kajita, Kameda, Kasuga,
  {et~al.}}]{mnu_atmospheric_oscillation_SK_1998}
Fukuda, Y., Hayakawa, T., Ichihara, E., {et~al.} 1998, \prl, 81, 1562

\bibitem[{Fukugita \& Yanagida(1986)}]{neutrino_baryogenesis_fukugita_1986}
Fukugita, M., \& Yanagida, T. 1986, Phys. Lett. B, 174, 45

\bibitem[{Gair {et~al.}(2013)Gair, Vallisneri, Larson, \&
  Baker}]{review_GR_tests_with_GWs_LISA_Gair_2013}
Gair, J.~R., Vallisneri, M., Larson, S.~L., \& Baker, J.~G. 2013, Living Rev.
  Relativ., 16, 7

\bibitem[{Galama {et~al.}(1998)Galama, Vreeswijk, Van~Paradijs, Kouveliotou,
  Augusteijn, B{\"o}hnhardt, Brewer, Doublier, Gonzalez, Leibundgut,
  {et~al.}}]{galama1998unusual}
Galama, T.~J., Vreeswijk, P., Van~Paradijs, J., {et~al.} 1998, Nature, 395, 670

\bibitem[{Gambini \& Pullin(1999)}]{QG_LQG_gambini_1999}
Gambini, R., \& Pullin, J. 1999, \prd, 59, 124021

\bibitem[{{Gao} {et~al.}(2019){Gao}, {Ai}, {Cao}, {Zhang}, {Zhu}, {Li},
  {Zhang}, \& {Bauswein}}]{gao2019relation}
{Gao}, H., {Ai}, S.-K., {Cao}, Z.-J., {et~al.} 2019, arXiv e-prints,
  arXiv:1905.03784

\bibitem[{{Gao} {et~al.}(2013){Gao}, {Ding}, {Wu}, {Zhang}, \& {Dai}}]{Bing_7}
{Gao}, H., {Ding}, X., {Wu}, X.-F., {Zhang}, B., \& {Dai}, Z.-G. 2013, \apj,
  771, 86

\bibitem[{Gao {et~al.}(2015)Gao, Wu, \& M{\'e}sz{\'a}ros}]{WEP_GRBs_Gao_2015}
Gao, H., Wu, X.-F., \& M{\'e}sz{\'a}ros, P. 2015, Astrophys. J., 810, 121

\bibitem[{Gao \& Wald(2000)}]{gao2000theorems}
Gao, S., \& Wald, R.~M. 2000, Class. Quantum Grav., 17, 4999

\bibitem[{Gao \& Fan(2006)}]{gao2006short}
Gao, W.-H., \& Fan, Y.-Z. 2006, Chinese J. Astron. Astrophys., 6, 513

\bibitem[{Garwin {et~al.}(1957)Garwin, Lederman, \&
  Weinrich}]{garwin1957observations}
Garwin, R.~L., Lederman, L.~M., \& Weinrich, M. 1957, Phys. Rev., 105, 1415

\bibitem[{Gehrels {et~al.}(2016)Gehrels, Cannizzo, Kanner, Kasliwal, Nissanke,
  \& Singer}]{gehrels2016galaxy}
Gehrels, N., Cannizzo, J.~K., Kanner, J., {et~al.} 2016, Astrophys. J., 820,
  136

\bibitem[{Gehrels {et~al.}(2015)Gehrels, Spergel, SDT,
  {et~al.}}]{cosmo_gehrels_2015}
Gehrels, N., Spergel, D., SDT, W., {et~al.} 2015, J. Phys.: Conf. Ser., 610,
  012007

\bibitem[{Gehrels {et~al.}(2005)Gehrels, Sarazin, O'Brien, Zhang, Barbier,
  Barthelmy, Blustin, Burrows, Cannizzo, Cummings, {et~al.}}]{gehrels2005short}
Gehrels, N., Sarazin, C., O'Brien, P.~T., {et~al.} 2005, Nature, 437, 851

\bibitem[{Gendreau {et~al.}(2012)Gendreau, Arzoumanian, \&
  Okajima}]{gendreau2012neutron}
Gendreau, K.~C., Arzoumanian, Z., \& Okajima, T. 2012, in Space Telescopes and
  Instrumentation 2012: Ultraviolet to Gamma Ray, Vol. 8443, International
  Society for Optics and Photonics, 844313

\bibitem[{Geng {et~al.}(2019)Geng, Zhang, K{\"o}lligan, Kuiper, \&
  Huang}]{geng2019propagation}
Geng, J.-J., Zhang, B., K{\"o}lligan, A., Kuiper, R., \& Huang, Y.-F. 2019,
  \apjl, 877, L40

\bibitem[{{Geng} {et~al.}(2019){Geng}, {Zhang}, {K{\"o}lligan}, {Kuiper}, \&
  {Huang}}]{jet_time_3}
{Geng}, J.-J., {Zhang}, B., {K{\"o}lligan}, A., {Kuiper}, R., \& {Huang}, Y.-F.
  2019, \apjl, 877, L40

\bibitem[{Ghirlanda {et~al.}(2016)Ghirlanda, Salafia, Pescalli, Ghisellini,
  Salvaterra, Chassande-Mottin, Colpi, Nappo, D’Avanzo, Melandri,
  {et~al.}}]{ghirlanda2016short}
Ghirlanda, G., Salafia, O.~S., Pescalli, A., {et~al.} 2016, Astronomy \&
  Astrophysics, 594, A84

\bibitem[{Ghirlanda {et~al.}(2018)Ghirlanda, Nappo, Ghisellini, Melandri,
  Marcarini, Nava, Salafia, Campana, \& Salvaterra}]{ghirlanda2018bulk}
Ghirlanda, G., Nappo, F., Ghisellini, G., {et~al.} 2018, Astron. Astrophys.,
  609, A112

\bibitem[{Ghirlanda {et~al.}(2019)Ghirlanda, Salafia, Paragi, Giroletti, Yang,
  Marcote, Blanchard, Agudo, An, Bernardini, {et~al.}}]{ghirlanda2019compact}
Ghirlanda, G., Salafia, O.~S., Paragi, Z., {et~al.} 2019, Science, 363, 968

\bibitem[{Giacomazzo \& Perna(2013)}]{giacomazzo2013formation}
Giacomazzo, B., \& Perna, R. 2013, \apjl, 771, L26

\bibitem[{Giacomazzo {et~al.}(2015)Giacomazzo, Zrake, Duffell, MacFadyen, \&
  Perna}]{giacomazzo2015producing}
Giacomazzo, B., Zrake, J., Duffell, P.~C., MacFadyen, A.~I., \& Perna, R. 2015,
  \apj, 809, 39

\bibitem[{Gill {et~al.}(2019)Gill, Granot, De~Colle, \&
  Urrutia}]{gill2019numerical}
Gill, R., Granot, J., De~Colle, F., \& Urrutia, G. 2019, \apj, 883, 15

\bibitem[{Gleyzes {et~al.}(2015)Gleyzes, Langlois, Piazza, \&
  Vernizzi}]{cgw_dark_energy_Gleyzes_2015}
Gleyzes, J., Langlois, D., Piazza, F., \& Vernizzi, F. 2015, JCAP, 2015, 018

\bibitem[{{Goldstein} {et~al.}(2016){Goldstein}, {Burns}, {Hamburg},
  {Connaughton}, {Veres}, {Briggs}, {Hui}, \& {The GBM-LIGO
  Collaboration}}]{Goldstein2016}
{Goldstein}, A., {Burns}, E., {Hamburg}, R., {et~al.} 2016, ArXiv e-prints,
  arXiv:1612.02395

\bibitem[{Goldstein {et~al.}(2017)Goldstein, Veres, Burns, Briggs, Hamburg,
  Kocevski, Wilson-Hodge, Preece, Poolakkil, Roberts,
  {et~al.}}]{GBM_only_paper}
Goldstein, A., Veres, P., Burns, E., {et~al.} 2017, \apjl, 848, L14

\bibitem[{{Goldstein} {et~al.}(2017){Goldstein}, {Veres}, {Burns}, {Briggs},
  {Hamburg}, {Kocevski}, {Wilson-Hodge}, {Preece}, {Poolakkil}, {Roberts},
  {Hui}, {Connaughton}, {Racusin}, {von Kienlin}, {Dal Canton}, {Christensen},
  {Littenberg}, {Siellez}, {Blackburn}, {Broida}, {Bissaldi}, {Cleveland},
  {Gibby}, {Giles}, {Kippen}, {McBreen}, {McEnery}, {Meegan}, {Paciesas}, \&
  {Stanbro}}]{Goldstein2017}
{Goldstein}, A., {Veres}, P., {Burns}, E., {et~al.} 2017, \apjl, 848, L14

\bibitem[{Gompertz {et~al.}(2013)Gompertz, O'Brien, \&
  Wynn}]{gompertz2013magnetar}
Gompertz, B.~P., O'Brien, P.~T., \& Wynn, G.~A. 2013, Mon. Not. R. Astron.
  Soc., 438, 240

\bibitem[{Gompertz {et~al.}(2018)Gompertz, Levan, Tanvir, Hjorth, Covino,
  Evans, Fruchter, {Gonz{\'a}lez-Fern{\'a}ndez}, Jin, Lyman,
  {et~al.}}]{gompertz2018diversity}
Gompertz, B.~P., Levan, A.~J., Tanvir, N.~R., {et~al.} 2018, Astrophys. J.,
  860, 62

\bibitem[{Goobar {et~al.}(2017)Goobar, Amanullah, Kulkarni, Nugent, Johansson,
  Steidel, Law, M{\"o}rtsell, Quimby, Blagorodnova, {et~al.}}]{lensed_typeIa}
Goobar, A., Amanullah, R., Kulkarni, S.~R., {et~al.} 2017, Science, 356, 291

\bibitem[{{Goodman}(1986)}]{SGRB_progenitor_NS_Goodman_1986}
{Goodman}, J. 1986, \apjl, 308, L47

\bibitem[{{Goodman}(1997)}]{goodman1997radio}
---. 1997, \na, 2, 449

\bibitem[{Gorski {et~al.}(2005)Gorski, Hivon, Banday, Wandelt, Hansen,
  Reinecke, \& Bartelmann}]{gorski2005healpix}
Gorski, K.~M., Hivon, E., Banday, A.~J., {et~al.} 2005, Astrophys. J., 622, 759

\bibitem[{Gottlieb {et~al.}(2018)Gottlieb, Nakar, Piran, \&
  Hotokezaka}]{gottlieb2018cocoon}
Gottlieb, O., Nakar, E., Piran, T., \& Hotokezaka, K. 2018, Mon. Not. R.
  Astron. Soc., 479, 588

\bibitem[{{Graham} {et~al.}(2017){Graham}, {Hogan}, {Kasevich}, {Rajendran}, \&
  {Romani}}]{graham2017mid}
{Graham}, P.~W., {Hogan}, J.~M., {Kasevich}, M.~A., {Rajendran}, S., \&
  {Romani}, R.~W. 2017, arXiv e-prints, arXiv:1711.02225

\bibitem[{{Granot} {et~al.}(2017){Granot}, {Guetta}, \&
  {Gill}}]{granot2017lessons}
{Granot}, J., {Guetta}, D., \& {Gill}, R. 2017, \apjl, 850, L24

\bibitem[{Granot \& Kumar(2003)}]{granot2003constraining}
Granot, J., \& Kumar, P. 2003, Astrophys. J., 591, 1086

\bibitem[{Grove {et~al.}(2019)Grove, Cheung, Kerr, Mitchell, Phlips, Woolf,
  Wulf, Wilson-Hodge, Kocevski, Briggs, {et~al.}}]{grove2019glowbug}
Grove, J.~E., Cheung, C.~C., Kerr, M., {et~al.} 2019, AAS/High Energy
  Astrophysics Division, AAS/High Energy Astrophysics Division, 109

\bibitem[{{Guidorzi} {et~al.}(2017){Guidorzi}, {Margutti}, {Brout}, {Scolnic},
  {Fong}, {Alexander}, {Cowperthwaite}, {Annis}, {Berger}, {Blanchard},
  {Chornock}, {Coppejans}, {Eftekhari}, {Frieman}, {Huterer}, {Nicholl},
  {Soares-Santos}, {Terreran}, {Villar}, \& {Williams}}]{guidorzi2017improved}
{Guidorzi}, C., {Margutti}, R., {Brout}, D., {et~al.} 2017, \apjl, 851, L36

\bibitem[{Guiriec {et~al.}(2010)Guiriec, Briggs, Connaugthon, Kara, Daigne,
  Kouveliotou, Van~der Horst, Paciesas, Meegan, Bhat,
  {et~al.}}]{guiriec2010time}
Guiriec, S., Briggs, M.~S., Connaugthon, V., {et~al.} 2010, Astrophys. J., 725,
  225

\bibitem[{Guiriec {et~al.}(2011)Guiriec, Connaughton, Briggs, Burgess, Ryde,
  Daigne, M{\'e}sz{\'a}ros, Goldstein, McEnery, Omodei,
  {et~al.}}]{guiriec2011detection}
Guiriec, S., Connaughton, V., Briggs, M.~S., {et~al.} 2011, \apjl, 727, L33

\bibitem[{Guiriec {et~al.}(2016)Guiriec, Kouveliotou, Hartmann, Granot, Asano,
  M{\'e}sz{\'a}ros, Gill, Gehrels, \& McEnery}]{guiriec2016unified}
Guiriec, S., Kouveliotou, C., Hartmann, D.~H., {et~al.} 2016, \apjl, 831, L8

\bibitem[{Gupta \& Desai(2018)}]{gupta2018limit}
Gupta, S., \& Desai, S. 2018, Ann. Phys., 399, 85

\bibitem[{Guralnik {et~al.}(1964)Guralnik, Hagen, \& Kibble}]{Higgs_3}
Guralnik, G.~S., Hagen, C.~R., \& Kibble, T.~W. 1964, \prl, 13, 585

\bibitem[{{Guth}(1981)}]{cosmo_inflation_1981}
{Guth}, A.~H. 1981, \prd, 23, 347

\bibitem[{Gyulassy \& McLerran(2005)}]{gyulassy2005new}
Gyulassy, M., \& McLerran, L. 2005, Nuclear Physics A, 750, 30

\bibitem[{Haggard {et~al.}(2017)Haggard, Nynka, Ruan, Kalogera, Cenko, Evans,
  \& Kennea}]{haggard2017deep}
Haggard, D., Nynka, M., Ruan, J.~J., {et~al.} 2017, \apjl, 848, L25

\bibitem[{{Hajela} {et~al.}(2019)}]{gcn_170817_latettime}
{Hajela}, A., {et~al.} 2019, {Further Chandra observations of GW170817 ~581-583
  days since merger}, Tech. rep., GSFC

\bibitem[{Hajela {et~al.}(2019)Hajela, Margutti, Alexander, Kathirgamaraju,
  Baldeschi, Guidorzi, Giannios, Fong, Wu, MacFadyen, {et~al.}}]{hajela2019two}
Hajela, A., Margutti, R., Alexander, K., {et~al.} 2019, \apjl, 886, L17

\bibitem[{Halzen \& Hooper(2002)}]{halzen2002high}
Halzen, F., \& Hooper, D. 2002, Rep. Progr. Phys., 65, 1025

\bibitem[{Hansen \& Lyutikov(2001)}]{precursors_hansen_2010}
Hansen, B. M.~S., \& Lyutikov, M. 2001, Mon. Not. R. Astron. Soc., 322, 695

\bibitem[{Harrison {et~al.}(1999)Harrison, Bloom, Frail, Sari, Kulkarni,
  Djorgovski, Axelrod, Mould, Schmidt, Wieringa,
  {et~al.}}]{harrison1999optical}
Harrison, F., Bloom, J., Frail, D.~A., {et~al.} 1999, \apjl, 523, L121

\bibitem[{Hasco{\"e}t {et~al.}(2012)Hasco{\"e}t, Daigne, Mochkovitch, \&
  Vennin}]{hascoet2012fermi}
Hasco{\"e}t, R., Daigne, F., Mochkovitch, R., \& Vennin, V. 2012, Mon. Not. R.
  Astron. Soc., 421, 525

\bibitem[{Heavens {et~al.}(2014)Heavens, Jimenez, \&
  Verde}]{heavens2014standard}
Heavens, A., Jimenez, R., \& Verde, L. 2014, \prl, 113, 241302

\bibitem[{Hess(1912)}]{hess1912beobachtungen}
Hess, V.~F. 1912, Phys. Z., 13, 1084

\bibitem[{Higgs(1964)}]{Higgs_1}
Higgs, P.~W. 1964, \prl, 13, 508

\bibitem[{Hinderer {et~al.}(2019)Hinderer, Nissanke, Foucart, Hotokezaka,
  Vincent, Kasliwal, Schmidt, Williamson, Nichols, Duez,
  {et~al.}}]{hinderer2019distinguishing}
Hinderer, T., Nissanke, S., Foucart, F., {et~al.} 2019, \prd, 100, 063021

\bibitem[{Hirata {et~al.}(1987)Hirata, Kajita, Koshiba, Nakahata, Oyama, Sato,
  Suzuki, Takita, Totsuka, Kifune, {et~al.}}]{SN1987A_neutrinos_Hirata_1987}
Hirata, K., Kajita, T., Koshiba, M., {et~al.} 1987, \prl, 58, 1490

\bibitem[{Hopkins \& Beacom(2006)}]{cosmic_SFR_Hopkins_2006}
Hopkins, A.~M., \& Beacom, J.~F. 2006, Astrophys. J., 651, 142

\bibitem[{Horowitz {et~al.}(2014)Horowitz, Kumar, \&
  Michaels}]{horowitz2014electroweak}
Horowitz, C.~J., Kumar, K.~S., \& Michaels, R. 2014, Eur. Phys. J. A, 50, 48

\bibitem[{Horowitz \& Piekarewicz(2001)}]{horowitz2001neutron}
Horowitz, C.~J., \& Piekarewicz, J. 2001, \prl, 86, 5647

\bibitem[{Hotokezaka {et~al.}(2018)Hotokezaka, Beniamini, \&
  Piran}]{bns_rates_meta_2018}
Hotokezaka, K., Beniamini, P., \& Piran, T. 2018, arXiv e-prints, 1801.01141

\bibitem[{Hotokezaka {et~al.}(2013)Hotokezaka, Kiuchi, Kyutoku, Okawa,
  Sekiguchi, Shibata, \& Taniguchi}]{hotokezaka2013mass}
Hotokezaka, K., Kiuchi, K., Kyutoku, K., {et~al.} 2013, \prd, 87, 024001

\bibitem[{{Hotokezaka} {et~al.}(2019){Hotokezaka}, {Nakar}, {Gottlieb},
  {Nissanke}, {Masuda}, {Hallinan}, {Mooley}, \& {Deller}}]{H0_170817_VLBI}
{Hotokezaka}, K., {Nakar}, E., {Gottlieb}, O., {et~al.} 2019, Nature Astron.,
  3, 940

\bibitem[{{Hotokezaka} {et~al.}(2016){Hotokezaka}, {Nissanke}, {Hallinan},
  {Lazio}, {Nakar}, \& {Piran}}]{omni_radio_optimization_Hotokezaka_2016}
{Hotokezaka}, K., {Nissanke}, S., {Hallinan}, G., {et~al.} 2016, \apj, 831, 190

\bibitem[{{Hotokezaka} \& {Piran}(2015)}]{omni_radio_general_Hotokezaka_2015}
{Hotokezaka}, K., \& {Piran}, T. 2015, \mnras, 450, 1430

\bibitem[{Hotokezaka {et~al.}(2016)Hotokezaka, Wanajo, Tanaka, Bamba, Terada,
  \& Piran}]{kilonova_gamma_rays_hotokezaka_2016}
Hotokezaka, K., Wanajo, S., Tanaka, M., {et~al.} 2016, Mon. Not. R. Astron.
  Soc., 459, 35

\bibitem[{Howell {et~al.}(2019)Howell, Ackley, Rowlinson, \&
  Coward}]{source_evo_Howell_2018}
Howell, E.~J., Ackley, K., Rowlinson, A., \& Coward, D. 2019, Mon. Not. R.
  Astron. Soc., 485, 1435

\bibitem[{Hubble(1929{\natexlab{a}})}]{cosmo_hubble_1929}
Hubble, E. 1929{\natexlab{a}}, Proc. Natl. Acad. Sci. USA, 15, 168

\bibitem[{Hubble(1925)}]{cosmo_hubble_1925}
Hubble, E.~P. 1925, Astrophys. J., 62

\bibitem[{Hubble(1929{\natexlab{b}})}]{cosmo_hubble_m31_1929}
---. 1929{\natexlab{b}}, Astrophys. J., 69

\bibitem[{{Hulse} \& {Taylor}(1975)}]{Hulse_Taylor_pulsar_discovery}
{Hulse}, R.~A., \& {Taylor}, J.~H. 1975, \apjl, 195, L51

\bibitem[{Hurley {et~al.}(1999)Hurley, Cline, Mazets, Barthelmy, Butterworth,
  Marshall, Palmer, Aptekar, Golenetskii, Il'Inskii,
  {et~al.}}]{hurley1999giant}
Hurley, K., Cline, T., Mazets, E., {et~al.} 1999, Nature, 397, 41

\bibitem[{Hurley {et~al.}(2005)Hurley, Boggs, Smith, Duncan, Lin, Zoglauer,
  Krucker, Hurford, Hudson, Wigger, {et~al.}}]{hurley2005exceptionally}
Hurley, K., Boggs, S., Smith, D., {et~al.} 2005, Nature, 434, 1098

\bibitem[{{Hurley} {et~al.}(2011){Hurley}, {Golenetskii}, {Aptekar}, {Mazets},
  {Pal'Shin}, {Frederiks}, {Mitrofanov}, {Golovin}, {Kozyrev}, {Litvak},
  {Sanin}, {Boynton}, {Fellows}, {Harshman}, {Starr}, {von Kienlin}, {Rau},
  {Yamaoka}, {Ohno}, {Fukazawa}, {Takahashi}, {Tashiro}, {Terada}, {Murakami},
  {Makishima}, {Barthelmy}, {Cummings}, {Gehrels}, {Krimm}, {Cline},
  {Goldsten}, {Del Monte}, {Feroci}, {Marisaldi}, {Briggs}, {Connaughton},
  {Meegan}, {Smith}, {Wigger}, \& {Hajdas}}]{hurley2010third}
{Hurley}, K., {Golenetskii}, S., {Aptekar}, R., {et~al.} 2011, in American
  Institute of Physics Conference Series, Vol. 1358, American Institute of
  Physics Conference Series, ed. J.~E. {McEnery}, J.~L. {Racusin}, \&
  N.~{Gehrels}, 385--388

\bibitem[{{Hyper-Kamiokande Proto-Collaboration}
  {et~al.}(2018){Hyper-Kamiokande Proto-Collaboration}, {:}, {Abe}, {Abe},
  {Aihara}, {Aimi}, {Akutsu}, {Andreopoulos}, {Anghel}, {Anthony},
  {et~al.}}]{mnu_hyperkamiokande_design_report_2018}
{Hyper-Kamiokande Proto-Collaboration}, {:}, {Abe}, K., {et~al.} 2018, arXiv
  e-prints, arXiv:1805.04163

\bibitem[{Ioka {et~al.}(2016)Ioka, Hotokezaka, \& Piran}]{ioka2016ultra}
Ioka, K., Hotokezaka, K., \& Piran, T. 2016, \apj, 833, 110

\bibitem[{Isi {et~al.}(2017)Isi, Pitkin, \& Weinstein}]{isi2017probing}
Isi, M., Pitkin, M., \& Weinstein, A.~J. 2017, \prd, 96, 042001

\bibitem[{Ivezi{\'c} {et~al.}(2019)Ivezi{\'c}, Kahn, Tyson, Abel, Acosta,
  Allsman, Alonso, AlSayyad, Anderson, Andrew,
  {et~al.}}]{lsst_drivers_Ivezic_2008}
Ivezi{\'c}, {\v{Z}}., Kahn, S.~M., Tyson, J.~A., {et~al.} 2019, Astrophys. J.,
  873, 111

\bibitem[{Iyer {et~al.}(2011)Iyer, Souradeep, Unnikrishnan, Dhurandhar, Raja,
  \& Sengupta}]{LIGO_India}
Iyer, B., Souradeep, T., Unnikrishnan, C.~S., {et~al.} 2011, {LIGO-India,
  Proposal of the Consortium for Indian Initiative in Gravitational-wave
  Observations (IndIGO)}, Tech. rep., LIGO.
\newblock \url{https://dcc.ligo.org/LIGOM1100296/public}

\bibitem[{Jacob \& Piran(2008)}]{jacob2008lorentz}
Jacob, U., \& Piran, T. 2008, JCAP, 2008, 031

\bibitem[{Jacobson {et~al.}(2006)Jacobson, Liberati, \&
  Mattingly}]{QG_LIV_jacobson_2006}
Jacobson, T., Liberati, S., \& Mattingly, D. 2006, Annals of Physics, 321, 150

\bibitem[{Jacobson \& Mattingly(2004)}]{jacobson2004einstein}
Jacobson, T., \& Mattingly, D. 2004, \prd, 70, 024003

\bibitem[{Ji {et~al.}(2016)Ji, Frebel, Chiti, \& Simon}]{ji2016r}
Ji, A.~P., Frebel, A., Chiti, A., \& Simon, J.~D. 2016, Nature, 531, 610

\bibitem[{Jim{\'e}nez {et~al.}(2016)Jim{\'e}nez, Piazza, \&
  Velten}]{speed_of_gravity_pulsars_jimenez_2016}
Jim{\'e}nez, J.~B., Piazza, F., \& Velten, H. 2016, \prl, 116, 061101

\bibitem[{Jin {et~al.}(2016)Jin, Hotokezaka, Li, Tanaka, D'Avanzo, Fan, Covino,
  Wei, \& Piran}]{jin2016macronova}
Jin, Z.-P., Hotokezaka, K., Li, X., {et~al.} 2016, Nature Commun., 7, 12898

\bibitem[{Johnson(2019)}]{johnson2019populating}
Johnson, J.~A. 2019, Science, 363, 474

\bibitem[{{Johnson} {et~al.}(2019){Johnson}, {Zasowski}, {Weinberg}, {Ting},
  {Sobeck}, {Smith}, {Silva Aguirre}, {Nataf}, {Lucatello}, {Kollmeier},
  {Hekker}, {Cunha}, {Chiappini}, {Carlberg}, {Bird}, {Basu}, \&
  {Anguiano}}]{johnson2019origin}
{Johnson}, J.~A., {Zasowski}, G., {Weinberg}, D., {et~al.} 2019, arXiv
  e-prints, arXiv:1907.04388

\bibitem[{Just {et~al.}(2015)Just, Bauswein, Pulpillo, Goriely, \&
  Janka}]{just2015comprehensive}
Just, O., Bauswein, A., Pulpillo, R.~A., Goriely, S., \& Janka, H.-T. 2015,
  Mon. Not. R. Astron. Soc., 448, 541

\bibitem[{Just {et~al.}(2016)Just, Obergaulinger, Janka, Bauswein, \&
  Schwarz}]{just2016neutron}
Just, O., Obergaulinger, M., Janka, H.-T., Bauswein, A., \& Schwarz, N. 2016,
  \apjl, 816, L30

\bibitem[{Kahya(2011)}]{EP_dark_matter_emulators_Kahya_2011}
Kahya, E.~O. 2011, Phys. Lett. B, 701, 291

\bibitem[{Kahya \& Desai(2016)}]{WEP_GWs_kahya_2016}
Kahya, E.~O., \& Desai, S. 2016, Phys. Lett. B, 756, 265

\bibitem[{{Kalogera} {et~al.}(2007){Kalogera}, {Belczynski}, {Kim},
  {O'Shaughnessy}, \& {Willems}}]{review_CO_formation_Kalogera_2007}
{Kalogera}, V., {Belczynski}, K., {Kim}, C., {O'Shaughnessy}, R., \& {Willems},
  B. 2007, \physrep, 442, 75

\bibitem[{Kanner {et~al.}(2012)Kanner, Camp, Racusin, Gehrels, \&
  White}]{kanner2012seeking}
Kanner, J., Camp, J., Racusin, J., Gehrels, N., \& White, D. 2012, Astrophys.
  J., 759, 22

\bibitem[{Karki {et~al.}(2016)Karki, Tuyenbayev, Kandhasamy, Abbott, Abbott,
  Anders, Berliner, Betzwieser, Cahillane, Canete,
  {et~al.}}]{ALIGO_photon_calibrators}
Karki, S., Tuyenbayev, D., Kandhasamy, S., {et~al.} 2016, Rev. Sci.
  Instruments, 87, 114503

\bibitem[{{Kasen} {et~al.}(2013){Kasen}, {Badnell}, \&
  {Barnes}}]{KN_correct_opacities_Kasen_2013}
{Kasen}, D., {Badnell}, N.~R., \& {Barnes}, J. 2013, \apj, 774, 25

\bibitem[{Kasen {et~al.}(2017)Kasen, Metzger, Barnes, Quataert, \&
  Ramirez-Ruiz}]{kasen2017origin}
Kasen, D., Metzger, B., Barnes, J., Quataert, E., \& Ramirez-Ruiz, E. 2017,
  Nature, 551, 80

\bibitem[{Kasliwal {et~al.}(2017)Kasliwal, Nakar, Singer, Kaplan, Cook,
  Van~Sistine, Lau, Fremling, Gottlieb, Jencson,
  {et~al.}}]{kasliwal2017illuminating}
Kasliwal, M.~M., Nakar, E., Singer, L.~P., {et~al.} 2017, Science, 358, 1559

\bibitem[{Kasliwal {et~al.}(2019)Kasliwal, Kasen, Lau, Perley, Rosswog, Ofek,
  Hotokezaka, Chary, Sollerman, Goobar, {et~al.}}]{kasliwal2019spitzer}
Kasliwal, M.~M., Kasen, D., Lau, R.~M., {et~al.} 2019, Mon. Not. R. Astron.
  Soc. Lett.

\bibitem[{Kathirgamaraju {et~al.}(2019)Kathirgamaraju, Tchekhovskoy, Giannios,
  \& Barniol~Duran}]{kathirgamaraju2019counterparts}
Kathirgamaraju, A., Tchekhovskoy, A., Giannios, D., \& Barniol~Duran, R. 2019,
  Mon. Not. R. Astron. Soc. Lett., 484, L98

\bibitem[{Katz \& Canel(1996)}]{katz1996long}
Katz, J.~I., \& Canel, L.~M. 1996, Astrophys. J., 471, 915

\bibitem[{{Kawaguchi} {et~al.}(2019){Kawaguchi}, {Shibata}, \&
  {Tanaka}}]{kawaguchi_kilonova_remnants}
{Kawaguchi}, K., {Shibata}, M., \& {Tanaka}, M. 2019, arXiv e-prints,
  arXiv:1908.05815

\bibitem[{Kawamura {et~al.}(2011)Kawamura, Ando, Seto, Sato, Nakamura, Tsubono,
  Kanda, Tanaka, Yokoyama, Funaki, {et~al.}}]{kawamura2011japanese}
Kawamura, S., Ando, M., Seto, N., {et~al.} 2011, Class. Quantum Grav., 28,
  094011

\bibitem[{Keppel \& Ajith(2010)}]{GR_mg_Keppel_2010}
Keppel, D., \& Ajith, P. 2010, \prd, 82, 122001

\bibitem[{Kimura {et~al.}(2018)Kimura, Murase, Bartos, Ioka, Heng, \&
  M{\'e}sz{\'a}ros}]{kimura2018transejecta}
Kimura, S.~S., Murase, K., Bartos, I., {et~al.} 2018, \prd, 98, 043020

\bibitem[{Kimura {et~al.}(2017)Kimura, Murase, M{\'e}sz{\'a}ros, \&
  Kiuchi}]{kimura2017high}
Kimura, S.~S., Murase, K., M{\'e}sz{\'a}ros, P., \& Kiuchi, K. 2017, \apjl,
  848, L4

\bibitem[{{Kisaka} \& {Ioka}(2015)}]{2015ApJ...804L..16K}
{Kisaka}, S., \& {Ioka}, K. 2015, \apjl, 804, L16

\bibitem[{Kiuchi {et~al.}(2014)Kiuchi, Kyutoku, Sekiguchi, Shibata, \&
  Wada}]{kiuchi2014high}
Kiuchi, K., Kyutoku, K., Sekiguchi, Y., Shibata, M., \& Wada, T. 2014, \prd,
  90, 041502

\bibitem[{Kiuchi {et~al.}(2019)Kiuchi, Kyutoku, Shibata, \&
  Taniguchi}]{kiuchi2019revisiting}
Kiuchi, K., Kyutoku, K., Shibata, M., \& Taniguchi, K. 2019, \apjl, 876, L31

\bibitem[{Kiuchi {et~al.}(2015)Kiuchi, Sekiguchi, Kyutoku, Shibata, Taniguchi,
  \& Wada}]{kiuchi2015high}
Kiuchi, K., Sekiguchi, Y., Kyutoku, K., {et~al.} 2015, \prd, 92, 064034

\bibitem[{{Klebesadel} {et~al.}(1973){Klebesadel}, {Strong}, \&
  {Olson}}]{GRBs_astrophysical_origin_1973}
{Klebesadel}, R.~W., {Strong}, I.~B., \& {Olson}, R.~A. 1973, \apjl, 182, L85

\bibitem[{Knox(2006)}]{curvature_knox}
Knox, L. 2006, \prd, 73, 023503

\bibitem[{Kobayashi \& Zhang(2007)}]{kobayashi2007onset}
Kobayashi, S., \& Zhang, B. 2007, Astrophys. J., 655, 973

\bibitem[{Kocevski {et~al.}(2007)Kocevski, Butler, \&
  Bloom}]{kocevski2007pulse}
Kocevski, D., Butler, N., \& Bloom, J.~S. 2007, Astrophys. J., 667, 1024

\bibitem[{{Kocevski} {et~al.}(2018){Kocevski}, {Burns}, {Goldstein}, {Dal
  Canton}, {Briggs}, {Blackburn}, {Veres}, {Hui}, {Hamburg}, {Roberts},
  {Wilson-Hodge}, {Connaughton}, {Racusin}, {Littenberg}, {von Kienlin}, \&
  {Bissaldi}}]{Targeted_Search_Kocevski_2018}
{Kocevski}, D., {Burns}, E., {Goldstein}, A., {et~al.} 2018, \apj, 862, 152

\bibitem[{K{\"o}ppel {et~al.}(2019)K{\"o}ppel, Bovard, \&
  Rezzolla}]{koppel2019general}
K{\"o}ppel, S., Bovard, L., \& Rezzolla, L. 2019, \apjl, 872, L16

\bibitem[{Korobkin {et~al.}(2019)Korobkin, Hungerford, Fryer, Mumpower, Misch,
  Sprouse, Lippuner, Surman, Couture, Bloser, {et~al.}}]{korobkin2019gamma}
Korobkin, O., Hungerford, A.~M., Fryer, C.~L., {et~al.} 2019, arXiv e-prints,
  1905.05089

\bibitem[{Kostelecky \& Mewes(2012)}]{LIV_neutrinos_Kostelecky_2011}
Kostelecky, A., \& Mewes, M. 2012, Phys. Rev. D, 85, 096005

\bibitem[{Kosteleck{\`y} {et~al.}(2003)Kosteleck{\`y}, Lehnert, \&
  Perry}]{LIV_spactime_varying_kostelecky_2003}
Kosteleck{\`y}, V.~A., Lehnert, R., \& Perry, M.~J. 2003, \prd, 68, 123511

\bibitem[{Kosteleck{\`y} \& Mewes(2008)}]{kostelecky2008astrophysical}
Kosteleck{\`y}, V.~A., \& Mewes, M. 2008, \apjl, 689, L1

\bibitem[{Kosteleck{\`y} \& Russell(2011)}]{SME_summary}
Kosteleck{\`y}, V.~A., \& Russell, N. 2011, Rev. Mod. Phys., 83, 11

\bibitem[{Kosteleck{\`y} \& Samuel(1989)}]{QG_string_kostelecky_1989}
Kosteleck{\`y}, V.~A., \& Samuel, S. 1989, \prd, 39, 683

\bibitem[{{Kouveliotou} {et~al.}(1993){Kouveliotou}, {Meegan}, {Fishman},
  {Bhat}, {Briggs}, {Koshut}, {Paciesas}, \& {Pendleton}}]{Kouveliotou1993}
{Kouveliotou}, C., {Meegan}, C.~A., {Fishman}, G.~J., {et~al.} 1993, \apjl,
  413, L101

\bibitem[{Kraus \& Tomboulis(2002)}]{LIV_EGB_kraus_2002}
Kraus, P., \& Tomboulis, E.~T. 2002, \prd, 66, 045015

\bibitem[{Krauss \& Tremaine(1988)}]{WEP_1987A_Krass_1988}
Krauss, L.~M., \& Tremaine, S. 1988, \prl, 60, 176

\bibitem[{{Kruckow} {et~al.}(2018){Kruckow}, {Tauris}, {Langer}, {Kramer}, \&
  {Izzard}}]{kurckow2018}
{Kruckow}, M.~U., {Tauris}, T.~M., {Langer}, N., {Kramer}, M., \& {Izzard},
  R.~G. 2018, \mnras, 481, 1908

\bibitem[{Kulkarni {et~al.}(1999)Kulkarni, Djorgovski, Odewahn, Bloom, Gal,
  Koresko, Harrison, Lubin, Armus, Sari, {et~al.}}]{kulkarni1999afterglow}
Kulkarni, S., Djorgovski, S., Odewahn, S., {et~al.} 1999, Nature, 398, 389

\bibitem[{Kumar \& Granot(2003)}]{kumar2003evolution}
Kumar, P., \& Granot, J. 2003, \apj, 591, 1075

\bibitem[{Kumar \& Zhang(2015)}]{kumar2015physics}
Kumar, P., \& Zhang, B. 2015, Phys. Rep., 561, 1

\bibitem[{{Kuns} {et~al.}(2019){Kuns}, {Yu}, {Chen}, \&
  {Adhikari}}]{kuns2019astrophysics}
{Kuns}, K.~A., {Yu}, H., {Chen}, Y., \& {Adhikari}, R.~X. 2019, arXiv e-prints,
  arXiv:1908.06004

\bibitem[{Kuzmin {et~al.}(1985)Kuzmin, Rubakov, \&
  Shaposhnikov}]{neutrino_baryogenesis_kuzmin_1985}
Kuzmin, V.~A., Rubakov, V.~A., \& Shaposhnikov, M.~E. 1985, Phys. Lett. B, 155,
  36

\bibitem[{Kyutoku \& Seto(2017)}]{kyutoku2017gravitational}
Kyutoku, K., \& Seto, N. 2017, \prd, 95, 083525

\bibitem[{{Lamb} {et~al.}(2019){Lamb}, {Tanvir}, {Levan}, {de Ugarte Postigo},
  {Kawaguchi}, {Corsi}, {Evans}, {Gompertz}, {Malesani}, {Page}, {Wiersema},
  {Rosswog}, {Shibata}, {Tanaka}, {van der Horst}, {Cano}, {Fynbo}, {Fruchter},
  {Greiner}, {Heintz}, {Higgins}, {Hjorth}, {Izzo}, {Jakobsson}, {Kann},
  {O'Brien}, {Perley}, {Pian}, {Pugliese}, {Starling}, {Th{\"o}ne}, {Watson},
  {Wijers}, \& {Xu}}]{lamb2019short}
{Lamb}, G.~P., {Tanvir}, N.~R., {Levan}, A.~J., {et~al.} 2019, \apj, 883, 48

\bibitem[{Langer \& Norman(2006)}]{langer2006collapsar}
Langer, N., \& Norman, C.~A. 2006, \apjl, 638, L63

\bibitem[{Lattimer(2012)}]{lattimer2012nuclear}
Lattimer, J.~M. 2012, Annu. Rev. Nucl. Part. Sci., 62, 485

\bibitem[{Lattimer \& Prakash(2000)}]{lattimer2000nuclear}
Lattimer, J.~M., \& Prakash, M. 2000, Phys. Rep., 333, 121

\bibitem[{{Lattimer} \& {Schramm}(1974)}]{KN_r_process_Lattimer_1974}
{Lattimer}, J.~M., \& {Schramm}, D.~N. 1974, \apjl, 192, L145

\bibitem[{Lazzati {et~al.}(2017)Lazzati, Deich, Morsony, \&
  Workman}]{lazzati2017off}
Lazzati, D., Deich, A., Morsony, B.~J., \& Workman, J.~C. 2017, Mon. Not. R.
  Astron. Soc., 471, 1652

\bibitem[{Lazzati {et~al.}(2005)Lazzati, Ghirlanda, \&
  Ghisellini}]{lazzati2005soft}
Lazzati, D., Ghirlanda, G., \& Ghisellini, G. 2005, Mon. Not. R. Astron. Soc.
  Lett., 362, L8

\bibitem[{Lazzati {et~al.}(2018)Lazzati, Perna, Morsony, Lopez-Camara,
  Cantiello, Ciolfi, Giacomazzo, \& Workman}]{lazzati2018late}
Lazzati, D., Perna, R., Morsony, B.~J., {et~al.} 2018, \prl, 120, 241103

\bibitem[{Lazzati {et~al.}(2001)Lazzati, Ramirez-Ruiz, \&
  Ghisellini}]{extended_emission_2001_lazzati}
Lazzati, D., Ramirez-Ruiz, E., \& Ghisellini, G. 2001, Astron. Astrophys., 379,
  L39

\bibitem[{Le~Floc'h {et~al.}(2003)Le~Floc'h, Duc, Mirabel, Sanders, Bosch,
  Diaz, Donzelli, Rodrigues, Greiner, Mereghetti, {et~al.}}]{le2003hosts}
Le~Floc'h, E., Duc, P.-A., Mirabel, I., {et~al.} 2003, Astron. Astrophys., 400,
  499

\bibitem[{Leavitt(1908)}]{cosmo_leavitt_1908}
Leavitt, H.~S. 1908, Ann. Harvard College Obs., 60, 87

\bibitem[{Leavitt \& Pickering(1912)}]{cosmo_leavitt_1912}
Leavitt, H.~S., \& Pickering, E.~C. 1912, Harvard College Obs. Circ., 173, 1

\bibitem[{Lee {et~al.}(2000)Lee, Wijers, \& Brown}]{lee2000blandford}
Lee, H.~K., Wijers, R. A. M.~J., \& Brown, G.~E. 2000, Physics Reports, 325, 83

\bibitem[{Lee \& Yang(1956)}]{lee1956question}
Lee, T.-D., \& Yang, C.-N. 1956, Phys. Rev., 104, 254

\bibitem[{Lee \& Ramirez-Ruiz(2007)}]{lee2007progenitors}
Lee, W.~H., \& Ramirez-Ruiz, E. 2007, New J. Phys., 9, 17

\bibitem[{Lei {et~al.}(2013)Lei, Zhang, \& Liang}]{lei2013hyperaccreting}
Lei, W.-H., Zhang, B., \& Liang, E.-W. 2013, Astrophys. J., 765, 125

\bibitem[{Lei {et~al.}(2017)Lei, Zhang, Wu, \& Liang}]{lei2017hyperaccreting}
Lei, W.-H., Zhang, B., Wu, X.-F., \& Liang, E.-W. 2017, \apj, 849, 47

\bibitem[{{Leibler} \& {Berger}(2010)}]{2010ApJ...725.1202L}
{Leibler}, C.~N., \& {Berger}, E. 2010, \apj, 725, 1202

\bibitem[{Lema{\^\i}tre(1927)}]{cosmo_lemaitre_1927}
Lema{\^\i}tre, G. 1927, Ann. Soc. Sci. Bruxelles, 47, 49

\bibitem[{Lema{\^\i}tre(1931)}]{cosmo_lemaitre_1931}
---. 1931, Mon. Not. R. Astron. Soc., 91, 483

\bibitem[{Li {et~al.}(2019)Li, Krastev, Wen, \& Zhang}]{li2019towards}
Li, B.-A., Krastev, P.~G., Wen, D.-H., \& Zhang, N.-B. 2019, European Phys. J.
  A, 55, 117

\bibitem[{{Li}(2000)}]{2000PhRvD..61h4016L}
{Li}, L.-X. 2000, \prd, 61, 084016

\bibitem[{Li {et~al.}(2018)Li, Xiong, Zhang, Lu, Song, Cao, Chang, Chen, Chen,
  Chen, {et~al.}}]{li2018insight}
Li, T., Xiong, S., Zhang, S., {et~al.} 2018, Science China: Phys. Mech.
  Astron., 61, 031011

\bibitem[{Li {et~al.}(2011)Li, Chornock, Leaman, Filippenko, Poznanski, Wang,
  Ganeshalingam, \& Mannucci}]{li2011nearby}
Li, W., Chornock, R., Leaman, J., {et~al.} 2011, \mnras, 412, 1473

\bibitem[{Li {et~al.}(2016)Li, Hu, Fan, \& Wei}]{li2016grb}
Li, X., Hu, Y.-M., Fan, Y.-Z., \& Wei, D.-M. 2016, Astrophys. J., 827, 75

\bibitem[{{Lien} {et~al.}(2016){Lien}, {Sakamoto}, {Barthelmy}, {Baumgartner},
  {Cannizzo}, {Chen}, {Collins}, {Cummings}, {Gehrels}, {Krimm}, {Markwardt},
  {Palmer}, {Stamatikos}, {Troja}, \& {Ukwatta}}]{Lien2016}
{Lien}, A., {Sakamoto}, T., {Barthelmy}, S.~D., {et~al.} 2016, \apj, 829, 7

\bibitem[{{Linde}(1982)}]{cosmo_inflation_1982}
{Linde}, A.~D. 1982, Phys. Lett. B, 108, 389

\bibitem[{Linder(2003)}]{expansion_linder_2003}
Linder, E.~V. 2003, \prl, 90, 091301

\bibitem[{Linder(2005)}]{curved_linder}
---. 2005, Astropart. Phys., 24, 391

\bibitem[{Linder(2011)}]{lens_linder}
---. 2011, \prd, 84, 123529

\bibitem[{Lippuner {et~al.}(2017)Lippuner, Fern{\'a}ndez, Roberts, Foucart,
  Kasen, Metzger, \& Ott}]{lippuner2017signatures}
Lippuner, J., Fern{\'a}ndez, R., Roberts, L.~F., {et~al.} 2017, Mon. Not. R.
  Astron. Soc., 472, 904

\bibitem[{Lipunov {et~al.}(2017)Lipunov, Gorbovskoy, Kornilov, Tyurina,
  Balanutsa, Kuznetsov, Vlasenko, Kuvshinov, Gorbunov, Buckley,
  {et~al.}}]{lipunov2017master}
Lipunov, V.~M., Gorbovskoy, E., Kornilov, V.~G., {et~al.} 2017, \apjl, 850, L1

\bibitem[{Liska {et~al.}(2017)Liska, Hesp, Tchekhovskoy, Ingram, van~der Klis,
  \& Markoff}]{liska2017formation}
Liska, M., Hesp, C., Tchekhovskoy, A., {et~al.} 2017, Mon. Not. R. Astron. Soc.
  Lett., 474, L81

\bibitem[{Liska {et~al.}(2019)Liska, Tchekhovskoy, Ingram, \& Van
  Der~Klis}]{liska2019bardeen}
Liska, M., Tchekhovskoy, A., Ingram, A., \& Van Der~Klis, M. 2019, Mon. Not. R.
  Astron. Soc., 487, 550

\bibitem[{Lithwick \& Sari(2001)}]{lithwick2001lower}
Lithwick, Y., \& Sari, R. 2001, Astrophys. J., 555, 540

\bibitem[{Littenberg {et~al.}(2015)Littenberg, Farr, Coughlin, Kalogera, \&
  Holz}]{mass_gap_littenberg_2015}
Littenberg, T.~B., Farr, B., Coughlin, S., Kalogera, V., \& Holz, D.~E. 2015,
  \apjl, 807, L24

\bibitem[{Longo(1987)}]{neutrinos_speed_1987A_longo_1987}
Longo, M.~J. 1987, \prd, 36, 3276

\bibitem[{Longo(1988)}]{WEP_1987A_Longo_1988}
---. 1988, \prl, 60, 173

\bibitem[{Lorimer(2008)}]{review_binary_NS_lorimer_2008}
Lorimer, D.~R. 2008, Living Rev. Relativ., 11, 8

\bibitem[{L{\"u} \& Zhang(2014)}]{lu2014test}
L{\"u}, H.-J., \& Zhang, B. 2014, \apj, 785, 74

\bibitem[{L{\"u} {et~al.}(2015)L{\"u}, Zhang, Lei, Li, \&
  Lasky}]{lu2015millisecond}
L{\"u}, H.-J., Zhang, B., Lei, W.-H., Li, Y., \& Lasky, P.~D. 2015, Astrophys.
  J., 805, 89

\bibitem[{LVC(2017{\natexlab{a}})}]{GCN_LVC_190425z}
LVC. 2017{\natexlab{a}}, {LIGO/Virgo S190425z: Updated localization from LIGO
  and Virgo data}, Tech. rep., GSFC

\bibitem[{LVC(2017{\natexlab{b}})}]{gcnLVCGW170817_0}
---. 2017{\natexlab{b}}, {LIGO/Virgo G298048: Fermi GBM trigger
  524666471/170817529: LIGO/Virgo Identification of a possible
  gravitational-wave counterpart}, Tech. rep., GSFC

\bibitem[{LVC(2017{\natexlab{c}})}]{gcnLVCGW170817_1}
---. 2017{\natexlab{c}}, {LIGO/Virgo Identification of a binary neutron star
  candidate coincident with Fermi GBM trigger 524666471/170817529}, Tech. rep.,
  GSFC

\bibitem[{LVC(2017{\natexlab{d}})}]{gcnLVCGW170817_2}
---. 2017{\natexlab{d}}, {LIGO/Virgo G298048: Further analysis of a binary
  neutron star candidate with updated sky localization}, Tech. rep., GSFC

\bibitem[{LVC(2017{\natexlab{e}})}]{GCN_LVC_S190718y}
---. 2017{\natexlab{e}}, {LIGO/Virgo S190718y: Identification of a GW compact
  binary merger candidate}, Tech. rep., GSFC

\bibitem[{LVC(2019)}]{gcn_LVC_S190425z}
---. 2019, {{LIGO}/{Virgo} {S190425z}: Identification of a GW compact binary
  merger candidate}, Tech. rep., GSFC

\bibitem[{LVC(2020)}]{gcn_GRB_200415A_IPN}
---. 2020, GRB Coordinates Network, 27585

\bibitem[{Lynch {et~al.}(2018)Lynch, Coughlin, Vitale, Stubbs, \&
  Katsavounidis}]{lynch2018observational}
Lynch, R., Coughlin, M., Vitale, S., Stubbs, C.~W., \& Katsavounidis, E. 2018,
  \apjl, 861, L24

\bibitem[{Lyutikov {et~al.}(2003)Lyutikov, Pariev, \&
  Blandford}]{lyutikov2003polarization}
Lyutikov, M., Pariev, V.~I., \& Blandford, R.~D. 2003, Astrophys. J., 597, 998

\bibitem[{Macias \& Ramirez-Ruiz(2019)}]{macias2019constraining}
Macias, P., \& Ramirez-Ruiz, E. 2019, \apjl, 877, L24

\bibitem[{Madau \& Dickinson(2014{\natexlab{a}})}]{cosmic_SFR_Madau_2014}
Madau, P., \& Dickinson, M. 2014{\natexlab{a}}, Annu. Rev. Astron. Astrophys.,
  52, 415

\bibitem[{Madau \& Dickinson(2014{\natexlab{b}})}]{madau2014cosmic}
---. 2014{\natexlab{b}}, Annu. Rev. Astron. Astrophys., 52, 415

\bibitem[{Magueijo(2003)}]{LIV_varying_c_magueijo_2003}
Magueijo, J. 2003, Rep. Prog. Phys., 66, 2025

\bibitem[{Maione {et~al.}(2017)Maione, De~Pietri, Feo, \&
  L{\"o}ffler}]{maione2017spectral}
Maione, F., De~Pietri, R., Feo, A., \& L{\"o}ffler, F. 2017, \prd, 96, 063011

\bibitem[{Majorana(1937)}]{neutrino_majorana_1937}
Majorana, E. 1937, Il Nuovo Cimento (1924-1942), 14, 171

\bibitem[{Maki {et~al.}(1962)Maki, Nakagawa, \& Sakata}]{neutrino_maki_1962}
Maki, Z., Nakagawa, M., \& Sakata, S. 1962, Prog. Theor. Phys., 28, 870

\bibitem[{Mandhai {et~al.}(2018)Mandhai, Tanvir, Lamb, Levan, \&
  Tsang}]{mandhai2018rate}
Mandhai, S., Tanvir, N., Lamb, G., Levan, A., \& Tsang, D. 2018, Galaxies, 6,
  130

\bibitem[{Mao \& Paczynski(1992)}]{mao1992cosmological}
Mao, S., \& Paczynski, B. 1992, \apj, 388, L45

\bibitem[{Margalit \& Metzger(2019)}]{margalit2019multi}
Margalit, B., \& Metzger, B.~D. 2019, arXiv e-prints, 1904.11995

\bibitem[{Margutti {et~al.}(2017)Margutti, Berger, Fong, Guidorzi, Alexander,
  Metzger, Blanchard, Cowperthwaite, Chornock, Eftekhari,
  {et~al.}}]{margutti2017electromagnetic}
Margutti, R., Berger, E., Fong, W.-f., {et~al.} 2017, \apjl, 848, L20

\bibitem[{Margutti {et~al.}(2018)Margutti, Alexander, Xie, Sironi, Metzger,
  Kathirgamaraju, Fong, Blanchard, Berger, MacFadyen,
  {et~al.}}]{margutti2018binary}
Margutti, R., Alexander, K.~D., Xie, X., {et~al.} 2018, \apjl, 856, L18

\bibitem[{Mart{\'\i}nez-Pinedo {et~al.}(2012)Mart{\'\i}nez-Pinedo, Fischer,
  Lohs, \& Huther}]{martinez2012charged}
Mart{\'\i}nez-Pinedo, G., Fischer, T., Lohs, A., \& Huther, L. 2012, \prl, 109,
  251104

\bibitem[{Mather {et~al.}(1994)Mather, Cheng, Cottingham, Eplee~Jr, Fixsen,
  Hewagama, Isaacman, Jensen, Meyer, Noerdlinger, {et~al.}}]{cosmo_mather_1994}
Mather, J.~C., Cheng, E.~S., Cottingham, D.~A., {et~al.} 1994, Astrophys. J.,
  420, 439

\bibitem[{Matsumoto {et~al.}(2020)Matsumoto, Kimura, Murase, \&
  M{\'e}sz{\'a}ros}]{matsumoto2020linking}
Matsumoto, T., Kimura, S.~S., Murase, K., \& M{\'e}sz{\'a}ros, P. 2020, Mon.
  Not. R. Astron. Soc., 493, 783

\bibitem[{Mattingly(2005)}]{QG_LIV_Mattingly_2005}
Mattingly, D. 2005, Living Rev. Relativ., 8, 5

\bibitem[{Mazets {et~al.}(1979)Mazets, Golenetskii, Il'Inskii, Guryan,
  {et~al.}}]{GMF_790305_LMC}
Mazets, E., Golenetskii, S., Il'Inskii, V., Guryan, Y.~A., {et~al.} 1979,
  Nature, 282, 587

\bibitem[{Mazets {et~al.}(1981)Mazets, Golenetskii, Il'Inskii, Panov, Aptekar,
  Gur'Yan, Proskura, Sokolov, Sokolova, Kharitonova,
  {et~al.}}]{mazets1981catalog}
Mazets, E.~P., Golenetskii, S.~V., Il'Inskii, V.~N., {et~al.} 1981, Astrophys.
  Space Sci., 80, 3

\bibitem[{{Mazets} {et~al.}(2008){Mazets}, {Aptekar}, {Cline}, {Frederiks},
  {Goldsten}, {Golenetskii}, {Hurley}, {von Kienlin}, \&
  {Pal'shin}}]{GMF_070201_M31}
{Mazets}, E.~P., {Aptekar}, R.~L., {Cline}, T.~L., {et~al.} 2008, \apj, 680,
  545

\bibitem[{McClelland {et~al.}(2014)McClelland, Evans, Schnabel, Lantz, Martin,
  \& Quetschke}]{LIGO_Voyager}
McClelland, D., Evans, M., Schnabel, R., {et~al.} 2014, {Instrument Science
  White Paper}, Tech. rep., LIGO.
\newblock \url{https://dcc.ligo.org/LIGO-T1400316/public}

\bibitem[{McEnery {et~al.}(2019)McEnery, Barrio, Agudo, Ajello, {\'A}lvarez,
  Ansoldi, Anton, Auricchio, Stephen, Baldini, {et~al.}}]{mcenery2019all}
McEnery, J., Barrio, J.~A., Agudo, I., {et~al.} 2019, arXiv e-prints,
  1907.07558

\bibitem[{Meegan {et~al.}(1992)Meegan, Fishman, Wilson, Paciesas, Pendleton,
  Horack, Brock, \& Kouveliotou}]{meegan1992spatial}
Meegan, C., Fishman, G., Wilson, R., {et~al.} 1992, Nature, 355, 143

\bibitem[{{Meegan} {et~al.}(2009{\natexlab{a}}){Meegan}, {Lichti}, {Bhat},
  {Bissaldi}, {Briggs}, {Connaughton}, {Diehl}, {Fishman}, {Greiner}, {Hoover},
  {van der Horst}, {von Kienlin}, {Kippen}, {Kouveliotou}, {McBreen},
  {Paciesas}, {Preece}, {Steinle}, {Wallace}, {Wilson}, \&
  {Wilson-Hodge}}]{Meegan2009}
{Meegan}, C., {Lichti}, G., {Bhat}, P.~N., {et~al.} 2009{\natexlab{a}}, \apj,
  702, 791

\bibitem[{{Meegan} {et~al.}(2009{\natexlab{b}}){Meegan}, {Lichti}, {Bhat},
  {Bissaldi}, {Briggs}, {Connaughton}, {Diehl}, {Fishman}, {Greiner}, {Hoover},
  {van der Horst}, {von Kienlin}, {Kippen}, {Kouveliotou}, {McBreen},
  {Paciesas}, {Preece}, {Steinle}, {Wallace}, {Wilson}, \&
  {Wilson-Hodge}}]{Meegan09}
---. 2009{\natexlab{b}}, \apj, 702, 791

\bibitem[{Meng {et~al.}(2018)Meng, Geng, Zhang, Wei, Xiao, Liu, Gao, Wu, Liang,
  Huang, {et~al.}}]{meng2018origin}
Meng, Y.-Z., Geng, J.-J., Zhang, B.-B., {et~al.} 2018, Astrophys. J., 860, 72

\bibitem[{Mertens(2016)}]{mertens2016direct}
Mertens, S. 2016, J. Phys.: Conf. Ser., 718, 022013

\bibitem[{Messenger \& Read(2012)}]{cosmo_no_EM_messenger_2012}
Messenger, C., \& Read, J. 2012, \prl, 108, 091101

\bibitem[{Messick {et~al.}(2017)Messick, Blackburn, Brady, Brockill, Cannon,
  Cariou, Caudill, Chamberlin, Creighton, Everett,
  {et~al.}}]{messick2017analysis}
Messick, C., Blackburn, K., Brady, P., {et~al.} 2017, \prd, 95, 042001

\bibitem[{M{\'e}sz{\'a}ros \& Rees(1997)}]{meszaros1997poynting}
M{\'e}sz{\'a}ros, P., \& Rees, M.~J. 1997, \apjl, 482, L29

\bibitem[{M{\'e}sz{\'a}ros {et~al.}(1998)M{\'e}sz{\'a}ros, Rees, \&
  Wijers}]{meszaros1998viewing}
M{\'e}sz{\'a}ros, P., Rees, M.~J., \& Wijers, R.~A.~M.~J. 1998, \apj, 499, 301

\bibitem[{Meszaros \& Waxman(2001)}]{meszaros2001tev}
Meszaros, P., \& Waxman, E. 2001, \prl, 87, 171102

\bibitem[{Metzger {et~al.}(2011)Metzger, Giannios, Thompson, Bucciantini, \&
  Quataert}]{Metzger2011protomagnetar}
Metzger, B., Giannios, D., Thompson, T., Bucciantini, N., \& Quataert, E. 2011,
  \mnras, 413, 2031

\bibitem[{Metzger(2020)}]{metzger2020kilonovae}
Metzger, B.~D. 2020, Living Rev. Relativ., 23, 1

\bibitem[{Metzger {et~al.}(2014)Metzger, Bauswein, Goriely, \&
  Kasen}]{metzger2014neutron}
Metzger, B.~D., Bauswein, A., Goriely, S., \& Kasen, D. 2014, Mon. Not. R.
  Astron. Soc., 446, 1115

\bibitem[{Metzger \& Berger(2012)}]{metzger_berger_2012}
Metzger, B.~D., \& Berger, E. 2012, Astrophys. J., 746, 48

\bibitem[{Metzger \& Fern{\'a}ndez(2014)}]{metzger2014red}
Metzger, B.~D., \& Fern{\'a}ndez, R. 2014, Mon. Not. R. Astron. Soc., 441, 3444

\bibitem[{{Metzger} \& {Piro}(2014)}]{Bing_6}
{Metzger}, B.~D., \& {Piro}, A.~L. 2014, \mnras, 439, 3916

\bibitem[{Metzger {et~al.}(2008{\natexlab{a}})Metzger, Piro, \&
  Quataert}]{metzger2008time}
Metzger, B.~D., Piro, A.~L., \& Quataert, E. 2008{\natexlab{a}}, Mon. Not. R.
  Astron. Soc., 390, 781

\bibitem[{Metzger {et~al.}(2009)Metzger, Piro, \&
  Quataert}]{metzger2009neutron}
---. 2009, Mon. Not. R. Astron. Soc., 396, 304

\bibitem[{Metzger {et~al.}(2008{\natexlab{b}})Metzger, Quataert, \&
  Thompson}]{metzger2008short}
Metzger, B.~D., Quataert, E., \& Thompson, T.~A. 2008{\natexlab{b}}, Mon. Not.
  R. Astron. Soc., 385, 1455

\bibitem[{Metzger {et~al.}(2018)Metzger, Thompson, \&
  Quataert}]{metzger2018magnetar}
Metzger, B.~D., Thompson, T.~A., \& Quataert, E. 2018, Astrophys. J., 856, 101

\bibitem[{Metzger \& Zivancev(2016)}]{precursors_Metzger_2016}
Metzger, B.~D., \& Zivancev, C. 2016, Mon. Not. R. Astron. Soc., 461, 4435

\bibitem[{{Metzger} {et~al.}(2010){Metzger}, {Mart{\'{\i}}nez-Pinedo},
  {Darbha}, {Quataert}, {Arcones}, {Kasen}, {Thomas}, {Nugent}, {Panov}, \&
  {Zinner}}]{KN_first_named_Metzger_2010}
{Metzger}, B.~D., {Mart{\'{\i}}nez-Pinedo}, G., {Darbha}, S., {et~al.} 2010,
  \mnras, 406, 2650

\bibitem[{Mewes(2019)}]{dispersion_general_1}
Mewes, M. 2019, \prd, 99, 104062

\bibitem[{Meyer {et~al.}(1992)Meyer, Howard, Mathews, Hoffman, \&
  Woosley}]{meyer1992r}
Meyer, B.~S., Howard, W.~M., Mathews, G.~J., Hoffman, R., \& Woosley, S.~E.
  1992, Unstable Nuclei in Astrophysics, 37

\bibitem[{Miller {et~al.}(2019{\natexlab{a}})Miller, Ryan, Dolence, Burrows,
  Fontes, Fryer, Korobkin, Lippuner, Mumpower, \& Wollaeger}]{miller2019full}
Miller, J.~M., Ryan, B.~R., Dolence, J.~C., {et~al.} 2019{\natexlab{a}}, \prd,
  100, 023008

\bibitem[{Miller {et~al.}(2019{\natexlab{b}})Miller, Lamb, Dittmann, Bogdanov,
  Arzoumanian, Gendreau, Guillot, Harding, Ho, Lattimer,
  {et~al.}}]{miller2019psr}
Miller, M.~C., Lamb, F.~K., Dittmann, A.~J., {et~al.} 2019{\natexlab{b}},
  \apjl, 887, L24

\bibitem[{Minaev \& Pozanenko(2017)}]{SGRB_precursors_INTEGRAL_2017}
Minaev, P.~Y., \& Pozanenko, A.~S. 2017, Astron. Lett., 43, 1

\bibitem[{{Minazzoli}(2019)}]{minazzoli2019strong}
{Minazzoli}, O. 2019, arXiv e-prints, arXiv:1912.06891

\bibitem[{Minazzoli {et~al.}(2019)Minazzoli, Johnson-McDaniel, \&
  Sakellariadou}]{minazzoli2019shortcomings}
Minazzoli, O., Johnson-McDaniel, N.~K., \& Sakellariadou, M. 2019, arXiv
  e-prints, 1907.12453

\bibitem[{Mirshekari {et~al.}(2012)Mirshekari, Yunes, \&
  Will}]{mirshekari2012constraining}
Mirshekari, S., Yunes, N., \& Will, C.~M. 2012, \prd, 85, 024041

\bibitem[{Mirzoyan {et~al.}(2019)Mirzoyan, Noda, Moretti, Berti, Nigro, Hoang,
  Micanovic, Takahashi, Chai, Moralejo, {et~al.}}]{MAGIC_190114C}
Mirzoyan, R., Noda, K., Moretti, E., {et~al.} 2019, {MAGIC detects the GRB
  190114C in the TeV energy domain}, Tech. rep., GSFC

\bibitem[{Mochkovitch {et~al.}(1993)Mochkovitch, Hernanz, Isern, \&
  Martin}]{mochkovitch1993gamma}
Mochkovitch, R., Hernanz, M., Isern, J., \& Martin, X. 1993, Nature, 361, 236

\bibitem[{Moffat(1993)}]{LIV_varying_c_moffat_1993}
Moffat, J.~W. 1993, Int. J. Mod. Phys. D, 2, 351

\bibitem[{Mogushi {et~al.}(2019)Mogushi, Cavagli{\`a}, \&
  Siellez}]{mogushi2019jet}
Mogushi, K., Cavagli{\`a}, M., \& Siellez, K. 2019, Astrophys. J., 880, 55

\bibitem[{Mohapatra \& Smirnov(2006)}]{mnu_review_2006}
Mohapatra, R.~N., \& Smirnov, A.~Y. 2006, Annu. Rev. Nucl. Part. Sci., 56, 569

\bibitem[{Mooley {et~al.}(2018)Mooley, Deller, Gottlieb, Nakar, Hallinan,
  Bourke, Frail, Horesh, Corsi, \& Hotokezaka}]{mooley2018superluminal}
Mooley, K.~P., Deller, A.~T., Gottlieb, O., {et~al.} 2018, Nature, 561, 355

\bibitem[{Moore \& Nelson(2001)}]{GR_cgw_CR_limit_Moore_2001}
Moore, G.~D., \& Nelson, A.~E. 2001, Journal of High Energy Physics, 2001, 023

\bibitem[{Most {et~al.}(2019)Most, Papenfort, Dexheimer, Hanauske, Schramm,
  St{\"o}cker, \& Rezzolla}]{most2019signatures}
Most, E.~R., Papenfort, L.~J., Dexheimer, V., {et~al.} 2019, \prl, 122, 061101

\bibitem[{Most {et~al.}(2018)Most, Weih, Rezzolla, \&
  Schaffner-Bielich}]{most2018new}
Most, E.~R., Weih, L.~R., Rezzolla, L., \& Schaffner-Bielich, J. 2018, \prl,
  120, 261103

\bibitem[{{Mueller} {et~al.}(2019){Mueller}, {Baker}, {Barke}, {Bender},
  {Berti}, {Caldwell}, {Conklin}, {Cornish}, {Ferrara}, {Holley-Bockelmann},
  {Kamai}, {Larson}, {Livas}, {McWilliams}, {Mueller}, {Natarajan}, {Rioux},
  {Sankar}, {Schnittman}, {Shoemaker}, {Slutsky}, {Stebbins}, {Thorpe}, \&
  {Ziemer}}]{baker2019space}
{Mueller}, G., {Baker}, J., {Barke}, S., {et~al.} 2019, \baas, 51, 243

\bibitem[{Murase {et~al.}(2009)Murase, M{\'e}sz{\'a}ros, \&
  Zhang}]{murase2009probing}
Murase, K., M{\'e}sz{\'a}ros, P., \& Zhang, B. 2009, \prd, 79, 103001

\bibitem[{Murase {et~al.}(2018)Murase, Toomey, Fang, Oikonomou, Kimura,
  Hotokezaka, Kashiyama, Ioka, \& M{\'e}sz{\'a}ros}]{murase2018double}
Murase, K., Toomey, M.~W., Fang, K., {et~al.} 2018, Astrophys. J., 854, 60

\bibitem[{Murayama \& Yanagida(2001)}]{LIV_SN1987A_Murayama_2001}
Murayama, H., \& Yanagida, T. 2001, Phys. Lett. B, 520, 263

\bibitem[{Murguia-Berthier {et~al.}(2014)Murguia-Berthier, Montes,
  Ramirez-Ruiz, De~Colle, \& Lee}]{murguia2014necessary}
Murguia-Berthier, A., Montes, G., Ramirez-Ruiz, E., De~Colle, F., \& Lee, W.~H.
  2014, \apjl, 788, L8

\bibitem[{Nagakura {et~al.}(2014)Nagakura, Hotokezaka, Sekiguchi, Shibata, \&
  Ioka}]{nagakura2014jet}
Nagakura, H., Hotokezaka, K., Sekiguchi, Y., Shibata, M., \& Ioka, K. 2014,
  \apjl, 784, L28

\bibitem[{Nagano \& Watson(2000)}]{nagano2000observations}
Nagano, M., \& Watson, A.~A. 2000, Rev. Mod. Phys., 72, 689

\bibitem[{{Nakar} \& {Piran}(2011)}]{omni_radio_nakar_2011}
{Nakar}, E., \& {Piran}, T. 2011, \nat, 478, 82

\bibitem[{{Nakar} \& {Sari}(2012)}]{cocoon_breakout_nakar_2012}
{Nakar}, E., \& {Sari}, R. 2012, \apj, 747, 88

\bibitem[{Nakar {et~al.}(2012)}]{nakar2012relativistic}
Nakar, E., {et~al.} 2012, Astrophys. J., 747, 88

\bibitem[{Nishizawa \& Kobayashi(2018)}]{nishizawa2018parity}
Nishizawa, A., \& Kobayashi, T. 2018, \prd, 98, 124018

\bibitem[{Nishizawa \& Nakamura(2014)}]{nishizawa_2014_speed_of_gravity}
Nishizawa, A., \& Nakamura, T. 2014, \prd, 90, 044048

\bibitem[{{Nissanke} {et~al.}(2013){Nissanke}, {Holz}, {Dalal}, {Hughes},
  {Sievers}, \& {Hirata}}]{H0-Nissanke-2013}
{Nissanke}, S., {Holz}, D.~E., {Dalal}, N., {et~al.} 2013, arXiv e-prints,
  arXiv:1307.2638

\bibitem[{Norris \& Bonnell(2006)}]{extended_emission_2006_norris}
Norris, J.~P., \& Bonnell, J.~T. 2006, Astrophys. J., 643, 266

\bibitem[{Nouri {et~al.}(2018)Nouri, Duez, Foucart, Deaton, Haas, Haddadi,
  Kidder, Ott, Pfeiffer, Scheel, {et~al.}}]{nouri2018evolution}
Nouri, F.~H., Duez, M.~D., Foucart, F., {et~al.} 2018, \prd, 97, 083014

\bibitem[{Nynka {et~al.}(2018)Nynka, Ruan, Haggard, \& Evans}]{nynka2018fading}
Nynka, M., Ruan, J.~J., Haggard, D., \& Evans, P.~A. 2018, \apjl, 862, L19

\bibitem[{Oechslin \& Janka(2006)}]{oechslin2006torus}
Oechslin, R., \& Janka, H.-T. 2006, Mon. Not. R. Astron. Soc., 368, 1489

\bibitem[{Oechslin \& Janka(2007)}]{oechslin2007gravitational}
---. 2007, \prl, 99, 121102

\bibitem[{Ofek(2007)}]{ofek2007soft}
Ofek, E.~O. 2007, \apj, 659, 339

\bibitem[{Oganesyan {et~al.}(2020)Oganesyan, Ascenzi, Branchesi, Salafia,
  Dall’Osso, \& Ghirlanda}]{oganesyan2020structured}
Oganesyan, G., Ascenzi, S., Branchesi, M., {et~al.} 2020, The Astrophysical
  Journal, 893, 88

\bibitem[{{Oppenheimer} \&
  {Volkoff}(1939)}]{NS_first_models_Oppenheimer_Volkoff_1939}
{Oppenheimer}, J.~R., \& {Volkoff}, G.~M. 1939, Phys. Rev., 55, 374

\bibitem[{{\"O}zel \& Freire(2016)}]{NS-EOS-Ozel}
{\"O}zel, F., \& Freire, P. 2016, Annu. Rev. Astron. Astrophys., 54, 401

\bibitem[{{Paczynski}(1986)}]{Paczynski1986}
{Paczynski}, B. 1986, \apjl, 308, L43

\bibitem[{Paczynski \& Rhoads(1993)}]{paczynski1993radio}
Paczynski, B., \& Rhoads, J.~E. 1993, arXiv e-prints, astro-ph/9307024

\bibitem[{Palatiello {et~al.}(2017)Palatiello, Noda, Inoue, Colin, Moretti, \&
  Longo}]{MAGIC_160821B}
Palatiello, M., Noda, K., Inoue, S., {et~al.} 2017, in Proceedings of the 7th
  International Fermi Symposium, held 15-20 October 2017, in
  Garmisch-Partenkirchen, Germany (IFS2017)

\bibitem[{Palmer {et~al.}(2005)Palmer, Barthelmy, Gehrels, Kippen, Cayton,
  Kouveliotou, Eichler, Wijers, Woods, Granot,
  {et~al.}}]{GMF_041227_SGR1806_Palmer}
Palmer, D.~M., Barthelmy, S., Gehrels, N., {et~al.} 2005, Nature, 434, 1107

\bibitem[{Panaitescu(2005{\natexlab{a}})}]{panaitescu2005jets}
Panaitescu, A. 2005{\natexlab{a}}, \mnras, 363, 1409

\bibitem[{Panaitescu(2005{\natexlab{b}})}]{panaitescu2005models}
---. 2005{\natexlab{b}}, Mon. Not. R. Astron. Soc., 362, 921

\bibitem[{Panaitescu \& Kumar(2001)}]{panaitescu2001jet}
Panaitescu, A., \& Kumar, P. 2001, \apj, 554, 667

\bibitem[{Pannarale {et~al.}(2015)Pannarale, Berti, Kyutoku, Lackey, \&
  Shibata}]{pannarale2015gravitational}
Pannarale, F., Berti, E., Kyutoku, K., Lackey, B.~D., \& Shibata, M. 2015,
  \prd, 92, 081504

\bibitem[{Pardo {et~al.}(2018)Pardo, Fishbach, Holz, \&
  Spergel}]{large_dim_pardo_2018}
Pardo, K., Fishbach, M., Holz, D.~E., \& Spergel, D.~N. 2018, JCAP, 2018, 048

\bibitem[{Pauli(1930)}]{neutrino_pauli_1930}
Pauli, W. 1930, Phys. Today, 31, 27

\bibitem[{Penzias \& Wilson(1965)}]{cosmo_CMB_1965}
Penzias, A.~A., \& Wilson, R.~W. 1965, Astrophys. J., 142, 419

\bibitem[{Perley {et~al.}(2009)Perley, Metzger, Granot, Butler, Sakamoto,
  Ramirez-Ruiz, Levan, Bloom, Miller, Bunker, {et~al.}}]{perley2009grb}
Perley, D.~A., Metzger, B.~D., Granot, J., {et~al.} 2009, Astrophys. J., 696,
  1871

\bibitem[{Perlmutter {et~al.}(1999)Perlmutter, Aldering, Goldhaber, Knop,
  Nugent, Castro, Deustua, Fabbro, Goobar, Groom,
  {et~al.}}]{Universe_expanding_Perlmutter_1999}
Perlmutter, S., Aldering, G., Goldhaber, G., {et~al.} 1999, Astrophys. J., 517,
  565

\bibitem[{Perna \& Keeton(2009)}]{lens_anisotropic_PErna_2009}
Perna, R., \& Keeton, C.~R. 2009, Mon. Not. R. Astron. Soc., 397, 1084

\bibitem[{Perna {et~al.}(2003)Perna, Sari, \& Frail}]{perna2003jets}
Perna, R., Sari, R., \& Frail, D. 2003, \apj, 594, 379

\bibitem[{Peters \& Mathews(1963)}]{peters1963gravitational}
Peters, P.~C., \& Mathews, J. 1963, Phys. Rev., 131, 435

\bibitem[{Piran(1999)}]{piran1999gamma}
Piran, T. 1999, Phys. Rep., 314, 575

\bibitem[{{Piran} {et~al.}(2013){Piran}, {Nakar}, \&
  {Rosswog}}]{omni_radio_KN_Piran_2012}
{Piran}, T., {Nakar}, E., \& {Rosswog}, S. 2013, \mnras, 430, 2121

\bibitem[{{Planck Collaboration}(2018)}]{Planck_2018_cosmo}
{Planck Collaboration}. 2018, ArXiv e-prints, arXiv:1807.06209

\bibitem[{Poisson \& Will(2014)}]{poisson2014gravity}
Poisson, E., \& Will, C.~M. 2014, {Gravity: Newtonian, post-newtonian,
  relativistic} (Cambridge University Press)

\bibitem[{Pontecorvo(1957)}]{neutrino_pontecorvo_1957}
Pontecorvo, B. 1957, Zhur. Eksp. Teor. Fiz., 33

\bibitem[{Pontecorvo(1958)}]{neutrino_pontecorvo_1958}
---. 1958, Zhur. Eksp. Teor. Fiz., 34

\bibitem[{Pontecorvo(1967)}]{neutrino_pontecorvo_1968}
---. 1967, Zhurnal Eksperimental'noi i Teoreticheskoi Fiziki, 53, 1717

\bibitem[{Pontecorvo(1968)}]{mnu_oscillations_original_pontecorvo_1968}
---. 1968, Sov. Phys. JETP, 26, 165

\bibitem[{Popham {et~al.}(1999)Popham, Woosley, \&
  Fryer}]{popham1999hyperaccreting}
Popham, R., Woosley, S.~E., \& Fryer, C. 1999, \apj, 518, 356

\bibitem[{{Postnov} \& {Yungelson}(2014)}]{review_CO_evolution_Postnov_2014}
{Postnov}, K.~A., \& {Yungelson}, L.~R. 2014, Living Rev. Relativ., 17, 3

\bibitem[{Pruet {et~al.}(2004)Pruet, Thompson, \&
  Hoffman}]{pruet2004nucleosynthesis}
Pruet, J., Thompson, T.~A., \& Hoffman, R.~D. 2004, Astrophys. J., 606, 1006

\bibitem[{Punturo {et~al.}(2010)Punturo, Abernathy, Acernese, Allen, Andersson,
  Arun, Barone, Barr, Barsuglia, Beker, {et~al.}}]{Einstein_Telescope}
Punturo, M., Abernathy, M., Acernese, F., {et~al.} 2010, Class. Quantum Grav.,
  27, 194002

\bibitem[{Qian \& Vogel(2015)}]{mnu_mass_hierarchy_review_2015}
Qian, X., \& Vogel, P. 2015, Progr. Part. Nucl. Phys., 83, 1

\bibitem[{Qian(2000)}]{qian2000supernovae}
Qian, Y.-Z. 2000, \apjl, 534, L67

\bibitem[{Rachen \& M{\'e}sz{\'a}ros(1998)}]{rachen1998photohadronic}
Rachen, J.~P., \& M{\'e}sz{\'a}ros, P. 1998, \prd, 58, 123005

\bibitem[{Racusin {et~al.}(2008)Racusin, Karpov, Sokolowski, Granot, Wu,
  Pal'Shin, Covino, Van Der~Horst, Oates, Schady,
  {et~al.}}]{racusin2008broadband}
Racusin, J., Karpov, S., Sokolowski, M., {et~al.} 2008, Nature, 455, 183

\bibitem[{{Racusin} {et~al.}(2017){Racusin}, {Perkins}, {Briggs}, {de Nolfo},
  {Krizmanic}, {Caputo}, {McEnery}, {Shawhan}, {Morris}, {Connaughton},
  {Kocevski}, {Wilson-Hodge}, {Hui}, {Mitchell}, \&
  {McBreen}}]{racusin2017burstcube}
{Racusin}, J., {Perkins}, J.~S., {Briggs}, M.~S., {et~al.} 2017, arXiv
  e-prints, arXiv:1708.09292

\bibitem[{Racusin {et~al.}(2009)Racusin, Liang, Burrows, Falcone, Sakamoto,
  Zhang, Zhang, Evans, \& Osborne}]{racusin2009jet}
Racusin, J.~L., Liang, E.~W., Burrows, D.~N., {et~al.} 2009, Astrophys. J.,
  698, 43

\bibitem[{Radice {et~al.}(2018)Radice, Perego, Zappa, \&
  Bernuzzi}]{radice2018gw170817}
Radice, D., Perego, A., Zappa, F., \& Bernuzzi, S. 2018, \apjl, 852, L29

\bibitem[{Raithel(2019)}]{raithel2019constraints}
Raithel, C.~A. 2019, Eur. Phys. J. A, 55, 80

\bibitem[{Raithel {et~al.}(2018)Raithel, {\"O}zel, \&
  Psaltis}]{raithel2018tidal}
Raithel, C.~A., {\"O}zel, F., \& Psaltis, D. 2018, \apjl, 857, L23

\bibitem[{Ratra \& Peebles(1988)}]{quintessence_1}
Ratra, B., \& Peebles, P. J.~E. 1988, \prd, 37, 3406

\bibitem[{Ravasio {et~al.}(2018)Ravasio, Oganesyan, Ghirlanda, Nava,
  Ghisellini, Pescalli, \& Celotti}]{ravasio2018consistency}
Ravasio, M.~E., Oganesyan, G., Ghirlanda, G., {et~al.} 2018, Astron.
  Astrophys., 613, A16

\bibitem[{Ravi \& Lasky(2014)}]{ravi2014birth}
Ravi, V., \& Lasky, P.~D. 2014, Mon. Not. R. Astron. Soc., 441, 2433

\bibitem[{Read {et~al.}(2013)Read, Baiotti, Creighton, Friedman, Giacomazzo,
  Kyutoku, Markakis, Rezzolla, Shibata, \& Taniguchi}]{read2013matter}
Read, J.~S., Baiotti, L., Creighton, J.~D., {et~al.} 2013, \prd, 88, 044042

\bibitem[{Rees \& M{\'e}sz{\'a}ros(1994)}]{rees1994unsteady}
Rees, M.~J., \& M{\'e}sz{\'a}ros, P. 1994, arXiv e-prints, astro-ph/9404038

\bibitem[{Refsdal(1964)}]{lens_Refsdal_1964}
Refsdal, S. 1964, Mon. Not. R. Astron. Soc., 128, 307

\bibitem[{Reichart(1998)}]{reichart1998redshift}
Reichart, D.~E. 1998, \apjl, 495, L99

\bibitem[{Reines \& Cowan~Jr(1953)}]{reines_cowan_1953}
Reines, F., \& Cowan~Jr, C.~L. 1953, Phys. Rev., 92, 830

\bibitem[{{Reitze} {et~al.}(2019){Reitze}, {Adhikari}, {Ballmer}, {Barish},
  {Barsotti}, {Billingsley}, {Brown}, {Chen}, {Coyne}, {Eisenstein}, {Evans},
  {Fritschel}, {Hall}, {Lazzarini}, {Lovelace}, {Read}, {Sathyaprakash},
  {Shoemaker}, {Smith}, {Torrie}, {Vitale}, {Weiss}, {Wipf}, \&
  {Zucker}}]{reitze2019cosmic}
{Reitze}, D., {Adhikari}, R.~X., {Ballmer}, S., {et~al.} 2019, \baas, 51, 35

\bibitem[{{Rezzolla} {et~al.}(2011){Rezzolla}, {Giacomazzo}, {Baiotti},
  {Granot}, {Kouveliotou}, \& {Aloy}}]{rezzolla2011}
{Rezzolla}, L., {Giacomazzo}, B., {Baiotti}, L., {et~al.} 2011, \apjl, 732, L6

\bibitem[{Rezzolla {et~al.}(2018)Rezzolla, Most, \& Weih}]{rezzolla2018using}
Rezzolla, L., Most, E.~R., \& Weih, L.~R. 2018, \apjl, 852, L25

\bibitem[{{Rezzolla} \& {Takami}(2016)}]{2016PhRvD..93l4051R}
{Rezzolla}, L., \& {Takami}, K. 2016, \prd, 93, 124051

\bibitem[{Rhoads(1997)}]{rhoads1997tell}
Rhoads, J.~E. 1997, \apjl, 487, L1

\bibitem[{{Riess} {et~al.}(2019){Riess}, {Casertano}, {Yuan}, {Macri}, \&
  {Scolnic}}]{H0-Riess_2019}
{Riess}, A.~G., {Casertano}, S., {Yuan}, W., {Macri}, L.~M., \& {Scolnic}, D.
  2019, \apj, 876, 85

\bibitem[{Riess {et~al.}(1998)Riess, Filippenko, Challis, Clocchiatti, Diercks,
  Garnavich, Gilliland, Hogan, Jha, Kirshner,
  {et~al.}}]{Universe_expanding_Riess_1998}
Riess, A.~G., Filippenko, A.~V., Challis, P., {et~al.} 1998, Astron. J., 116,
  1009

\bibitem[{Riess {et~al.}(2016)Riess, Macri, Hoffmann, Scolnic, Casertano,
  Filippenko, Tucker, Reid, Jones, Silverman, {et~al.}}]{riess2016}
Riess, A.~G., Macri, L.~M., Hoffmann, S.~L., {et~al.} 2016, Astrophys. J., 826,
  56

\bibitem[{Riess {et~al.}(2018)Riess, Casertano, Yuan, Macri, Bucciarelli,
  Lattanzi, MacKenty, Bowers, Zheng, Filippenko, {et~al.}}]{H0-Riess_2018}
Riess, A.~G., Casertano, S., Yuan, W., {et~al.} 2018, Astrophys. J., 861, 126

\bibitem[{Riley {et~al.}(2019)Riley, Watts, Bogdanov, Ray, Ludlam, Guillot,
  Arzoumanian, Baker, Bilous, Chakrabarty, {et~al.}}]{riley2019nicer}
Riley, T.~E., Watts, A.~L., Bogdanov, S., {et~al.} 2019, \apjl, 887, L21

\bibitem[{Roberts {et~al.}(2012)Roberts, Reddy, \& Shen}]{roberts2012medium}
Roberts, L.~F., Reddy, S., \& Shen, G. 2012, Physical Review C, 86, 065803

\bibitem[{Rossi {et~al.}(2002)Rossi, Lazzati, \& Rees}]{rossi2002afterglow}
Rossi, E., Lazzati, D., \& Rees, M.~J. 2002, \mnras, 332, 945

\bibitem[{Rosswog(2007)}]{rosswog2007fallback}
Rosswog, S. 2007, Mon. Not. R. Astron. Soc. Lett., 376, L48

\bibitem[{Rovelli(2008)}]{QG_LQG_rovelli_2008}
Rovelli, C. 2008, Living Rev. Relativ., 11, 5

\bibitem[{Rowlinson {et~al.}(2013)Rowlinson, {O'Brien}, Metzger, Tanvir, \&
  Levan}]{rowlinson2013signatures}
Rowlinson, A., {O'Brien}, P.~T., Metzger, B.~D., Tanvir, N.~R., \& Levan, A.~J.
  2013, Mon. Not. R. Astron. Soc., 430, 1061

\bibitem[{Rowlinson {et~al.}(2010)Rowlinson, O'Brien, Tanvir, Zhang, Evans,
  Lyons, Levan, Willingale, Page, Onal, {et~al.}}]{rowlinson2010unusual}
Rowlinson, A., O'Brien, P.~T., Tanvir, N.~R., {et~al.} 2010, Mon. Not. R.
  Astron. Soc., 409, 531

\bibitem[{Rubin {et~al.}(1980)Rubin, Ford~Jr, \& Thonnard}]{cosmo_rubin_1980}
Rubin, V.~C., Ford~Jr, W.~K., \& Thonnard, N. 1980, Astrophys. J., 238, 471

\bibitem[{Rubin {et~al.}(1978)Rubin, Thonnard, \& Ford~Jr}]{cosmo_rubin_1978}
Rubin, V.~C., Thonnard, N., \& Ford~Jr, W.~K. 1978, Astrophys. J., 225, L107

\bibitem[{{Ruffert} \& {Janka}(1998)}]{ruffert_janka}
{Ruffert}, M., \& {Janka}, H.-T. 1998, \aap, 338, 535

\bibitem[{Ruiz \& Shapiro(2017)}]{ruiz2017general}
Ruiz, M., \& Shapiro, S.~L. 2017, \prd, 96, 084063

\bibitem[{{Ryan} {et~al.}(2019){Ryan}, {van Eerten}, {Piro}, \&
  {Troja}}]{ryan2019gamma}
{Ryan}, G., {van Eerten}, H., {Piro}, L., \& {Troja}, E. 2019, arXiv e-prints,
  arXiv:1909.11691

\bibitem[{Sadowski {et~al.}(2008)Sadowski, Belczynski, Bulik, Ivanova, Rasio,
  \& O'Shaughnessy}]{review_merger_rate_compact_sadowski_2008}
Sadowski, A., Belczynski, K., Bulik, T., {et~al.} 2008, Astrophys. J., 676,
  1162

\bibitem[{Safarzadeh {et~al.}(2019)Safarzadeh, Berger, Leja, \&
  Speagle}]{safarzadeh2019measuring}
Safarzadeh, M., Berger, E., Leja, J., \& Speagle, J.~S. 2019, \apjl, 878, L14

\bibitem[{Sakstein \& Jain(2017)}]{sakstein2017implications}
Sakstein, J., \& Jain, B. 2017, \prl, 119, 251303

\bibitem[{Salafia \& Giacomazzo(2020)}]{salafia2020accretion}
Salafia, O.~S., \& Giacomazzo, B. 2020, arXiv e-prints, 2006.07376

\bibitem[{Salam(1968)}]{SM_2}
Salam, A. 1968, in Prog. Of the Nobel Symposium, 1968, Stockholm, Sweden, Vol.
  367

\bibitem[{Samajdar \& Arun(2017)}]{GW_dispersion_future_Samajdar_2017}
Samajdar, A., \& Arun, K.~G. 2017, \prd, 96, 104027

\bibitem[{Sari {et~al.}(1999)Sari, Piran, \& Halpern}]{sari1999jets}
Sari, R., Piran, T., \& Halpern, J.~P. 1999, \apjl, 519, L17

\bibitem[{{Sari} {et~al.}(1998){Sari}, {Piran}, \&
  {Narayan}}]{sari1998synchrotron}
{Sari}, R., {Piran}, T., \& {Narayan}, R. 1998, \apjl, 497, L17

\bibitem[{Sathyaprakash {et~al.}(2012)Sathyaprakash, Abernathy, Acernese,
  Ajith, Allen, Amaro-Seoane, Andersson, Aoudia, Arun, Astone,
  {et~al.}}]{ET_science_objectives_sathyaprakash_2012}
Sathyaprakash, B., Abernathy, M., Acernese, F., {et~al.} 2012, Class. Quantum
  Grav., 29, 124013

\bibitem[{Sathyaprakash {et~al.}(2010)Sathyaprakash, Schutz, \& Van
  Den~Broeck}]{et_2010_cosmography}
Sathyaprakash, B.~S., Schutz, B.~F., \& Van Den~Broeck, C. 2010, Class. Quantum
  Grav., 27, 215006

\bibitem[{{Savchenko} {et~al.}(2017){Savchenko}, {Ferrigno}, {Kuulkers},
  {Bazzano}, {Bozzo}, {Brandt}, {Chenevez}, {Courvoisier}, {Diehl}, {Domingo},
  {Hanlon}, {Jourdain}, {von Kienlin}, {Laurent}, {Lebrun}, {Lutovinov},
  {Martin-Carrillo}, {Mereghetti}, {Natalucci}, {Rodi}, {Roques}, {Sunyaev}, \&
  {Ubertini}}]{Savchenko2017}
{Savchenko}, V., {Ferrigno}, C., {Kuulkers}, E., {et~al.} 2017, \apjl, 848, L15

\bibitem[{Schnittman {et~al.}(2018)Schnittman, Dal~Canton, Camp, Tsang, \&
  Kelly}]{EM_chirp_schnittman_2018}
Schnittman, J.~D., Dal~Canton, T., Camp, J., Tsang, D., \& Kelly, B.~J. 2018,
  Astrophys. J., 853, 123

\bibitem[{Schussler {et~al.}(2019)}]{GCN_HESS_190829A}
Schussler, F., {et~al.} 2019, {{GRB190829A}: Detection of VHE gamma-ray
  emission with H.E.S.S.}, Tech. rep., GSFC

\bibitem[{Schutz(1986)}]{H0-Schutz-1986}
Schutz, B.~F. 1986, Nature, 323, 310

\bibitem[{Schutz(2002)}]{schutz2002lighthouses}
---. 2002, in Lighthouses of the Universe: The Most Luminous Celestial Objects
  and Their Use for Cosmology (Springer), 207--224

\bibitem[{Schutz(2011)}]{GW_patterns_Schutz_2011}
---. 2011, Class. Quantum Grav., 28, 125023

\bibitem[{Sekiguchi {et~al.}(2011)Sekiguchi, Kiuchi, Kyutoku, \&
  Shibata}]{GWs_neutrino_BNS_sekiguchi_2011}
Sekiguchi, Y., Kiuchi, K., Kyutoku, K., \& Shibata, M. 2011, \prl, 107, 051102

\bibitem[{Senno {et~al.}(2016)Senno, Murase, \&
  M{\'e}sz{\'a}ros}]{senno2016choked}
Senno, N., Murase, K., \& M{\'e}sz{\'a}ros, P. 2016, \prd, 93, 083003

\bibitem[{Shapiro(1964)}]{shapiro_delay_1964}
Shapiro, I.~I. 1964, \prl, 13, 789

\bibitem[{Shapley(1918)}]{cosmo_shapley_1918}
Shapley, H. 1918, Astrophys. J., 48

\bibitem[{Shibata \& Taniguchi(2006)}]{shibata2006merger}
Shibata, M., \& Taniguchi, K. 2006, \prd, 73, 064027

\bibitem[{Shibata \& Ury{\=u}(2000)}]{shibata2000simulation}
Shibata, M., \& Ury{\=u}, K. 2000, \prd, 61, 064001

\bibitem[{Shibata {et~al.}(2019)Shibata, Zhou, Kiuchi, \&
  Fujibayashi}]{shibata2019constraint}
Shibata, M., Zhou, E., Kiuchi, K., \& Fujibayashi, S. 2019, arXiv e-prints,
  1905.03656

\bibitem[{Shoemaker \& Murase(2018)}]{GW_GRB_WEP_Shoemaker_2018}
Shoemaker, I.~M., \& Murase, K. 2018, \prd, 97, 083013

\bibitem[{Shull {et~al.}(2012)Shull, Smith, \&
  Danforth}]{cosmo_missing_baryon_2012}
Shull, J.~M., Smith, B.~D., \& Danforth, C.~W. 2012, Astrophys. J., 759, 23

\bibitem[{Siegel {et~al.}(2019)Siegel, Barnes, \&
  Metzger}]{siegel2019collapsars}
Siegel, D.~M., Barnes, J., \& Metzger, B.~D. 2019, Nature, 569, 241

\bibitem[{Siegel \& Metzger(2017)}]{siegel2017three}
Siegel, D.~M., \& Metzger, B.~D. 2017, Phys. Rev. Let., 119, 231102

\bibitem[{Singer \& Price(2016)}]{singer2016rapid}
Singer, L.~P., \& Price, L.~R. 2016, \prd, 93, 024013

\bibitem[{Singer {et~al.}(2014)Singer, Price, Farr, Urban, Pankow, Vitale,
  Veitch, Farr, Hanna, Cannon, {et~al.}}]{singer2014first}
Singer, L.~P., Price, L.~R., Farr, B., {et~al.} 2014, Astrophys. J., 795, 105

\bibitem[{Smolin(2008)}]{QG_general_Smolin_2008}
Smolin, L. 2008, {Three roads to quantum gravity} (Basic books)

\bibitem[{Smoot {et~al.}(1992)Smoot, Bennett, Kogut, Wright, Aymon, Boggess,
  Cheng, De~Amici, Gulkis, Hauser, {et~al.}}]{cosmo_smoot_1992}
Smoot, G.~F., Bennett, C.~L., Kogut, A., {et~al.} 1992, Astrophys. J., 396, L1

\bibitem[{Soares-Santos {et~al.}(2017)Soares-Santos, Holz, Annis, Chornock,
  Herner, Berger, Brout, Chen, Kessler, Sako,
  {et~al.}}]{soares2017electromagnetic}
Soares-Santos, M., Holz, D.~E., Annis, J., {et~al.} 2017, \apjl, 848, L16

\bibitem[{Sokolsky(2018)}]{sokolsky2018introduction}
Sokolsky, P. 2018, {Introduction to ultrahigh energy cosmic ray physics} (CRC
  Press)

\bibitem[{Solodukhin(2011)}]{LIV_BH_solodukhin_2011}
Solodukhin, S.~N. 2011, Living Rev. Relativ., 14, 8

\bibitem[{Song {et~al.}(2019)Song, Ai, Wang, Xing, Gao, \&
  Zhang}]{song2019viewing}
Song, H.-R., Ai, S.-K., Wang, M.-H., {et~al.} 2019, \apjl, 881, L40

\bibitem[{Starling {et~al.}(2005)Starling, Wijers, Hughes, Tanvir, Vreeswijk,
  Rol, \& Salamanca}]{starling2005spectroscopy}
Starling, R.~L.~C., Wijers, R.~A.~M.~J., Hughes, M.~A., {et~al.} 2005, \mnras,
  360, 305

\bibitem[{{Starobinski{\v i}}(1979)}]{cosmo_inflation_1979}
{Starobinski{\v i}}, A.~A. 1979, Sov. J. Exp. Theor. Phys. Lett., 30, 682

\bibitem[{Stodolsky(1988)}]{neutrinos_speed_1987A_stodolsky_1988}
Stodolsky, L. 1988, Phys. Lett. B, 201, 353

\bibitem[{Stone {et~al.}(2013)Stone, Loeb, \& Berger}]{stone2013pulsations}
Stone, N., Loeb, A., \& Berger, E. 2013, \prd, 87, 084053

\bibitem[{Strong {et~al.}(1974)Strong, Klebesadel, \&
  Olson}]{strong1974preliminary}
Strong, I.~B., Klebesadel, R.~W., \& Olson, R.~A. 1974, \apj, 188, L1

\bibitem[{{Sun} {et~al.}(2019){Sun}, {Li}, {Zhang}, {Zhang}, {Bauer}, {Xue}, \&
  {Yuan}}]{Bing_4}
{Sun}, H., {Li}, Y., {Zhang}, B., {et~al.} 2019, arXiv e-prints,
  arXiv:1908.01107

\bibitem[{{Sun} {et~al.}(2017){Sun}, {Zhang}, \& {Gao}}]{Bing_2}
{Sun}, H., {Zhang}, B., \& {Gao}, H. 2017, \apj, 835, 7

\bibitem[{{Svinkin} {et~al.}(2015){Svinkin}, {Hurley}, {Aptekar},
  {Golenetskii}, \& {Frederiks}}]{GMF_IPN_Dmitry_search}
{Svinkin}, D.~S., {Hurley}, K., {Aptekar}, R.~L., {Golenetskii}, S.~V., \&
  {Frederiks}, D.~D. 2015, \mnras, 447, 1028

\bibitem[{Symbalisty \& Schramm(1982)}]{symbalisty1982neutron}
Symbalisty, E., \& Schramm, D.~N. 1982, Astrophys. Lett., 22, 143

\bibitem[{Tak {et~al.}(2019)Tak, Guiriec, Uhm, Yassine, Omodei, \&
  McEnery}]{tak2019multiple}
Tak, D., Guiriec, S., Uhm, Z.~L., {et~al.} 2019, Astrophys. J., 876, 76

\bibitem[{Takeda {et~al.}(2018)Takeda, Nishizawa, Michimura, Nagano, Komori,
  Ando, \& Hayama}]{GW_polarization_Takeda_2018}
Takeda, H., Nishizawa, A., Michimura, Y., {et~al.} 2018, \prd, 98, 022008

\bibitem[{Tamanini {et~al.}(2016)Tamanini, Caprini, Barausse, Sesana, Klein, \&
  Petiteau}]{h0_LISA_2016}
Tamanini, N., Caprini, C., Barausse, E., {et~al.} 2016, JCAP, 2016, 002

\bibitem[{Tanabashi {et~al.}(2018)Tanabashi, Hagiwara, Hikasa, Nakamura,
  Sumino, Takahashi, Tanaka, Agashe, Aielli, Amsler,
  {et~al.}}]{review_particle_physics_Tanabashi_2018}
Tanabashi, M., Hagiwara, K., Hikasa, K., {et~al.} 2018, \prd, 98, 030001

\bibitem[{Tanaka(2016)}]{review_kilonova_tanaka_2016}
Tanaka, M. 2016, Advances in Astronomy, 2016

\bibitem[{{Tanaka} \& {Hotokezaka}(2013)}]{KN_correct_opacities_Tanaka_2013}
{Tanaka}, M., \& {Hotokezaka}, K. 2013, \apj, 775, 113

\bibitem[{Tanvir {et~al.}(2013)Tanvir, Levan, Fruchter, Hjorth, Hounsell,
  Wiersema, \& Tunnicliffe}]{tanvir2013kilonova}
Tanvir, N.~R., Levan, A.~J., Fruchter, A.~S., {et~al.} 2013, Nature, 500, 547

\bibitem[{Tanvir {et~al.}(2017)Tanvir, Levan, Gonz{\'a}lez-Fern{\'a}ndez,
  Korobkin, Mandel, Rosswog, Hjorth, D'Avanzo, Fruchter, Fryer,
  {et~al.}}]{gw170817_kilonova_tanvir}
Tanvir, N.~R., Levan, A.~J., Gonz{\'a}lez-Fern{\'a}ndez, C., {et~al.} 2017,
  \apjl, 848, L27

\bibitem[{Tauris {et~al.}(2017)Tauris, Kramer, Freire, Wex, Janka, Langer,
  Podsiadlowski, Bozzo, Chaty, Kruckow, {et~al.}}]{Tauris-DNS-Systems}
Tauris, T.~M., Kramer, M., Freire, P.~C.~C., {et~al.} 2017, Astrophys. J., 846,
  170

\bibitem[{Taveras \& Yunes(2008)}]{taveras2008barbero}
Taveras, V., \& Yunes, N. 2008, \prd, 78, 064070

\bibitem[{Taylor {et~al.}(2004)Taylor, Frail, Berger, \&
  Kulkarni}]{taylor2004angular}
Taylor, G.~B., Frail, D.~A., Berger, E., \& Kulkarni, S.~R. 2004, \apjl, 609,
  L1

\bibitem[{{Taylor} \& {Weisberg}(1982)}]{1982ApJ...253..908T}
{Taylor}, J.~H., \& {Weisberg}, J.~M. 1982, \apj, 253, 908

\bibitem[{Taylor \& Gair(2012)}]{cosmo_no_EM_2012}
Taylor, S.~R., \& Gair, J.~R. 2012, \prd, 86, 023502

\bibitem[{Thompson(1994)}]{thompson1994model}
Thompson, C. 1994, \mnras, 270, 480

\bibitem[{Timmes {et~al.}(2019)Timmes, Fryer, Timmes, Hungerford, Couture,
  Adams, Aoki, Arcones, Arnett, Auchettl, {et~al.}}]{timmes2019catching}
Timmes, F., Fryer, C., Timmes, F., {et~al.} 2019, BAAS, 51, 2

\bibitem[{Tolman(1939)}]{tolman1939static}
Tolman, R.~C. 1939, Phys. Rev., 55, 364

\bibitem[{Toma {et~al.}(2009)Toma, Sakamoto, Zhang, Hill, McConnell, Bloser,
  Yamazaki, Ioka, \& Nakamura}]{toma2009statistical}
Toma, K., Sakamoto, T., Zhang, B., {et~al.} 2009, Astrophys. J., 698, 1042

\bibitem[{Troja {et~al.}(2010)Troja, Rosswog, \& Gehrels}]{troja2010precursors}
Troja, E., Rosswog, S., \& Gehrels, N. 2010, Astrophys. J., 723, 1711

\bibitem[{{Troja} {et~al.}(2007){Troja}, {Cusumano}, {O'Brien}, {Zhang},
  {Sbarufatti}, {Mangano}, {Willingale}, {Chincarini}, {Osborne}, {Marshall},
  {Burrows}, {Campana}, {Gehrels}, {Guidorzi}, {Krimm}, {La Parola}, {Liang},
  {Mineo}, {Moretti}, {Page}, {Romano}, {Tagliaferri}, {Zhang}, {Page}, \&
  {Schady}}]{2007ApJ...665..599T}
{Troja}, E., {Cusumano}, G., {O'Brien}, P.~T., {et~al.} 2007, \apj, 665, 599

\bibitem[{Troja {et~al.}(2017{\natexlab{a}})Troja, Piro, van Eerten, Wollaeger,
  Im, Fox, Butler, Cenko, Sakamoto, Fryer,
  {et~al.}}]{GW170817_Troja_Xray_discovery}
Troja, E., Piro, L., van Eerten, H., {et~al.} 2017{\natexlab{a}}, Nature, 551,
  71

\bibitem[{Troja {et~al.}(2017{\natexlab{b}})Troja, Lipunov, Mundell, Butler,
  Watson, Kobayashi, Cenko, Marshall, Ricci, Fruchter,
  {et~al.}}]{troja2017significant}
Troja, E., Lipunov, V.~M., Mundell, C.~G., {et~al.} 2017{\natexlab{b}}, Nature,
  547, 425

\bibitem[{Troja {et~al.}(2019)Troja, Van~Eerten, Ryan, Ricci, Burgess,
  Wieringa, Piro, Cenko, \& Sakamoto}]{troja2019year}
Troja, E., Van~Eerten, H., Ryan, G., {et~al.} 2019, Mon. Not. R. Astro. Soc.,
  489, 1919

\bibitem[{Tsang {et~al.}(2012)Tsang, Read, Hinderer, Piro, \&
  Bondarescu}]{precursors_tsang_resonant}
Tsang, D., Read, J.~S., Hinderer, T., Piro, A.~L., \& Bondarescu, R. 2012,
  \prl, 108, 011102

\bibitem[{Tso {et~al.}(2017)Tso, Isi, Chen, \& Stein}]{dispersion_general_2}
Tso, R., Isi, M., Chen, Y., \& Stein, L. 2017, in Proceedings of the Seventh
  Meeting on CPT and {Lorentz} Symmetry, World Scientific, 205--208

\bibitem[{Tunnicliffe {et~al.}(2013)Tunnicliffe, Levan, Tanvir, Rowlinson,
  Perley, Bloom, Cenko, O'Brien, Cobb, Wiersema,
  {et~al.}}]{SGRB_hostless_offset_Tunnicliffe_2013}
Tunnicliffe, R.~L., Levan, A.~J., Tanvir, N.~R., {et~al.} 2013, Mon. Not. R.
  Astron. Soc., 437, 1495

\bibitem[{Usman {et~al.}(2016)Usman, Nitz, Harry, Biwer, Brown, Cabero, Capano,
  Dal~Canton, Dent, Fairhurst, {et~al.}}]{usman2016pycbc}
Usman, S.~A., Nitz, A.~H., Harry, I.~W., {et~al.} 2016, Class. Quantum Grav.,
  33, 215004

\bibitem[{Usov(1992)}]{uso1992millisecond}
Usov, V. 1992, Nature, 357, 472

\bibitem[{Valenti {et~al.}(2017)Valenti, David, Yang, Cappellaro, Tartaglia,
  Corsi, Jha, Reichart, Haislip, \& Kouprianov}]{valenti2017discovery}
Valenti, S., David, J., Yang, S., {et~al.} 2017, \apjl, 848, L24

\bibitem[{{van de Voort} {et~al.}(2019){van de Voort}, {Pakmor}, {Grand },
  {Springel}, {G{\'o}mez}, \& {Marinacci}}]{Voort2019}
{van de Voort}, F., {Pakmor}, R., {Grand }, R. J.~J., {et~al.} 2019, arXiv
  e-prints, arXiv:1907.01557

\bibitem[{Van~Elewyck {et~al.}(2009)Van~Elewyck, Ando, Aso, Baret, Barsuglia,
  Bartos, Chassande-Mottin, Di~Palma, Dwyer, Finley, {et~al.}}]{van2009joint}
Van~Elewyck, V., Ando, S., Aso, Y., {et~al.} 2009, Int. J. Mod. Phys. D, 18,
  1655

\bibitem[{Van~Paradijs {et~al.}(1997)Van~Paradijs, Groot, Galama, Kouveliotou,
  Strom, Telting, Rutten, Fishman, Meegan, Pettini,
  {et~al.}}]{van1997transient}
Van~Paradijs, J., Groot, P.~J., Galama, T., {et~al.} 1997, Nature, 386, 686

\bibitem[{Vasileiou {et~al.}(2013)Vasileiou, Jacholkowska, Piron, Bolmont,
  Couturier, Granot, Stecker, Cohen-Tanugi, \&
  Longo}]{LIV_090510_Vasileiou_2013}
Vasileiou, V., Jacholkowska, A., Piron, F., {et~al.} 2013, \prd, 87, 122001

\bibitem[{Velasco(2019)}]{HESS_180720B}
Velasco, R. 2019, 1st International Cherenkov Telescope Array Symposium.
\newblock
  \url{https://indico.cta-observatory.org/event/1946/contributions/19893/}

\bibitem[{Veres {et~al.}(2018)Veres, M{\'e}sz{\'a}ros, Goldstein, Fraija,
  Connaughton, Burns, Preece, Hamburg, Wilson-Hodge, Briggs,
  {et~al.}}]{veres2018gamma}
Veres, P., M{\'e}sz{\'a}ros, P., Goldstein, A., {et~al.} 2018, arXiv e-prints,
  1802.07328

\bibitem[{Vietri(1995)}]{vietri1995acceleration}
Vietri, M. 1995, arXiv e-prints, astro-ph/9506081

\bibitem[{{Villar} {et~al.}(2017){Villar}, {Guillochon}, {Berger}, {Metzger},
  {Cowperthwaite}, {Nicholl}, {Alexand er}, {Blanchard}, {Chornock},
  {Eftekhari}, {Fong}, {Margutti}, \& {Williams}}]{villar2017combined}
{Villar}, V.~A., {Guillochon}, J., {Berger}, E., {et~al.} 2017, \apjl, 851, L21

\bibitem[{{Vitale} \& {Chen}(2018)}]{H0_vitale_NSBH}
{Vitale}, S., \& {Chen}, H.-Y. 2018, \prl, 121, 021303

\bibitem[{Voges {et~al.}(2000)Voges, Aschenbach, Boller, Brauninger, Briel,
  Burkert, Dennerl, Englhauser, Gruber, Haberl, {et~al.}}]{voges2000rosat}
Voges, W., Aschenbach, B., Boller, T., {et~al.} 2000, IAUC, 7432, 3

\bibitem[{Volovik(2001)}]{LIV_superfluid_analogues_volovik_2001}
Volovik, G.~E. 2001, Phys. Rep., 351, 195

\bibitem[{{von Kienlin} {et~al.}(2003){von Kienlin}, {Beckmann}, {Rau},
  {Arend}, {Bennett}, {McBreen}, {Connell}, {Deluit}, {Hanlon}, {Hurley},
  {Kippen}, {Lichti}, {Moran}, {Preece}, {Roques}, {Sch{\"o}nfelder},
  {Skinner}, {Strong}, \& {Williams}}]{2003A&A...411L.299V}
{von Kienlin}, A., {Beckmann}, V., {Rau}, A., {et~al.} 2003, \aap, 411, L299

\bibitem[{von Kienlin {et~al.}(2020)von Kienlin, Meegan, Paciesas, Bhat,
  Bissaldi, Briggs, Burns, Cleveland, Gibby, Giles, {et~al.}}]{von2020fourth}
von Kienlin, A., Meegan, C.~A., Paciesas, W.~S., {et~al.} 2020, \apj, 893, 46

\bibitem[{{von Kienlin, A.} {et~al.}(2003){von Kienlin, A.}, {Beckmann, V.},
  {Rau, A.}, {Arend, N.}, {Bennett, K.}, {McBreen, B.}, {Connell, P.}, {Deluit,
  S.}, {Hanlon, L.}, {Hurley, K.}, {Kippen, M.}, {Lichti, G. G.}, {Moran, L.},
  {Preece, R.}, {Roques, J.-P.}, {Sch{\"o}nfelder, V.}, {Skinner, G.}, {Strong,
  A.}, \& {Williams, R.}}]{SPIACS}
{von Kienlin, A.}, {Beckmann, V.}, {Rau, A.}, {et~al.} 2003, Astron.
  Astrophys., 411, L299

\bibitem[{Wallner {et~al.}(2015)Wallner, Faestermann, Feige, Feldstein, Knie,
  Korschinek, Kutschera, Ofan, Paul, Quinto, {et~al.}}]{wallner2015abundance}
Wallner, A., Faestermann, T., Feige, J., {et~al.} 2015, Nature Communications,
  6, 5956

\bibitem[{Wanajo(2013)}]{wanajo2013r}
Wanajo, S. 2013, \apjl, 770, L22

\bibitem[{Wanajo {et~al.}(2014)Wanajo, Sekiguchi, Nishimura, Kiuchi, Kyutoku,
  \& Shibata}]{wanajo2014production}
Wanajo, S., Sekiguchi, Y., Nishimura, N., {et~al.} 2014, \apjl, 789, L39

\bibitem[{Wanderman \& Piran(2010)}]{wanderman2010luminosity}
Wanderman, D., \& Piran, T. 2010, Mon. Not. R. Astron. Soc., 406, 1944

\bibitem[{Wanderman \& Piran(2015)}]{wanderman2015rate}
---. 2015, Mon. Not. R. Astron. Soc., 448, 3026

\bibitem[{Wang {et~al.}(2018)Wang, Peng, Wu, \& Dai}]{wang2018pre}
Wang, J.-S., Peng, F.-K., Wu, K., \& Dai, Z.-G. 2018, Astrophys. J., 868, 19

\bibitem[{{Wang} {et~al.}(2019){Wang}, {Liu}, {Zhang}, {Xi}, \&
  {Zhang}}]{SSC_190114C_3}
{Wang}, X.-Y., {Liu}, R.-Y., {Zhang}, H.-M., {Xi}, S.-Q., \& {Zhang}, B. 2019,
  arXiv e-prints, arXiv:1905.11312

\bibitem[{Waxman(1995)}]{waxman1995cosmological}
Waxman, E. 1995, \prl, 75, 386

\bibitem[{Waxman \& Bahcall(1997)}]{waxman1997high}
Waxman, E., \& Bahcall, J. 1997, \prl, 78, 2292

\bibitem[{Wei {et~al.}(2015)Wei, Gao, Wu, \&
  M{\'e}sz{\'a}ros}]{WEP_FRBs_Wei_2015}
Wei, J.-J., Gao, H., Wu, X.-F., \& M{\'e}sz{\'a}ros, P. 2015, \prl, 115, 261101

\bibitem[{Wei {et~al.}(2016)Wei, Wang, Gao, \& Wu}]{WEP_blazars_wei_2016}
Wei, J.-J., Wang, J.-S., Gao, H., \& Wu, X.-F. 2016, \apjl, 818, L2

\bibitem[{{Wei} {et~al.}(2017){Wei}, {Zhang}, {Wu}, {Gao}, {M{\'e}sz{\'a}ros},
  {Zhang}, {Dai}, {Zhang}, \& {Zhu}}]{GW_GRB_WEP_Wei_2017}
{Wei}, J.-J., {Zhang}, B.-B., {Wu}, X.-F., {et~al.} 2017, JCAP, 2017, 035

\bibitem[{Wei {et~al.}(2019)Wei, Zhang, Shao, Gao, Li, Yin, Wu, Wang, Zhang, \&
  Dai}]{blazar_WEP_2019a}
Wei, J.-J., Zhang, B.-B., Shao, L., {et~al.} 2019, J. High Energy Astrophys.,
  22, 1

\bibitem[{Weinberg {et~al.}(2013)Weinberg, Mortonson, Eisenstein, Hirata,
  Riess, \& Rozo}]{cosmo_vary_paper}
Weinberg, D.~H., Mortonson, M.~J., Eisenstein, D.~J., {et~al.} 2013, Phys.
  Rep., 530, 87

\bibitem[{Weinberg(1967)}]{SM_1}
Weinberg, S. 1967, \prl, 19, 1264

\bibitem[{Weinberg(2008)}]{weinberg2008effective}
---. 2008, \prd, 77, 123541

\bibitem[{Will(1998)}]{GR_mg_Will_1998}
Will, C.~M. 1998, \prd, 57, 2061

\bibitem[{Will(2014)}]{Review_Tests_of_GR_overall_Will_2014}
---. 2014, Living Rev. Relativ., 17

\bibitem[{Will(2018)}]{will2018theory}
---. 2018, {Theory and experiment in gravitational physics} (Cambridge
  University Press)

\bibitem[{Williams {et~al.}(2018)Williams, Clark, Williamson, \&
  Heng}]{williams2018constraints}
Williams, D., Clark, J.~A., Williamson, A.~R., \& Heng, I.~S. 2018, Astrophys.
  J., 858, 79

\bibitem[{Williamson {et~al.}(2014)Williamson, Biwer, Fairhurst, Harry,
  Macdonald, Macleod, \& Predoi}]{Joint-Williamson-2014}
Williamson, A.~R., Biwer, C., Fairhurst, S., {et~al.} 2014, \prd, 90, 122004

\bibitem[{Willingale {et~al.}(2007)Willingale, {O'Brien}, Osborne, Godet, Page,
  Goad, Burrows, Zhang, Rol, Gehrels, {et~al.}}]{willingale2007testing}
Willingale, R., {O'Brien}, P.~T., Osborne, J.~P., {et~al.} 2007, Astrophys. J.,
  662, 1093

\bibitem[{Wollaeger {et~al.}(2018)Wollaeger, Korobkin, Fontes, Rosswog, Even,
  Fryer, Sollerman, Hungerford, Van~Rossum, \& Wollaber}]{wollaeger2018impact}
Wollaeger, R.~T., Korobkin, O., Fontes, C.~J., {et~al.} 2018, Mon. Not. R.
  Astron. Soc., 478, 3298

\bibitem[{{Wood}(2016)}]{HAWC_2}
{Wood}, J. 2016, PhD thesis, University of Maryland, College Park,
  arXiv:1801.01550

\bibitem[{Woosley(1993)}]{woosley1993gamma}
Woosley, S.~E. 1993, Astrophys. J., 405, 273

\bibitem[{Woosley \& Bloom(2006)}]{woosley2006supernova}
Woosley, S.~E., \& Bloom, J.~S. 2006, Annu. Rev. Astron. Astrophys., 44, 507

\bibitem[{Woosley \& Hoffman(1992)}]{woosley1992alpha}
Woosley, S.~E., \& Hoffman, R.~D. 1992, Astrophys. J., 395, 202

\bibitem[{Woosley {et~al.}(1994)Woosley, Wilson, Mathews, Hoffman, \&
  Meyer}]{woosley1994r}
Woosley, S.~E., Wilson, J.~R., Mathews, G.~J., Hoffman, R.~D., \& Meyer, B.~S.
  1994, Astrophys. J., 433, 229

\bibitem[{Wu {et~al.}(1957)Wu, Ambler, Hayward, Hoppes, \&
  Hudson}]{wu1957experimental}
Wu, C.-S., Ambler, E., Hayward, R.~W., Hoppes, D.~D., \& Hudson, R.~P. 1957,
  Phys. Rev., 105, 1413

\bibitem[{Wu {et~al.}(2019)Wu, Banerjee, Metzger, Mart{\'\i}nez-Pinedo,
  Aramaki, Burns, Hailey, Barnes, \& Karagiorgi}]{wu2019finding}
Wu, M.-R., Banerjee, P., Metzger, B.~D., {et~al.} 2019, arXiv e-prints,
  1905.03793

\bibitem[{Xie {et~al.}(2018)Xie, Zrake, \& MacFadyen}]{xie2018numerical}
Xie, X., Zrake, J., \& MacFadyen, A. 2018, \apj, 863, 58

\bibitem[{{Xie} {et~al.}(2018){Xie}, {Zrake}, \& {MacFadyen}}]{jet_time_1}
{Xie}, X., {Zrake}, J., \& {MacFadyen}, A. 2018, \apj, 863, 58

\bibitem[{{Xue} {et~al.}(2019){Xue}, {Zheng}, {Li}, {Brandt}, {Zhang}, {Luo},
  {Zhang}, {Bauer}, {Sun}, {Lehmer}, {Wu}, {Yang}, {Kong}, {Li}, {Sun}, {Wang},
  \& {Vito}}]{Bing_3}
{Xue}, Y.~Q., {Zheng}, X.~C., {Li}, Y., {et~al.} 2019, \nat, 568, 198

\bibitem[{Yagi \& Yunes(2013)}]{yagi2013love}
Yagi, K., \& Yunes, N. 2013, Science, 341, 365

\bibitem[{Yang {et~al.}(2015)Yang, Jin, Li, Covino, Zheng, Hotokezaka, Fan,
  Piran, \& Wei}]{yang2015possible}
Yang, B., Jin, Z.-P., Li, X., {et~al.} 2015, Nature Communications, 6, 7323

\bibitem[{Yang {et~al.}(2017)}]{GCN_DLT_last_obs}
Yang, S., {et~al.} 2017, {LIGO/Virgo G298048: Continued observation for
  DLT17ck}, Tech. rep., GSFC

\bibitem[{Yonetoku {et~al.}(2012)Yonetoku, Murakami, Gunji, Mihara, Toma,
  Morihara, Takahashi, Wakashima, Yonemochi, Sakashita,
  {et~al.}}]{yonetoku2012magnetic}
Yonetoku, D., Murakami, T., Gunji, S., {et~al.} 2012, \apjl, 758, L1

\bibitem[{Yost {et~al.}(2003)Yost, Harrison, Sari, \& Frail}]{yost2003study}
Yost, S.~A., Harrison, F.~A., Sari, R., \& Frail, D.~A. 2003, Astrophys. J.,
  597, 459

\bibitem[{{Yu} {et~al.}(2013){Yu}, {Zhang}, \& {Gao}}]{Bing_5}
{Yu}, Y.-W., {Zhang}, B., \& {Gao}, H. 2013, \apjl, 776, L40

\bibitem[{Yunes {et~al.}(2010)Yunes, {O'Shaughnessy}, Owen, \&
  Alexander}]{yunes2010testing}
Yunes, N., {O'Shaughnessy}, R., Owen, B.~J., \& Alexander, S. 2010, \prd, 82,
  064017

\bibitem[{Yunes \& Siemens(2013)}]{review_GR_tests_with_GWs_yunes_2013}
Yunes, N., \& Siemens, X. 2013, Living Rev. Relativ., 16, 9

\bibitem[{Zalamea \& Beloborodov(2011)}]{zalamea2011neutrino}
Zalamea, I., \& Beloborodov, A.~M. 2011, Mon. Not. R. Astron. Soc., 410, 2302

\bibitem[{Zappa {et~al.}(2018)Zappa, Bernuzzi, Radice, Perego, \&
  Dietrich}]{zappa2018gravitational}
Zappa, F., Bernuzzi, S., Radice, D., Perego, A., \& Dietrich, T. 2018, \prl,
  120, 111101

\bibitem[{{Zhang}(2018)}]{zhang2018physics}
{Zhang}, B. 2018, {The Physics of Gamma-Ray Bursts} (Cambridge University
  Press), doi:10.1017/9781139226530

\bibitem[{{Zhang}(2019)}]{zhang2019delay}
---. 2019, Frontiers Phys., 14, 64402

\bibitem[{{Zhang} \& {Kumar}(2013)}]{ICMART_2}
{Zhang}, B., \& {Kumar}, P. 2013, \prl, 110, 121101

\bibitem[{Zhang \& M{\'e}sz{\'a}ros(2001)}]{zhang2001gamma}
Zhang, B., \& M{\'e}sz{\'a}ros, P. 2001, \apjl, 552, L35

\bibitem[{{Zhang} \& {M{\'e}sz{\'a}ros}(2001)}]{2001ApJ...552L..35Z}
{Zhang}, B., \& {M{\'e}sz{\'a}ros}, P. 2001, \apjl, 552, L35

\bibitem[{{Zhang} \& {M{\'e}sz{\'a}ros}(2002)}]{structure_2}
---. 2002, \apj, 571, 876

\bibitem[{Zhang \& Yan(2010)}]{zhang2010internal}
Zhang, B., \& Yan, H. 2010, Astrophys. J., 726, 90

\bibitem[{Zhang {et~al.}(2007)Zhang, Zhang, Liang, Gehrels, Burrows, \&
  M{\'e}sz{\'a}ros}]{zhang2007making}
Zhang, B., Zhang, B.-B., Liang, E.-W., {et~al.} 2007, \apjl, 655, L25

\bibitem[{Zhang {et~al.}(2018{\natexlab{a}})Zhang, Zhang, Sun, Lei, Gao, Li,
  Shao, Zhao, Hu, L{\"u}, {et~al.}}]{zhang2018peculiar}
Zhang, B.-B., Zhang, B., Sun, H., {et~al.} 2018{\natexlab{a}}, Nature Commun.,
  9, 447

\bibitem[{Zhang {et~al.}(2018{\natexlab{b}})Zhang, Li, \&
  Xu}]{zhang2018combined}
Zhang, N.-B., Li, B.-A., \& Xu, J. 2018{\natexlab{b}}, Astrophys. J., 859, 90

\bibitem[{Zhang {et~al.}(2019)Zhang, Kole, Bao, Batsch, Bernasconi, Cadoux,
  Chai, Dai, Dong, Gauvin, {et~al.}}]{zhang2019detailed}
Zhang, S.-N., Kole, M., Bao, T.-W., {et~al.} 2019, Nature Astron., 3, 258

\bibitem[{Zhu(2015)}]{Zhu_dissertation}
Zhu, S. 2015, PhD thesis, University of Maryland, College Park.
\newblock \url{http://hdl.handle.net/1903/17258}

\bibitem[{Zhuge {et~al.}(1994)Zhuge, Centrella, \&
  McMillan}]{zhuge1994gravitational}
Zhuge, X., Centrella, J.~M., \& McMillan, S. L.~W. 1994, \prd, 50, 6247

\bibitem[{Zucker {et~al.}(2016)Zucker, Vitale, \& Sigg}]{LIGO_Aplus}
Zucker, M., Vitale, S., \& Sigg, D. 2016, {Getting an A+: Enhancing Advanced
  LIGO}, Tech. rep., LIGO.
\newblock \url{https://dcc.ligo.org/LIGO-G1601435/public}

\bibitem[{Zwicky(1933)}]{cosmo_zwicky_1933}
Zwicky, F. 1933, Helv. Phys. Acta, 6, 110

\bibitem[{Zwicky(1937)}]{cosmo_zwicky_1937}
---. 1937, Astrophys. J., 86, 217

\end{thebibliography}
